**University of Bucharest**
**Faculty of Physics**

# Theoretical description of prompt fission neutron multiplicity and spectra

**by**

**Cristian MANAILESCU**

**A dissertation submitted in partial fulfillment of the requirements for the degree of**

**Doctor in Physics**

**Thesis coordinators:**

**Prof. Dr. Anabella TUDORA**
**(University of Bucharest, Faculty of Physics)**

**Dr. Olivier SEROT**
**(Directeur de Recherche, CEA–Cadarache)**

**2012**

*To my family…*

## Acknowledgements

Undertaking this PhD has been a truly life-changing experience for me and it would not have been possible to do without the support and guidance that I received from many people.

I would like to first say a big thank you to my advisor from the University of Bucharest, Prof. Dr. Anabella Tudora for all the support and encouragement she gave me during the whole PhD period. Without her guidance and constant feedback this PhD would not have been achievable.

I would also like to express my sincere gratitude to my advisor from CEA Cadarache, Dr. Olivier Serot for his support, patience and help he gave me during the course of my PhD study. His continuous guidance helped me in all the time of research and writing of this thesis.

My thanks also goes to Olivier Litaize from who's advanced knowledge on C++ programming skills I benefited on numerous occasions.

Besides my advisors, I would like to thank the rest of my thesis committee: Dr. Mihail Mirea and Dr. Gilles Noguere for taking their time to read the manuscript and for their useful comments and suggestions.

Last but not least I wish to thank my family for their love, understanding, patience and endless support.

# Contents

# Chapter I

## Introduction

For the design of nuclear reactors and for other energetic and non–energetic applications it is important to know, with a high accuracy, the prompt fission neutron spectrum as a function of both the fissioning nucleus and its excitation energy, and the average number of prompt neutrons emitted in the fission process.

Concerning the average number of prompt neutron emitted in fission, until 20 years ago, the most important role in evaluations was played by the experimental measurements. Different systematics (such as of Howerton, of Blois and Fréhaut and so on), based exclusively on the behaviour of experimental multiplicity data were also used in the evaluations. Obviously, this kind of systematics is not able to assure prompt neutron multiplicity predictions.

For the evaluation of the prompt fission neutron spectrum, two types of representations were used (Maxwell and Watt) depending on one or two free parameters adjusted in order to reproduce the experimental data for a given fissioning nucleus at a given excitation energy. For this reason early spectrum Maxwell and Watt representations have not predictive power.

The modelisation of prompt neutron and gammas emission in fission (post-scission part) has been developed later than the nuclear interaction models (pre-scission part of fission). In the 90's, important steps were made by Märten and his co–workers (Märten and Seelinger, 1984), known as the Dresda research group, who tried to make a very detailed description of the post-scission process, by the development of the Complex Cascade Evaporation Model (CCEM), a model that requires a substantial amount of experimental information.

The cascade evaporation model accounts for the distribution of the fission fragment excitation energy in each step of the cascade and for other important physical effects like the energy dependence of the inverse process of compound nucleus formation, the center-of-mass motion of the fission fragments, the anisotropy effects in the center of mass system, the complete fission fission–fragment mass and kinetic energy distribution and also the semi–empirical fission–fragment nuclear level densities

Another important characteristic was to consider the competition between the emission of neutrons and gammas for each fission fragment as it was made in the Hauser–Feschbach statistical model (Hauser and Feshbach, 1952). Computer codes (STATIS, GNASH, TALYS) including the Hauser–Feshbach formalism can perform such a calculation for any nuclei that appears as a fission fragments needing as an input the excitation energy. Such attempts were performed few years ago by the research group of Bruyères-le-Châtel, France, by modifying the TALYS code.

Because of the huge amount of calculation and the large number of parameters needed, both CCEM and Hauser–Feshbach models are not appropriate for the evaluation purpose of the prompt emission data.

Since the early 90's the model proposed by Madland and Nix (Madland and Nix, 1982) known as the "Los Alamos model" was successfully used. This model account for some important physical ingredients like the distribution of the fission–fragment excitation energy, the energy dependence of the inverse process of compound nucleus formations, the center–of–mass motion of the fission–fragments and also the multiple fission chances at high incident neutron energy.

The initial model proposed by Madland and Nix has taken into account only one fragmentation, the so-called most probable fragmentation, determined by weighting over fission fragment distributions (by taking into account the entire fragment range covered by the distributions or only a subset, the so-called 7 point approximation). This model, with only one fragmentation, needs as input only a few parameters, taken as average values, usually obtained from experimental data or independent models. For these reasons, the Los Alamos model (with only one fragmentation and average values of model parameters) is today largely used for prompt emission data evaluations, especially at higher incident energies where multiple fission chances are involved.

Basic features of this model made possible the development of other models, considering the multi–modal concept (taking into account one fragmentation, the most probable one, for each fission mode) or by considering all fragmentations covered by the mass and kinetic energy distributions as is made in the Point by Point model (Vladuca and Tudora, 2000a,b, 2001a,b,c; Hambsch et al., 2003; Tudora et al., 2005; Tudora, 2006, 2008, 2009, 2010a,b).

Subsequent improvements of the Point by Point model concerned the treatment of the level density of fission fragments, prompt gammas–rays energy, total excitation energy partition between fully-accelerated fission fragments by using parameterizations or by taking into account what is happening at scission and so on.

Along with the deterministic treatment of the PbP model, in the recent years the probabilistic Monte Carlo treatment of the prompt neutron emission has known important developments. The most important feature of this treatment is the possibility to take into account the sequential emission of prompt neutrons and gamma-rays. As for today Monte-Carlo treatments of prompt emission were developed by three groups: Litaize, Serot and Regnier from CEA–Cadarache (Litaize and Serot, 2010, 2011; Regnier et al., 2012a,b), Talou and Lemaire from Los Alamos National Laboratory (Lemaire et al., 2005) and Randrup and Vogt from Livermore National Laboratory (Randrup and Vogt, 2009; Vogt et al., 2009).

Nowadays both treatments, the deterministic PbP and the probabilistic Monte Carlo, are able to provide all quantities characterizing the prompt neutron and gamma-ray emission: quantities as a function of fragment, average quantities as a function of TKE or total average ones.

The Monte–Carlo treatment is very useful to obtain nuclear data predictions also in the high energy range ($10^{16}$-$10^{18}$ eV), where experimental data are difficult to obtain (Rebel et al., 2008; Sima et al., 2011).

The present work concerns two of successful models used today: PbP and the Monte Carlo approaches. Therefore the thesis is structured as follows:

The description of the Point by Point model and of the extended Los Alamos model for higher energies that takes into account the secondary chains and ways is given in Chapter II. In this chapter are given also examples of PbP and most probable fragmentation approach calculations for various quantities which characterize prompt emission: multi–parametric matrices (meaning different quantities as a function of fragment and of TKE), quantities as a function of fragment mass, quantities as a function of the TKE and total average quantities, for different spontaneous and neutron induced fissioning systems.

Special care was given to the TXE partition between the fully accelerated fission fragments, two partition methods used in the PbP model being discussed in details.

In Chapter III is given the description of the Monte Carlo treatment included in the FIFRELIN code. Only those aspects that differ from the PbP treatment are emphasized, namely the treatment of the moment of inertia entering the rotational energy calculation and the TXE partition method based on a mass dependent temperature ratio law. A special attention is given to the latest developments of the code concerning the inclusion of the energy dependent compound nucleus cross–section of the inverse process of neutron evaporation from fragments. In this chapter examples of calculation with the FIFRELIN code for the case of the standard fissioning system $^{252}$Cf(SF) are given.

Original results for several plutonium spontaneous fissioning systems ($^{236,238,240,242,244}Pu$) and one neutron induced fissioning system ($^{239}Pu(n_{th},f)$) obtained with both PbP and Monte-Carlo treatments are given in Chapter IV. The comparison between the results obtained from both types of treatments allows the each other validation and also lead to some improvements that can be brought to the TXE partition in the FIFRELIN code.

The last chapter includes an overview of the most important conclusions resulting from the inter–comparison of the results obtained with both treatments in chapter IV.

# Chapter II

## Description of the Point by Point model and of the most probable fragmentation approach (Los Alamos model) extended to take into account secondary nucleus chains and paths

As for today, the Los Alamos model with subsequent improvements is considered an agreed model for evaluation purposes. This is due not only to the better physical ingredients compared to other models, but also to the fact that in the case of only one fragmentation only few input model parameters taken as averaging quantities coming from experimental data or independent models are required as follows: the energy released in fission ($<E_r>$), the total kinetic energy of the fission–fragments ($<TKE>$), the neutron separation energy from the fission–fragments ($<S_n>$), the average prompt gamma–ray energy ($<E_\gamma>$) and the average level density parameter $<a>$ (parameterized as $<C>=A/<a>$, where $A$ is the mass number of the fissioning nucleus).

For the case of the spontaneous fission or neutron induced fission in the energy range where only one compound nucleus undergoing fission is formed (first fission chance), the basic features of the Los Alamos model can be used in three ways:

i) Taking into account only one fragmentation (most probable fragmentation), this being the treatment used in the basic work of Madland and Nix (Madland and Nix, 1982). In this case the model parameters $\langle E_r \rangle$, $\langle TKE \rangle$, $\langle S_n \rangle$, $\langle E\gamma \rangle$ and $\langle C \rangle = A/\langle a \rangle$ are taken as average values.

ii) Taking into account the multi–modal concept of fission, considering a fragmentation and corresponding average parameters for each mode "*m*" ($\langle E_r \rangle_m$, $\langle TKE \rangle_m$, etc). The total prompt fission neutron spectrum, the total prompt neutron multiplicity as well as other quantities characterizing the prompt neutron and gamma–ray emission are calculated as a superposition of the respective quantity corresponding to each fission mode weighted with the modal branching ratios (Hambsch et al., 2002; Hambsch et al., 2003).

iii) Taking into account the entire fission–fragments range, the so–called "Point by Point" (PbP) approach (Tudora et al., 2005; Vladuca et al., 2006; Tudora, 2006, 2008, 2009, 2010a,b,c). In this case the total average quantities characterizing the prompt neutron and

gamma–ray emission are calculated as a superposition of the respective quantity corresponding to a fission–fragment pair weighted with the charge and mass distributions.

At higher incident neutron energies, when more than one fission chances are involved, neither the multi–modal nor the Point by Point approaches is used. This is due to the lack of experimental data regarding the mass and charge distributions corresponding to each compound nucleus undergoing fission, and the difficulties encountered by taking into account many fragmentations of the secondary fissioning nuclei having a continuous range of excitation energies.

Therefore for prompt neutron data evaluation purposes the remaining case is to take into account only one fragmentation (the most probable fragmentation) for each compound nucleus and to use average values for the input model parameters (Madland and Nix, 1982). The average model parameters are dependent on the excitation energy of the respective fissioning nucleus, details about these dependences can be found in (Vladuca and Tudora, 2000a; 2001a,b,c).

### II.1 Basic features of the Point by Point model

In order to calculate the neutron energy spectrum in the center–of–mass system of a given fission fragment, the standard nuclear evaporation theory is used, and than a transformation to the laboratory system is performed. The prompt neutron spectrum in the center–of–mass (CMS) system corresponding to an individual fission fragment (indexed "i") is given by (Madland and Nix, 1982):

$$\Phi_i(\varepsilon) = \sigma_{ci}(\varepsilon) \int_0^{Tmi} K_i(T) P(T) \exp(-\varepsilon/T) dT \tag{2.1}$$

with the normalization constant:

$$K_i(T) = \left[ \int_0^{\infty} \sigma_{ci}(\varepsilon) \exp(-\varepsilon/T) d\varepsilon \right]^{-1} \tag{2.2}$$

where $\varepsilon$ is the neutron energy in the center of mass system. The Boltzmann's constant was included in the nuclear temperature, so that it has energetic units. In the PbP model $\sigma_c(\varepsilon)$ is obtained from optical model calculations using phenomenological potential parameterizations appropriate for nuclei appearing as fission fragments.

In equation (2.1), *P(T)* is the residual temperature distribution of a fragment given by the following expression (Hambsch et al., 2005):

$$P(T) = \begin{cases} \dfrac{2}{T_{mi}^2} \dfrac{(s+1)^2}{4s} T, & 0 \le T \le \dfrac{2T_m i}{s+1} \\ \dfrac{2}{T_{mi}^2} \dfrac{s+1}{2(s-1)} \left( -\dfrac{s+1}{2s} T + T_{mi} \right), & \dfrac{2T_{mi}}{s+1} < T \le \dfrac{2sT_{mi}}{s+1} \end{cases} \tag{2.3}$$

where $T_{mi}$ is the maximum value of this distribution and $s \geq 1$. Obviously, when s is taken equal to 1, the traditional triangular distribution proposed by Madland and Nix (Madland and Nix, 1982) is obtained. An example of such distributions is given in **Fig. 2.1** for the case of the spontaneous fission of $^{252}$Cf.

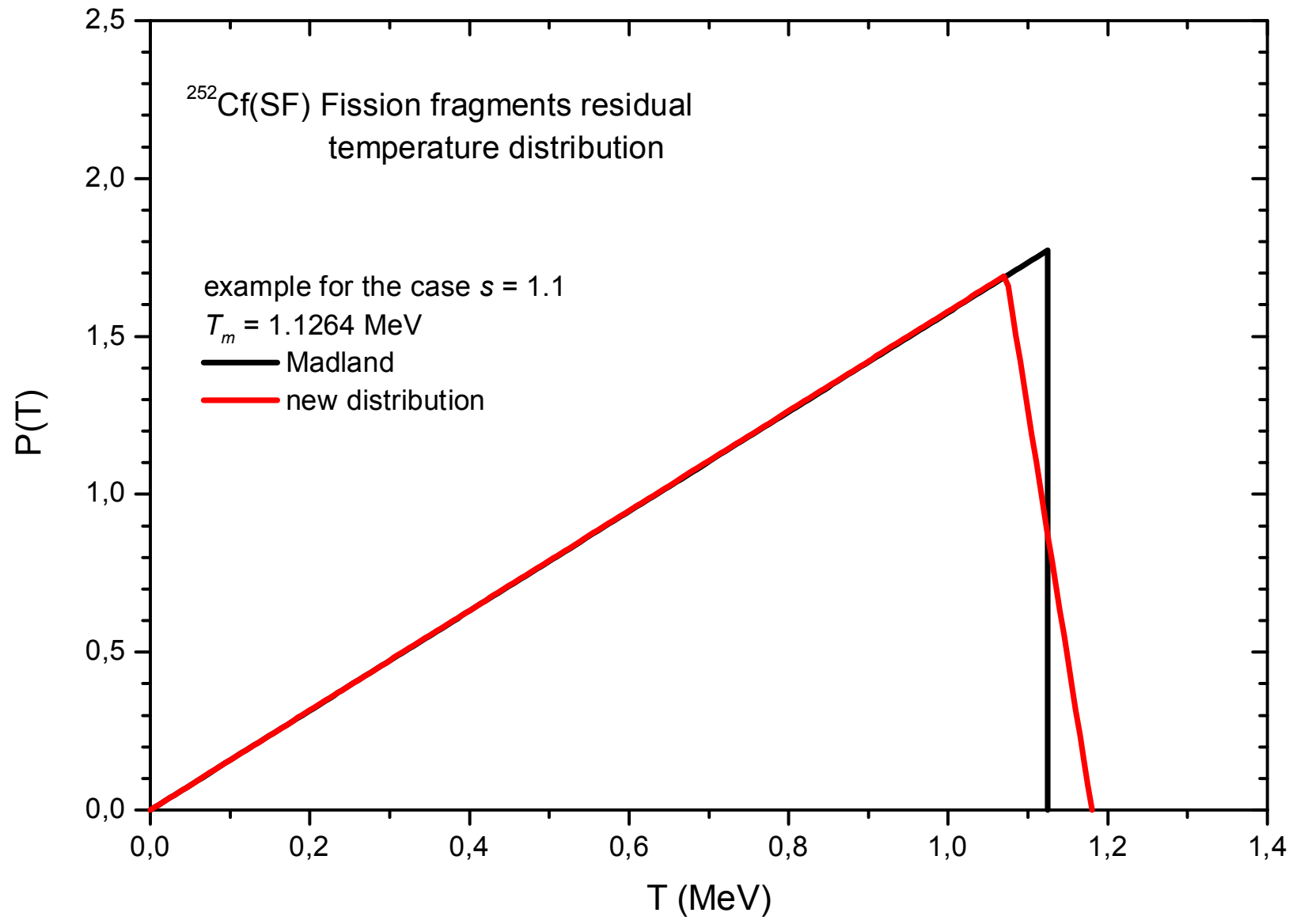

**Fig.2.1** Residual nuclear temperature distribution for $^{252}$Cf(SF) $T_m$=1.1264 MeV, classical version (cutoff at well-defined temperature) and the new version (cutoff in a narrow temperature range)

In the PbP model $\sigma_c(\varepsilon)$ is needed for all fission–fragments forming the fragmentation range. In the case of the most probable fragmentation approach, $\sigma_c(\varepsilon)$ is needed only for the light and heavy fragments forming the most probable fragmentation of each compound nucleusundergoing fission (each fission chance). In the case of the multi–modal fission, $\sigma_c(\varepsilon)$ is needed for the most probable fragmentation corresponding to each fission mode. The best case would be to obtain the inverse cross–sections from an experimental dataset for all the fission–fragments involved. Unfortunately such experiments are not possible because these types of nuclei, with neutron excesses as large as those encountered in fission–fragments are very unstable. Consequently, the inverse cross–section must be provided by the optical model calculations using the SCAT2 code (Bersillon, 1991) with phenomenological optical potentials with isospin dependences adequate for the nuclei appearing as fission–fragment.

Several phenomenological optical potentials can be used, such as: Becchetti–Greenless, Wilmore–Hodgson or the new optical model parameterization of Koning–Delaroche taken from (RIPL–3, 2012a).

A simplified formula for the calculation of the inverse cross–section was proposed by Iwamoto (Iwamoto, 2008), with not encouraging results, the advantage being only the time consumption of the calculations (see **Appendix 1**).

According to Terrel (Terrel, 1959) and Hambsch et al. (Hambsch et al., 2005) the anisotropy of neutrons emission, if present, is symmetrical about 90º and the fission fragment prompt neutron spectrum in the center–of–mass could be described by the following equation:

$$\Phi(\varepsilon,\theta_{CM}) = \Phi(\varepsilon)\frac{1+b\cos^2\theta_{CM}}{1+b/3} \tag{2.4}$$

where *Φ(ε)* is the center–of–mass neutron spectrum from (2.1) and *b* is the anisotropy parameter.

If a possible anisotropy effect is taken into account, then the neutron energy spectrum in the laboratory system for an individual fission–fragment becomes:

$$N(E,E_f,\sigma_c^f) = \frac{1}{2T_m^2\sqrt{EF_i}}\int_{(\sqrt{E}-\sqrt{EF_i})^2}^{(\sqrt{E}+\sqrt{EF_i})^2}\sigma_c^f(\varepsilon)\sqrt{\varepsilon}\left(\frac{1}{1+b/3}+\frac{b(E-\varepsilon-EF_i)^2}{4\varepsilon EF_i(1+b/3)}\right)I(\varepsilon)d\varepsilon \tag{2.5}$$

$$\text{with } I(\varepsilon) = \begin{cases} \frac{1}{4s}\int_0^a K_i(T)(s+1)^2 T\exp(-\varepsilon/T)dT \\ \frac{s+1}{2(s-1)}\int_a^{sa} K_i(T)\left(-\frac{s+1}{2s}T+T_{mi}\right)\exp(-\varepsilon/T)dT \end{cases},\quad a=\frac{2T_{mi}}{s+1} \tag{2.6}$$

In the above expressions $EF_i$ is the average kinetic energy per nucleon obtained from momentum conservation, calculated for a given pair of fragments as following:

$$EF_i = \frac{A_{H,L}}{A_{L,H}}\frac{TKE}{A_0} \tag{2.7}$$

where $A_L$, $A_H$ are the mass numbers of complementary light and heavy fragments, *TKE* is the average total kinetic energy of the respective pair of fission fragments and $A_0$ is the mass number of the fissioning nucleus.

The prompt neutron energy spectrum for a pair of fission–fragment in the laboratory system is given by:

$$N(E) = \frac{r}{r+1}N_L(E)+\frac{1}{r+1}N_H(E) \tag{2.8}$$

where $N_{L,H}(E)$ are calculated according to equations (2.5)–(2.6) and *r* is the ratio of neutrons emitted by the light and by the heavy complementary fragments:

$$r = \frac{\nu_L}{\nu_H} \tag{2.9}$$

All existing experimental data of *ν(A)* showed that equal number of neutrons is emitted by fragment pairs with $A_H$ around 140 (which is usually one of the most probable fragments). Therefore, for this case the ratio from equation (2.9) is taken equal to one. A ratio not equal to 1 in the case of the most probable fragmentation approach, as made by several authors is wrong contradicting the experimental *ν(A)* behavior. This assumption, of a ratio not equal to one must be considered when more than one fragmentation are considered as in the case of the PbP or the multi–modal approaches.

Another source of non–isotropic neutrons can be neutron emission at the moment of scission, the so–called scission neutrons. Kornilov (Kornilov et al., 2001) has re–analyzed three independent experiments on spontaneous fission of $^{252}$Cf in order to obtain a better description of the neutron evaporation from the fragments. He showed that a good agreement exists between these experiments, and that a neutron surplus of (30±5) % exists at about 90º relative to the direction of the moving fragments. These neutrons do not originate from fully accelerated fission–fragments, and would represent about 10 % of the total fission yield. A similar conclusion has been drawn by Franklyn (Franklyn et al., 1978) from angular distribution of neutrons for the neutron induced fission of $^{235}$U. This angular distribution could be very well fitted only by considering a contribution of the scission neutrons of 20 % to the total neutron yield. Consequently, there is a clear need of more precise experimental data regarding the scission neutrons.

In the new version of the computer code including the PbP, multi–modal and most probable approaches, (Vladuca and Tudora, 2000b), the emission of the so–called scission neutrons can be also taken into account (Visan and Tudora, 2011). The scission neutron spectrum being considered as a Weisskopf–Ewing evaporation spectrum from the fissioning nucleus:

$$N_{sciz}(E) = k(\theta_{sciz}, \sigma_c)\sigma_c(E)E\exp(-E/\theta_{sciz}) \tag{2.10}$$

with the normalization constant given by:

$$k(\theta_{sciz}, \sigma_c) = \left[\int_0^\infty \sigma_c(E)E\exp(-E/\theta_{sciz})dE\right]^{-1} \tag{2.11}$$

In eq.(2.10) $\sigma_c(E)$ is the compound nucleus cross–section of the inverse process of neutron evaporation from the fissioning nucleus. Taking into account that this nucleus is a permanent deformed one, the direct interaction mechanism has to be treated by the coupled channel method. Consequently $\sigma_c(E)$ is provided by the ECIS code (Raynal, 1994) using deformed optical model parameterizations for actinides such as Vladuca et. al. (Vladuca et al., 1996), Capote et. al. (Capote et al., 2008), Soukhovitskii (Soukhovitskii et al., 2004).

The scission neutrons temperature $\theta_{sciz}$ and the amount of the scission neutrons $w$ are input parameters. In the end, the total spectrum is obtain as:

$$N_{total}(E) = (1-w)N(E) + wN_{sciz} \tag{2.12}$$

The maximum value of $P(T)$ distribution is obtained from $E_i^* = a_i T_{mi}^2$ where $E_i^*$ is the fragment excitation energy at full acceleration and $a_i$ is the level density parameter of the respective fragment (calculated at the energy $E_i^*$ in the frame of the generalized super–fluid model (Ignatiuk, 1998), see **Appendix 2**). The excitation energies of complementary fully accelerated fission fragments are obtained using different methods of total excitation energy partition. In the Point by Point treatment two methods are used. These methods will be described in details in a following section.

For each fission–fragment pair the total excitation energy at full acceleration is calculated as:

$$TXE = E_r + B_n + E_n - TKE \tag{2.13}$$

where $E_n$ is the incident neutron energy, $B_n$ is the neutron binding energy in the compound nucleus and $E_r$ is the energy release of the respective fragmentation (Q–value, calculated using mass excess taken from nuclear data libraries (RIPL–3, 2012b). In the case of spontaneous fission both $B_n$ and $E_n$ are set equal to zero in equation (2.13). The total kinetic energy is usually provided by experimental $TKE(A)$ data, but in the absence of the experimental data it can be calculated using available models. A simple approach based on the electrostatic repulsion between the fission–fragments connected by a neck in the pre–scission configuration was proposed (Manea and Tudora, 2011).

From energy conservation the prompt neutron multiplicity corresponding to a fragment pair is obtained as:

$$TXE = \nu_{pair}\left(<\varepsilon> + <S_n>\right) + E_\gamma \tag{2.14}$$

where $<\varepsilon>$ is the average prompt neutron energy in the center of mass system (the first order momentum of prompt neutron spectra of complementary fragments according to equation (2.1), $E_\gamma$ is the average prompt gamma–ray energy and $<S_n>$ is the average neutron separation energy from complementary fission–fragments given by:

$$<S_n> = \frac{1}{2}\left(\overline{S_{nL}} + \overline{S_{nH}}\right) \tag{2.15}$$

where the average neutron separation energy from each fragment is taken as $S_{nx}/x$ (with x=1,2,3,… depending on the amount of fragment excitation energy, accounting the sequential emission).

Using one of the two methods, the level density parameter and the maximum values of the residual temperature distribution at full acceleration for each fission–fragment can be determined. With these parameters, the prompt neutron multiplicities, spectra as well as other quantities characterizing the prompt neutron emission can be calculated for each fragment or pair of fragments according to the equation (2.1)–(2.14).

### II.1.1 Total excitation energy partition based on the $\nu_H/\nu_{pair}$ parameterization

The use of the *TXE* partition according to the ratio of prompt neutron number emitted by complementary fragments in the frame of the Point by Point model was possible due to the systematic behavior of $\nu_H/(\nu_L+\nu_H)$ as a function of the heavy mass $A_H$ deduced exclusively from te behaviour of experimental *ν(A)* data. The experimental *ν(A)* were represented as $\nu_H/(\nu_L+\nu_H)$ versus $A_H$ (Tudora, 2006; Manailescu et al., 2011). This representation was preferred over the traditional *ν(A)* because the nuclei forming the heavy fragment group are not changing significantly from one fissioning system to another. For all fissioning systems where experimental on *ν(A)* exists, the quantity $\nu_H/\nu_{pair}$ as a function of $A_H$ exhibits a systematic behavior as follows (see **Figs. 2.2–2.4**) (Manailescu et al., 2011):

- a minimum in $\nu_H/\nu_{pair}$ around the mass $A_H$=130 occurs, driven by the magic numbers Z=50 and N=82 and by the very large negative values of the shell corrections

- the complementary fragments emit almost an equal number of neutrons around the mass number 140

- the light fragment emits more neutrons than the heavy fragment only in the range $A_H<140$ ($\nu_H/\nu_{pair} < 0.5$) while above this mass the heavy fragment emits more than the light fragment ($\nu_H/\nu_{pair} > 0.5$).

Because of the scarcity of the experimental *ν(A)* data, only six fissioning systems were studied in (Manailescu et al., 2011): three thermal neutron induced fissioning systems ($^{233,235}$U($n_{th}$,f), $^{239}$Pu($n_{th}$,f)), one neutron induced fissioning system at a higher incident neutron energy ($^{237}$Np(n,f) at 0.8 and 5.5 MeV) and two spontaneous fissioning systems ($^{248}$Cm(SF), $^{252}$Cf(SF)). For all the studied cases the fission–fragments range was chosen according to the Point by Point treatment as follows: the entire fragment mass range covered by the experimental fission fragment distribution with a step of 1 mass unit. For each mass unit, 4 charge numbers Z were taken as the nearest integer values above and below the most probable charge obtained from the unchanged charge distribution corrected with a possible charge polarization.

As can be seen in **Figs. 2.2–2.4**, for all the cases $\nu_H/\nu_{pair}$ shows a minimum at $A_H$=130 (except for the spontaneous fission of $^{248}$Cm, where the minimum is shifted with few mass units below this value). In the case of neutron induced fissioning systems, in the heavy mass range of few mass units around $A_H$=140, an equal number of neutrons are emitted, while for the spontaneous fission, this range is narrower and shifted above $A_H$=140. It is important to mention that for these spontaneous fissioning systems, the experimental data regarding the mass yields presents maxima at few mass units above $A_H$=140.

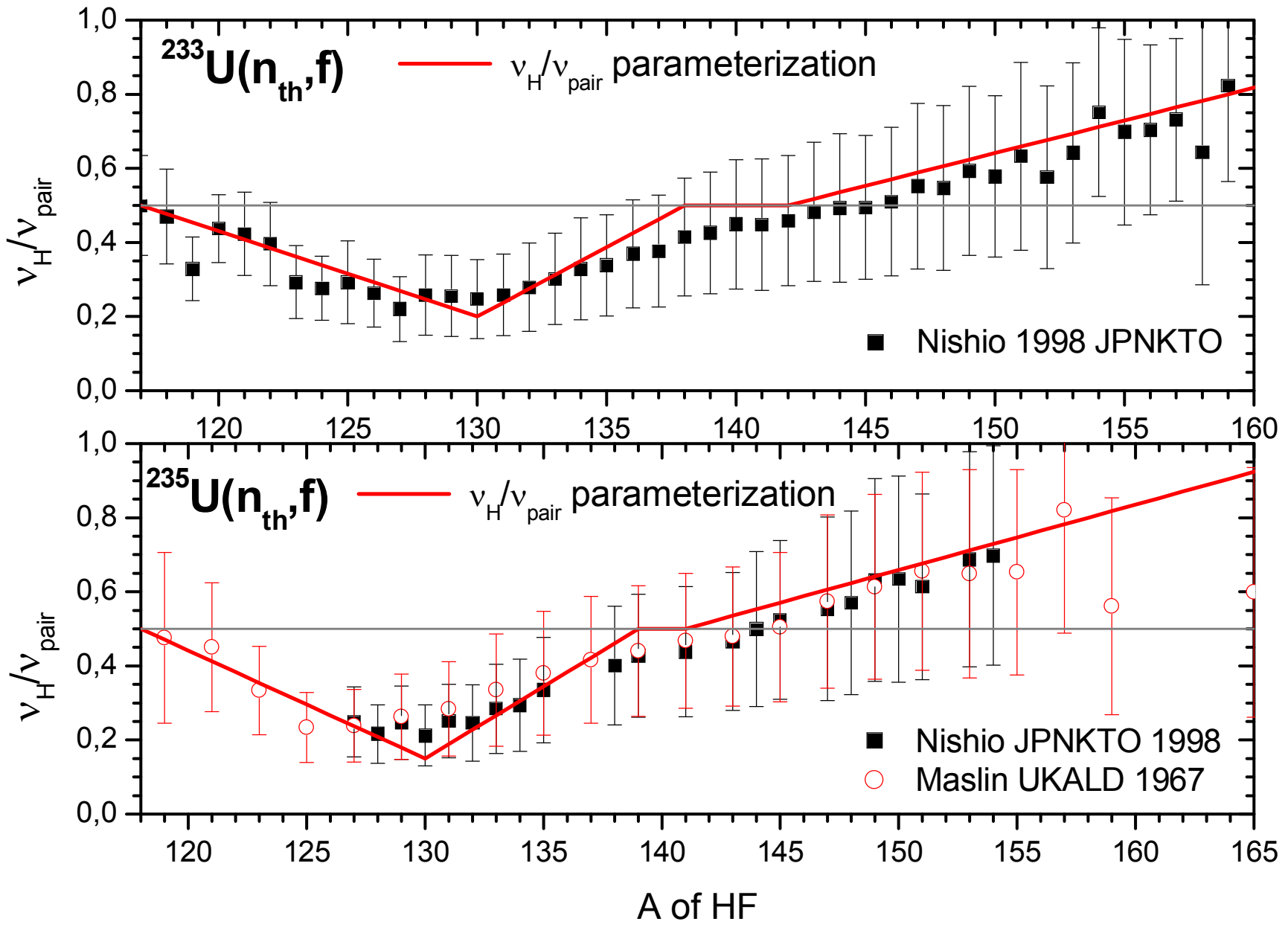

**Fig. 2.2** $\nu_H/\nu_{pair}$ parameterization in comparison with the experimental data for $^{233}$U (upper part) and for $^{235}$U (lower part) (Manailescu et al., 2011)

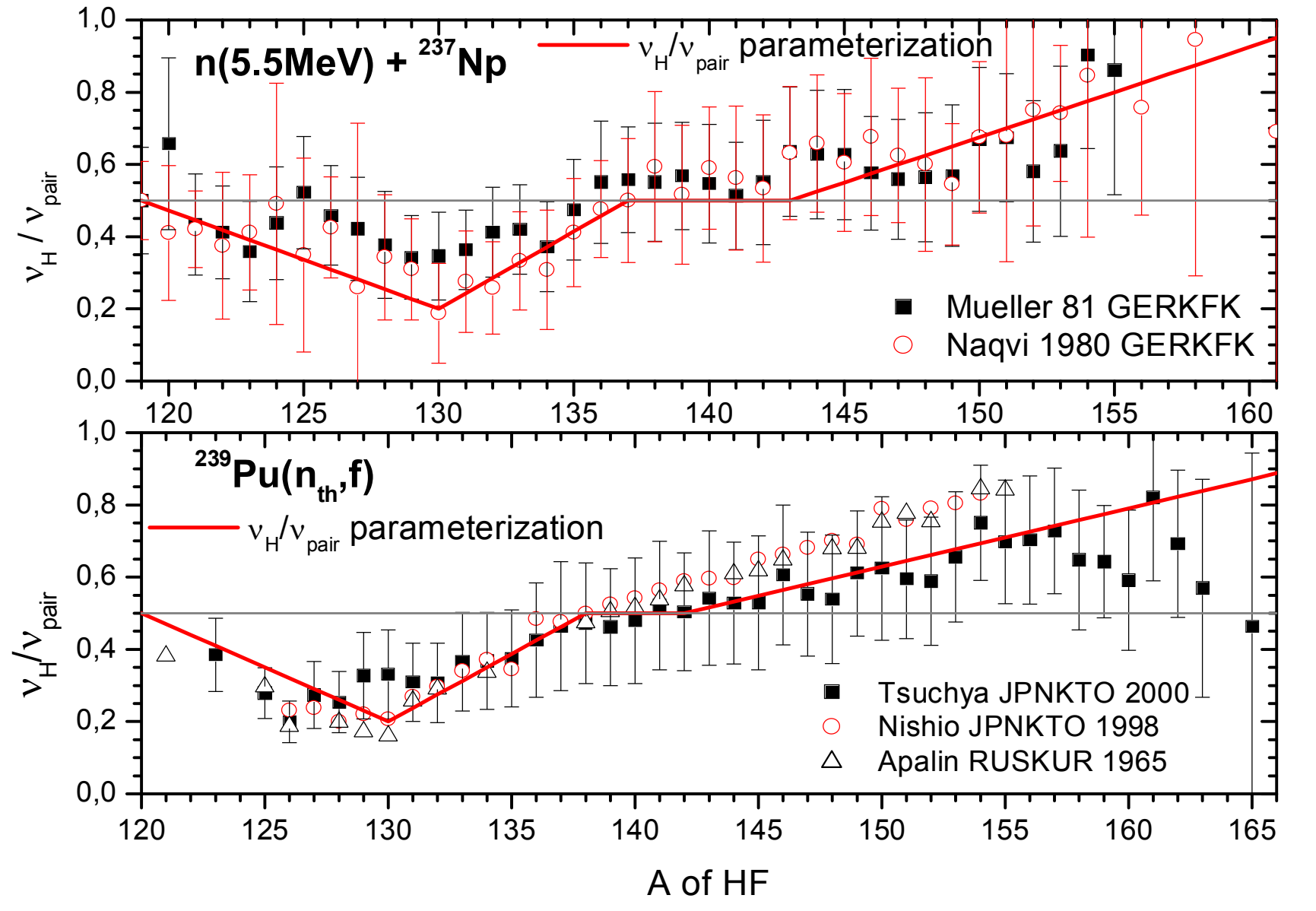

**Fig. 2.3** $\nu_H/\nu_{pair}$ parameterization in comparison with the experimental data for $^{237}$Np (upper part) and for $^{239}$Pu (lower part) (Manailescu et al., 2011)

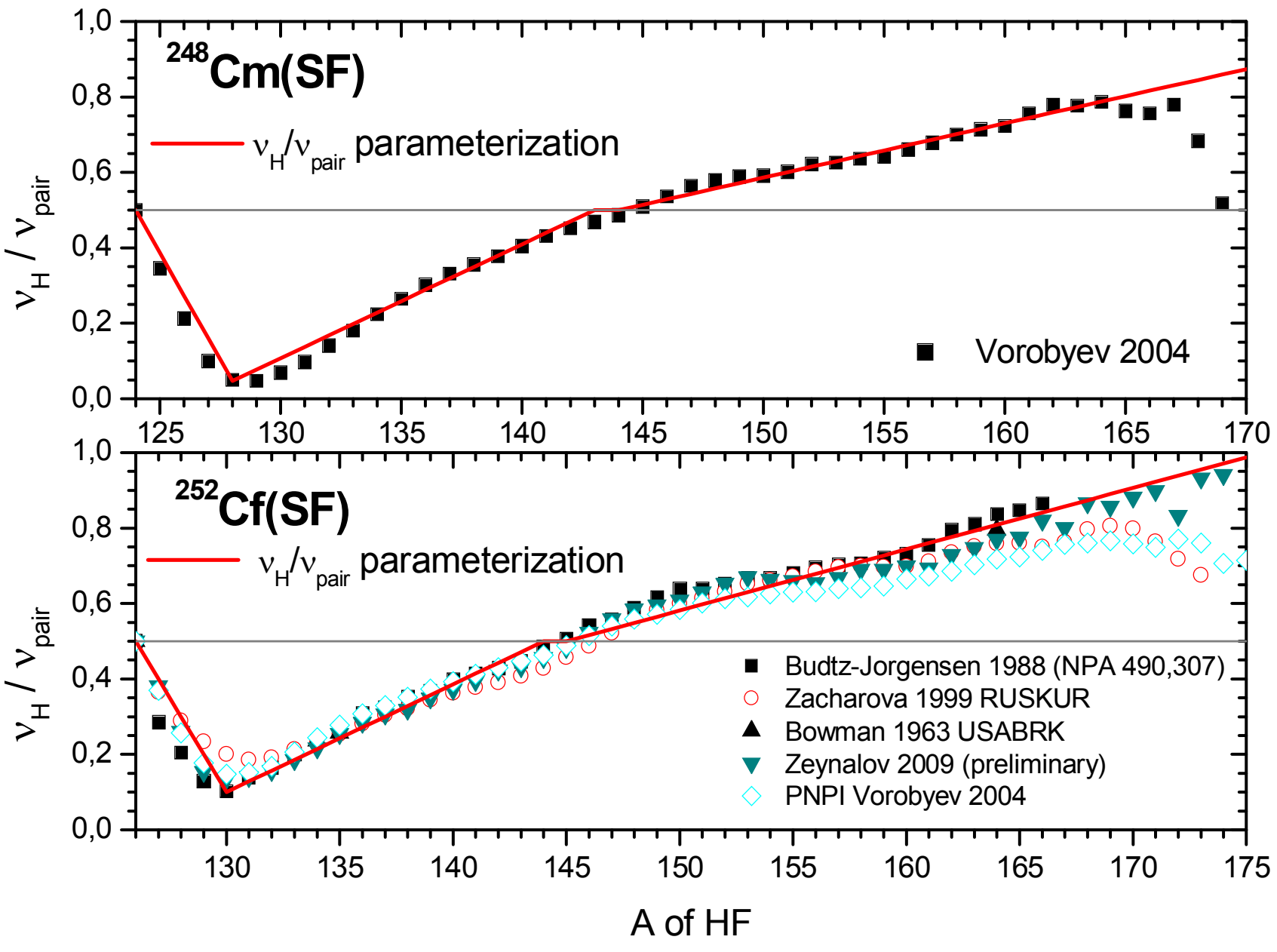

**Fig. 2.4** $\nu_H/\nu_{pair}$ parameterization in comparison with the experimental data for $^{248}Cm$ (upper part) and for $^{252}Cf$ (lower part) (Manailescu et al., 2011)

Using the $\nu_H/\nu_{pair}(A_H)$ parameterization plotted with lines in **Figs. 2.2–2.4**, for each pair of complementary fission–fragments the total excitation energy TXE is shared according to the following relations:

$$TXE = E_L^* + E_H^* \qquad \frac{E_L^*}{E_H^*} = \frac{\nu_L}{\nu_H} \tag{2.16}$$

giving:

$$E_H^* = \frac{\nu_H}{\nu_{pair}} TXE \qquad E_L^* = \left(1 - \frac{\nu_H}{\nu_{pair}}\right) TXE \tag{2.17}$$

The level density parameter of each fragment is calculated in the frame of the generalized super–fluid model of Ignatiuk (Ignatiuk, 1998):

$$a(Z,A,E^*) = \begin{cases} \tilde{a}(A)\left(1 + \frac{\delta W(Z,A)}{U^*}\left(1 - \exp(-\gamma U^*)\right)\right) & ,U^* \geq U_{cr} \\ a_{cr} & ,U^* < U_{cr} \end{cases} \qquad U^* = E^* - E_{cond} \tag{2.18}$$

where $E_{cond}$ is the condensation energy, $a_{cr}$ is the critical level density parameter, $\tilde{a}(A)$ is the asymptotic value of the level density parameter, $\gamma$ is the damping parameter and $\delta W$ is the shell correction taken from Moller and Nix (RIPL–3, 2012d). Details about the generalized super–fluid model can be found in **Appendix 2**.

For each pair of fragments, the obtained $E^*_{L,H}$ according to equation (2.17) enter equation (2.18) in order to provide $a_{L,H}$.

Using the parameterization $\nu_H/\nu_{pair}$ for each fissioning system it was possible to obtain a parameterization of the ratio of the residual temperatures ($RT=T_L/T_H$) as a function of the heavy mass. This ratio is plotted in **Fig. 2.5** as a function of the heavy mass for the neutron induced fissioning systems studied in (Manailescu et al., 2011).

An interesting behaviour of the temperature ratio $RT(A_H)$ can be observed as following: *i)* a maximum occurs at $A_H=130$ and is around 1.5–1.6 , *ii)* in the $A_H$ range between *135–145* the temperature ratio is approximately equal to 1 (practically the same temperature for HF and LF) and *iii)* for $A_H>145$ the decrease of RT is almost linear and the slope does not differ very much from one neutron induced fissioning system to onother. This systematic trend allows the parameterizations of $RT(A_H)$ which are also plotted with different line styles in **Fig 2.5**. The slopes and the intercepts of the $RT$ parameterization versus the fissility parameter are plotted in **Fig. 2.6**. The almost constant values suggest that an unique $RT$ parameterization for the neutron induced fissioning systems can be deduced, represented by the dashed lines in **Fig 2.6**.

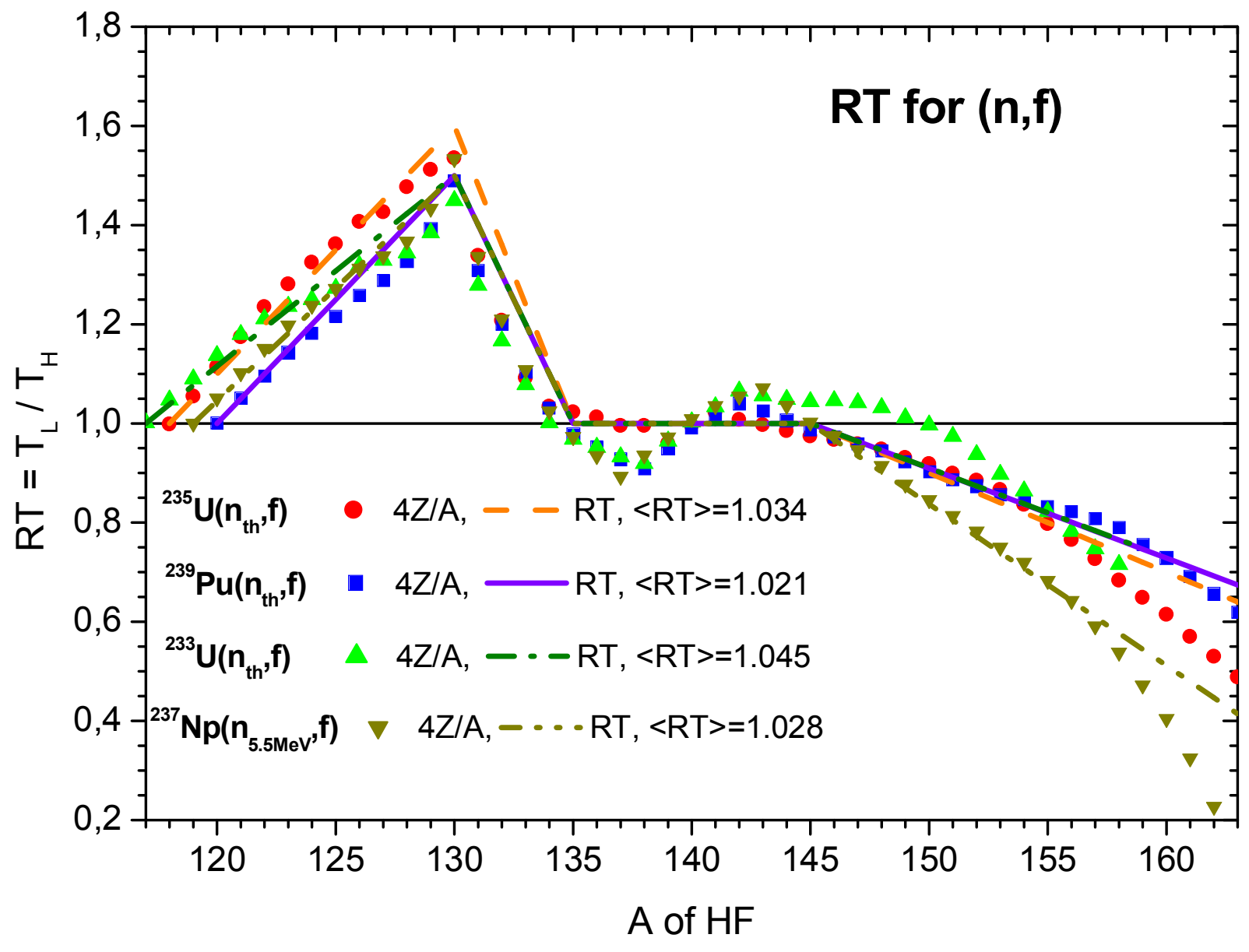

**Fig. 2.5** Fission–fragment ratios and parameterizations for neutron induced fission

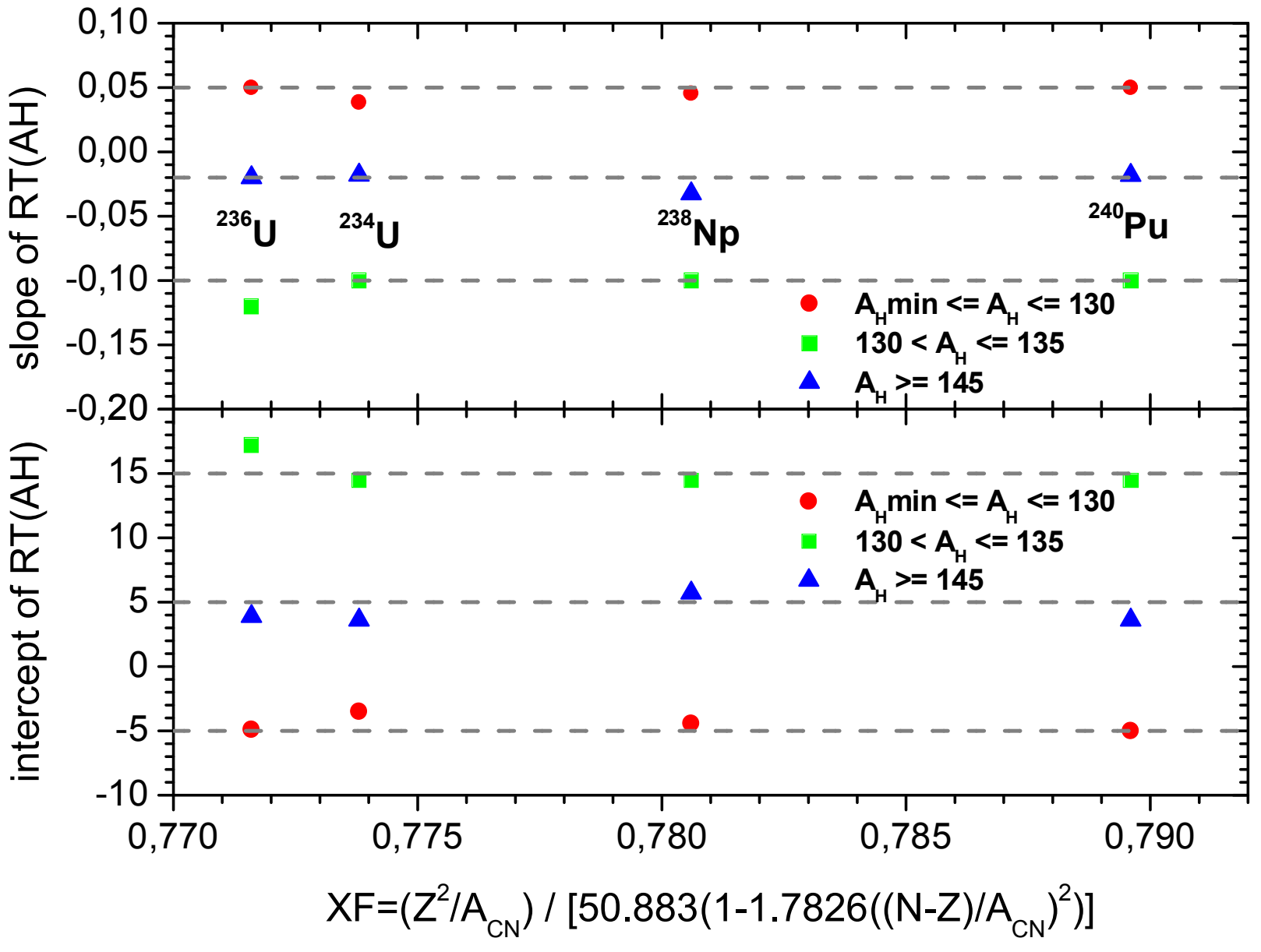

**Fig. 2.6** Slope and intercept of the RT parameterizations for neutron induced fission

For the spontaneous fissioning systems studied in (Manailescu et al., 2011), it was observed a different behaviour of the $RT(A_H)$ compared to the neutron induced fission ones as follows: *i)* the maximum of $RT$ (around $A_H=130$) is higher than in the case of (n,f), *ii)* the $A_H$ range where $RT$ is equal to 1 is limited to 1÷2 mass units around $A_H=145$. For both systems, practically the same $RT$ values in the $A_H$ range above 134 was observed. The visible differences in the region $A_H<134$ are mainly due to the shifted minimum of the experimental $\nu_H/\nu_{pair}$ data in the case of $^{248}$Cm(SF).

This TXE partition method is not depending on what is going on during the scission process (pre–saddle, saddle–to–scission), avoiding the ambiguities regarding the models, the parameters and the assumptions at scission. As a disadvantage, the parameterization $\nu_H/\nu_{pair}$ or $RT(A_H)$ should be used with caution for fissioning systems far from the studied ones.

### II.1.2 TXE partition including extra–deformation of fragments and excitation energies at scission

This method does not need any experimental data and parameter adjustments and can be applied to any fissioning system (spontaneous or at moderate incident neutron energy). The method is based on the following physical assumptions and models: statistical equilibrium at scission is reached, the Fermi–gas description of fragment level densities in the range of available excitation energies at scission is appropriate and the generalized super–fluid model of Ignatiuk (Ignatiuk, 1998) for the level density parameters of fragments can be applied. As in other papers (Ruben et al, 1991; Maerten et al., 1989; Terrel, 1965; Kildir and Aras, 1982), the extra deformation energies of nascent fragments are calculated considering the fragments as rotational ellipsoids nearly touching at the scission point, with the deformability determined from liquid drop model with the shell corrections taken into account.

According to Ruben and co–workers (Ruben et al., 1991; Maerten et al., 1989), the energy conservation for each pair of nascent fragments at scission is given by:

$$E_r + B_n + E_n = E_{pre} + E_{coul} + E_{def} + E_{sc}^* \tag{2.19}$$

where $E_r$ is the energy release (usually calculated using mass excesses from nuclear data tables), $B_n$ and $E_n$ are the neutron binding energy in the fissioning nucleus and the incident neutron energy. For the spontaneous fission case, both $E_n$ and $B_n$ are taken equal to zero. $E_{pre}$ and $E_{coul}$ are the pre–scission kinetic energy and the Coulomb repulsion energy between the two nascent fragments, respectively. $E_{def}$ is the sum of the extra-deformation energies ($E_{def} = E_{def}^L + E_{def}^H$) of complementary nascent fragments. $E_{sc}^*$ is the available excitation energy (collective and intrinsic) at scission. According to (Ruben et al., 1991) this energy can be described by the dissipative and heating energies.

After the full acceleration of the fission fragments, the total kinetic energy is given by:

$$TKE = E_{pre} + E_{coul} \tag{2.20}$$

Therefore, the total excitation energy of the fully–accelerated complementary fission–fragments can be written as (Morariu et al., 2012):

$$TXE = E_{def}^L + E_{def}^H + E_{sc}^* = E_L^* + E_H^* \tag{2.21}$$

In order to obtain the excitation energy of each fully accelerated fission fragment, two steps are necessary: (1) the calculation of the fragment extra–deformation energies at scission and (2) the partition of the total available excitation energy at scission $E_{sc}^*$ obtained by subtracting the extra–deformation energies at scission from the TXE between the two nascent fragments:

$$E^{*}_{sc} = TXE - \left(E^{L}_{def} + E^{H}_{def}\right) = E^{L}_{sc} + E^{H}_{sc} \tag{2.22}$$

The extra–deformation energy $E^{L,H}_{def}$ entering equations (2.21) and (2.22) must be understood as an additional deformation at scission which will be relaxed into excitation at full acceleration. Therefore, compared to the nascent fragments at scission, the well–separated fully accelerated fission–fragments are less deformed.

As shown by Terrel (Terrel, 1965), the deformability α is related to the stiffness parameter C2 (quadrupole deformation). The nuclear stiffness initially provided by the liquid drop model is strongly influenced by shell effects. The deformability α can be expressed by a semi–empirical relation as follows: $\alpha=\alpha_{LDM}(K-\delta W)/(K+\delta W)$ including shell corrections δW as proposed by Kildir and Aras (Kildir and Aras, 1982). In Ref. (Ruben et al, 1991; Maerten et al., 1989; Kildir and Aras, 1982) the shell correction of Myers–Swiatecki (RIPL–3, 2012d) have been used and the best fit of the stiffness data was obtain for K=8 MeV. In the case of the more recent shell correction data of Moller et al., as given in the Reference Input Parameter Library (RIPL–3, 2012c) a good fit of the experimental stiffness data was obtained for K=12 MeV.

Few examples of deformability calculation are given in **Fig. 2.7** as following: the liquid drop model deformability provided by the relation of Terrel (Terrel, 1965) (full circles) and by the relation proposed by Kildir and Aras (Kildir and Aras, 1982) (open circles). as can be seen in the figure, the two relations for $\alpha_{LDM}$ show very similar results. In the figure is also represented the deformability with shell effects correction of Myers–Swiatecki (down triangles) and K=8 MeV and the ones of Moller et al and K=12 MeV (stars) taken from (RIPL–3, 2012d). Due to the shell corrections, pronounced differences between the two cases appear only for the heavy fragments with $A_H$ above 160.

The total excitation energy magnitude is in the order of 25–50 MeV for different spontaneous or neutron–induced fissioning systems at low and moderate incident energy. The extra-deformation energy of a nascent fragment at scission relative to the fragment deformation at full acceleration does not exceed 10 MeV. Then it is reasonably to assume that the values of the available excitation energy at scission remain sufficiently high that the level density of the fragments can be described by Fermi–gas type functions (Morariu et al., 2012).

Consequently, the fragment excitation energy at scission can be expressed as a function of the level density parameter and the nuclear temperature as $E_{sc}^{L,H} = a_{sc}^{L,H} \tau_{L,H}^2$. Assuming statistical equilibrium at scission (equal nuclear temperature $\tau_L = \tau_H$ of complementary nascent fragments) and the level density description by the Fermi–gas model, the available excitation energy at scission is shared between the complementary nascent fragments in the same ratio as their level density parameters:

$$\frac{E_{sc}^{L}}{E_{sc}^{H}} = \frac{a_{sc}^{L}}{a_{sc}^{H}} \qquad (2.23)$$

where $a_{sc}^{L,H}$ must be understood as effective level density parameters accounting for collective and intrinsic/single–particle excitations. Usually, the effective level density parameters are obtained from the straightforward procedure of s–wave average spacing $\langle D_0 \rangle$ at binding neutron energy.

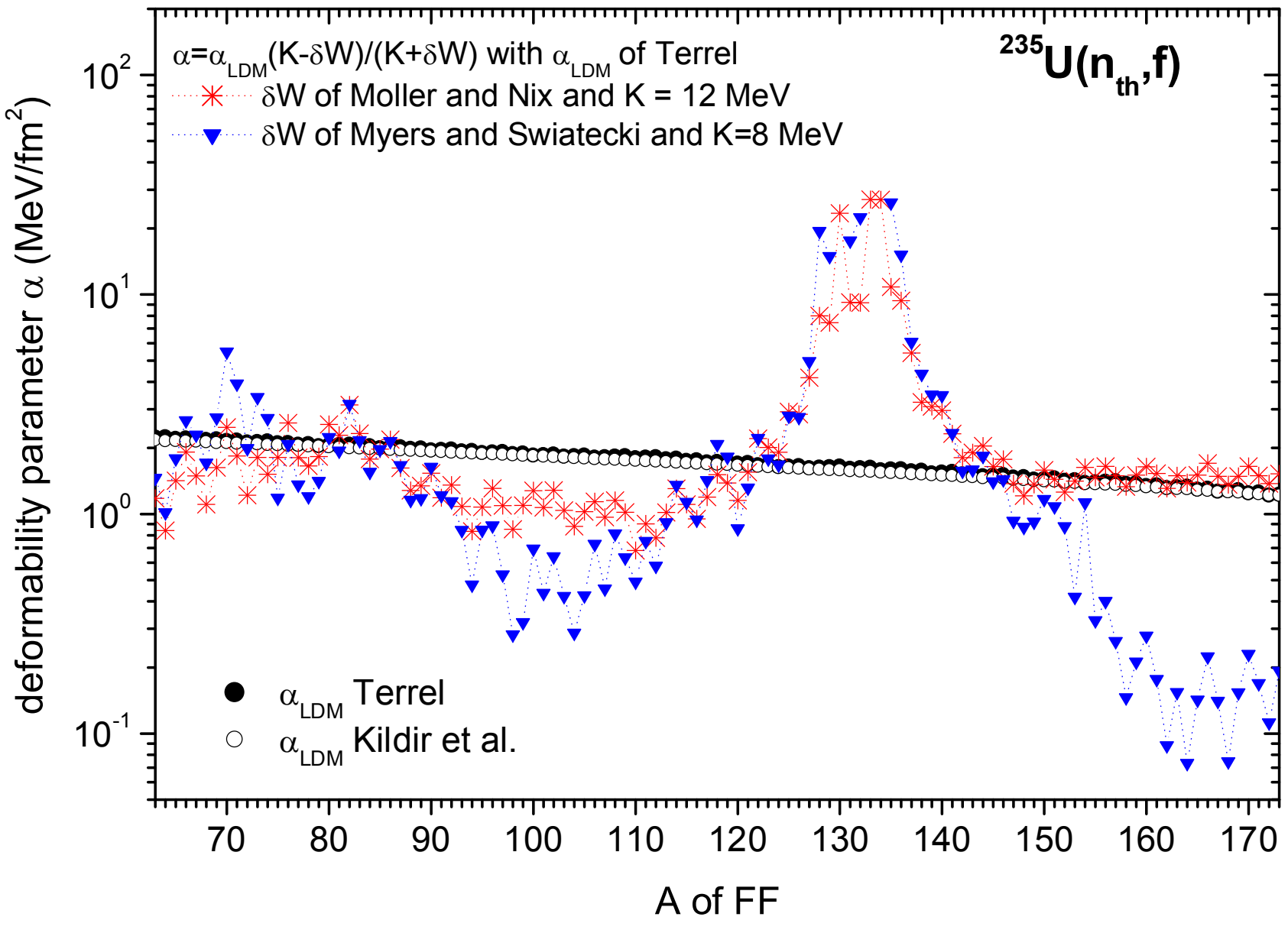

**Fig. 2.7** Example of deformability parameter calculation for the fission –fragment range of $^{235}$U($n_{th}$,f) (Morariu et al., 2012)

The super–fluid model (Ignatiuk, 1998) can be used in the above formula because the ratio of the level density parameters of complementary fragments provided by its use is practically the same as the ratio of the effective level density parameters. In order to support this fact an example is given in **Fig. 2.8**, where the level density parameter provided by the super–fluid model is plotted for two cases of asymptotic $\tilde{a}$ parameterizations (the one of (Ignatiuk, 1998) (open circles) and the one of (Egidy and Bucurescu, 2005) (stars)).

In the figure the $a_L/ a_H$ ratio provided by the systematic of Gilbert and Cameron (Gilbert and Cameron, 1965) is given too (squares) (see **Appendix 2**).

The values of the excitation energy at scission $E_{sc}^{L,H}$ and the level density parameter $a_{sc}^{L,H}$ of complementary nascent fragments described by the equations (2.22)–(2.23) and by the super–fluid model respectively (see **Appendix 2**) are obtained simultaneously by an iterative procedure.

Finally the excitation energy of each fully accelerated fragment is obtained as a sum of extra-deformation and excitation energy at scission:

$$E_{L,H}^{*} = E_{def}^{L,H} + E_{sc}^{L,H} \tag{2.24}$$

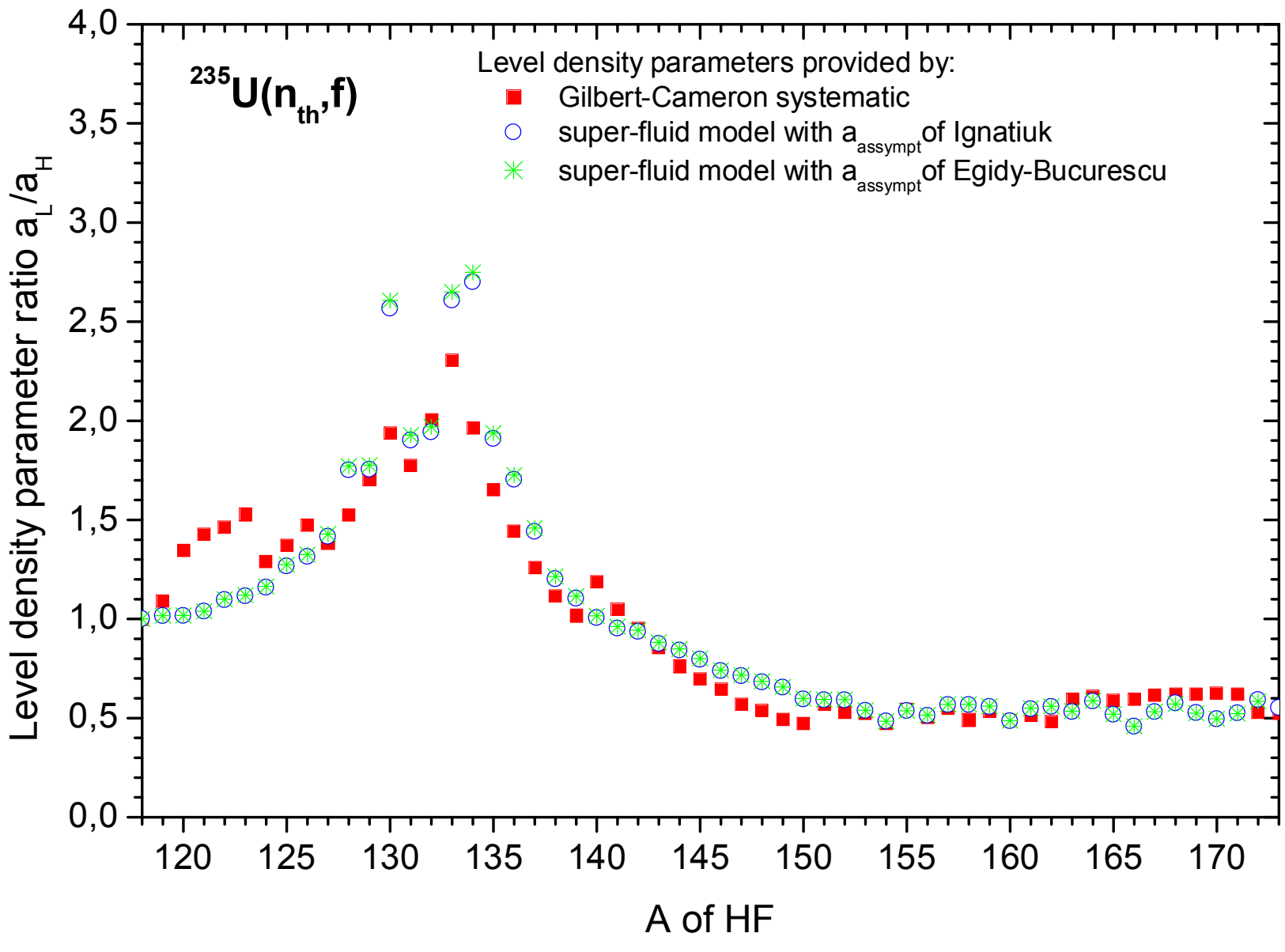

**Fig. 2.8** Level density parameter ratio of complementary fragments provided by the super–fluid model and by the effective level density parameter systematic of Gilbert–Cameron (Morariu et al, 2012)

In order to validate the described *TXE* partition procedure, in the following the fragment excitation energy is compared to the data obtained directly from experimental neutron multiplicity $\nu(A)$. Under the straightforward assumption that almost the entire prompt neutron emission takes place from fully accelerated fission fragments, the *TXE* given by equation (2.13) can be partitioned according to the ratio $\nu_L/\nu_H$ obtained from available experimental $\nu(A)$ data.

For the case of $^{235}U(n_{th},f)$, $E^{*}(A)$ data obtained from experimental $\nu(A)$ data sets of Nishio and Maslin are plotted with open symbols in figure **2.9**. As it can be seen in the figure, the calculated $E^*(A)$ describe very well the experimental data. For the calculation of $E^*(A)$, the total excitation energy was obtained according to equation (2.13) using the experimental TKE(A) of (Straede et al., 1987). In the figure $E^*(A)$ obtained by using for the extra deformation energies the $\beta_2$ parameterization of HFB 14 (Morariu et al., 2012) and from the parameterization of (Schmidt and Jurado, 2012) are plotted with full circles and stars respectively. The good agreement obtained by using $\beta_2$ of Schmidt encouraged the use of this simple parameterization as an appropriate global description of the $\beta_2$ deformation parameters of nascent fragments.

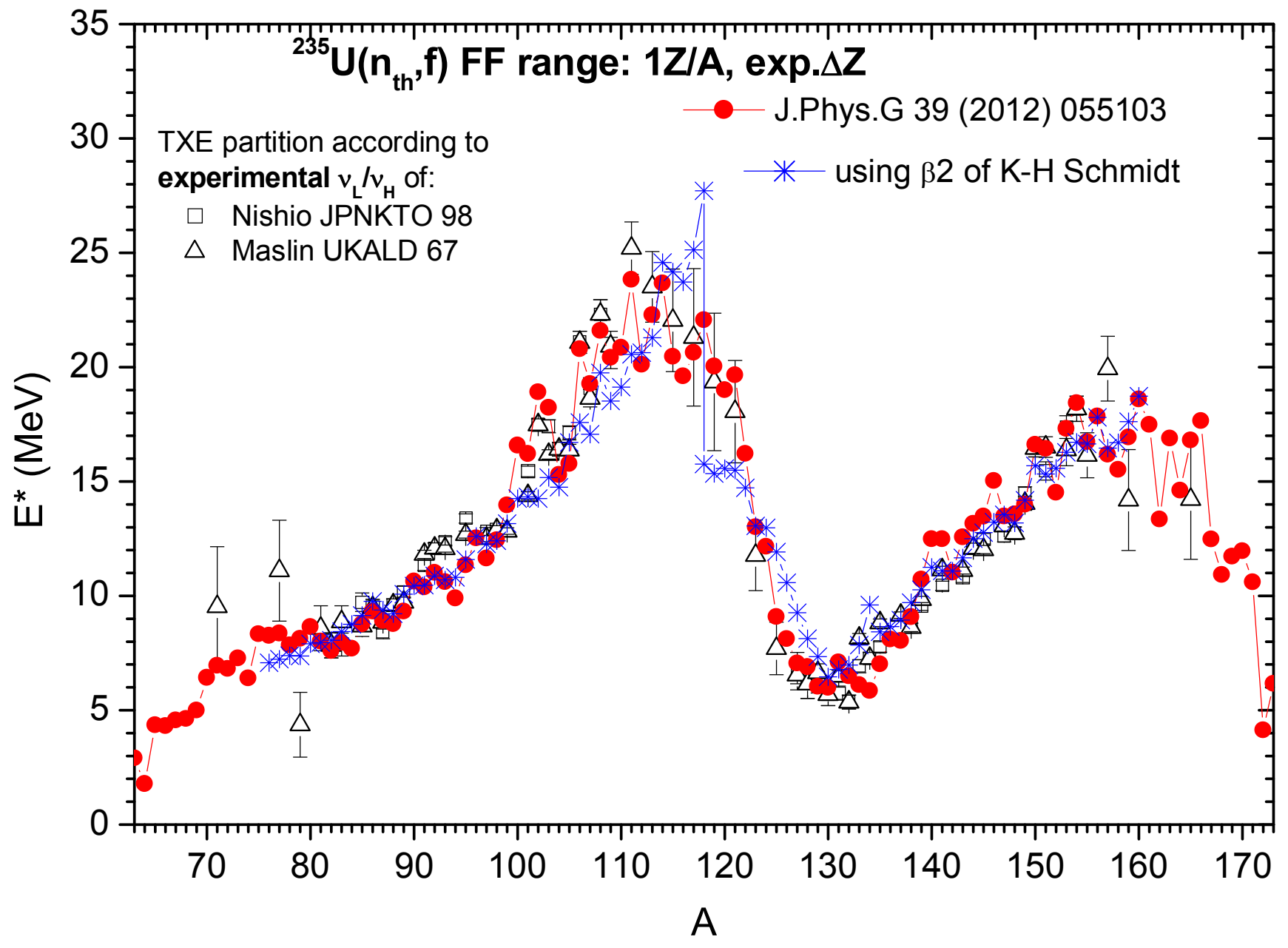

**Fig. 2.9** $^{235}U(n_{th},f)$: $E^*(A)$ obtained by using $\beta_2$ parameterization of (Schmidt and Jurado, 2012) in the frame of the same method as in (Morariu et al., 2012)in comparison with $E^*(A)$ reported in (Morariu et al., 2012) and with "indirect" experimental data obtained as in (Morariu et al., 2012)

The same procedure was applied for the case of indiced fission $^{237}Np$ at En=0.8 MeV and 5.5 MeV. The calculated $E^*(A)$ obtained by using the $\beta_2$ are plotted with magenta and cyan symbols connected with lines in **Fig. 2.10** in comparison with the $E^*(A)$ reported in (Morariu et al., 2012) (red and blue symbols) and the "indirect" experimental data of Naqvi and Muller (different black and grey smbols). Especially for En= 0.8 MeV, the use of these global $\beta_2$ leads to $E^*(A)$ results close to the ones previously reported (the red and the magenta points connected with lines are very close each other in the mass ranges $A_L<105$ and $A_H>130$).

As it can be seen, in the case of present E*(A) results the increase with the incident energy is visible only for the heavy fragments. Though an increase of $E^*(A)$ is observed for light fragments with $A_L$ between 100÷119, for the other light fragments with $A_L$<100 *E*(A)* can be considered practically constant with En. So, the use of $\beta_2$ global expressions of (Schmidt and Jurado, 2012) in the frame of the TXE partition method from (Morariu et al., 2012) can support the behaviour of experimental *ν(A)* at En = 0.8 and 5.5 MeV.

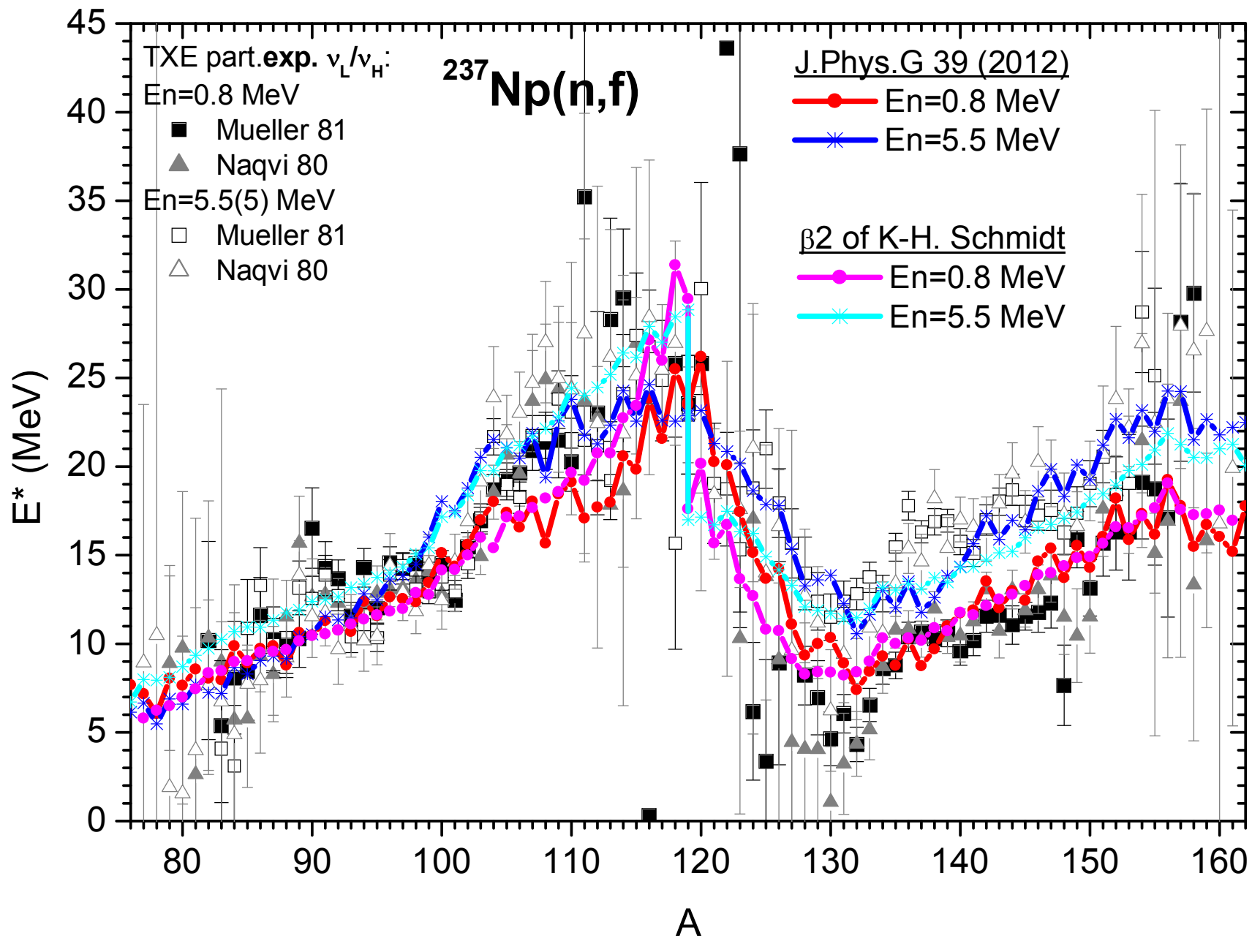

**Fig.2.10** $^{237}$Np(n,f): $E^*$(A) obtained by using $\beta_2$ parameterization of (Schmidt and Jurado, 2012) in the frame of the same method as in (Morariu et al., 2012)in comparison with $E^*$(A) reported in (Morariu et al., 2012) and with "indirect" experimental data obtained as in (Morariu et al., 2012)

In the case of $^{239}$Pu($n_{th}$,f), the E*(A) results obtained by using the global expression of $\beta_2$ proposed by Schmidt and Jurado in (Schmidt and Jurado, 2012) in the frame of the TXE partition method of (Morariu et al., 2012) are plotted with blue circles in **Fig. 2.11**. As it can be seen the present E*(A) results are close to the ones reported in (Morariu et al., 2012) (red circles) and describe rather well the "indirect" experimental data. Consequently, in the case of thermal induced fission of $^{239}$Pu, the $\beta_2$ parameterizations from (Schmidt and Jurado, 2012) are working well too.

For this reason, the same calculation was extended for the case of spontaneous fission of $^{240}$Pu, the obtained results being given with blue stars in the same figure. As observation, for both cases, $^{239}$Pu(nth,f) and $^{240}$Pu(SF), the same experimental TKE(A) distribution of Wagemans et al. was used. Taking into account that the fission fragments range is the same in both cases, the difference in TXE for each pair of fragments is given only by Bn.

Looking at the behaviour of the present $E^*(A)$ results given in **Fig. 2.11** (blue circle and stars), the increase of $E^*(A)$ with the excitation energy of the fissioning nucleus $^{240}$Pu is visible for both heavy and light fragments, being a little bit more pronounced for the heavy fragments, especially in the mass range $A_H>135$.

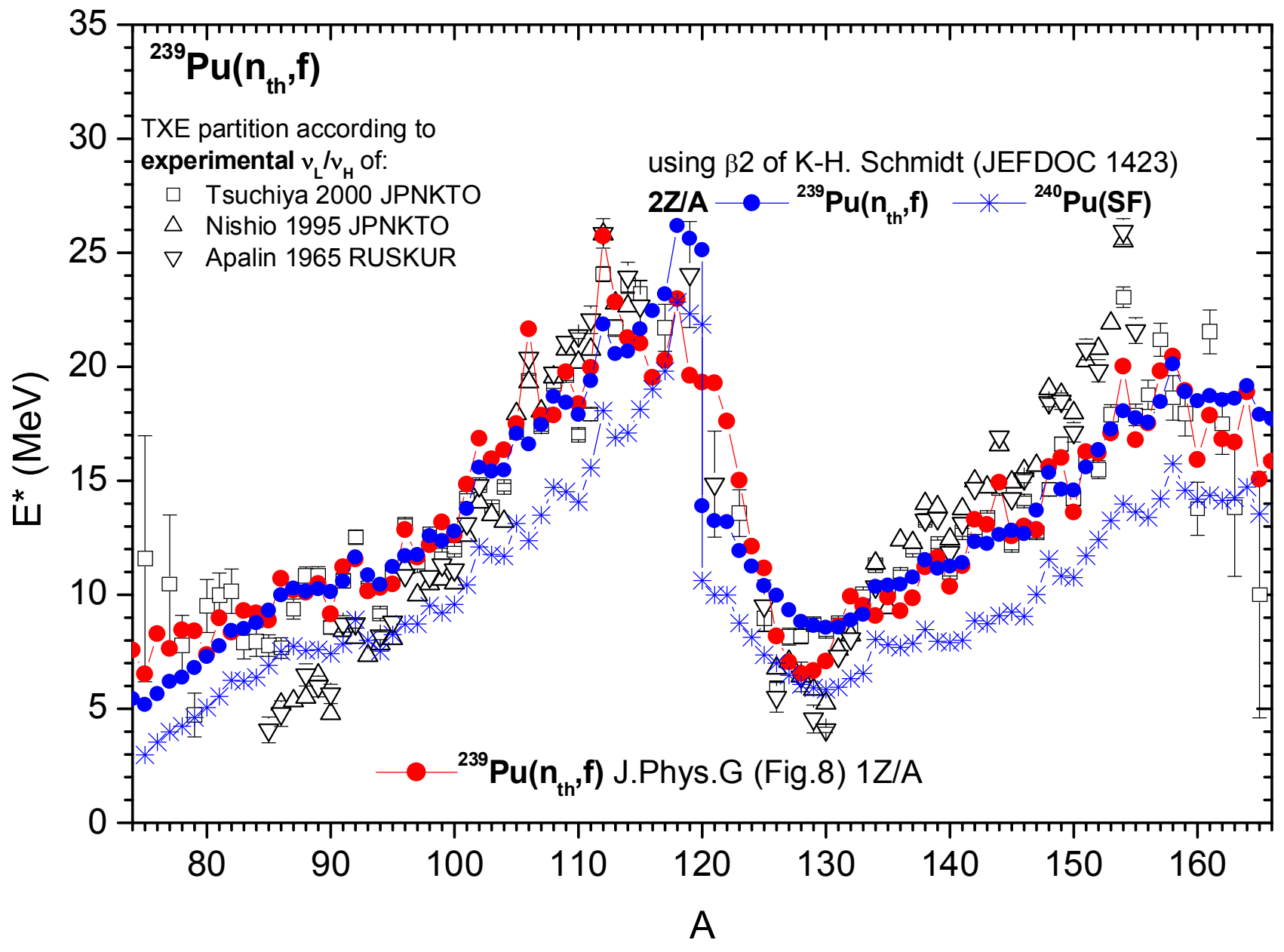

**Fig. 2.11** $^{239}$Pu($n_{th}$,f): E$^*$(A) obtained by using the $\beta_2$ of (Schmidt and Jurado, 2012) in the frame of the same method as in (Morariu et al., 2012) (blue circle) in comparison with the E$^*$(A) calculation from (Morariu et al., 2012) (red circle) and "indirect" experimental data as in (Morariu et al., 2012) (open symbols). E$^*$(A) of $^{240}$Pu(SF) obtained by using the $\beta_2$ of (Schmidt and Jurado, 2012) in the frame of the same method as in (Morariu et al., 2012) (blue stars)

The good agreement of the calculated $E^*(A)$ with the "indirect" experimental ones (obtained from experimental $\nu(A)$ data) can be considered as a validation of the above described TXE partition procedure. This fact proves again that the assumptions and models used are adequate.

This TXE partition method, based exclusively on models and straightforward assumptions and using as input only quantities available in the nuclear data libraries (such as mass excesses, shell corrections) can assure a better prediction of prompt emission data than other TXE partition methods

### II.1.3 Fragmentation range

In the Point by Point treatment (Tudora et al., 2005; Vladuca et al., 2006; Tudora, 2006, 2008, 2009, 2010a,b; Manea and Tudora, 2011; Manailescu et al, 2011; Morariu et al., 2012; Tudora et al, 2008, 2012a,), the primary multi–parametric matrices (e.g. ν(Z,A,TKE), ε(Z,A,TKE) and so on) are calculated as following:

*i)* the most important role is played by the choice of the fragmentation range, or the so–called fission–fragment range. This range is usually generated by taking into account all fragment mass pairs $\{A_L, A_H\}$ covered by an experimental or simulated mass distribution Y(A). For each pair of masses, two or more fragments are taken with the charge numbers Z as the nearest integer above and below the most probable charge that is considered UCD corrected with the possible charge polarization ΔZ. Finally the fragmentation range is $\{Z_{Lj}, A_{Lj}, Z_{Hj}, A_{Hj}\}$ with the index "j" running over the number of fragmentations

*ii)* for each pair of the fragmentation range mentioned on item *i)* the prompt neutron quantities(multiplicity, spectrum and so on) are calculated at a given TKE value (all fragment pairs receive the same TKE value)

*iii)* the calculation of item *ii)* is repeated for TKE values covering a convenient range.

In other words the PbP matrix ν(Z,A,TKE) means: $Z_j$, $A_j$ covering the fragmentation range (where Z and A are referring to light and heavy fragments with the index j from 1 up to the number of fragment pairs according to item *i)*) and $TKE_k$ covering the TKE range according to item *iii)* (for instance a TKE range from 130 MeV to 230 MeV with a step of 5 MeV).

### II.1.4 Fission fragment distributions

The fragment distributions used in the Point by Point treatment are the following:

a) the charge distribution of each fragment pair taken as a narrow Gaussian function according to Wagemans et al. (Wagemans, 1991):

$$P(Z) = \frac{1}{\sqrt{\pi c}} \exp\left(-(Z - Z_p)^2 / c\right), \quad c = 2(\sigma^2 + 1/12) \qquad (2.25)$$

where $Z_p$ is the most probable charge (UCD plus charge polarization)

b) the total kinetic energy distribution for each pair of fragments (also with a Gaussian shape):

$$p(A,TKE) = \frac{1}{\sigma_{TKE}(A)\sqrt{2\pi}} \exp\left( -\frac{(TKE - TKE(A))^2}{2(\sigma_{TKE}(A))^2} \right) \quad (2.26)$$

where TKE(A) and $\sigma_{TKE}$(A) are experimental data.

c) the fragment mass distribution Y(A). Obviously experimental Y(A) data are preferred, but in the absence of the experimental distributions, simulations of Y(A) can be used too. Details about can be found in (Tudora, 2010a).

Other distributions needed in Point by Point calculations can be easily obtained from the experimental single distributions Y(A), TKE(A) and $\sigma_{TKE}$(A), too. For instance the double distribution Y(A,TKE) can be re–constructed as $Y(A,TKE) = Y(A) \cdot p(A,TKE)$ with *p(A,TKE)* given by equation (2.26). The distribution Y(TKE) can be obtained as $Y(TKE) = \sum_A Y(A,TKE) \Big/ \sum_A Y(A)$.

As observation: taking into account that the Point by Point multi–parametric matrices are generated for all Z, A covering the fragmentation range at each TKE value (with TKE covering a chosen range with an usual step of 5 MeV) according to the computational scheme of the code, it is convenient to use double distributions Y(A,TKE) re–constructed from the single ones. But primary experimental double distributions Y(A,TKE) can be also used without any problem (needing only a simple change of the format of experimental matrix files).

## II.2 Average quantities provided by the Point by Point model

The Point by Point model is able to provide almost all quantities related to the prompt neutron emission. These quantities can be categorized according to the averaging manner of the multi–parametric matrices as following:

a) quantities which are a function of the fragment mass (such as ν(A), ε(A), $E_\gamma$(A), *param*(A), where *param* is: *Er*, *TKE*, *Sn*, and *a*) are obtained by averaging the corresponding matrix over Z and TKE.

As an example, the prompt neutron multiplicity as a function of fragment mass number (usually named sawtooth) is obtained using the following formula:

$$\bar{\nu}(A) = \sum_{Z,TKE} \nu(Z,A,TKE)p(Z,A)Y(A,TKE) \Big/ \sum_{Z,TKE} p(Z,A)Y(A,TKE) \qquad (2.27)$$

b) quantities which are a function of TKE (such as $\bar{\nu}(TKE)$, ε(TKE), *param*(TKE)) are obtained by averaging the corresponding matrix over Z and A:.

As an example:

$$\bar{\nu}(TKE) = \sum_{Z,A} \nu(Z,A,TKE)p(Z,A)Y(A,TKE) \Big/ \sum_{Z,A} p(Z,A)Y(A,TKE) \qquad (2.28)$$

c) total average quantities (for example, $<\nu_p>$, spectra, $<E_\gamma>$, and also the average values of the model parameters (*Er*, *TKE, Sn*, and *a,* generically labeled here *param*) are obtained by averaging the multi–parametric matrix over Z, A and TKE. The total average parameters (*<Er>*, *<TKE>, <Sn>*, and *<a>)* obtained in this manner are used in the most probable fragmentation approach.

### II.3 Most probable fragmentation approach

The Point by Point treatment is used to calculate the average values of the model parameters (<Er>, <TKE>, <Sn>, <a>) according to the following formula:

$$< param >= \frac{\sum_{Z,A,TKE} param(Z,A,TKE)\,p(Z,A)Y(Z,A,TKE)}{\sum_{Z,A,TKE} p(Z,A)Y(Z,A,TKE)} \tag{2.29}$$

where *param* is replaced by one of the model parameters mentioned above.

The total average parameters obtained by PbP treatment at different incident energies En showed a dependence on En which is taken polynomial in the most probable fragmentation approach as following (Vladuca and Tudora, 2001a,b,c):

$$< param > (En) =< param >_{th} + \alpha E_n + \beta E_n^2 \tag{2.30}$$

with $E_n = E^* - B_n$

where $<param>$ are, respectively, <Er>, <Sn>, <TKE>. The index “th” denotes to the average value of the model parameter at thermal incident neutron energy. More details can be found in (Tudora, 2009) and references therein.

Similarly, the average model parameters can be expressed as a function of the excitation energy of the fissioning nucleus:

$$< param > (E^*) =< param >_0 + aE^* + bE^{*2} \tag{2.31}$$

where $<param>_0$ is the average value of the model parameter at zero excitation energy (spontaneous fission). In the above relations, $\alpha$, $\beta$, $a$, $b$ are the coefficients expressing the first and the second polynomial order dependence on energy.

According to (Fréhaut, 1989), the average prompt gamma–ray energy can be considered as linearly dependent on the prompt neutron multiplicity:

$$< E_\gamma >= p < \nu_p > + q \tag{2.32}$$

with the slope and the intercept depending on the charge and the mass numbers of the fissioning nucleus (Vladuca and Tudora, 2001a,b):

$$p = 6.710 - 0.156\frac{Z^2}{A}$$
$$q = 0.750 + 0.088\frac{Z^2}{A} \quad (2.33)$$

For all the isotopes families (Th, Pa, U, Np, Pu, Am) studied in (Tudora, 2009), systematic behaviors were deduced for the thermal average model parameters $<E_r>_{th}$, $<TKE>_{th}$ and $<S_n>_{th}$ (for details see (Tudora, 2009)):

- linear behavior of the quantity $B_n + <E_r>_{th}$ versus the fissility parameter
- linear dependence of $<TKE>_{th}$ on the coulombian parameter $Z^2/A^{1/3}$ of the fissioning system
- parabolic dependence of $<Sn>_{th}$ on the fissility parameter:

$$XF = \left(Z^2/A\right)/\left(50.883\left(1 - 1.782\eta^2\right)\right) \quad (2.34)$$

where $\eta = (N - Z)/A$, with $Z$ and $A$ being the charge and respectively the mass number of the fissioning nucleus and $N=A-Z$.

More details regarding the systematic behaviour of the model parameters can be found in (Tudora, 2009).

## II.4 Multiple fission chances in the frame of the most probable fragmentation approach

For neutron incident energies up to 20 MeV only the fission of the nuclei formed by neutron evaporation from the precursor of the main chain is possible (named nuclei of the principal chain). At higher incident energies, fission of the nuclei formed by charged particle emission must be taken into account, too (see **Fig. 2.12**). Consequently, the prompt fission neutron spectrum, as well as the other quantities resulting from the fission of these nuclei must be also considered.

The extended model (Tudora et al., 2004) takes into account the prompt neutrons and gammas from the main and several secondary nucleus chains and ways, namely:

(1) the main chain (abbreviated *"n"*) – the fissioning nuclei formed by neutron evaporation;

The secondary ways are named as follows:

(2) *"protons"* (abbreviated *"p"*) – the fissioning nuclei formed by protons emission from nuclei of the main chain;
(3) *"neutrons via protons"* (abbreviated *"pn"*) – the fissioning nuclei formed by neutron evaporation from the nuclei formed by proton emission;
(4) *"deuterons"* (abbreviated *"d"*) – the fissioning nuclei formed by deuterons emission from nuclei of the main chain;
(5) *"alpha"* (abbreviated *"α"*) – the fissioning nuclei formed by alpha emission from nuclei of the main chain;
(6) *"neutrons via alpha"* (abbreviated *"αn"*) – the fissioning nuclei formed by neutron evaporation from the nuclei formed by alpha emission;

In the following, the above six ways will be indexed with $k=1(n)$, $2(p)$, $3(pn)$, $4(d)$, $5(\alpha)$, $6(\alpha n)$ (Tudora et al., 2004). The three nucleus chains according to the example of **Fig. 2.12** will be noted $c=(I)$ (the main chain), $c=(II)$ (the secondary chain of *Pa* nuclei) and $c=(III)$ (the secondary chain of *Th* nuclei). In the case of the main nucleus chain labeled I there is only one way $k=1$ and for this reason in the following the index 1 will be used for this chain and way.

The extended model (Tudora et al., 2004) can be also used for a variable cross section of the inverse process of compound nucleus formation, but because of the large amount of the calculations, for the nuclei of the secondary chains the consideration of a constant cross section is preferred.

The energetics of these reactions are defined as follows.

For the main chain /way (indexed 1) the excitation energy of the nuclei undergoing fission is given by:

$$
\begin{aligned}
E^{*(1)}_{1} &= En + Bn^{(1)}_{1} \\
E^{*(1)}_{i} &= E^{*(1)}_{i-1} - Bn^{(1)}_{i-1} - \langle \varepsilon_{ev} \rangle^{(1)}_{i-1}, \quad i = 2,\ldots,N^{(1)}
\end{aligned}
\tag{2.35}
$$

where $N^{(1)}$ is the number of nuclei (chances) of the main chain, $B^{(1)}_{n1}$ is the neutron binding energy from the $i$th nucleus of the main chain and $E_n$ is the incident neutron energy.

The quantity $\langle \varepsilon_{ev} \rangle^{(1)}_{i}$ is calculated using for the evaporation spectrum the Weisskopf–Ewing model:

$$
\begin{aligned}
\varphi^{(1)}_{i}(E) &= K(\theta^{(1)}_{i})\sigma_{c}(E)E\exp\left(-\frac{E}{\theta^{(1)}_{i}}\right) \\
K(\theta^{(1)}_{i}) &= \left(\int_{0}^{\infty}\sigma_{c}(E)E\exp\left(-\frac{E}{\theta^{(1)}_{i}}\right)dE\right)^{-1} \\
\langle \varepsilon_{ev} \rangle^{(1)}_{i} &= K(\theta^{(1)}_{i})\int_{0}^{\infty}E^{2}\sigma_{c}(E)E\exp\left(-\frac{E}{\theta^{(1)}_{i}}\right)dE
\end{aligned}
\tag{2.36}
$$

The evaporation temperature is given by:

$$
\theta^{(1)}_{i} = \sqrt{\frac{E^{*(1)}_{i} - Bn^{(1)}_{i}}{a^{(1)}_{i+1}}}, \quad i = 1,\ldots,(N^{(1)} - 1) \tag{2.37}
$$

where $a^{(1)}_{i}$ is the level density parameter of the residual nucleus $i$ of the main chain, obtained by the neutron evaporation.

The available excitation energies of the fissioning nuclei of the ways *k=2(p)* and *k=5(α)* are calculated using the recursive formula:

$$E^{*(k)}_{i} = E^{*(1)}_{i} - S^{(1)}_{ki} - \langle \varepsilon_{ev} \rangle_{ki}, \quad i = 1,\dots,N^{(1)} \tag{2.38}$$

where $N^{(k)}$ is the number of fissioning nuclei (chances) formed by $p$ or $\alpha$ emission. The quatities $E_i^{*(1)}$ and $S_{ki}^{(1)}$ are the excitation energy and respectively, the $p$ or $\alpha$ separation energy from the $i$th precursor nucleus of the main chain. $\langle \varepsilon_{ev} \rangle_i^{(1)}$ is calculated also by using for the evaporation spectrum the Weisskopf–Ewing model, where the evaporation temperature is given by:

$$\theta_{ki} = \sqrt{\frac{E^{*(1)}_{i} - S^{(1)}_{ki}}{a^{(c)}_{i}}}, \quad i = 1,\dots,N^{(k)} \tag{2.39}$$

with $a_i^{(c)}$ the level density parameter of the residual nucleus number $i$ formed by $p$ emission (nucleus chain c=II)) or $\alpha$ emission (nucleus chain c=III)).

In the case of the deuteron emission, (the way *k=4(d)*) the excitation energy and the evaporation temperature are:

$$\begin{aligned} E^{*(k)}_{i} &= E^{*(1)}_{i-1} - S^{(1)}_{ki-1} - \langle \varepsilon_{ev} \rangle_{ki-1}, \quad i = 2,\dots,N^{(k)} \\ \theta_{ki} &= \sqrt{\frac{E^{*(1)}_{i} - S^{(1)}_{ki}}{a^{(II)}_{i+1}}}, \quad i = 1,\dots.(N^{(k)} - 1) \end{aligned} \tag{2.40}$$

For the fissioning nuclei of the ways *k=3(pn)* and *k=6(αn)*, the excitation energy and the evaporation energy are calculated as:

$$\begin{aligned} E^{*(k)}_{i} &= E^{*(k)}_{i-1} - Bn^{(k)}_{i-1} - \langle \varepsilon_{ev} \rangle_{i-1}, \quad i = 2,\dots,N^{(k)} \\ E^{*(k)}_{1} &= E^{*(k-1)}_{1} \\ \theta^{(k)}_{i} &= \sqrt{\frac{E^{*(k)}_{i} - Bn^{(k)}_{i}}{a^{(c)}_{i+1}}}, \quad i = 1,\dots,(N^{(k)} - 1) \end{aligned} \tag{2.41}$$

where $Bn_i^{(k)}$ is the neutron separation energy from the nucleus indexed $i$ of the way *(k)* and $a_{i+1}^{(c)}$ refers to the nuclei of the chain $c=(II)$ for the case $k=3(pn)$ and respectively, of the chain $c=(III)$ for the case $k=6(\alpha n)$.

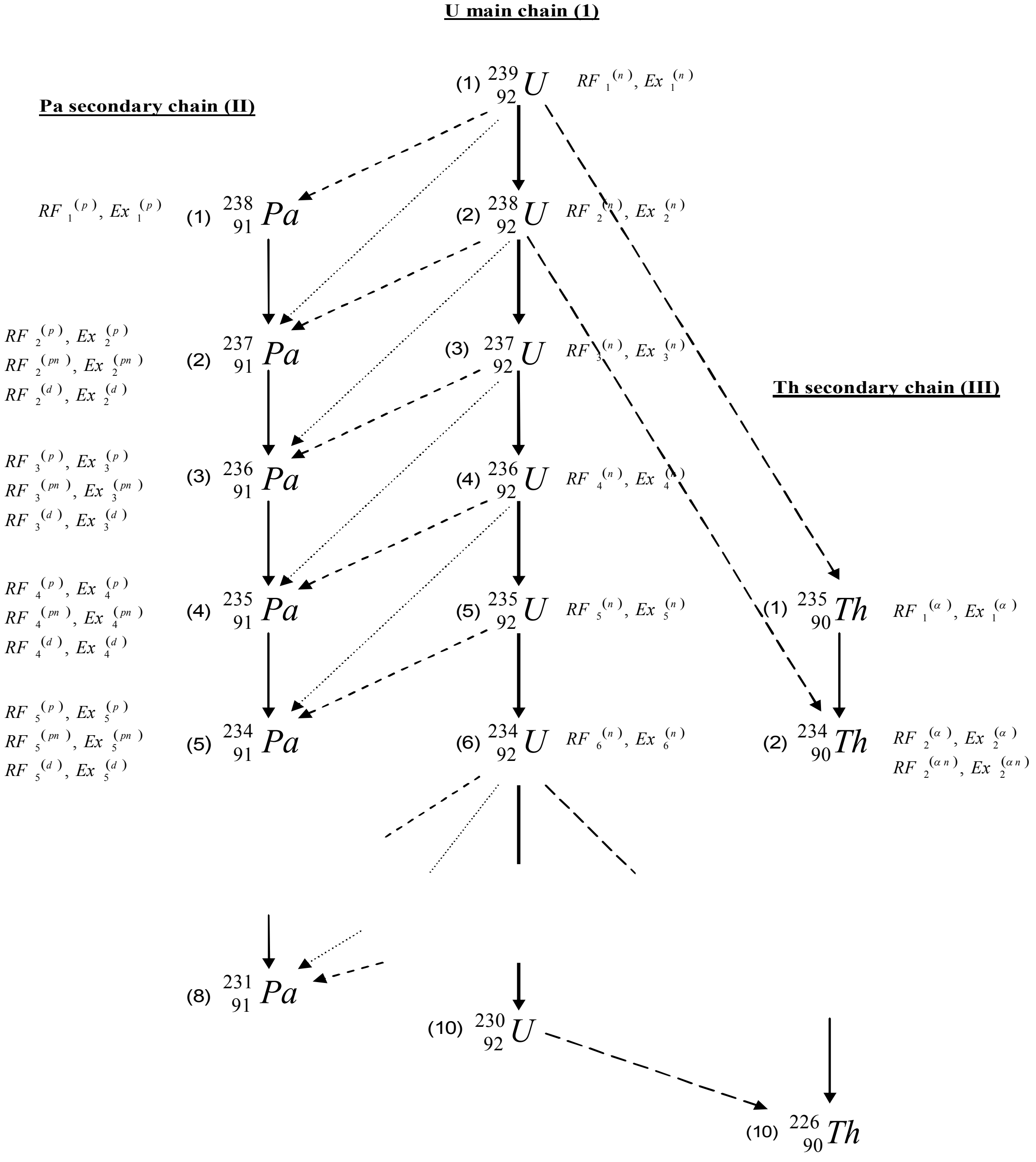

**Fig.2.12** Fission nucleus chain and ways, exemplified on the n+$^{238}$U reaction (Tudora et al., 2004)

The fission probability of each nucleus involved in the reaction is determined as fission cross–section ratio given by:

$$RF_i^{(c)} = \frac{\sigma_{fi}^{(c)}}{\sigma_f^{tot}} \qquad \text{c=I,II,III} \tag{2.42}$$

where $\sigma_f^{tot}$ and $\sigma_{fi}^{(c)}$ are the total fission cross–section and, respectively, the fission cross–section of the nucleus *i* of the chain *c*.

In order to obtain the partial fission cross–section ratio of the secondary chains (*(II)* and *(III)*), the production cross–sections of the *j*th secondary nucleus by protons $\sigma_{2j}$, neutrons from the chain *(II)* $\sigma_{3j}$, deuterons $\sigma_{4j}$, alpha particles $\sigma_{5j}$ and neutrons from the chain *(III)* $\sigma_{6j}$ are used:

$$RF_j^{(k)} = \frac{\sigma_{kj}}{\sum_{k'=2}^{4} \sigma_{k'f}} RF_j^{(II)} \qquad RF_j^{(k)} = \frac{\sigma_{kj}}{\sum_{k'=5}^{6} \sigma_{k'f}} RF_j^{(III)}$$
$$k = 2,3,4 \qquad\qquad k = 5,6 \tag{2.43}$$

The total prompt fission neutron multiplicity, the total prompt fission neutron spectra and the "*m*" order momentum of the spectrum are calculated as sum of multiplicities, spectra and "*m*" order moments of all fissioning nucleus ways:

$$\langle \nu \rangle_{tot} = \sum_{k=1}^{6} \langle \nu \rangle_{tot}^{(k)} \qquad N_{tot}(E) = \sum_{k=1}^{6} N_{tot}^{(k)}(E) \qquad \langle E^m \rangle_{tot} = \sum_{k=1}^{6} \langle E^m \rangle_{tot}^{(k)} \tag{2.44}$$

where for :

- *k=1(n), 3(pn), 6(αn)*

$$\langle \nu \rangle_{tot}^{(k)} = \sum_{i=n}^{N^{(k)}} RF_i^{(k)} \left( i - 1 + \langle \nu \rangle_i^{(k)} \right)$$

$$N_{tot}^{(k)}(E) = \sum_{i=n}^{N^{(k)}} \frac{RF_i^{(k)}}{\langle \nu \rangle_{tot}^{(k)}} \left( \sum_{j=1}^{i-1} \varphi_j^{(k)}(E) + \langle \nu \rangle_i^{(k)} N_i^{(k)}(E) \right) \qquad n = \begin{cases} 1 & k = 1 \\ 2 & k = 3,6 \end{cases}$$

$$\left\langle E^m \right\rangle_{tot}^{(k)} = \sum_{i=n}^{N^{(k)}} \frac{RF_i^{(k)}}{\left\langle \nu \right\rangle_{tot}^{(k)}} \left( \sum_{j=l}^{i-1} \left\langle \varepsilon_{ev}^m \right\rangle_j^{(k)} + \left\langle \nu \right\rangle_i^{(k)} \left\langle E^m \right\rangle_i^{(k)} \right)$$

- *k=2(p), 4(d) ,5(α)*

$$\left\langle \nu \right\rangle_{tot}^{(k)} = \sum_{i=n}^{N^{(k)}} RF_i^{(k)} \left\langle \nu \right\rangle_i^{(k)}$$

$$N_{tot}^{(k)}(E) = \sum_{i=n}^{N^{(k)}} \frac{RF_i^{(k)}}{\left\langle \nu \right\rangle_{tot}^{(k)}} \left\langle \nu \right\rangle_i^{(k)} N_i^{(k)}(E) \qquad n = \begin{cases} 1 & k = 2,5 \\ 2 & k = 4 \end{cases}$$

$$\left\langle E^m \right\rangle_{tot}^{(k)} = \sum_{i=n}^{N^{(k)}} \frac{RF_i^{(k)}}{\left\langle \nu \right\rangle_{tot}^{(k)}} \left\langle \nu \right\rangle_i^{(k)} \left\langle E^m \right\rangle_i^{(k)}$$

where $\left\langle \nu \right\rangle_i^{(k)}$ , $N_i^{(k)}(E)$ and $\left\langle E^m \right\rangle_i^{(k)}$ are the number, the individual spectrum and respectively the "*m*" order momentum of the emitted neutron by the nucleus *i* formed by way *k*.

Other quantities than can be calculated with this model (Tudora et al., 2004) are the average total gamma–ray energy $\left\langle E_\gamma \right\rangle_{tot}$ and the average total kinetic energy $\left\langle TKE \right\rangle$ of the fission–fragments:

$$\left\langle E_\gamma \right\rangle_{tot} = \sum_{k=1}^{6} \left\langle E_\gamma \right\rangle_{tot}^{(k)} \qquad \left\langle TKE \right\rangle_{tot} = \sum_{k=1}^{6} \left\langle TKE \right\rangle_{tot}^{(k)} \qquad n = \begin{cases} 1 & k = 1,2,5 \\ 2 & k = 3,4,6 \end{cases} \tag{2.45}$$

with

$$\left\langle E_\gamma \right\rangle_{tot}^{(k)} = \sum_{i=n}^{N(k)} RF_i^{(k)} \left\langle E_\gamma \right\rangle_i^{(k)} \qquad \left\langle TKE \right\rangle_{tot}^{(k)} = \sum_{i=n}^{N(k)} RF_i^{(k)} \left\langle TKE \right\rangle_i^{(k)}$$

## II.5 Examples of Point by Point model results

The experimental multi–parametric data concerning the prompt neutron multiplicity as a function of fission fragment mass and total kinetic energy as well as other experimental data like the prompt neutron average center–of–mass energy as a function of the fragment mass, the average multiplicity as a function of the fission–fragment total kinetic energy and the average prompt γ–ray energy versus the fragment mass, when they exists, allow a detailed verification of the model and parameters used to calculate the prompt emission data.

In the following examples of such calculations are presented according to the description from **Section II.2**.

Multi–parametric experimental data regarding the prompt neutron number emitted by a fission–fragment as a function of its mass number and as a function of the total kinetic energy are available for few fissioning systems.

An example of a multi-parametric matrix is given in Figs.**2.13** and **2.14**. The calculated $\nu_{pair}$ as a function of the heavy fragment mass number for different *TKE* values is shown in Fig. 2.13 for the case of the spontaneous fission of $^{252}$Cf. As it can be seen the obtained PbP results are in a good agreement with the experimental data of Zakharova and Bowman.

If we are looking to the fission–fragment pair multiplicity as a function of the total kinetic energy, a good agreement is also obtained; see for example Fig. 2.14 where $\nu_{pair}$ as a function of TKE for a few fragmentations is given for $^{252}$Cf(SF). Looking to this figure a linear decrease of the fission–fragments pair multiplicity with TKE is observed.

To better see the global good agreement between the Point by Point calculations and the experimental data, in **Fig. 2.15** the multi–parametric representation is given for $^{252}$Cf(SF). The available experimental data are plotted with open symbols and the calculations with full symbols. Except only for few spread experimental data, the Point by Point calculations are covering well the experimental points.

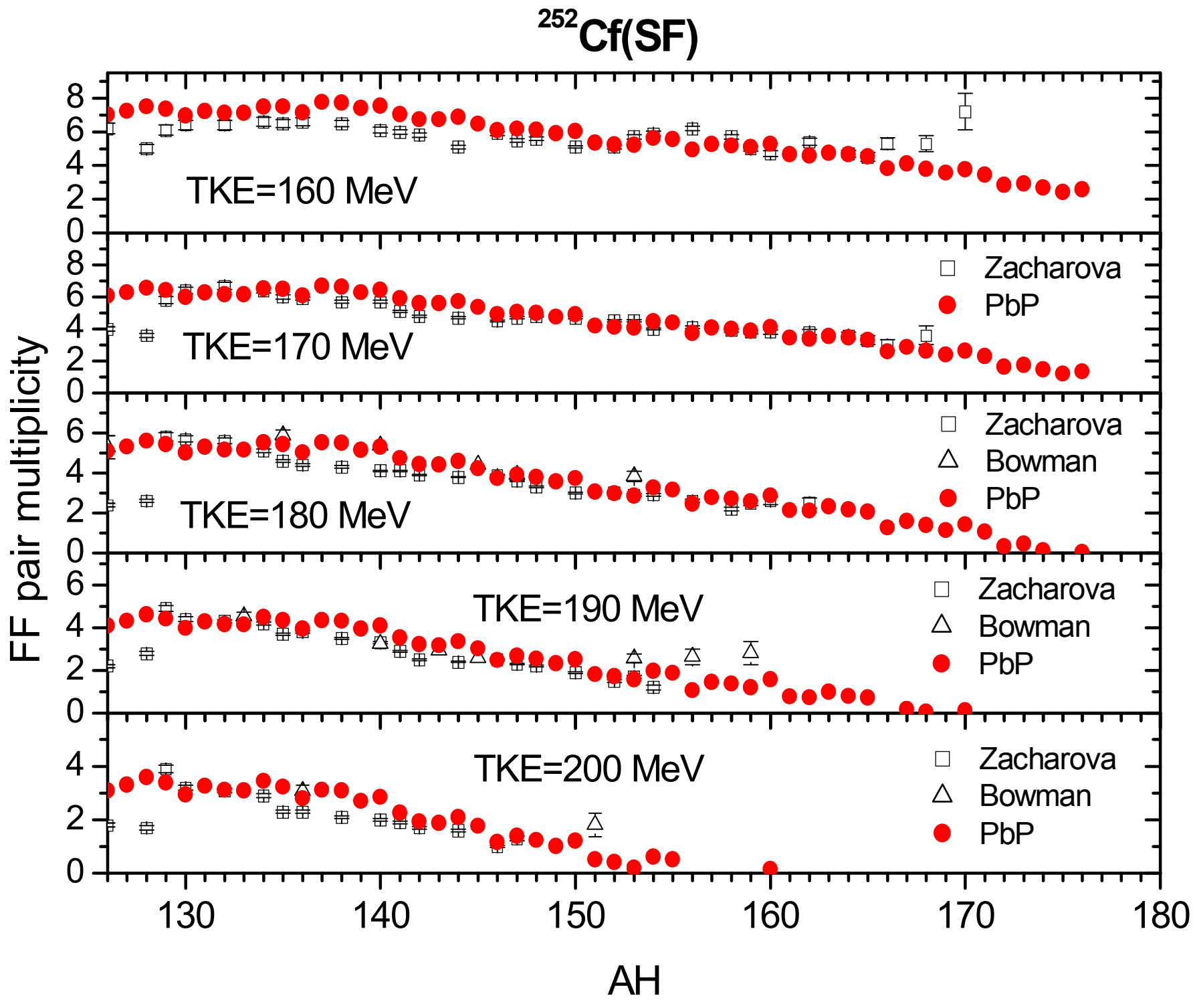

**Fig. 2.13** $^{252}$Cf(SF): Fission–fragment pair multiplicity for a given TKE versus the heavy fragment mass number (Tudora, 2008)

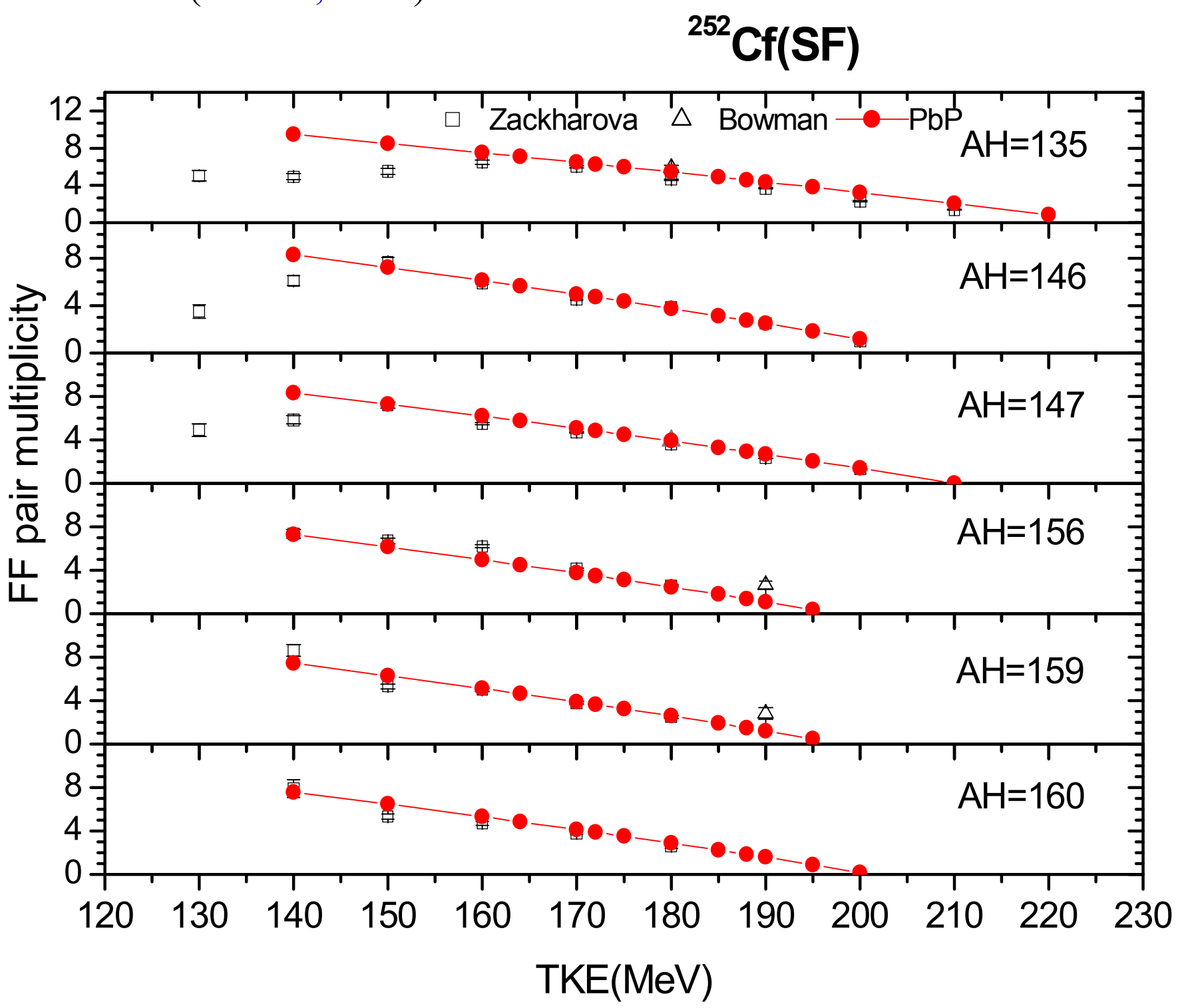

**Fig. 2.14** $^{252}$Cf(SF): Fission–fragments pair multiplicity as a function of TKE (Tudora, 2008)

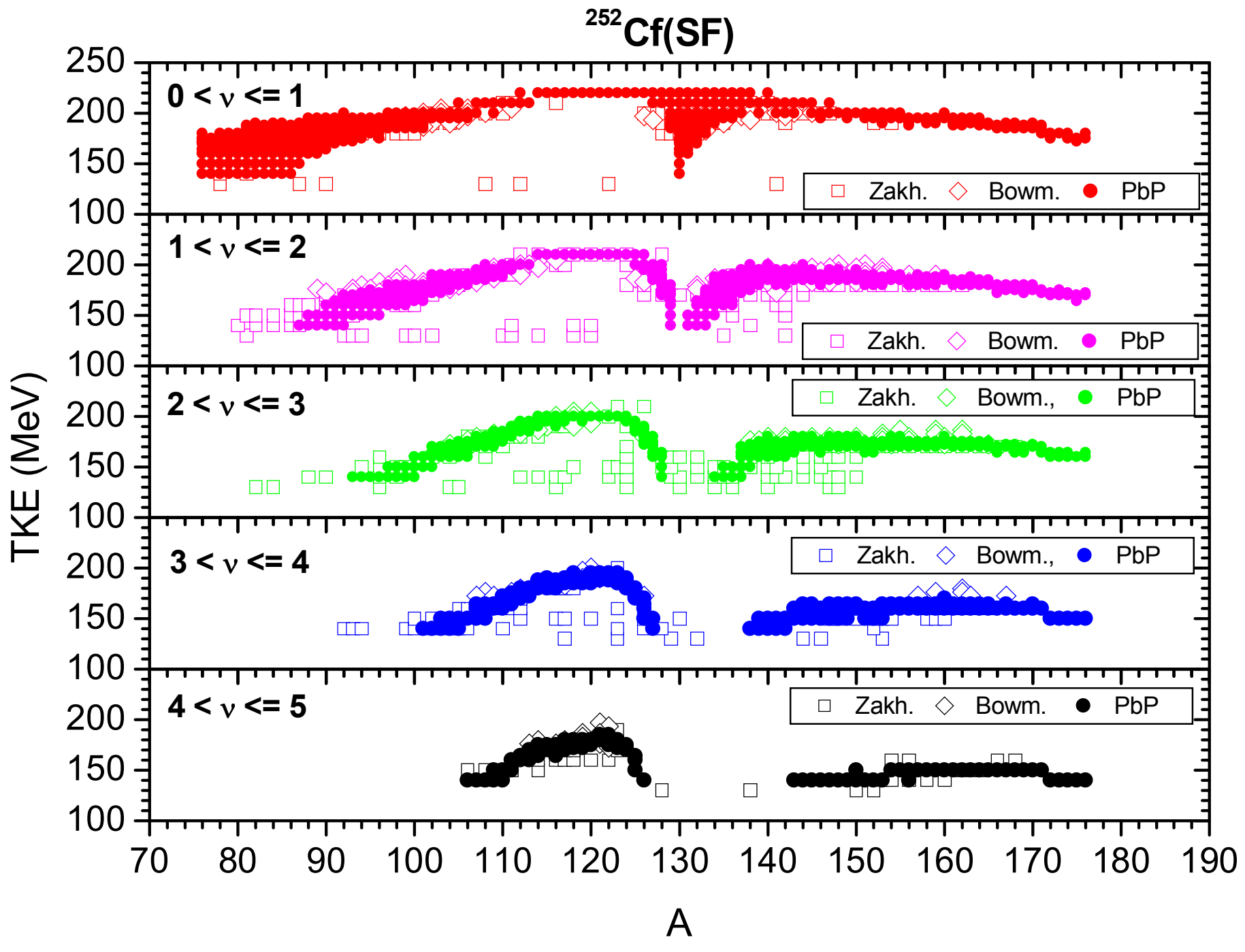

**Fig. 2.15** Multi–parametric representation of Point by Point calculation (full symbols) and experimental data (open symbols) for $^{252}$Cf(SF) (Tudora, 2008)

The Point by Point model itself as well as the model parameters corresponding to each fission–fragment pair can be verified by the supplementary test of the sawtooth (prompt neutron number emitted by the light or by the heavy fragment as a function of the fission–fragment mass number) which is the most sensitive quantity to the TXE partition. Unfortunately, the sawtooth experimental data are scarce, being available only for few neutron induced or spontaneous fissioning systems. The behaviour of the experimental sawtooth data shows an obvious dependence on the incident neutron energy but the majority of the existing data are for the thermal incident neutron energy. In **Fig. 2.16** the sawtooth for the thermal neutron induced fission of $^{235}$U is given. Here the TXE partition method described in **Section II.1.2** and also in (Manailescu et al, 2011) is used.

For the case of the neutron induced fission of $^{237}$Np, the Point by Point calculations of *ν(A)* at En=0.8 MeV and En=5.5 MeV, respectively, are plotted in **Fig. 2.17** in comparison with the two experimental data sets of Naqvi and Muller (black and grey symbols). The calculations were obtained using the TXE partition method based on $\nu_H/\nu_{pair}$ parameterization (Manailescu et al., 2011) (red full circles and blue stars connected with lines) and the one based on the calculation of the extra deformation energy at scission (Morariu et al., 2012) (magenta squares and green stars connected with lines). A good agreement of the *ν(A)* calculations with the experimental data is obtained, this being a validation of the TXE partition methods used.

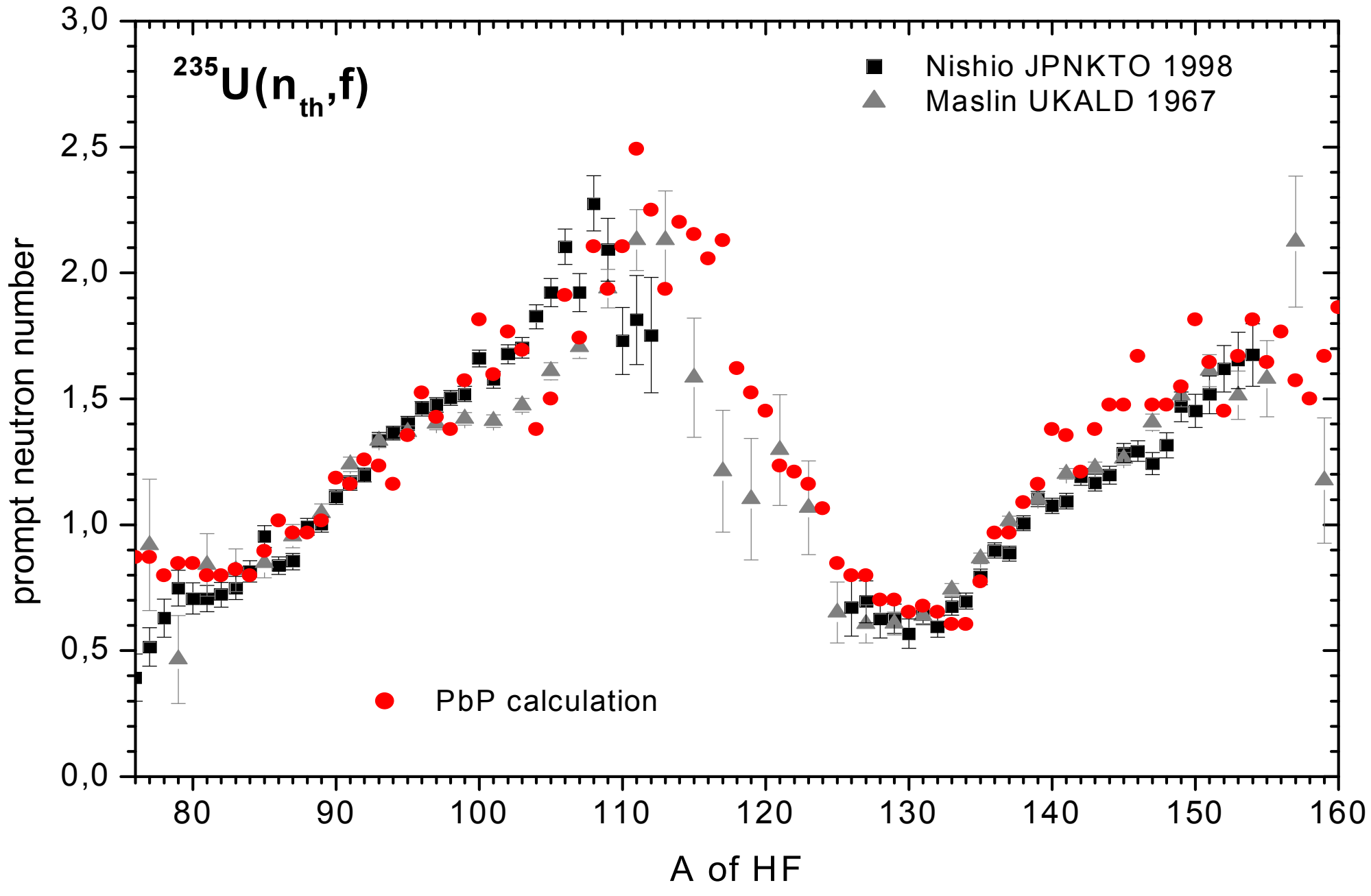

**Fig. 2.16** $^{235}U(n_{th},f)$: ν(A) calculation using the TXE partition method from (Manailescu et al, 2011) in comparison with experimental data (Morariu et al, 2012)

The comparison of the calculated average prompt neutron energies in the CMS with the available experimental data is given in **Fig. 2.18**. As it can be seen in the figure, good agreement with the experimental data was obtained for $^{252}Cf(SF)$ (upper part).

Good agreement was also obtained for $^{235}U(n_{th},f)$ except for the light fragment mass region were the calculation slightly underestimates the experimental data (middle part). In the $^{233}U(n_{th},f)$ case (lower part), the calculations do not agree well in the entire fission fragment mass range with the experimental data. Visible differences are in the mass range 116–130 where the experimental data of Nishio are slightly shifted with respect to the middle mass (symmetric fission.)

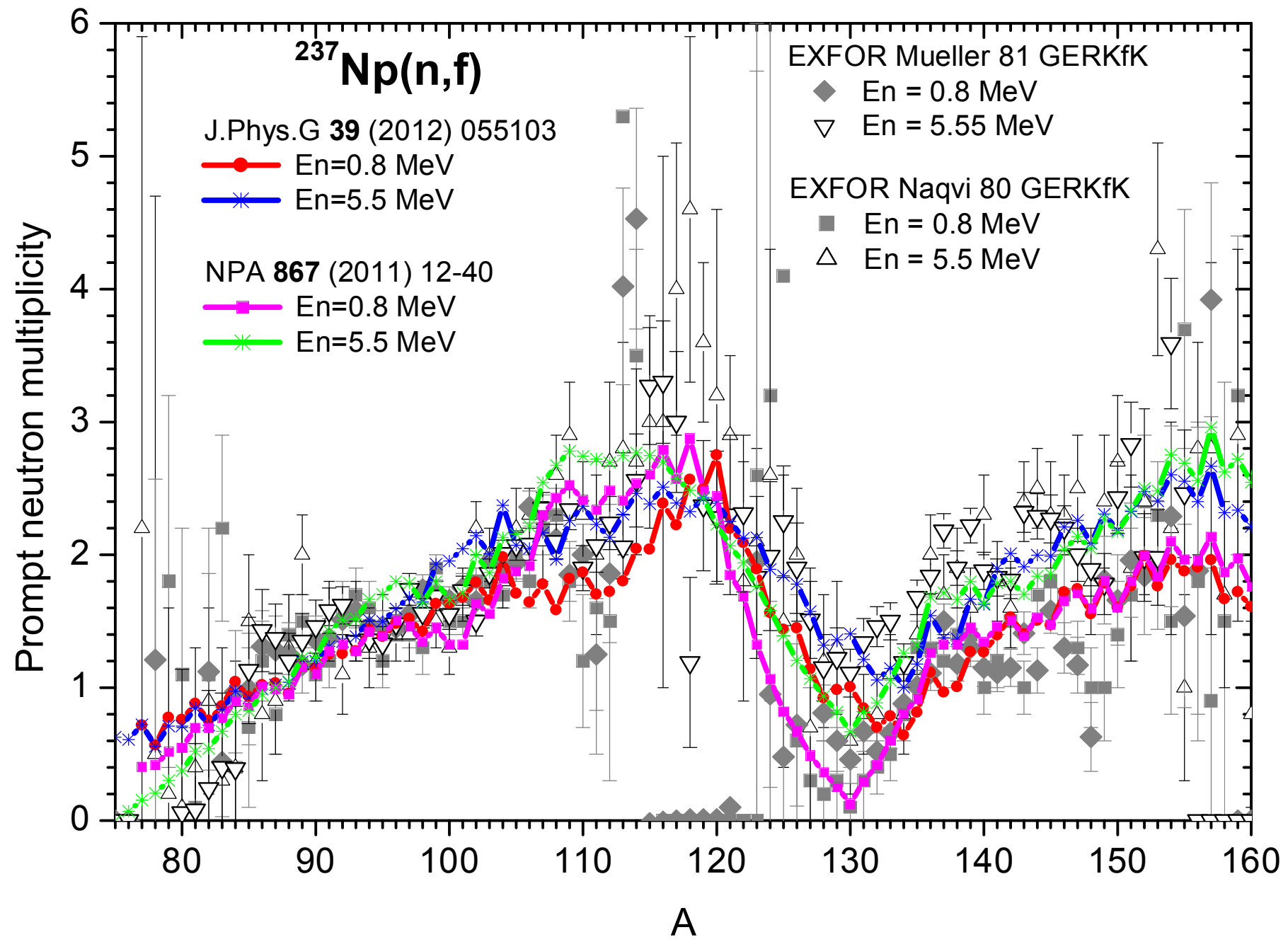

**Fig. 2.17** $^{237}$Np(n,f) PbP calculation of ν(A) at En=0.8MeV and En=5.5MeV in comparison with the experimental data

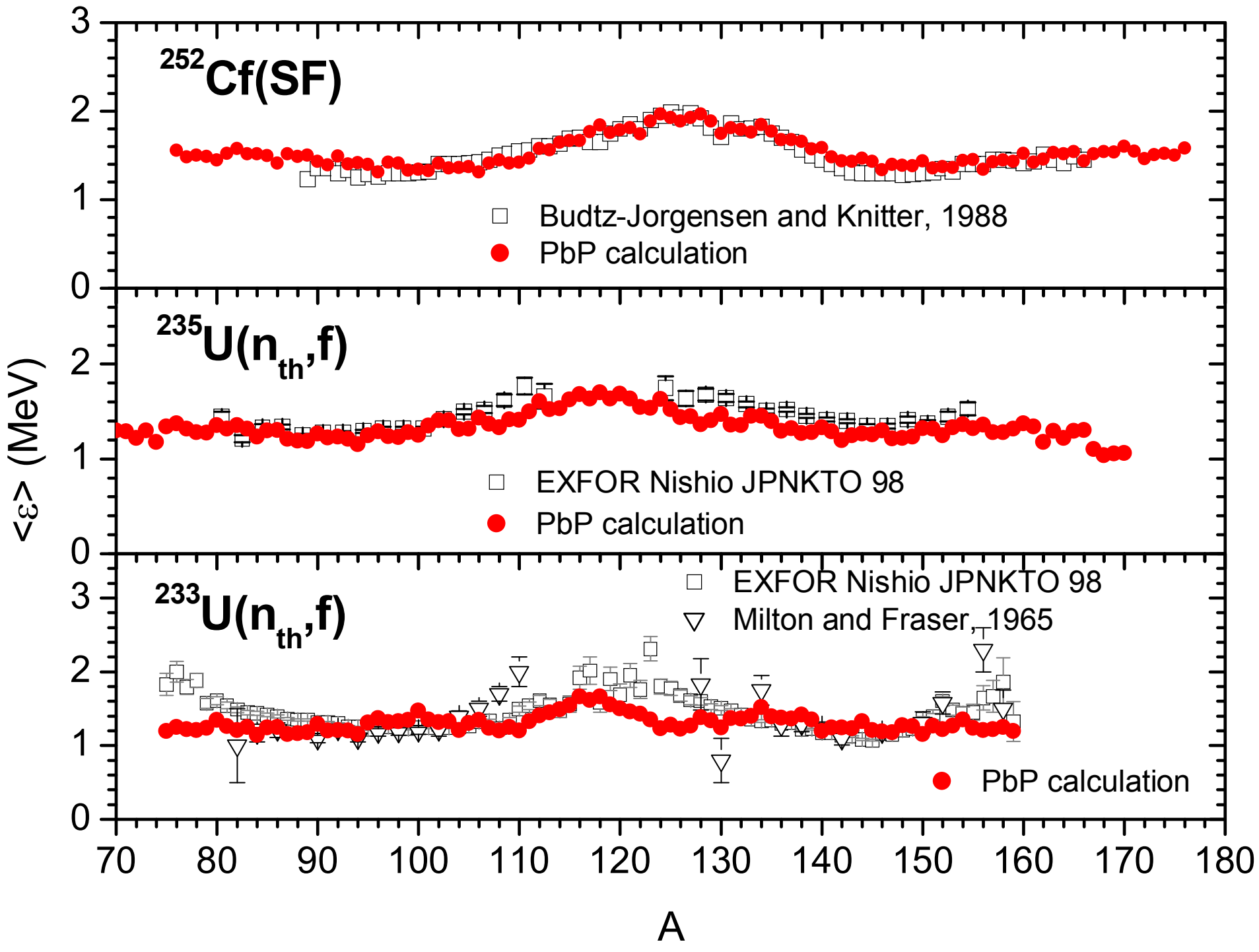

**Fig. 2.18** Average prompt neutron energy in CMS versus the fission–fragment mass number (Tudora, 2008)

The PbP model provides also the average prompt γ–ray energy (<$E_\gamma$>) of each fission–fragment pair. **In Fig. 2.19**, the calculated <$E_\gamma$>$_{pair}$ as a function of AL for $^{233}$U($n_{th}$,f) and $^{252}$Cf(SF) are given in comparison with the experimental data of Pleasonton and Nifenecker et al., respectively. As it can be seen the PbP results are in good agreement with the experimental data.

PbP calculations of $E_\gamma$(A) in comparison with the experimental data are plotted in **Fig. 2.20** for the case of the thermal induced fission of $^{235}$U. The results obtained using the TXE partition method from (Manailescu et al., 2011) are plotted with stars symbols and the results using the TXE partition method based on the calculation of the extra–deformation energy at scission (Morariu et al., 2012) are plotted with full circles.

As it can be seen the unique experimental data of prompt γ–ray energy as a function of fragment Eγ(A) measured for $^{235}$U(nth,f) are very well described by the Point by Point model results obtained with both methods of TXE partition.

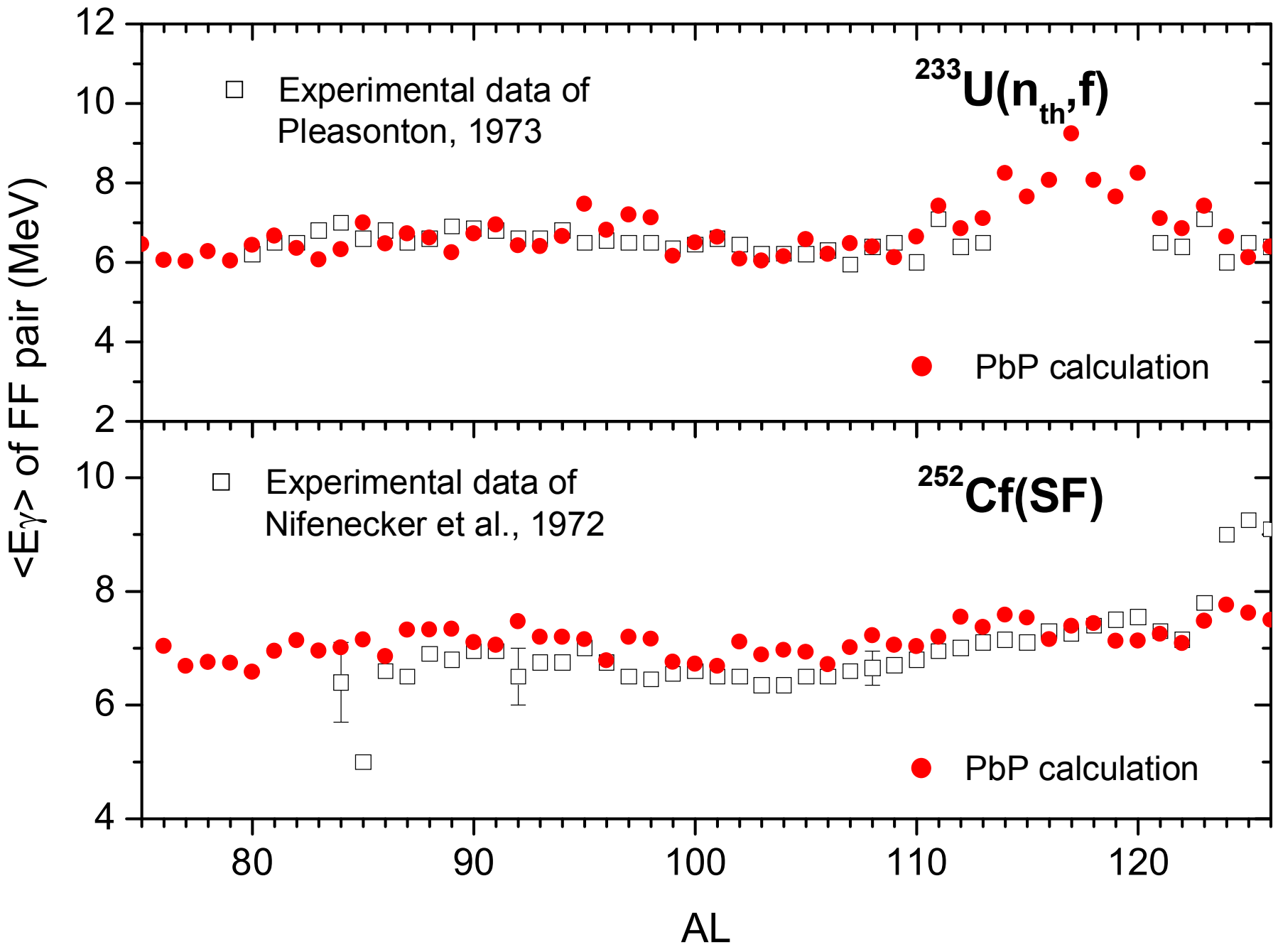

**Fig. 2.19** Average prompt γ–ray energy of the fission–fragment pair versus the light fragment mass number for $^{233}$U(nth,f) (upper part) and $^{252}$Cf(SF) (lower part)

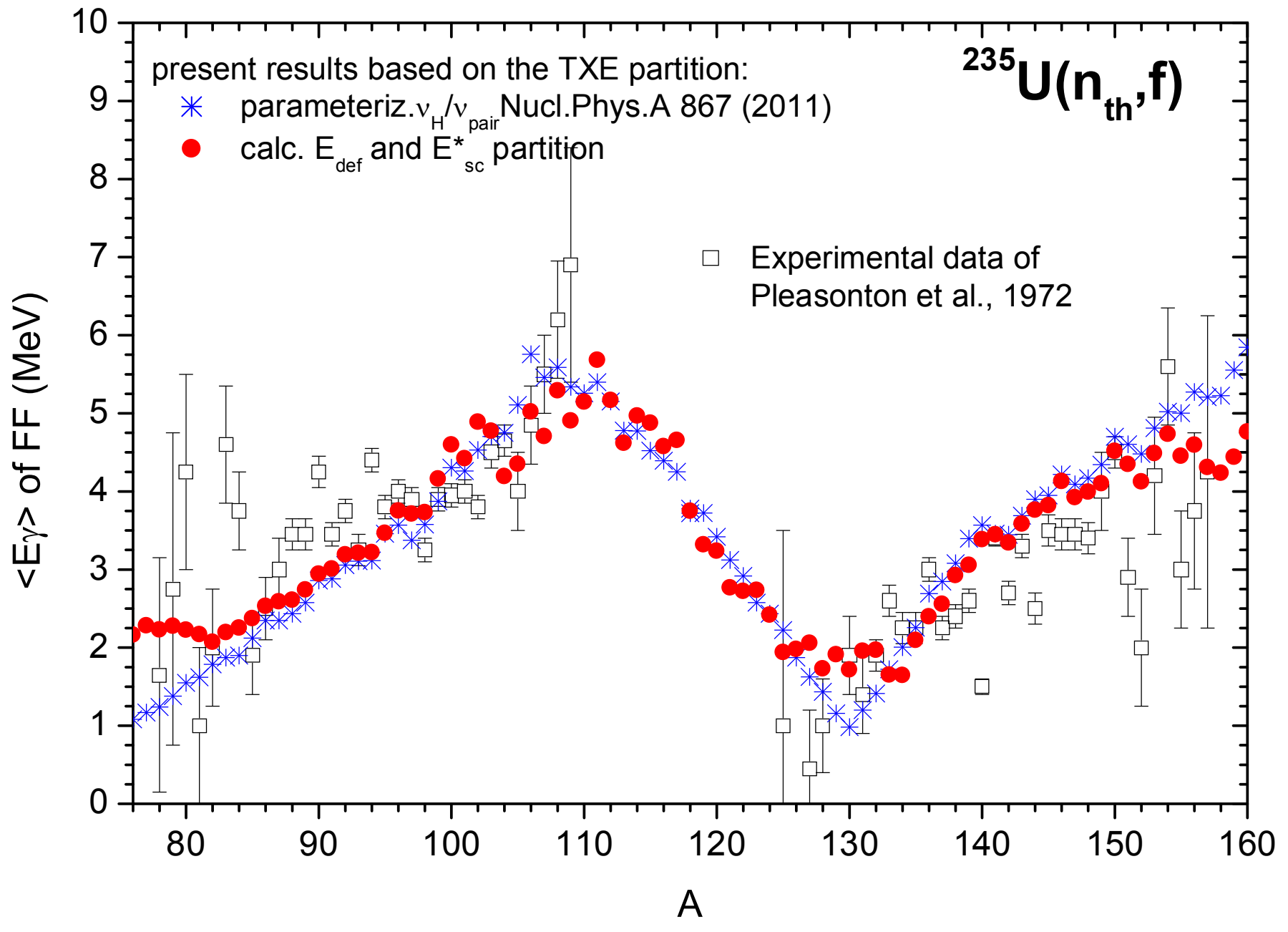

**Fig. 2.20** $^{235}$U(n$_{th}$,f) Average prompt γ–ray energy as a function of the fragment mass. Point by Point calculations using the TXE partition methods from (Manailescu et al., 2011) (stars symbols) and (Morariu et al., 2012) (full circles) in comparison with the experimental data (Tudora et al., 2012d)

Another type of prompt emission data are average quantities as a function of TKE. For instance in **Fig. 2.21** is given the average prompt neutron multiplicity as a function of the total kinetic energy of fission fragments <ν>(TKE) for the case of $^{252}$Cf(SF). For this reaction, benefiting of very accurate measurements of prompt emission data, the experimental <ν>(TKE) reported during the time by many research groups, using different methods, exhibit visible different behaviours consisting not only in different slopes dTKE/dν but also in a more or less visible flattening of <ν> at low TKE values.

As it can be seen in the figure, three experimental data sets (Budtz–Jorgensen (full black squares), Sing Shengyao (open down triangle) and Vorobyev (full gray diamonds)) exhibit almost the same slope and are close to each other for the TKE values between 165–220 MeV. The data of Budtz–Jorgensen and Vorobyev exhibit a visible flattening at TKE values lower than 160 MeV (Vorobyev data showing a pronounced dip below 150 MeV), while the Sing data maintain the same slope.

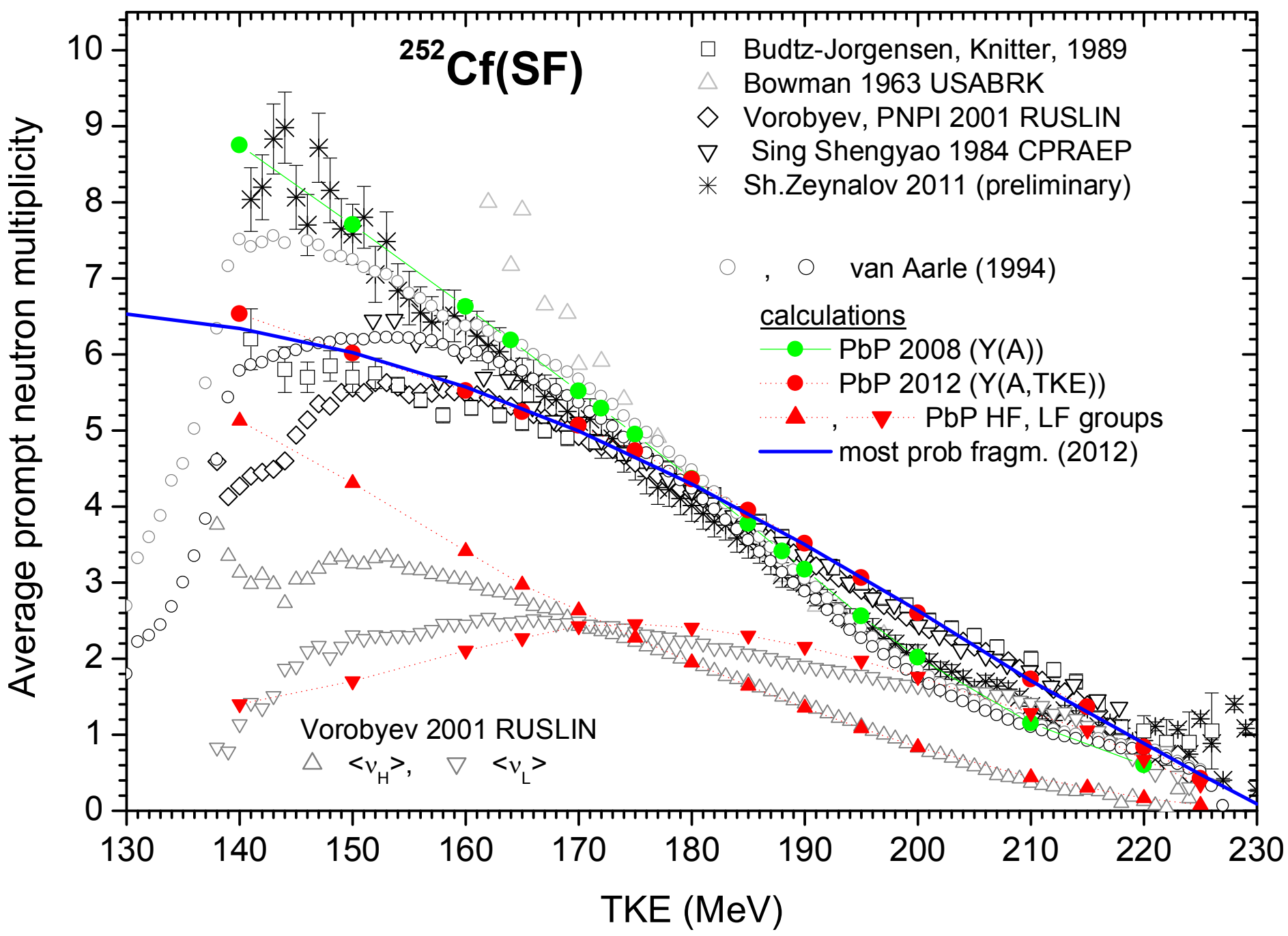

**Fig. 2.21** $^{252}$Cf(SF) Point by Point results obtained by averaging the matrix ν(Z,A,TKE) over Y(A,TKE) (red circles) and over Y(A) (green circles) in comparison with the experimental data. Most probable fragmentation result is plotted with solid blue (Tudora, 2012a).

The other data sets (Bowman (open up triangle), Zeynalov (stars) and van Aarle (open gray and black circles)) have a different slope compared to the three data sets mentioned above. The data sets labeled van Aarle were obtained by processing the experimental multi–parametric data in two manners: by taking into account the entire experimental matrix ν(A,TKE) (gray open circles) and by eliminating the very asymmetric fragmentations of the experimental distribution of van Aarle, in this case a flattening of <ν> at low TKE being obtained (black open circles). Both data sets of van Aarle exhibit the same slope dTKE/dν, close to the one of the Zeynalov data (Tudora, 2012a).

The previous PbP results reported in (Tudora, 2009), obtained by weighting the fission–fragment pair multiplicities over the mass distribution Y(A) (independent on TKE), are showing a linear decrease over the entire TKE range reproducing well the experimental data of Zeynalov.

The recent PbP results describe well the experimental data of Budtz–Jorgensen over the entire TKE range including the flattening of <ν> at low TKE values. Obviously the PbP calculation is in very good agreement with the data of Vorobyev and Sing, too, excepting at low TKE values where the Vorobyev data exhibit a pronounced dip and the Sing data maintain a linear behaviour (without any decrease) (Tudora, 2012a).

For both PbP calculations, the one reported in (Tudora, 2009) and the new one from (Tudora, 2012a)., the same multi–parametric matrix ν(Z,A,TKE) was used, the unique difference consisting in averaging over the double distribution Y(A,TKE) in the case of the results from (Tudora, 2012a) instead of the single distribution as it was done in (Tudora, 2009).

The experimental data regarding <ν>(TKE) for other fissioning systems exhibit a flat behaviour at low TKE values, too. An example is given in **Fig. 2.22** for the thermal neutron induced fission of $^{235}$U.

Similar Point by Point model calculation (weighting over Y(A) or over Y(A,TKE)) were done. <ν>(TKE) obtained by averaging over Y(A,TKE) describe well the experimental data of Vorobyev, Maslin and Nishio and also the flattening at low TKE values. <ν>(TKE) obtained by averaging over Y(A) exhibits a linear behaviour with a lower dTKE/dν slope and without flattening at low TKE.

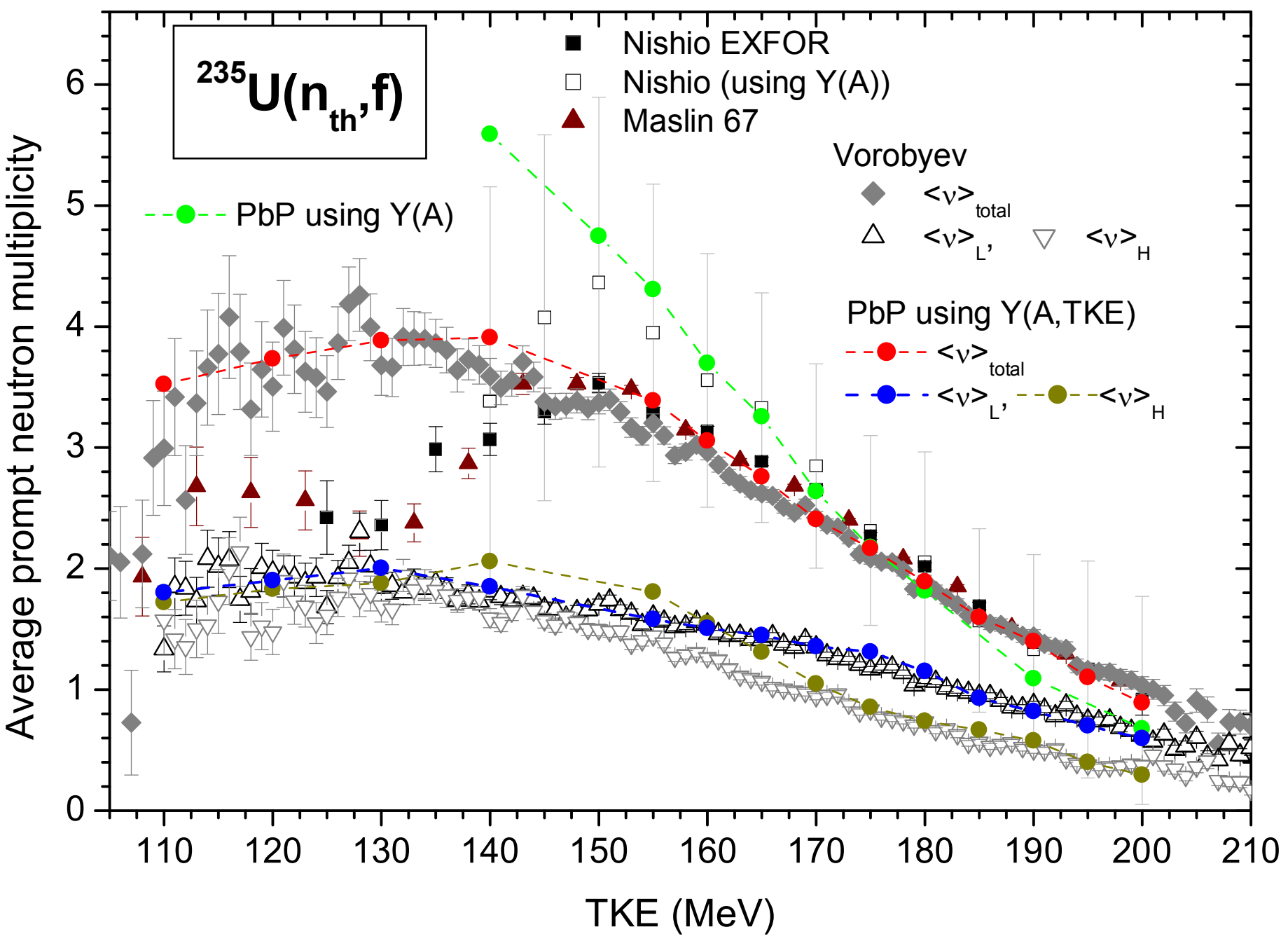

**Fig. 2.22** $^{235}$U(n$_{th}$,f) Point by Point calculatios using Y(A,TKE) reconstructed from experimental Y(A) and TKE(A) in comparison with the experimental data

In the case of $^{252}$Cf(SF), the average prompt neutron energy in the center–of–mass was also measured. The two existing experimental data plotted in **Fig 2.23** exhibit different slopes. The Point by Point results are obtained by averaging the multi–parametric matrix ε(A,Z,TKE) over the single distribution Y(A) (green circles) and over the reconstructed double distribution Y(A,TKE) (red circles). Reasonable good agreement was obtained for both types of calculations. <ε>(TKE) obtained by averaging over Y(A) seems to have a slope close to the Bowman data, while <ε>(TKE) obtained by averaging over Y(A,TKE) shows a slope close to the spread data of Nifenecker.

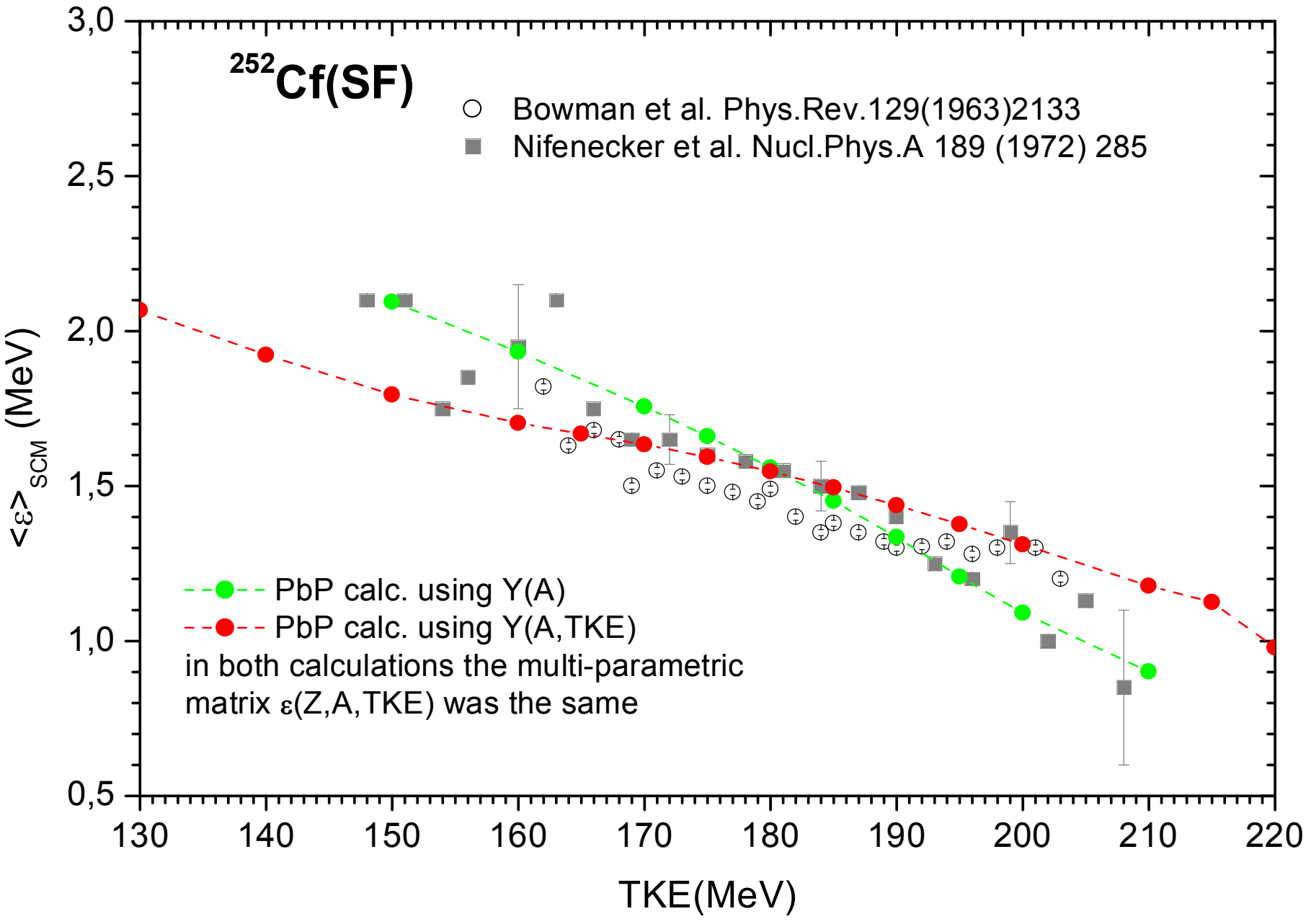

**Fig.2.23** $^{252}$Cf(SF): PbP calculations of average energy of prompt neutrons in CMS: using Y(A) (green circles) and Y(A,TKE) (red circles) in comparison with the experimental data (Tudora, 2012a)

The average values of model parameters: energy release <Er>, average neutron separation energy from fission–fragment <Sn> and average value of the level density parameter of fragments <a> (given as <C>=$A_0$/<a>, where $A_0$ is the mass number of the fissioning nucleus) can be obtained at a given value of TKE by averaging the parameters of fragment pairs calculated at the respective TKE value over the Y(A,TKE) distribution.

For instance, in **Fig. 2.24** the resulted <Er>, <Sn> and <C> as a function of TKE are plotted with full squares (upper, middle and lower part respectively) for the case of $^{252}$Cf(SF). Their nice and regular behaviour exhibited can be fitted very well. The appropriate fits are also plotted in the figure and the corresponding polynomial dependences on TKE are given in the figure's legend.

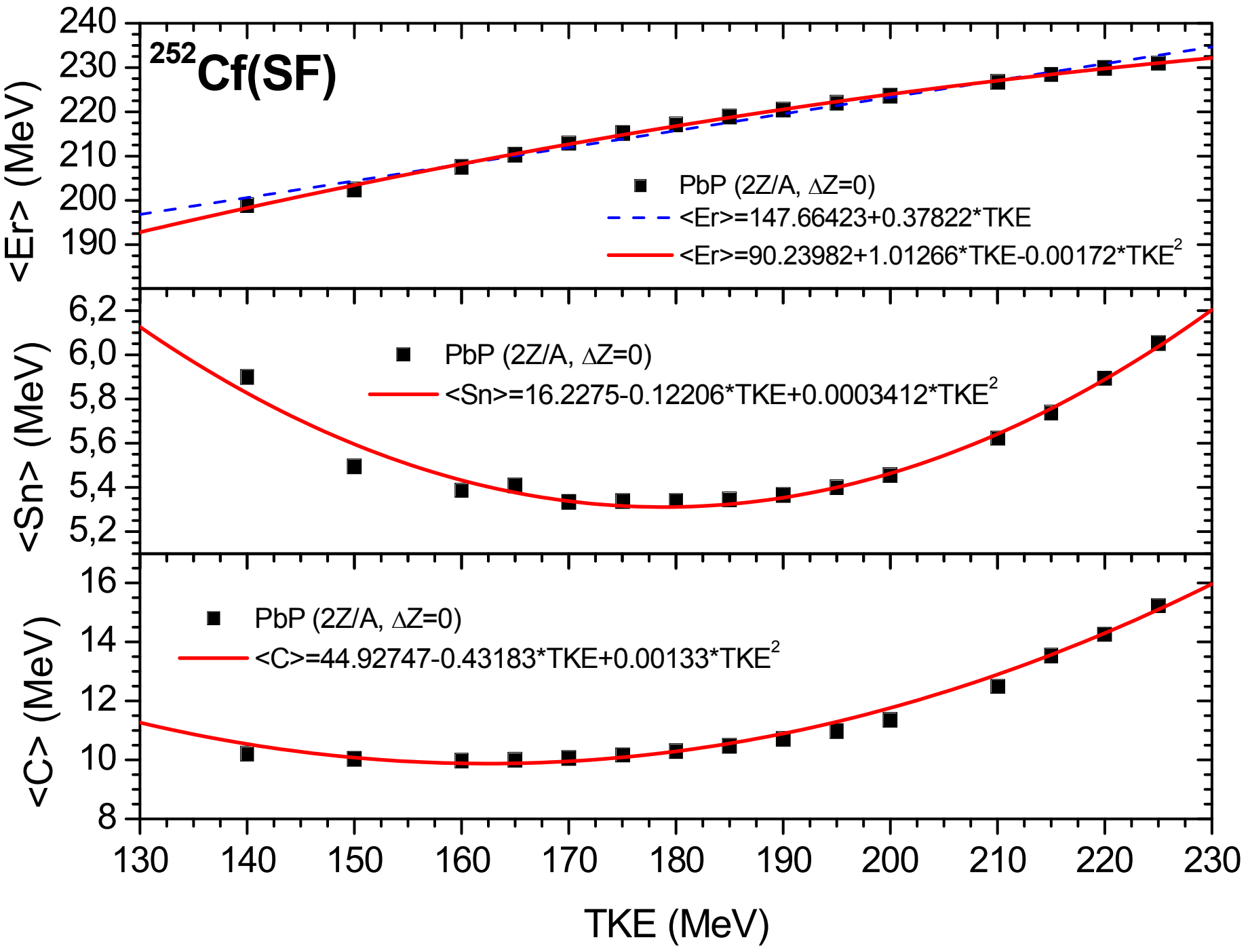

**Fig.2.24** $^{252}$Cf(SF) Average model parameters as a function of TKE obtained from the PbP treatment (full squares) and their appropriate fits (solid and dashed lines) (Tudora, 2012a)

The average model parameters depending on TKE allow the use of the most probable fragmentation approach, giving the <ν>(TKE) result plotted with blue continuous line in Fig.2.21. The possibility of use of the Los Alamos model with parameters depending on TKE has the advantage to provide <ν>(TKE) results at many TKE values in a very short computing time compared to other treatments (like PbP and Monte- Carlo).

Regarding the total average quantities, important for applications are especially the prompt neutron multiplicity and spectra.

An example of PbP spectrum calculation is given in **Fig. 2.25** as ratio to the Maxwellian for the case of thermal neutron induced fission of $^{233}$U. As it can be seen the PbP spectrum (red line) as well as the spectrum obtained in the frame of the most probable fragmentation approach with average parameters of the systematic (blue line) describes well the experimental data.

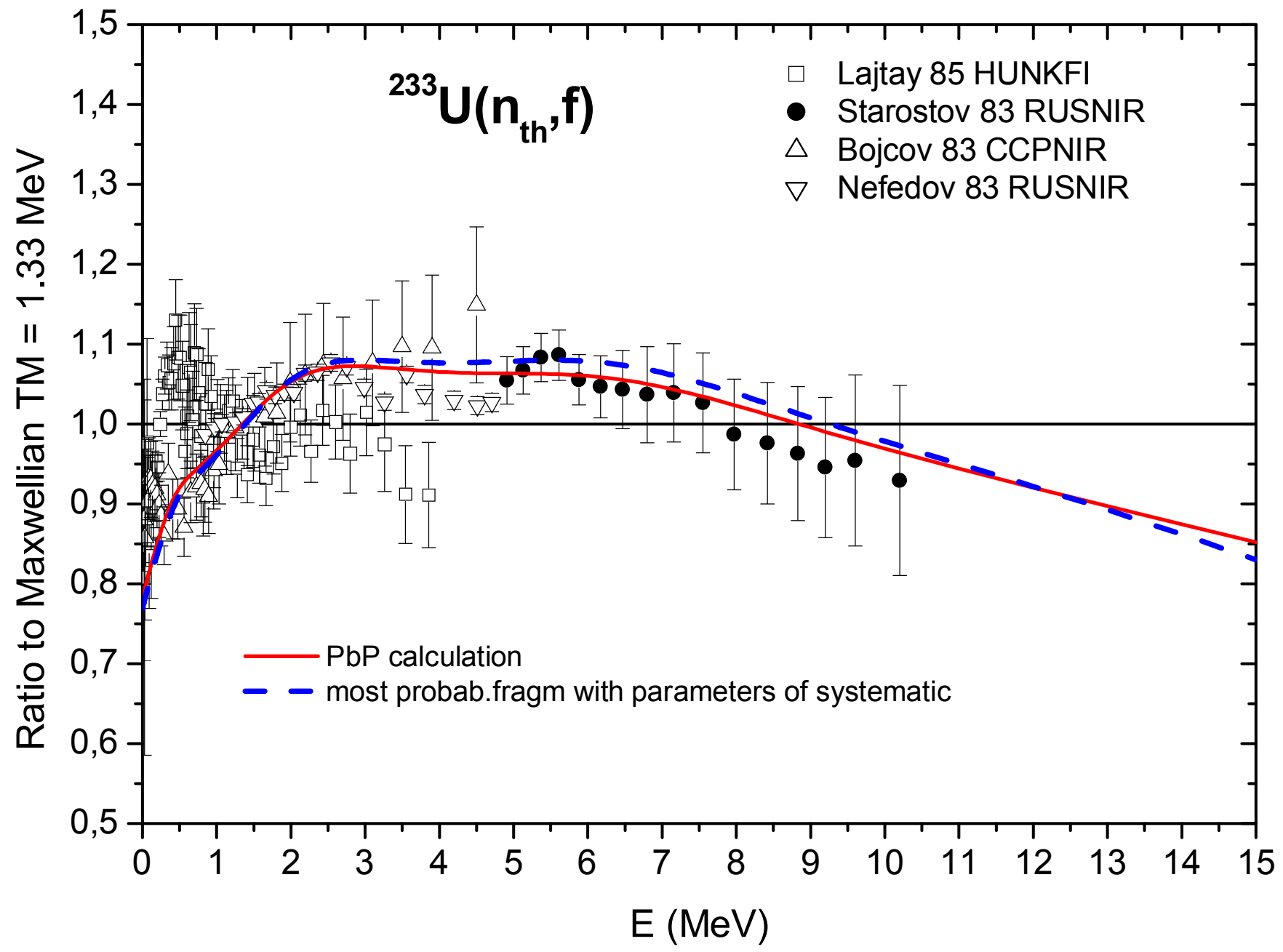

**Fig.2.25** $^{233}U(n_{th},f)$: Prompt fission neutron spectra calculation using the PbP and most probable fragmentation approaches in comparison with the experimental data

An other important quantity which can be obtained from the multi–parametric matrix provided by the Point by Point model is the prompt neutron multiplicity distribution *P(ν)*. This quantity is a very sensitive one depending on both, the mechanism of prompt neutron emission from the fission–fragments and the mass and kinetic energy distributions.

In **Fig. 2.26** is given an example for the case of spontaneous fission of $^{252}Cf$. As it can be seen, the PbP calculation results of *P(ν)* are in a very good agreement with the experimental data, overestimating a bit only the experimental data at ν=2.

Another example of *P(ν)* calculation is given in **Fig. 2.27** for $^{235}U(n_{th},f)$. The two experimental data sets (of Diven and Franklyn) are very close each other and very well decribed by the PbP result for ν = 4, 5, 6. For ν=1, 2, 3 the two experimental data sets differ to each other and the PbP result is placed between them.

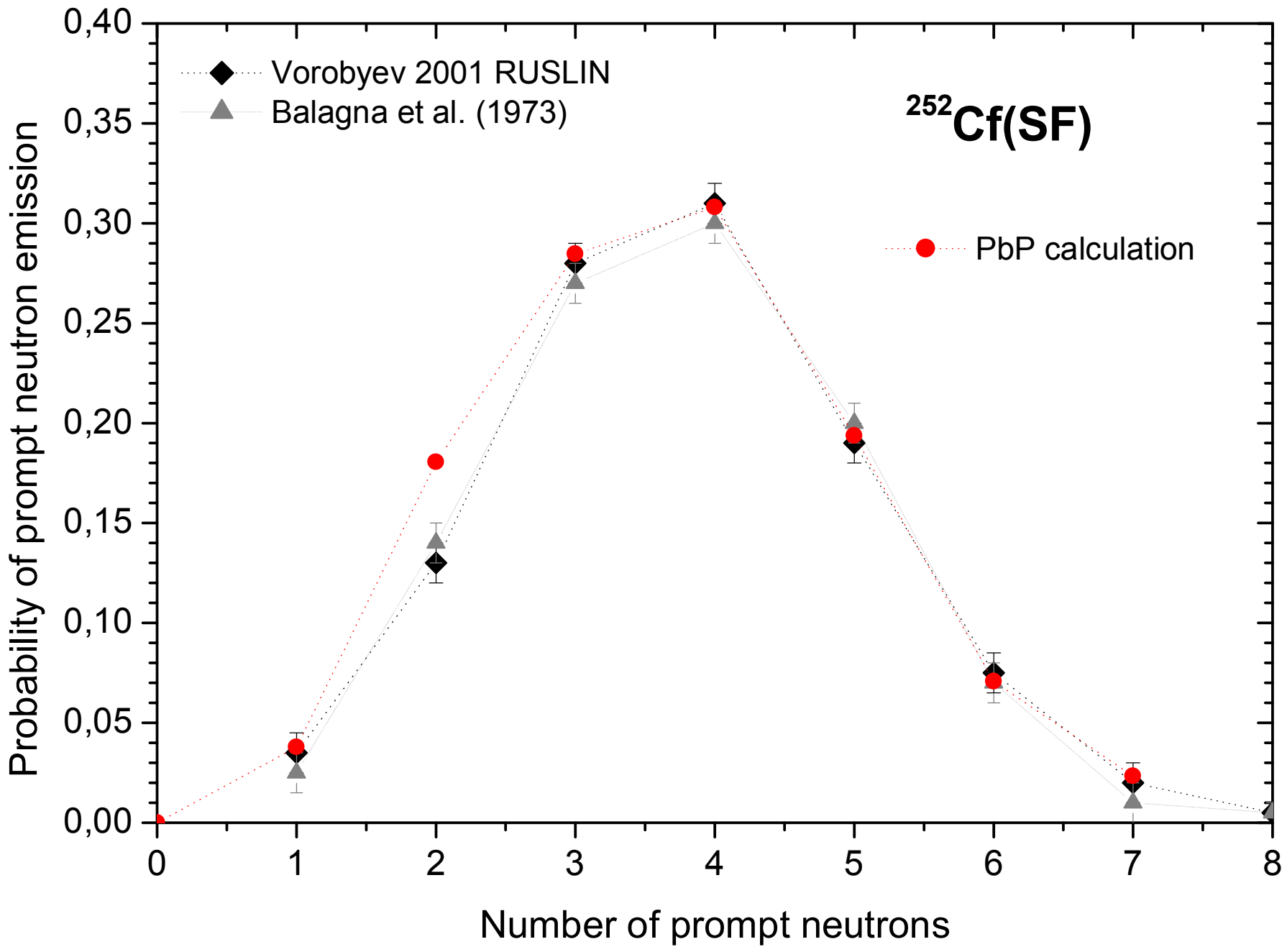

**Fig. 2.26** $^{252}$Cf(SF): Prompt neutron emission probability from Point by Point calculation in comparison with the experimental data (Tudora and Hambsch, 2010)

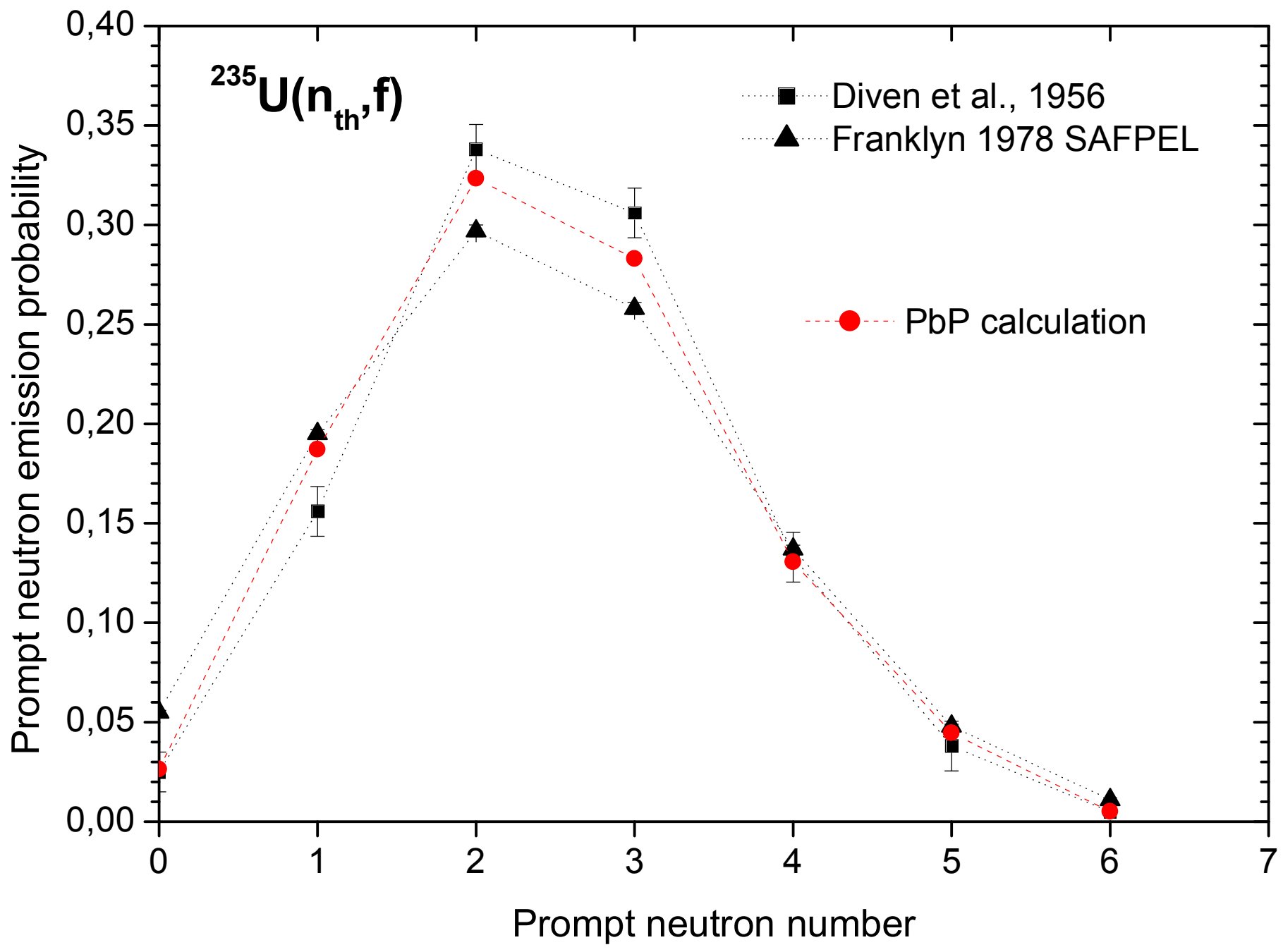

**Fig. 2.27** $^{235}$U($n_{th}$,f): Prompt neutron emission probability from Point by Point calculation in comparison with the experimental data (Tudora and Hambsch, 2010)

In the case of the Point by Point model, the total average quantities (such as $<\nu_p>$, spectra) calculated at different incident energies *En* reveal a dependence on *En* that is due to the fission–fragment distributions (in other words these distributions are depending on En). This fact is proven by the average parameters (<Er>, <TKE>, <Sn>) obtained by averaging over the fission–fragment distribution which are depending on En. Such calculations were already reported along the time (see for instance (Tudora et al., 2005; Vladuca et al., 2006; Tudora, 2006, Tudora et al., 2008 and references therein).

In **Figs. 2.28** and **2.29** two examples of prompt fission neutron multiplicity as a function of the incident neutron energy are given for the neutron induced fission of $^{237}$Np and $^{232}$Th, respectively.

For the case of $^{237}$Np(n,f) (**Fig. 2.28**) the Point by Point calculation were performed using two methods of TXE partition as follows: with red open circles are plotted the results using the TXE partition based on the $\nu_H/\nu_{pair}$ parameterization (Manailescu et al., 2011) and with red full circles the results using the TXE partition based on the calculation of the extra deformation energy at scission (Morariu et al., 2012). The most probable fragmentation result is plotted with red solid line. In the figure the ENDF/B–VII.1 and JENDL4 evaluations are also given for comparison.

In **Fig. 2.29 t**he total average prompt neutron multiplicity as a function of the incident energy for the neutron induced fission of $^{232}$Th is given up to 20 MeV in comparison with the experimental data. The Point by Point model results are plotted with full red circles, the most probable fragmentation results with red solid line (using the fission cross section from JEFF3.1) and with magenta dashed line (using the fission cross section from JENDL4). For comparison the evaluation from ENDF/B–VII, JEFF3.1 and JENDL4 are also plotted with different dash dotted lines.

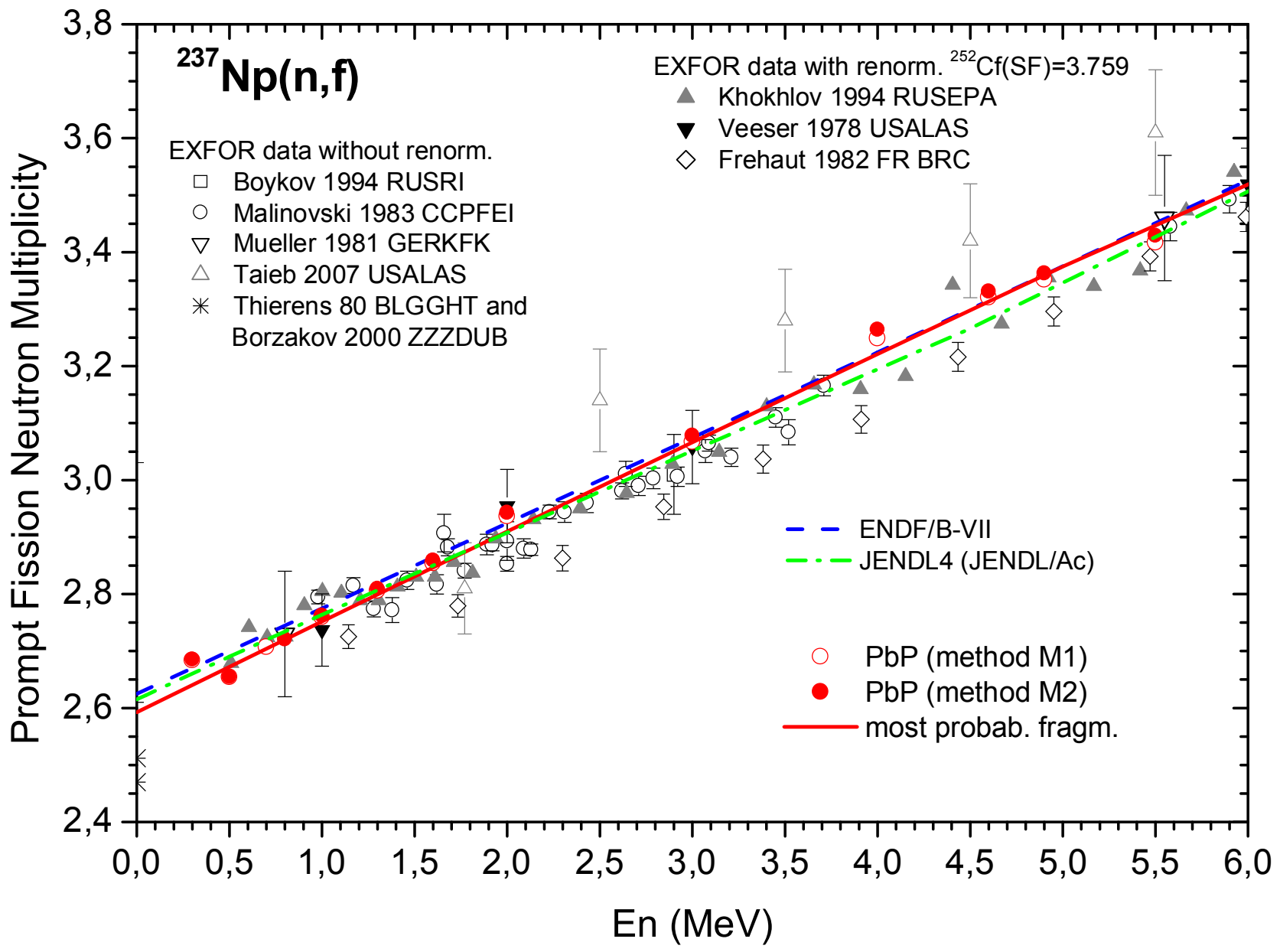

**Fig. 2.28** $^{237}$Np(n,f) Prompt fission neutron multiplicity as a function of the incident neutron energy in comparison with the experimental data (Tudora, 2012e)

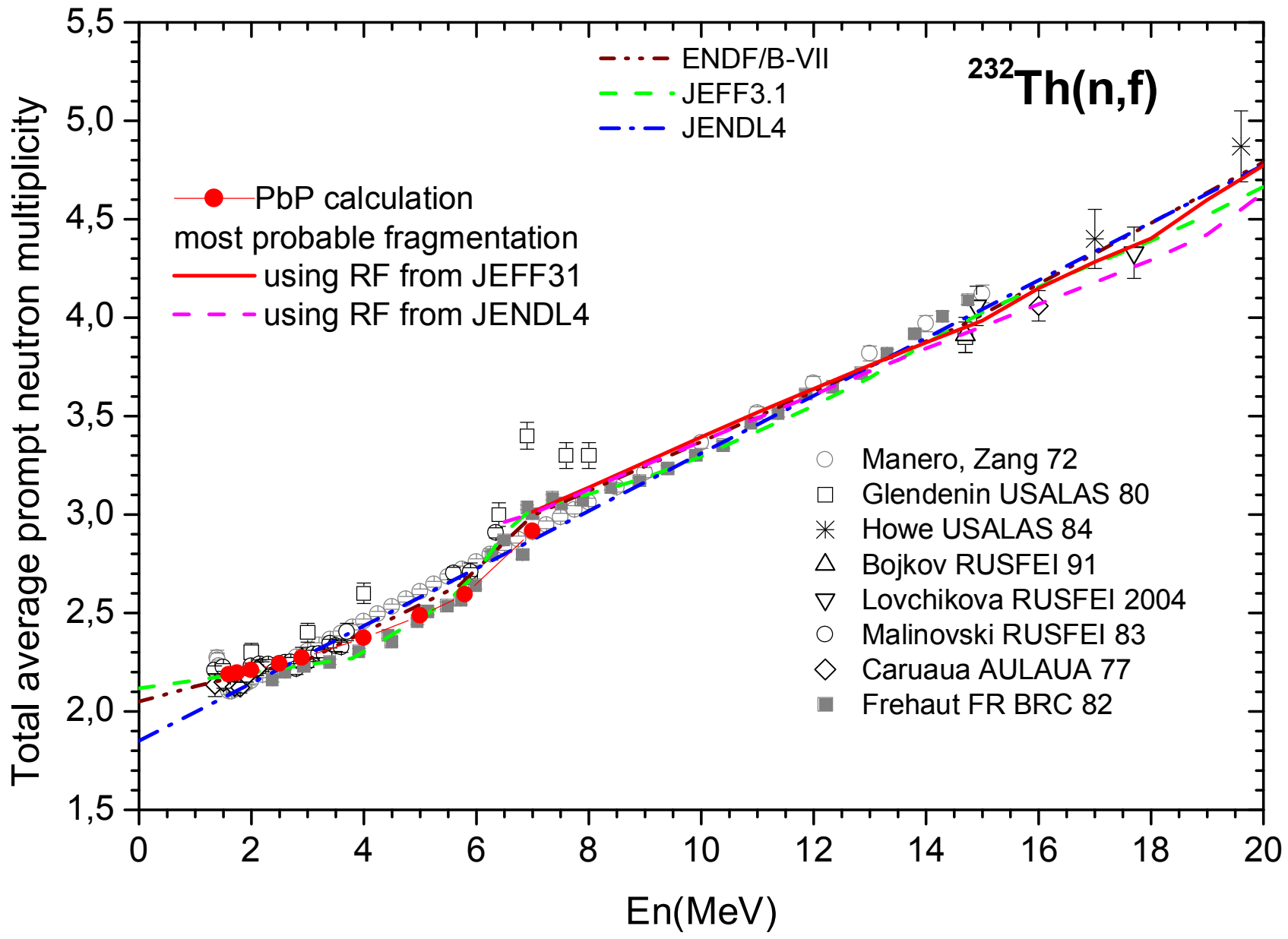

**Fig. 2.29** $^{232}$Th(n,f): Prompt fission neutron multiplicity as a function of the incident neutron energy in comparison with the experimental data (Tudora et al., 2012d)

In **Fig. 2.30** an example of calculated total average prompt γ–ray energy as a function of the incident energy is given in comparison with the experimental data of Fréhaut for $^{235}$U(n,f) (upper part) and $^{237}$Np(n,f) (lower part). The old results reported in (Vladuca and Tudora, 2000a, 2001a) are plotted in the figure with dash–dotted lines and the more recent results, with solid lines. These results are obtained by using average model parameters resulted from the Point by Point treatment (in the case of the main fissioning nuclei $^{236}$U and $^{238}$Np, respectively) and provided by the systematic of (Tudora, 2009) (in the case of secondary fission chances).

Another example of recent calculation of $<E_\gamma>$ as a function of the incident energy for $^{232}$Th(n,f) is given in **Fig. 2.31** in comparison with the experimental data of Fréhaut. As it can be seen in the figure, for both types of calculations, the Point by Point (red open circles) and most probable fragmentation approach (red solid line) give an overall good description of the experimental data.

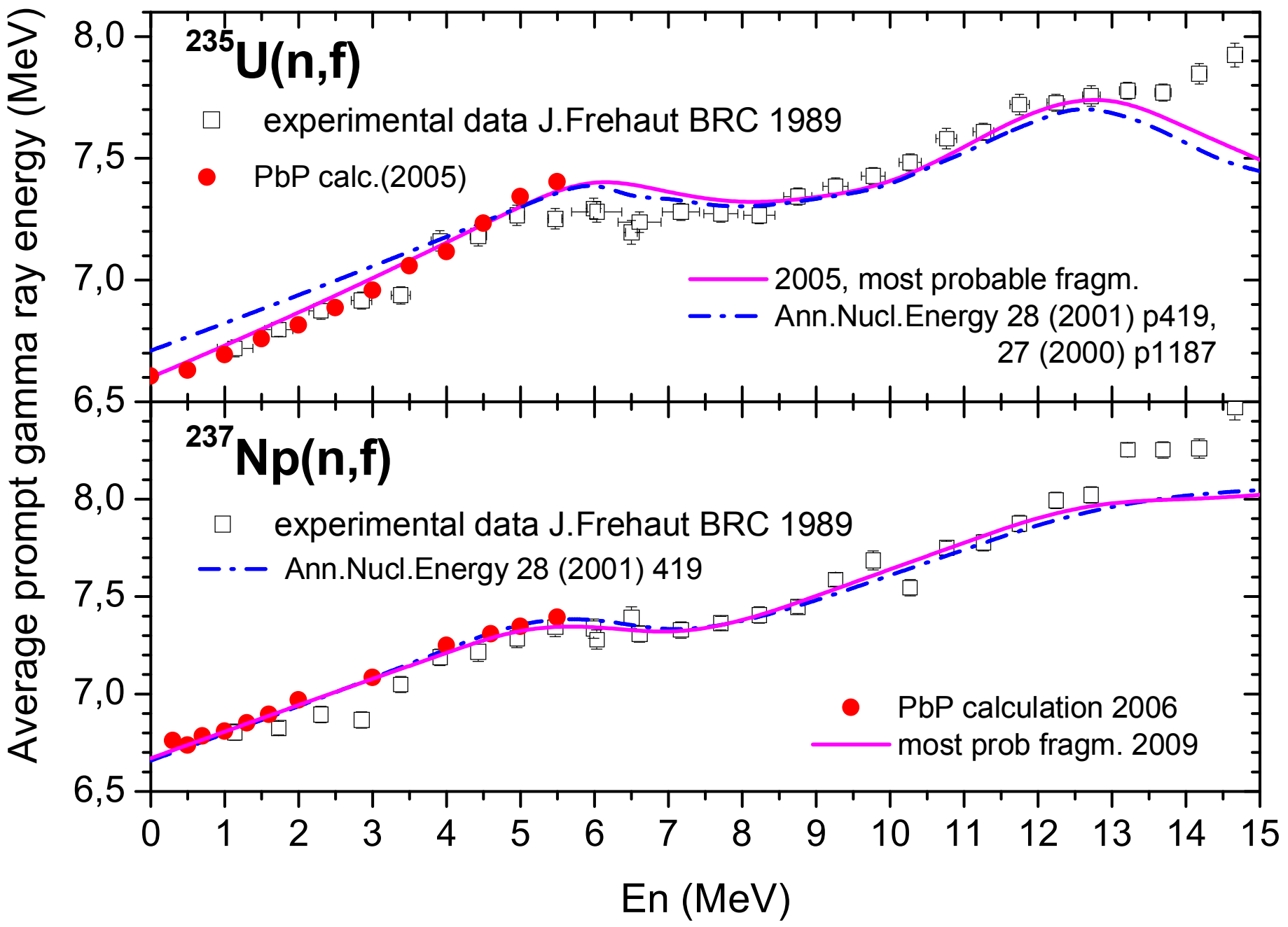

**Fig. 2.30** Average prompt γ–ray energy in comparison with experimental data for $^{235}$U(n,f) (upper part) and $^{237}$Np(n,f) (lower part) (Tudora et al., 2012d)

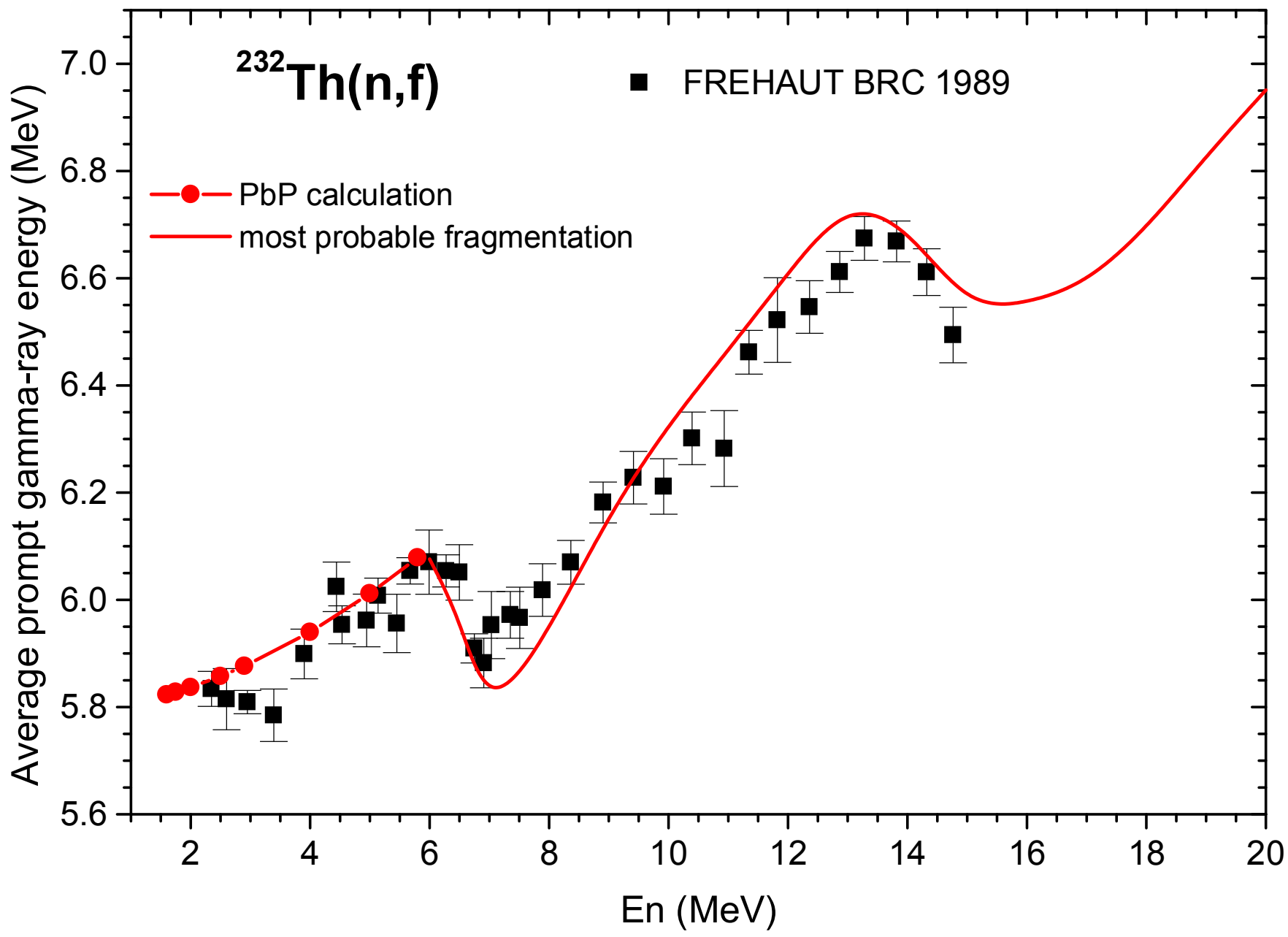

**Fig. 2.31** Average prompt γ–ray energy in comparison with experimental data for $^{232}$Th(n,f) (Tudora et al., 2012d)

# Chapter III

# Description of the FIFRELIN code

## III.1 Physical models

. In order to predict and analyze these various quantities characterizing the prompt emission in fission, a dedicated Monte Carlo code named FIFRELIN (Fission Fragment Evaporation Leading to an Investigation of Nuclear data) was developed.

The code goes a step beyond the Los Alamos model and looks in more details at the fission–fragment decay process by a Monte Carlo simulation of the sequential neutron emission. It contains various models mainly based on a mass–dependent temperature ratio of the fully accelerated complementary fragments and a spin–dependent excitation energy limit for neutron emission in order to improve de agreement with the experimental data (Litaize and Serot, 2010; 2011; Regnier et al., 2012a,b).

In order to perform a Monte Carlo simulation it is necessary to know the physical probability density functions. The most important probability functions concern the mass and the kinetic energy of the fission–fragments, together with the distribution needed for the selection of a nucleus which concerns the nuclear charge. For the mass and the kinetic energy experimental distributions are used. In order to take into account the different isobars within a mass chain, a Gaussian type nuclear charge distribution *P(Z)* was considered, as in the Point by Point treatment, described by equation (2.25), with the most probable charge for the light and for the heavy fragments obtained within the unchanged charge distribution.

### III.1.1 The distribution of the fission–fragment excitation energy

The decay path followed by a pair of fission–fragments depends on the available excitation energy. The total excitation energy *TXE* available for a given light and heavy pair $(A_L,Z_L)$, $(A_H,Z_H)$ is given in a similar manner as already described in Chapter II (see equation (2.13)). Similar to the Point by Point model, the total kinetic energy is taken from experimental data.

One of the long–standing questions about the nuclear fission process is related to the partition of the excitation energy between the two fragments. As it was written in (Litaize and Serot, 2010, 2011), at scission, the total excitation energy TXE is composed of intrinsic excitation energy $E_{sc}^{\text{int}r}$, deformation energy $E_{sc}^{def}$, and rotational energy $E_{sc}^{rot}$:

$$TXE = E_{sc}^{\text{int}r} + E_{sc}^{def} + E_{sc}^{rot} \tag{3.1}$$

At full acceleration (after the relaxation of the deformation energy), the fission–fragments are rotating and the total excitation energy is converted into intrinsic excitation energy and collective rotational energy:

$$TXE = E_L^* + E_H^* + E_L^{rot} + E_H^{rot} \tag{3.2}$$

where $E_L^{rot}$, $E_H^{rot}$ are the rotational energies of the light and heavy fragments and $E_L^*$, $E_H^*$ are their intrinsic excitation energies.

In the Monte Carlo treatment (FIFRELIN) the nucleus is considered as a Fermi–gas, therefore the excitation energy can be related to the nuclear temperature $T$ by:

$$E^* = aT^2 \tag{3.3}$$

where $a$ is the level density parameter. The determination of the level density parameter and the nuclear temperature is discussed in the next two sections.

### III.1.2 Effective level density parameter and nuclear temperature

The statistical properties of the excited nuclear levels have been a matter of concern and study for over fifty years. For the description of the level densities the Fermi–gas and the constant temperature models are used frequently with parameters obtained by fitting the experimental data.

The current understanding of the structure of low–lying nuclear levels is based on some important concepts including shell effects, pairing correlations and collective phenomena. All this concepts were incorporated into the generalized super–fluid model.

For practical applications of the statistical models, it is very important to obtain parameters of the level density description from reliable experimental data, the cumulative numbers of low–lying levels and the average distance between neutron resonances being used for this purpose.

Analysis of the experimental data was carried out initially by Gilbert and Cameron, and a rather simple systematic relationship has been proposed for the level density parameter (see **Appendix 2**).

One of the serious deficiencies of this systematic is the energy independence of the *a* parameter. The results of all consistent microscopic calculations of the nuclear level densities display damping of the shell effect at high excitation energies. In order to take into account the shell–effect damping, the level density parameter *a* should become energy dependent as approximated by the formula (Ig_1) (see **Appendix 2**).

Over the years, similar formulas were proposed, and the obtained differences between the corresponding level density parameters are mainly related to different shell corrections. If these corrections are taken together with the values of the *a* parameter for the neutron binding energies, then the asymptotic level density parameter can be derived on the basis of equation (Ig_1).

Inside the FIFRELIN code, different sets of shell and pairing energy corrections are implemented: Gilbert–Cameron, Myers–Swiatecki and Moller–Nix (RIPL–3, 2012d). Using the semi–classical formula for the asymptotic and damping parameter:

$$\widetilde{a} = \alpha A + \beta A^{2/3}, \qquad \gamma = \gamma_0 A^{-1/3} \tag{3.4}$$

and keeping in mind that the entire set of parameters ($a$, $\delta W$, $\Delta$, $\gamma$) entering equation (Ig_1) must be consistent, and considering the shell corrections from Myers–Swiatecki the following parameterizations for the asymptotic level density parameter and damping factor are obtained:

$$\widetilde{a} = 0.0959A + 0.1468A^{2/3}, \qquad \gamma = 0.325A^{-1/3} \tag{3.5}$$

Similar treatment for the level density parameter is made in the Point by Point model, but instead of the asymptotic level density parameter and the damping factor from equation (3.5) it is used a damping factor $\gamma = 0.4A^{-1/3}$ and an asymptotic level density parameterization proposed by Ignatiuk (Ignatiuk, 1998) or by Egidy and Bucurescu (Egidy and Bucurescu, 2005) given by:

$$\tilde{a} = 0.073A + 0.115A^{2/3} \quad \text{(Ignatiuk)}$$

and (3.6)

$$\tilde{a} = A(0.127 - 9.05 \cdot 10^{-5} A) \quad \text{(Egidy–Bucurescu)}$$

Regarding the nuclear temperature, a first assumption, already reported by Lemaire et al (Lemaire et al., 2005) was to consider that both fragments have the same temperature, the results obtained under this assumption were not reproducing the sawtooth shape of the prompt neutron multiplicity distribution ν(A).

Therefore, as originally proposed by Ohsawa (Ohsawa, 1991), a non–equitemperature model must be considered ($T_L = R_T T_H$).

The parameter $R_T$ was chosen to be constant or linearly dependent of the mass number and it will be discussed in a following section.

### III.1.3 Fission–fragment angular momentum

The main source of neutrons in low–energy fission is the emission from the primary fragments which have excitation energy larger than the neutron separation energy $S_n$. The neutron evaporation ends when the excitation energy becomes lower than a given limit. A first approximation is to consider that the neutron separation energy is the lower limit after which the gamma decay can start due to the available residual excitation energy. Unfortunately this approximation leads to an overestimation of the total average prompt neutron multiplicity $\bar{\nu}$ compared to the experimental data. Therefore, in order to reach a more consistent $\bar{\nu}$ value, a higher energy limit is required.

If the fission fragments are considered as rotating nuclei, then it is necessary to add a rotational energy to the neutron separation energy of the ground state:

$$E^*_{\lim} = S_n + E^{rot} \tag{3.7}$$

with the collective rotational energy approximated by the rotating liquid–drop model. Consequently, the rotational energy of a fission–fragment is given by:

$$E^{rot} = \frac{\hbar^2 J(J+1)}{2\Im} \tag{3.8}$$

where $J$ stands for the total angular momentum and $\Im$ stands for the moment of inertia. This formulation of the rotational energy is used in equation (3.2) for the partitioning of the excitation energy after full acceleration. In a quantum–mechanical description, spherical fission–fragments cannot exhibits collective rotation, therefore the above equation is no longer valid. The impact on the results is a negligible one, because the fission–fragments have a low initial excitation energy and they cannot emit as many neutrons no matter what excitation energy limited we consider, $S_n$ or $S_n+E_{rot}(J)$. Taking into consideration this simple assumption, the neutron evaporation occurs when the excitation energy is higher than the neutron separation energy plus the rotation energy:

$$E^* > S_n + E^{rot}(J) \tag{3.9}$$

When the condition from equation (3.9) is no longer satisfied, the gamma decay can start. In this excitation energy range, the state density of the residual nuclei is very high, so the gamma–rays are statistically emitted through a so–called “statistical cascade” which is dominated by dipolar gamma transitions.

The neutron emission is not expected to decrease the spin of the fragments by more than one unit of angular momentum and is as such of less importance in the determination of the initial fragment spins. On the contrary, the gamma emission is a very suitable tool in studying initial fragment spins because the emission time, the number, the energy, and the multipolarity of the gammas strongly depends on the value of the primary angular momentum.

The main conclusions of the experiments made over the last decades regarding the gamma emission were that the initial angular momentum of the fragments is large compared to the ground state spin and oriented perpendicular to the fission axis. If one assumes that the angular momentum is proportional to the number of emitted gamma rays, a saw–tooth–like behavior for the primary angular momentum as a function of the fragment mass is found.

Most of the calculations of initial fragment spin distribution made use of a desexcitation model developed by Huizenga and Vandenbosch (Huizenga and Vadenbosch, 1960; Vadenbosch and Huizenga, 1960). This model is based on the statistical model for the desexcitation of the fragments, in which no competition between neutron and gamma emission is considered. The model assumes that the distribution of levels with specific spin is given by:

$$P(J) \propto (2J+1)\exp\left(-(J+1/2)^2/2\sigma^2\right) \tag{3.10}$$

where *P(J)* is the probability distribution of levels with spin *J*, and $\sigma$ is the parameter which limits the population of high–spin levels. This parameter is in principal related to the moment of inertia and the temperature of the excited nucleus.

The deexcitation from a specific spin level by a transition is assumed to populate residual spin levels with a probability dependent on the availability of the specific levels given in equation (3.10). Following the neutron capture or after completion of the neutron evaporation a further assumption is made. This assumption is that before reaching the isomeric level or the ground state, the residual nucleus emits three *E1* gamma rays. Therefore, using a value of 3 or 4 for the $\sigma$ parameter a large variety of isomeric yield were empirically correlated. This method was applied by Warhanek and Vandenbosch in order to interpret the primary angular momentum of fission products (Wilhelmy et al., 1972). They assumed that the probability distribution of initial angular momentum states of the fragments could be represented by the following expression:

$$P(J) \propto (2J+1)\exp\left(\frac{(J+1/2)^2}{B^2}\right) \tag{3.11}$$

where *P(J)* is the probability distribution for each spin value *J* and *B* is almost equal to the root mean square value ($J_{rms}$) of $J+\frac{1}{2}$. *B* is related to the moment of inertia $\Im$ and the thermodynamic temperature *T* of the fragments by the following expression:

$$B^2 = 2\sigma^2 = 2\Im T/\hbar^2 \tag{3.12}$$

From the statistical correlation of the spin cutoff parameter $B$ with nuclear temperature $T$ of the fission fragments and the moment of inertia perpendicular to the fission axis, one expects an increase of $B$ or $J_{rms}$ at higher excitation energy. Higher $B$ values can result also from an increase of the moment of inertia due to the fragment deformability at higher excitation. On the other hand, if the neck oscillation is the origin of the fragment angular momentum, higher rotational motion of the compound nucleus system as a whole might restrict rapid neck oscillation, especially if the descent from saddle to scission is faster at higher energy. If the scission point deformation of the fragment does not change significantly, the $B$ value may now show substantial change with the initial excitation energy if the scission point deformation has the dominant influence on the $B$ value (Wagemans, 1991).

It should be pointed out that even for the spontaneous fission of $^{252}$Cf, which has an angular momentum of zero, the products do not have identical and canceling angular momentum. Whatever deviations that do exist between the two primary products can be made up by orbital angular momentum of the system. Based on study over 21 even–even deformed fragments, Wilhelmy (Wilhelmy et al., 1972) concluded that for the spontaneous fission of $^{252}$Cf case, the fission–fragments average angular momentum is around $(7\pm2)\hbar$ and the heavy fission–fragments have around 20% greater angular momentum than the light ones (Litaize and Serot, 2010.

### III.1.4 Fission–fragment moment of inertia

A very important characteristic of the deformed nuclei is the nuclear moment of inertia, which define the energetic spectrum of the rotational states. The experimental value of the moment of inertia, named “effective moment of inertia”, is usually determined based on the experimental value of the first rotational states. In particular, for the even–even nuclei, $\Im_{ef}$ is determined from the energy of the first rotational state as follows:

$$\Im_{ef} = \frac{3\lambda\!\!\!^{-}{}^2}{E_2} \tag{3.13}$$

where *E2* is the first excited state of spin equal 2.

The $\Im_{ef}$ "experimental" value is compared to the "theoretical" values obtained after some assumptions made for the nucleus.

The easiest hypothesis is to consider the nucleus as a deformed rigid body with a form of a rotational ellipsoid. In this case, the moment of inertia can be written as (Bohr and Mottelson, 1955):

$$\Im_{rigid} = \frac{2}{5} AMR^2 \left(1 + 0.31\beta + 0.44\beta^2\right) \tag{3.14}$$

where $A$ is the mass number, $M$ is the nucleon mass, $R$ is the radius ($R=R_0A^{1/3}$ *fm*, with $R_0=1.2$) and $\beta$ is the quadrupole deformation parameter in its ground state, which can be taken from databases (RIPL–3, 2012d).

The nucleus consists of nucleons with a high mobility one compare to the other, consequently the hypothesis of comparing it with a rigid body is a questionable one. Therefore it is assumed that for the spatial orientation of the deformed nucleus a small number of nucleons are participating, eventually the extra–core nucleon, valence nucleons in generally.

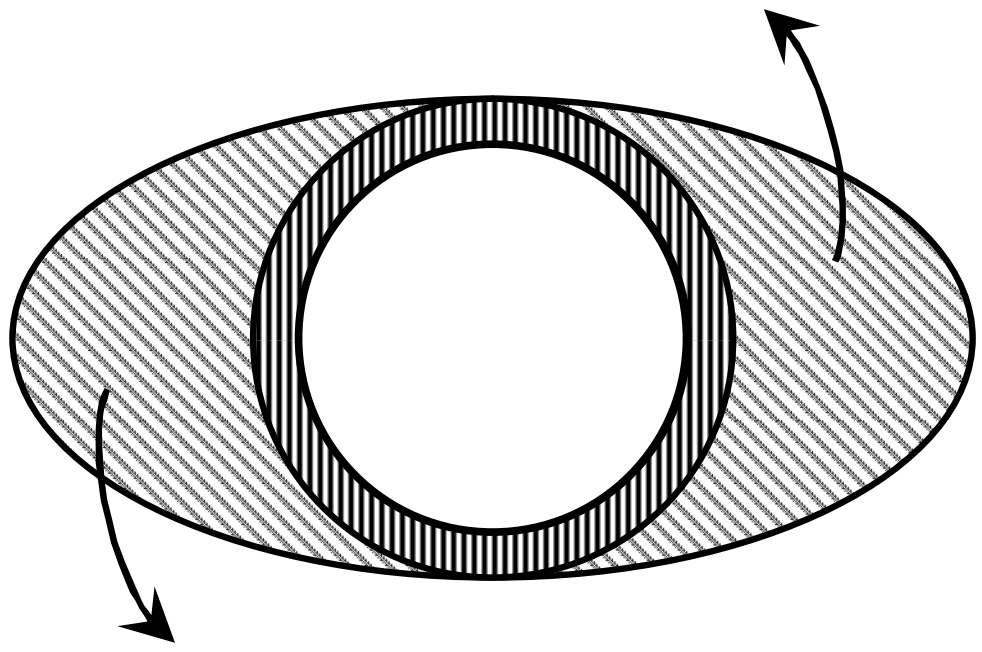

**Fig. 3.1** The moment of inertia defined by the valence nucleons

It's like these nucleons are separated from the spherical core nucleons by a super–fluid layer, which link the two types of nucleons (see figure 3.1). The collective motion of the valence nucleons is similar to a "wave" that propagates without friction (permanently) around the core generating permanent deformation of the nucleus. Considering this hypothesis, the so–called "hydrodynamic hypothesis", the moment of inertia has the following expression:

$$\Im_{liquid} = \frac{3}{5} AM\left((R_1 - R)^2 + (R_1 - R)^2 + (R_1 - R)^2\right) \tag{3.15}$$

where $R_1=R_2=b$ and $R_3=a$ stand for the semi–axes of the rotational ellipsoid given by:

$$a = R_0\left(1+\beta\sqrt{\frac{5}{4\pi}}\right) \qquad b = R_0\left(1-\frac{\beta}{2}\sqrt{\frac{5}{4\pi}}\right) \tag{3.16}$$

Therefore the moment of inertia from equation (3.15) becomes similar to the expression used by Inglis (Inglis, 1954):

$$\Im_{liquid} = \frac{3}{5}AM\left(2\cdot(b-R_0)^2+(a-R_0)^2\right) = \frac{9}{8\pi}AMR^2\beta^2 \tag{3.17}$$

It is well establish that a nucleus is neither a rigid body nor a fluid inside a rotating ellipsoid vessel: $\Im_{liquid} < \Im < \Im_{rigid}$. For instance, if we consider the first exited state $2^+$ of $^{164}$Er at 91.4 keV, from a numerical application of the equation (3.8) and using for the quadrupole deformation parameter $\beta \cong 0.3$, a value of $\hbar^2/2\Im$ =15 keV is obtained. Considering the moment of inertia for the rigid body model described by equation (3.14), we obtain $\hbar^2/2\Im_{rigid} = 6keV$, while for the fluid model case a value of $\hbar^2/2\Im_{liquid} = 90keV$ is obtained. Consequently the moment of inertia must have an intermediate value between these two extreme configurations.

### III.1.5 Temperature distribution

Over time different models were implemented in the FIFRELIN code. These models differ by the temperature ratio between the complementary fragments after full acceleration ($R_T=T_L/T_H$, where $T_L$ and $T_H$ are the temperature in the light and in the heavy fragments, respectively at full acceleration), the excitation energy limit for the neutron emission ($E^*_{\lim}$) and the moment of inertia $\Im$ involved in the rotational energy.

As already reported by Lemaire et al.( Lemaire et al., 2005), if we consider an equipartition of the temperature between the two fragments (in this case $R_T$=1) then the well–known sawtooth shape of the distribution of the multiplicity as a function of the mass number is no more reproduced. Also, Talou (Talou, 2010) mention that the ratio of the average neutron multiplicity for the light and, respectively, for the heavy fragment, $\overline{\nu_L}/\overline{\nu_H}$, is inverted and the average total neutron multiplicity $\overline{\nu}$ is overestimated.

Therefore, for the case of the spontaneous fission of $^{252}$Cf, on which the FIFRELIN code was initially tested, a different mass–dependent temperature ratio between the complementary was considered as follows (Litaize and Serot , 2010):

i) For the symmetric fission an equal temperature for both complementary fragments is expected and then the temperature ratio $R_T$=1.

ii) For the light mass number $A_L$=120, $R_T$ has a maximum value because the $^{252}$Cf case, the complementary heavy fragment is nearly spherical with 132 nucleons (corresponding to the closed proton shell Z=50 and the closed neutron shell N=82). Consequently the major part of the total excitation energy is gained by the light fragment leading to a higher value of the temperature compared to its double magic complementary partner.

iii) For the very asymmetric fission case, the heavy fragment is more deformed than the light fragment because the latter becomes shell stabilized. The light fragment can be almost spherical with a closed proton shell Z=28 and a closed neutron shell N=50, which lead to a lower temperature than the one's of the complementary heavy fragment. Therefore the ratio of the temperatures should have a value less than 1 ($R_T$<1).

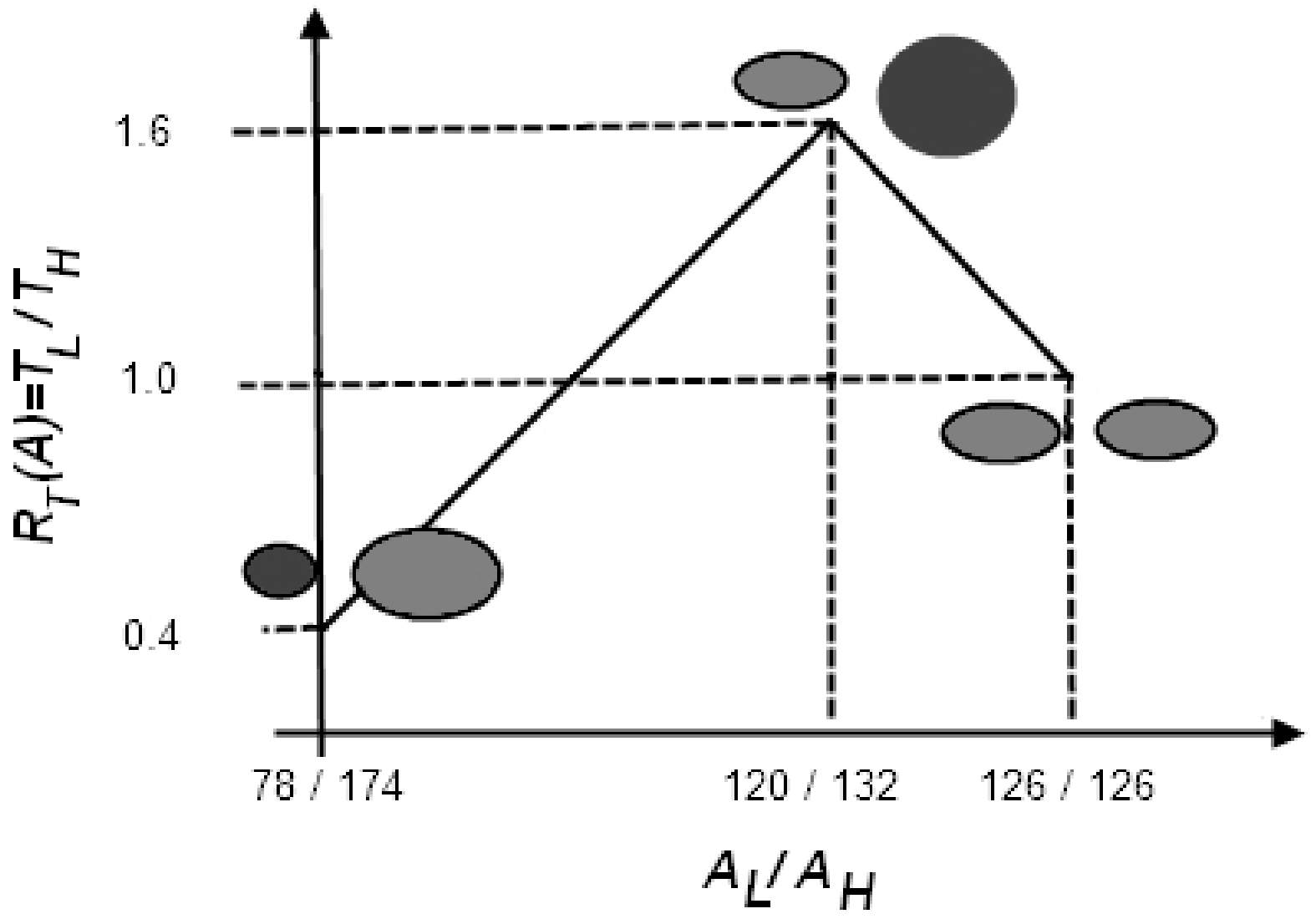

**Fig. 3.2** Temperature ratio law $R_T(A)$ for $^{252}$Cf(SF)

As can be seen in **Fig. 3.2**, between these three configurations, a simple composition of two linear laws is assumed. For mass split 78 / 174 the light fission –fragment is almost spherical and then, as it was mentioned before, its temperature is lower than its heavy complementary fragment, leading to a minimum value $R_T < 1$. The situation is reverse for the mass split 120 / 132 where the heavy fission–fragment is near spherical with a higher temperature than its light complementary fragment, leading to a maximum value $R_T > 1$. Finally, for the case of symmetric mass split 126 / 126, the temperature is the same for both light and heavy fragment and then the temperature ratio is equal to one, $R_T = 1$.

## III.2 New developments in the FIFRELIN code

### III.2.1 Energy dependent cross–section of the inverse process of neutron evaporation from fragments

Since the neutron spectrum of $^{252}$Cf is an international standard reference for metrological application, important efforts have been done in order to produce a revised evaluation (Manhart, 1987).

The Mannhart's evaluation was obtained from seven experiments based on the time–of–flight techniques. It is very well known that the most important neutron emission process is the evaporation from the fully accelerated fragments. Therefore, in order to reproduce with a higher accuracy the experimental data, statistical model approaches have been used to calculate the integral fission neutron spectrum, e. g., the fission neutron spectrum of the spontaneous fission of $^{252}$Cf by Browne and Dietrich (Browne and Dietrich, 1974), Gerasimenko and Rubchenya (Gerasimenko and Rubchenya, 1987), Märten and Seelinger (Märten and Seelinger, 1985), and Madland and Nix (Madland and Nix, 1982). The first three theoretical calculations are parameter free, the assumptions made by the different authors leading to differences between the results. For example, Browne and Dietrich (Browne and Dietrich, 1974), for simplification used only five fission fragment mass pair in order to represent the mass distribution. Gerasimenko and Rubchenya (Gerasimenko and Rubchenya, 1987) used the Hauser–Feschbach formalism and included the competition between the neutron and gamma–ray emission. Märten and Seelinger (Märten and Seelinger, 1985) applied the cascade emission model with the consideration of the anisotropy of the neutron emission in the center–of–mass system.

The Madland and Nix evaporation theory accounts for some important effects as the motion of the fission–fragments, the distribution of the fission–fragment residual temperature and especially the energy dependence of the compound nucleus cross–section of the inverse process of neutron evaporation from fragments, based on optical model calculations.

As the time being, results obtained with the FIFRELIN code were reported only for the case of the spontaneous fission of $^{252}$Cf (Litaize and Serot, 2010, 2011; Regnier et al., 2012a,b). All the results concerning the neutron spectrum showed a slightly underestimation below 500 keV and a slightly overestimation above 10 MeV. As already reported by Madland and Nix (Madland and Nix, 1982), these discrepancies between the calculated and the experimental neutron spectrum can be improved by taking into account an energy dependent cross–section for the inverse process of compound nucleus formation.

As described in **Chapter II**, the energy dependent compound nucleus cross–section of the inverse process of neutron evaporation from fragments appears in the expression of the prompt fission neutron spectrum corresponding to a fission–fragment in the center–of–mass (see equations (2.1), (2.2)).

The relative increase of the compound nucleus cross–section in a specific energy range determines a corresponding increase of the neutron spectrum because the neutron spectrum is mainly influenced by the shape and not by the absolute value of the cross–section. The fission–fragments are very unstable nuclei, which reduces the possibility of measuring the experimental values of the corresponding cross–sections. Therefore the values of the energy dependent cross–sections must be obtained from calculations.

The main used method to obtain the compound nucleus cross–section of the inverse process of neutron evaporation is the optical model calculation by using phenomenological potentials adequate for the region of nuclei appearing as fission fragments. Another manner is to use energy-dependent cross-sections given by global approximation analytical formulae (for instance the simplified formula proposed by Iwamoto). The energy dependent cross–section for the inverse process of compound nucleus formation, calculated by optical models in the same manner as in the case of the PbP model (see Chapter II) can be used in the case of FIFRELIN, too.

The most used phenomenological optical potential for the nuclei corresponding to the fission–fragments region is the Bechetti–Greenless potential (RIPL–3,2012a), this potential being use by Madland and Nix (Madland and Nix, 1982) and also in great part of the calculations performed with the PbP model. In order to generate de inverse cross–section for the fission–fragments the computer code SCAT2 (Bersillon, 1991) was used as in the case of the PbP model. This code allows the use of several optical potential parameterizations, including the Bechetti–Greenless's one.

Recently, A. J. Koning and J. P. Delaroche (RIPL–3,2012a), proposed a new parameterization of the spherical optical potential (for neutrons and protons) by adding an imaginary spin–orbit term and by modifying the energy dependence of the potential depths, this parameterization being also included in a new version of the SCAT2 code (Manea, 2010).

As example, in **Figs. 3.3** and **3.4** the compound nucleus cross–sections for the reactions involving the nuclei of the most probable fragmentation of $^{252}$Cf(SF) are plotted. It is important to notice that for both fragments, but especially for the case of the most probable light fragment (the $^{108}$Mo nucleus) the shape of the two inverse cross–sections are clear different one to the other, depending on the optical potential used, this meaning that the local extremes of the cross–sections calculated with the Koning–Delaroche optical potential (red line) are more visible than the ones calculated with the Becchetti–Greenless potential (black line). This fact has also an important influence on the prompt neutron emission results, especially on the spectrum shape.

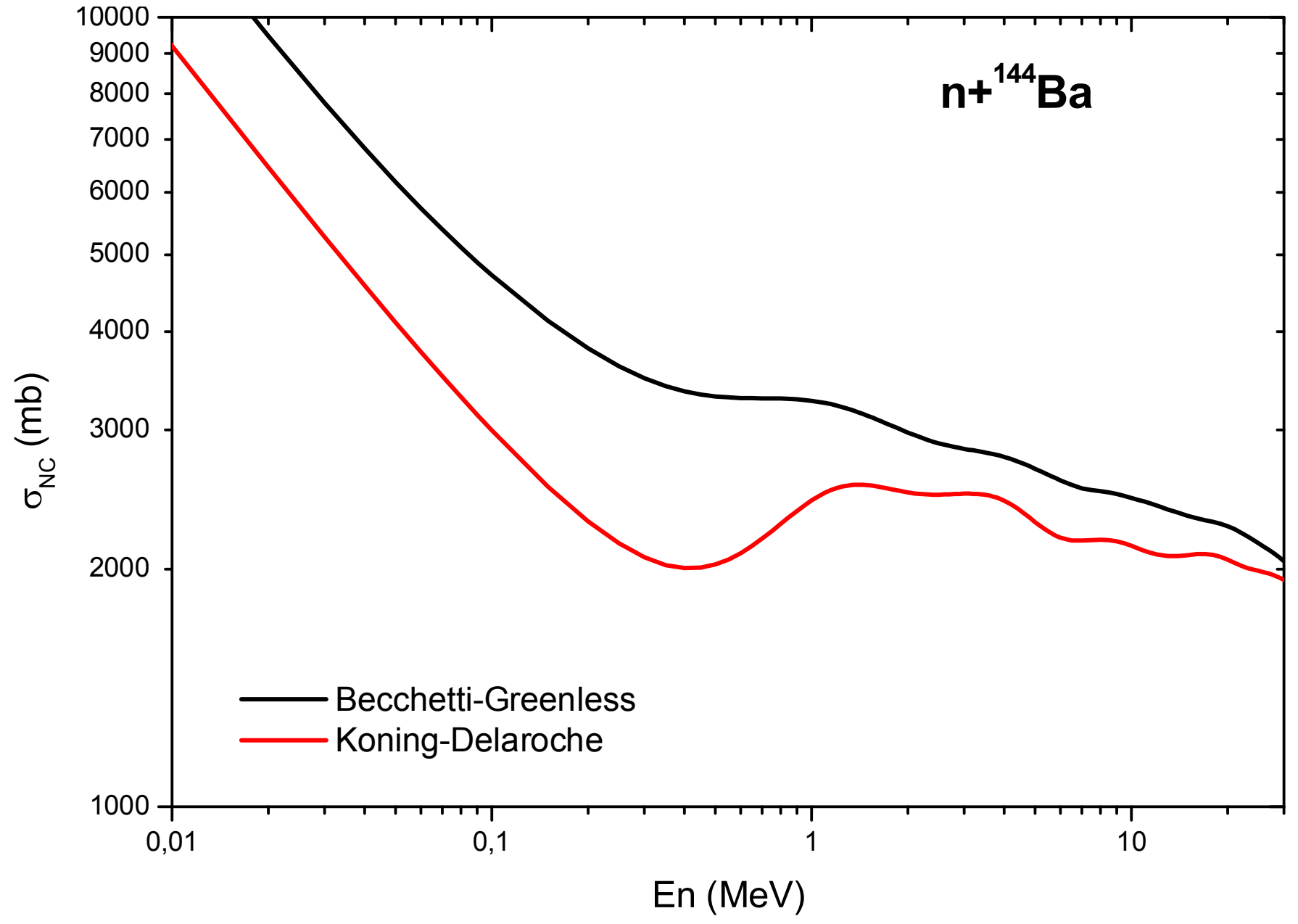

**Fig.3.3** Cross–sections for the n+$^{144}$Ba reaction using Becchetti–Greenless potential (black line) and Koning –Delaroche potential (red line)

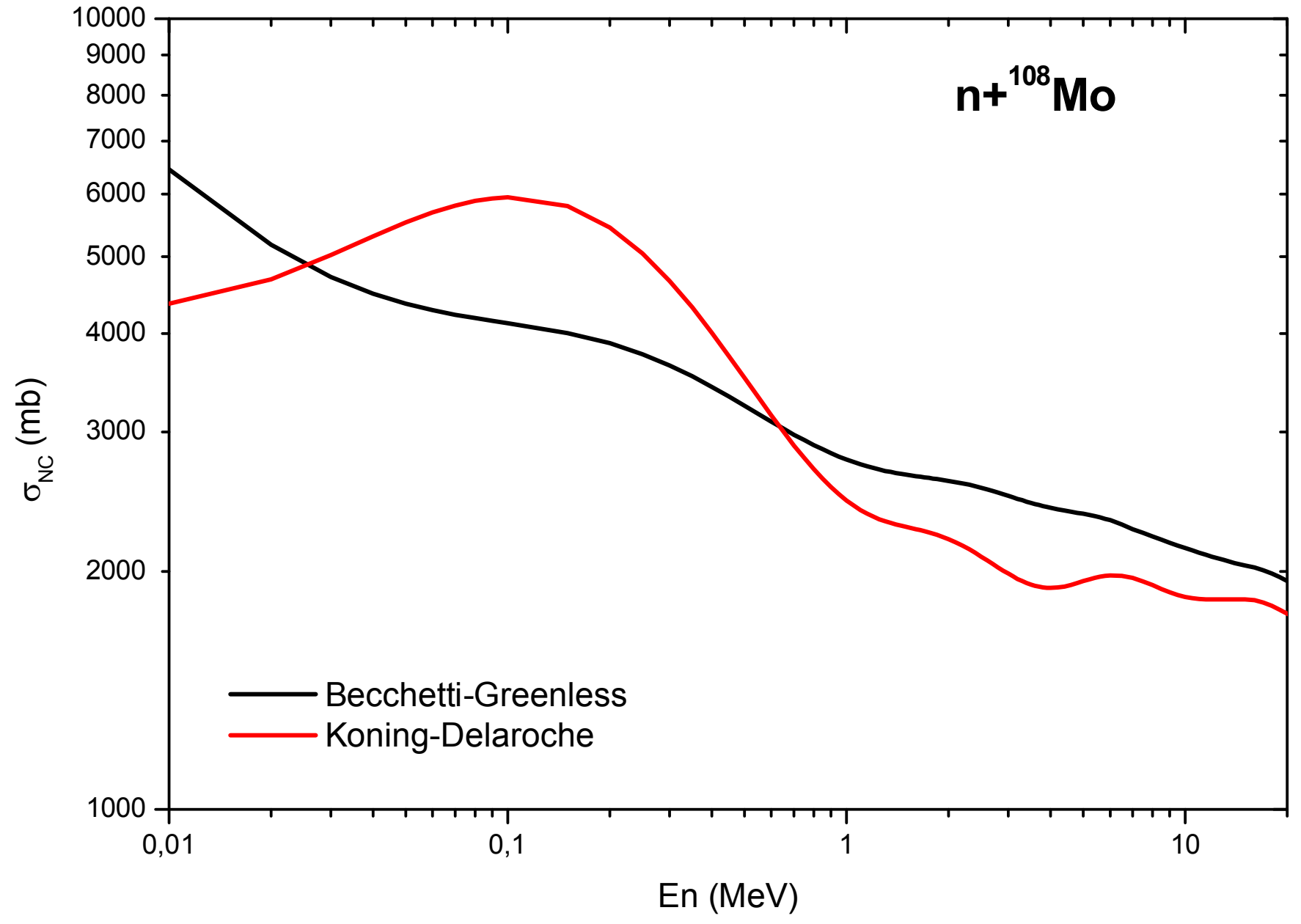

**Fig.3.4** Same as Fig. 3.3 for the n+$^{144}$Ba reaction

The visible differences between the shapes of the inverse cross–sections calculated with the two optical potentials are reflected in the different shapes of the calculated prompt neutron fission spectrum especially in the low energy region.

The PFNS in the laboratory system for the spontaneous fission of $^{252}$Cf for the two optical potentials and considering for the calculation of the moment of inertia a fraction of 0.4 of a rigid body is plotted in **Fig. 3.5** and **3.6**

In **Table 3.1** are given the values of the average total multiplicity for $^{252}$Cf(SF) obtained using the two optical potential for the inverse cross–section calculation. The results are coming from three different calculations, depending on the model chosen for the calculation of the moment of inertia: a fraction of 0.4 or 0.5 of a rigid body moment or the Inglis–Beliaev approximation by using the AMEDEE database (AMEDEE, 2011).

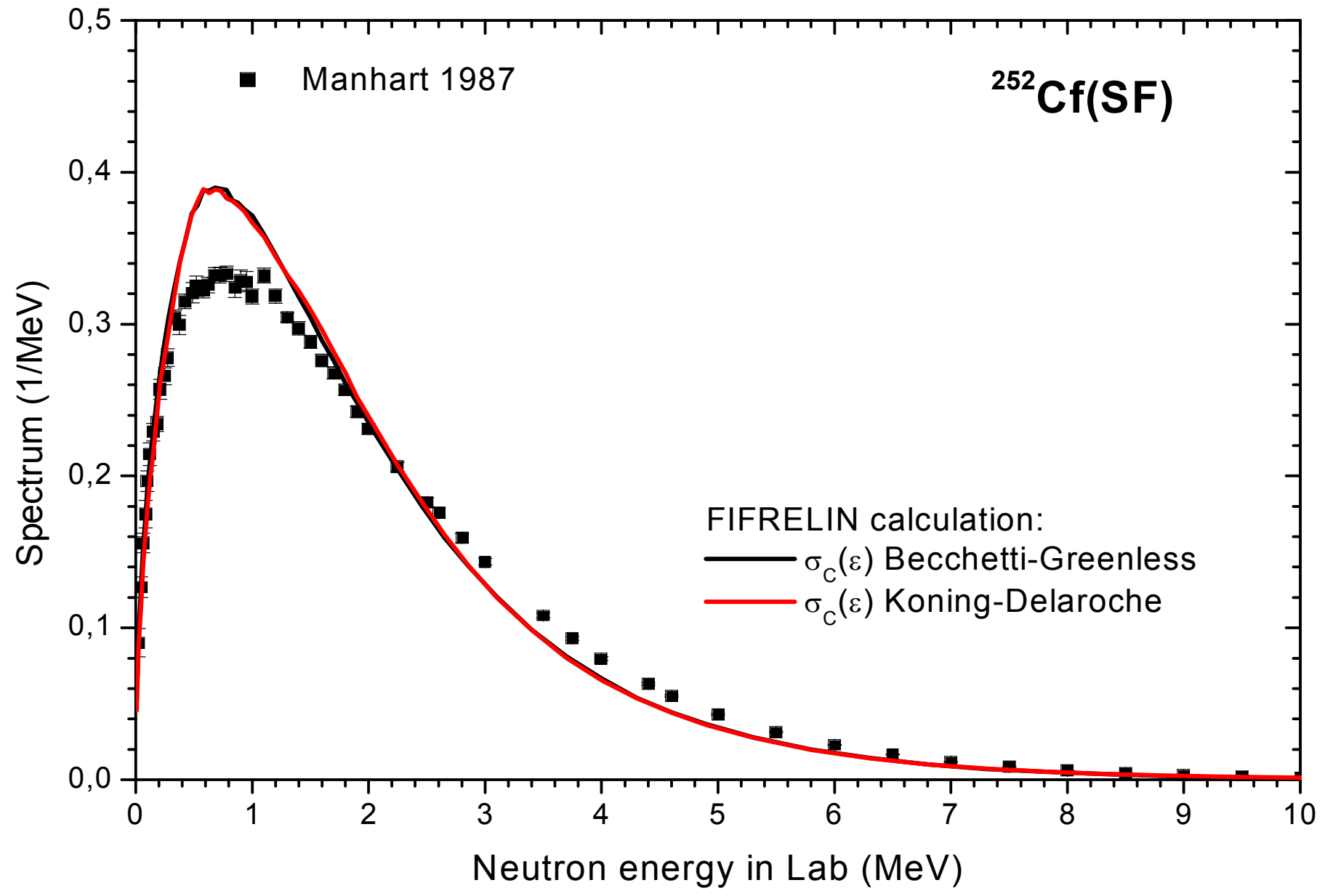

**Fig. 3.5** $^{252}$Cf(SF) Prompt fission neutron spectrum obtained using $\sigma_C(\varepsilon)$ calculated with Becchetti–Greenless potential (black line) and Koning–Delaroche potential (red line)

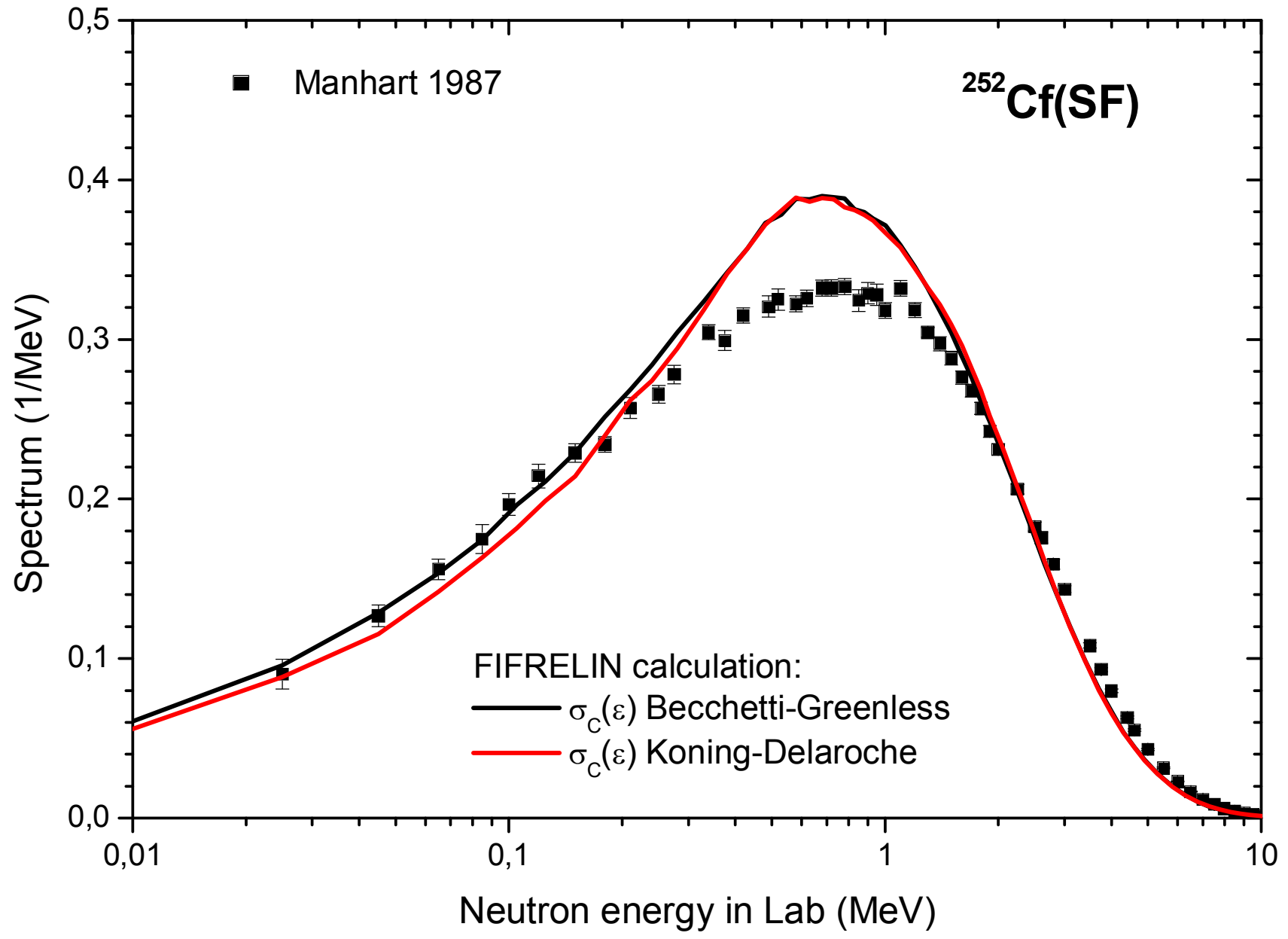

**Fig. 3.6** Same as Fig. 3.5 in a different scale

**Table 3.1** Total average prompt neutron multiplicities for $^{252}$Cf(SF) for various models

| Optical potential | $\Im$ | $\overline{\nu_L}$ | $\overline{\nu_H}$ | $\overline{\nu}$ |
|---|---|---|---|---|
| Becchetti–Greenless | $0.4 \cdot \Im_{rigid}$ | 2.17332 | 1.70884 | 3.88216 |
| | $0.5 \cdot \Im_{rigid}$ | 2.21450 | 1.74624 | 3.96074 |
| | Amedee | 2.17112 | 1.72482 | 3.89595 |
| Koning–Delaroche | $0.4 \cdot \Im_{rigid}$ | 2.18809 | 1.69337 | 3.88146 |
| | $0.5 \cdot \Im_{rigid}$ | 2.22934 | 1.73077 | 3.96012 |
| | Amedee | 2.18520 | 1.70997 | 3.89517 |

### III.2.2 New calculation of the moment of inertia

As mentioned in a previous section, one of the long–standing questions about the nuclear fission process is related to the excitation energy partition between the complementary fission–fragments forming a pair. After full acceleration, the total excitation energy of the rotating fission–fragments is described by the equation (3.2), with the rotational energy determined by the use of the liquid–drop model, where the rotational energy of a fission fragment is given by equation (3.8). In the case of $^{252}Cf$(SF) the best results with the FIFRELIN code were obtained by taking for the moment of inertia from the denominator of the equation (3.8) a fraction of 50% of a rigid body moment of inertia (Litaize and Serot, 2010).

A more realistic calculation is proposed by the use of the AMEDEE database (AMEDEE, 2011), containing the moments of inertia determined on the basis of an Inglis–Beliaev approximation (Inglis, 1954; Beliaev, 1961).

In its original version, available online from 2006, the AMEDEE database was covering only the mean field results for all the nuclei ranging from Carbon (Z=6) up to Darmstadtium (Z=110). In 2008 a first update was made, consisted in extending the database to cover the Super–Heavy mass region (up to Z=130). Large scale mean filed calculations from proton to neutron drip–lines have been performed using the Hartree–Fock–Bogoliubov method based on the Gogny nucleon–nucleon effective interaction. This extensive has shown the ability of the method to reproduce bulk nuclear structure data available experimentally. This includes nuclear masses, radii, matter densities, deformations, moment of inertia as well as collective mode (low energy and giant resonances).

As for today, the predictions of the database are available for more than 8000 nuclei. The latest update of the AMEDEE database includes nuclear matter densities calculated for the nucleus grand state deformation, single particle levels as function of the axial deformation as well as potential energy surfaces obtained within the triaxial symmetry framework, the latest providing useful information on the nuclear shapes, for instance when we are dealing with nucleus symmetries within the fission process.

The new version of FIRELIN has the possibility to use the moment of inertia from the AMEDEE database in order to calculate the rotational energy from eq (3.8).

### III.3 Results and discussion concerning FIFRELIN calculations for $^{252}$Cf(SF)

The emphasis of this part is to show the results obtained with the improved version of the Monte Carlo code FIFRELIN. As mentioned before the lately improvements of the code refers to the calculation of the moment of inertia and to the calculation of the prompt fission neutron spectra taking into account the energy–dependent compound nucleus cross section of the inverse process of neutron evaporation from fragments.

Up to now the code has been tested for the case of the spontaneous fission of $^{252}$Cf and the obtained results were in a good agreement with the experimental data regarding the prompt neutron multiplicity as a function of the fragment mass $\bar{\nu}(A)$ or as a function of the total kinetic energy $\bar{\nu}(TKE)$, the multiplicity distribution $P(\nu)$, energy spectra $N(E)$, and other quantities related to the prompt fission neutrons.

For all the performed simulations few different cases were considered:

- a fraction for the moment of inertia equal to 0.4 of the rigid body moment
- a fraction for the moment of inertia equal to 0.5 of the rigid body moment
- the new method for the calculation of the moment of inertia (using a Hartree-Fock-Bogoliubov formalisms, from the AMEDEE database (AMEDEE, 2011)
- taking/not taking into account the compound nucleus cross–section of the inverse process of neutron evaporation from fragments
- for the cases where the inverse cross–sections was taken into account, we considered two optical model parameterizations: Becchetti–Greenless and Koning–Delaroche (RIPL–3, 2012a).

The simulations were performed in two ways: using $10^7$ events for the cases where a constant cross section of the inverse process was considered and $10^6$ events for the cases where a dependence of the inverse cross section was taken into account. A sample of 97 masses ranging from 78 to 174 was considered and the experimental data of Varapai (Varapai, 2005) were used.

The experimental fragment distribution matrix f(A, KE) can be decomposed in a product of two normalized distribution functions, one of them being the pre-neutron mass yield Y(A) which is normalized to 2, and the other one, N(<KE>, $\sigma_{KE}$), is a mass–dependent Gaussian function of fragment kinetic energy characterized by a mean value <KE> and a standard deviation $\sigma_{KE}$, as can be seen in figures **3.7**.

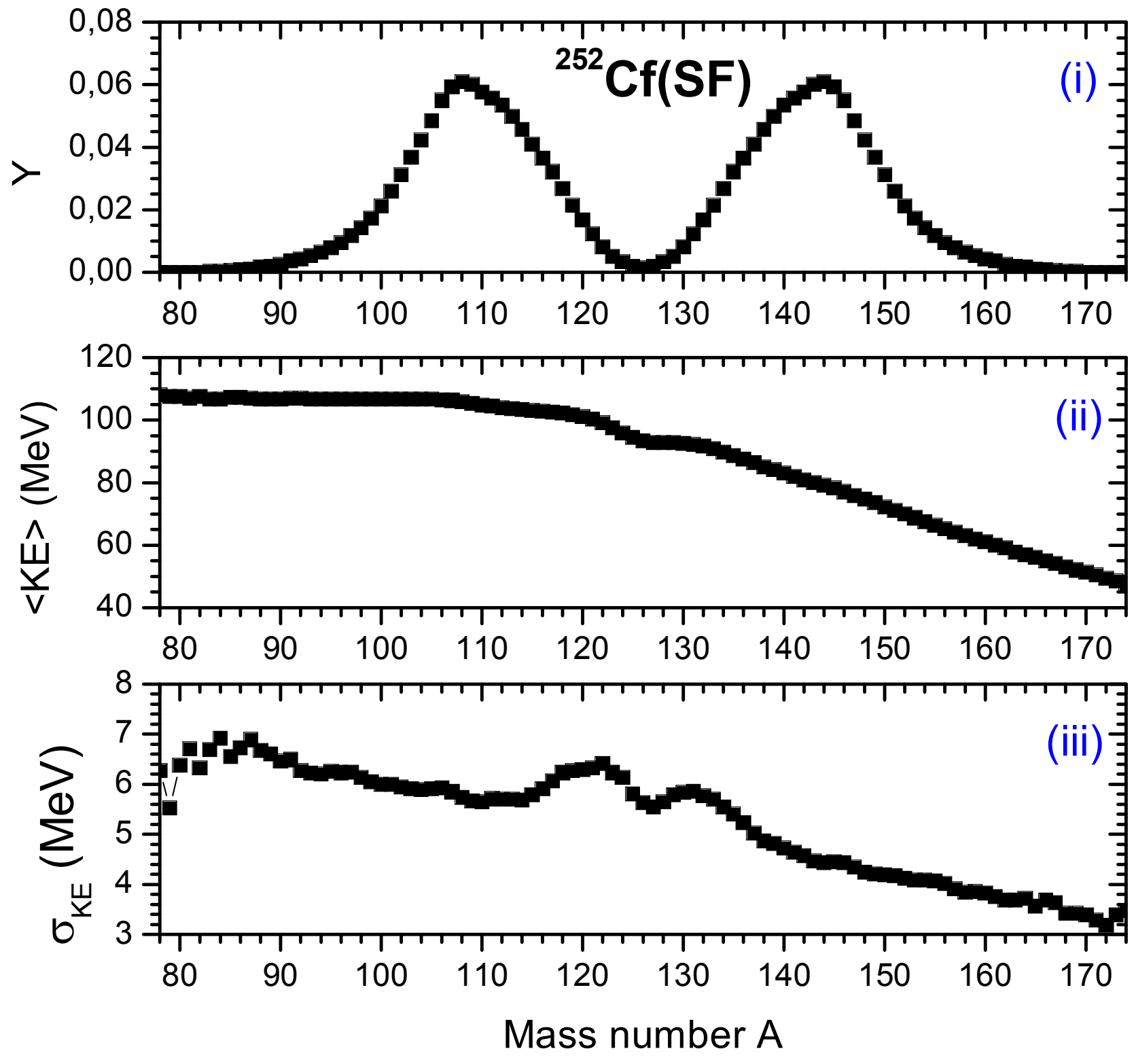

**Fig. 3.7** $^{252}$Cf experimental input data from (Varapai, 2005). Preneutron mass yield Y(A) **(i)** and kinetic energy [mean value <KE> **(ii)** and distribution width $\sigma_{KE}$ **(iii)**]

In order to distinguish between the different types of calculations, on the next figures are represented only the results for the cases where for calculation of the moment of inertia, a fraction of 0.4 of a rigid body moment of inertia was considered or the use of the AMEDEE database (AMEDEE, 2011), taking or not taking into account the energy dependence of the cross–section for the inverse process of compound nucleus formation. For the cases where the AMEDEE database was used three types of calculation are shown: for a constant inverse cross–section or a variable one calculated for the two types of optical model potential, Becchetti–Greenless and Koning–Delaroche (RIPL–3, 2012a). All these options can be found in **Table 3.2**.

In **Fig. 3.8** are given examples of calculations of the average neutron energy in the center of mass of the fission fragment pair as a function of the fragment mass in comparison with the data from (Budtz–Jorgensen and Knitter, 1988). As it can be seen, the PbP calculations are in a good agreement with the experimental data, while the FIFRELIN underestimate these data, the influence of the inverse cross–section being seen in the lower values of the calculations.

**Table 3.2** Input options for various models used in calculations

| Calculation | Moment of inertia | Inverse cross section |
|---|---|---|
| mom inertia =0.4*rigid+(xs const) | 0.4 from rigid body | constant |
| mom inertia =0.4*rigid + xs_BG | 0.4 from rigid body | Becchetti – Greenless |
| Amedee+(xs const) | Amedee | constant |
| Amedee +xs_BG | Amedee | Becchetti – Greenless |
| Amedee +xs_KD | Amedee | Koning – Delaroche |

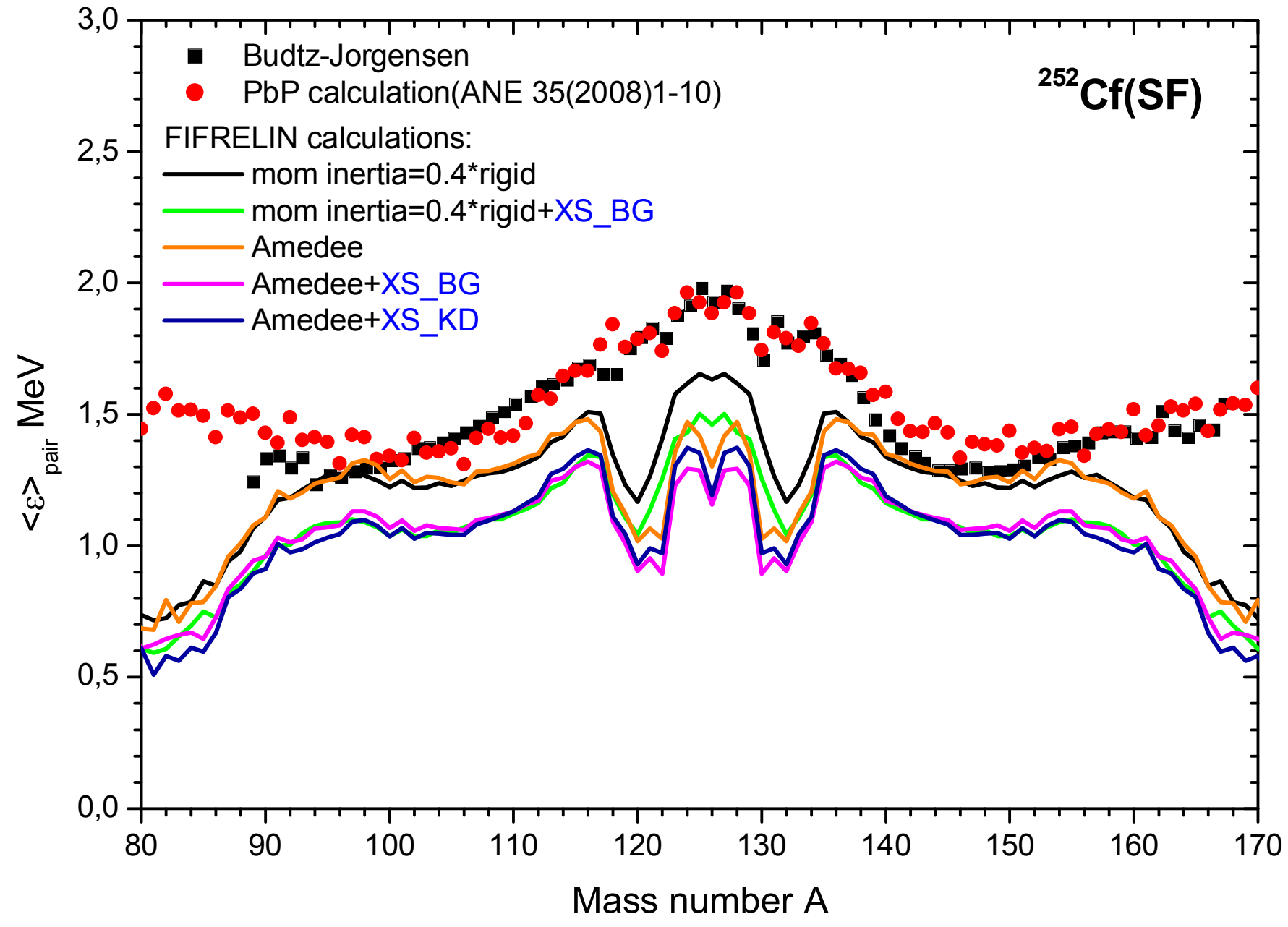

**Fig. 3.8** $^{252}$Cf(SF): Average center of mass neutron energy as a function of the mass number

The experimental data concerning the prompt fission neutron number as function of the fragment mass, when they exist, allow a more refined verification of the model's results and in the same time they can be used for validation of the used methods. In **Fig. 3.9** are given the sawtooth results (solid lines with different colors) for $^{252}$Cf(SF) in comparison with the experimental data of Bowman and Budtz–Jorgensen (black open squares and full triangles). Good agreement with the experimental data and with the Point by Point results (red full circles) was obtained for the entire mass region.

Another important parameter related to prompt fission neutron is the prompt neutron multiplicity distribution P(ν). This quantity is a very sensitive, depending on the prompt neutron emission mechanism and also on the fission–fragment mass and kinetic energy distributions. P(ν) results of FIFRELIN for the spontaneous fission of $^{252}$Cf is given in **Fig.3.10** in comparison with the experimental data of Vorobyev (black full squares) and Balagna (black open triangles). Except for the probability of emitting 4 neutrons where the calculations overestimates the experimental data, a nice agreement with the two sets of experimental data and also with the Point by Point results (red full circles) was obtained.

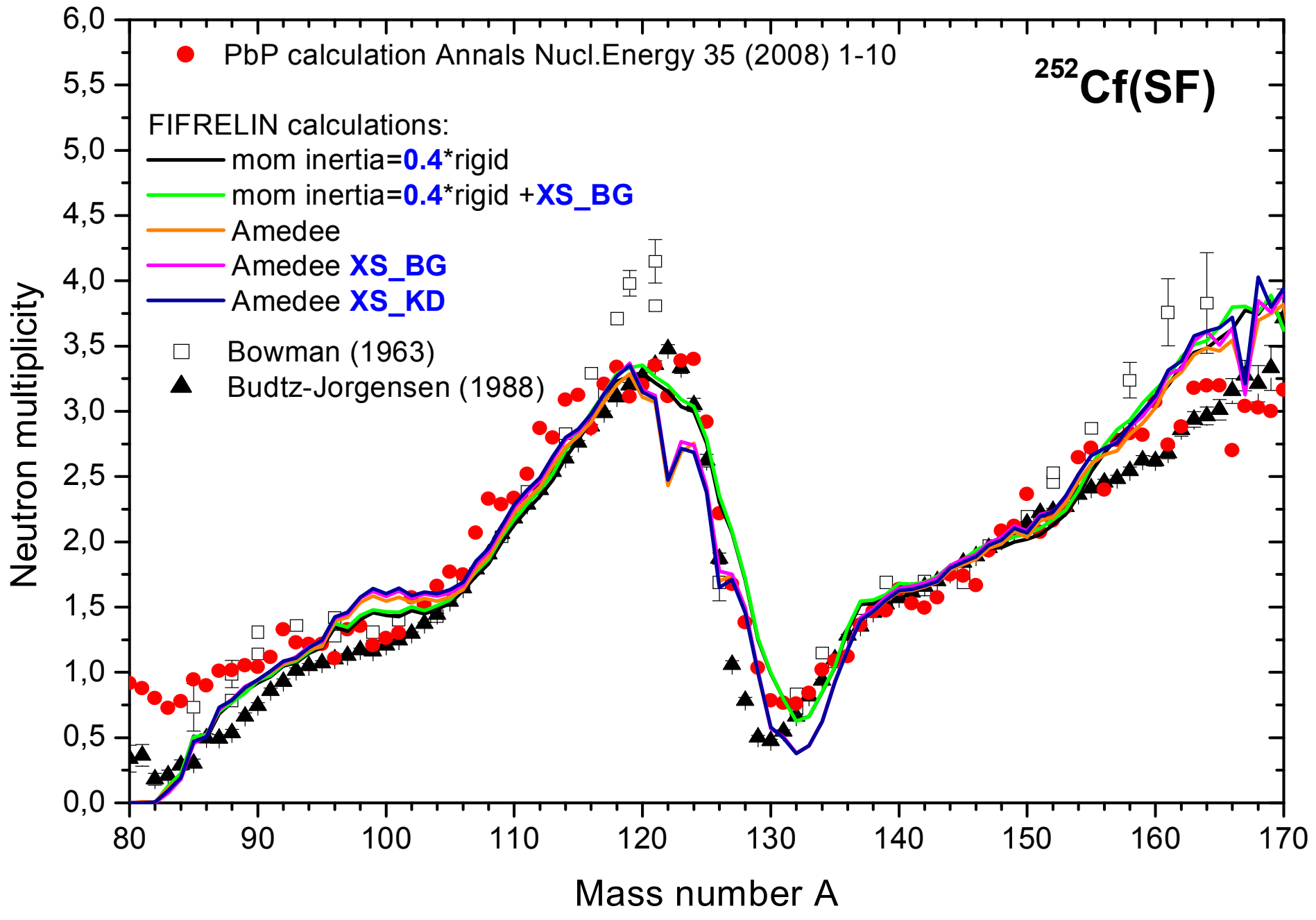

**Fig. 3.9** $^{252}$Cf(SF): Average prompt neutron multiplicity as a function of mass in comparison with the experimental data

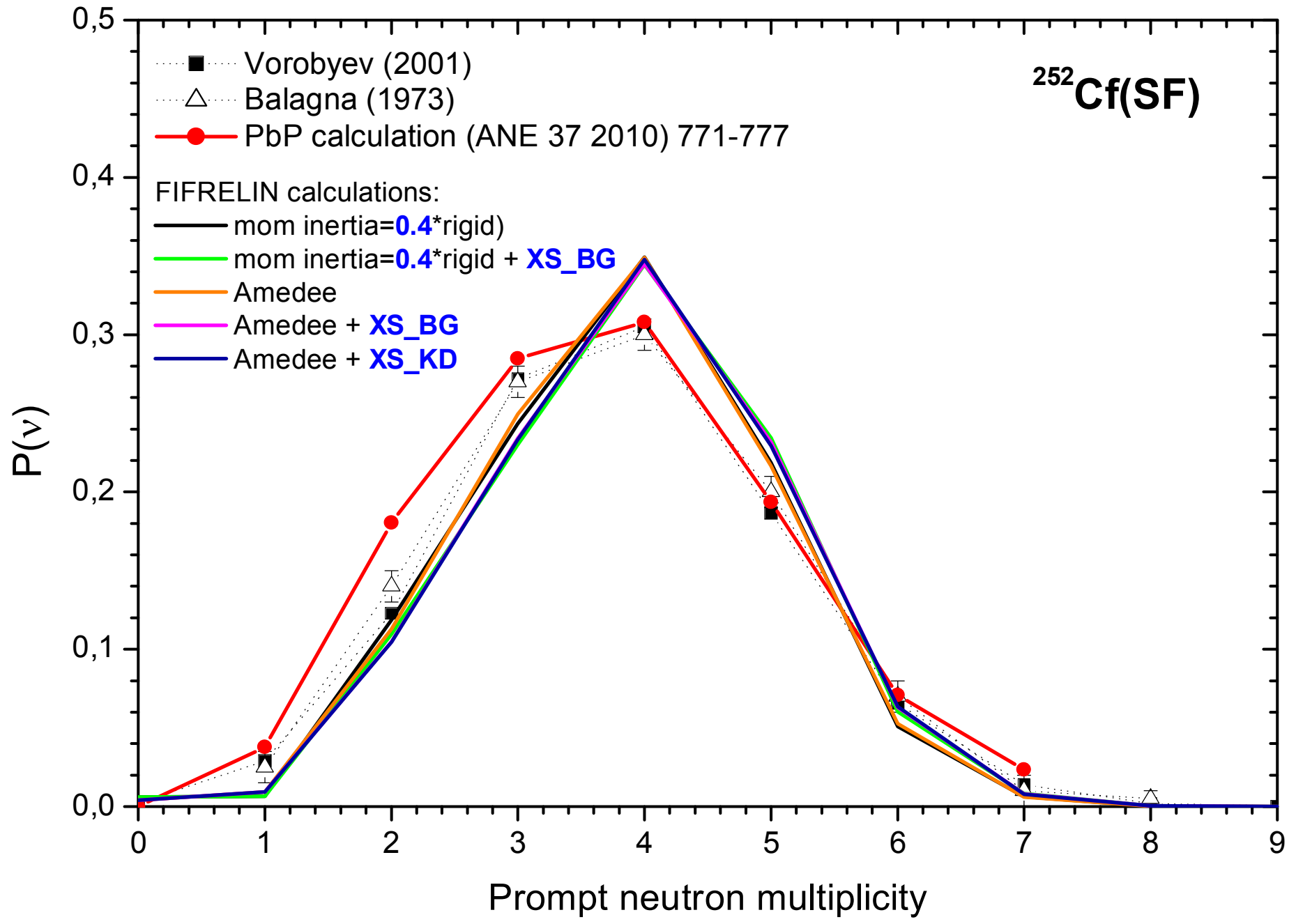

**Fig.3.10** $^{252}$Cf(SF): Prompt neutron multiplicity in comparison with the experimental data

The average prompt neutron multiplicity as a function of the total kinetic energy of fission fragments <ν>(TKE) for $^{252}$Cf(SF) is given in **Fig 3.11**. As already discussed in **Chapter II** the experimental data reported by the research groups exhibit visible different slopes dTKE/dν and also a more or less visible flattening of <ν> at low TKE values.

As it can be seen in the figure, the FIFRELIN results describe well the experimental data of Budtz–Jorgensen (black stars), Vorobyev (grey diamonds) and Sing (black full squares) over the entire TKE range including the flattening of <ν> at low TKE values and are in good agreement with the PbP calculation (cyan full circles connected with lines) obtained by averaging over the double distribution Y(A,TKE) reported in (Tudora, 2012a).

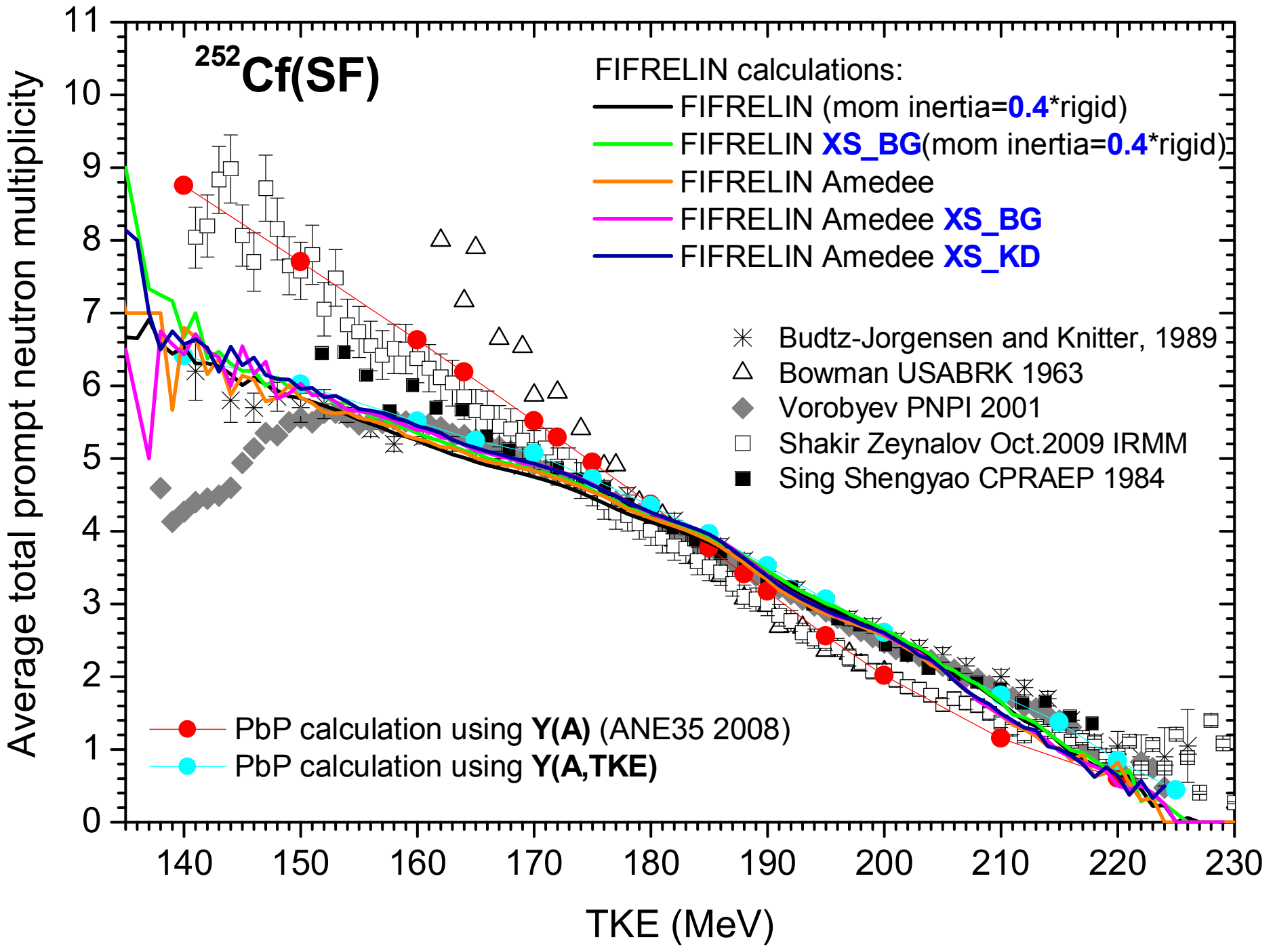

**Fig. 3.11** $^{252}$Cf(SF) FIFRELIN results of < ν>(TKE) in comparison with the experimental data (different black and grey symbols) and with the PbP calculations (red and cyan full circles connected with lines)

It was shown that estimated uncertainties in the prompt fission neutron spectrum from low–energy neutron–induced reactions can significantly impact the results of transport simulation for critical assemblies, uncertainties remaining below 500 keV and above 6 MeV. As example, the prompt neutron spectra for $^{252}$Cf(SF) is given in **Fig 3.12** as a ratio to a Maxwellian spectrum with TM=1.42 MeV.

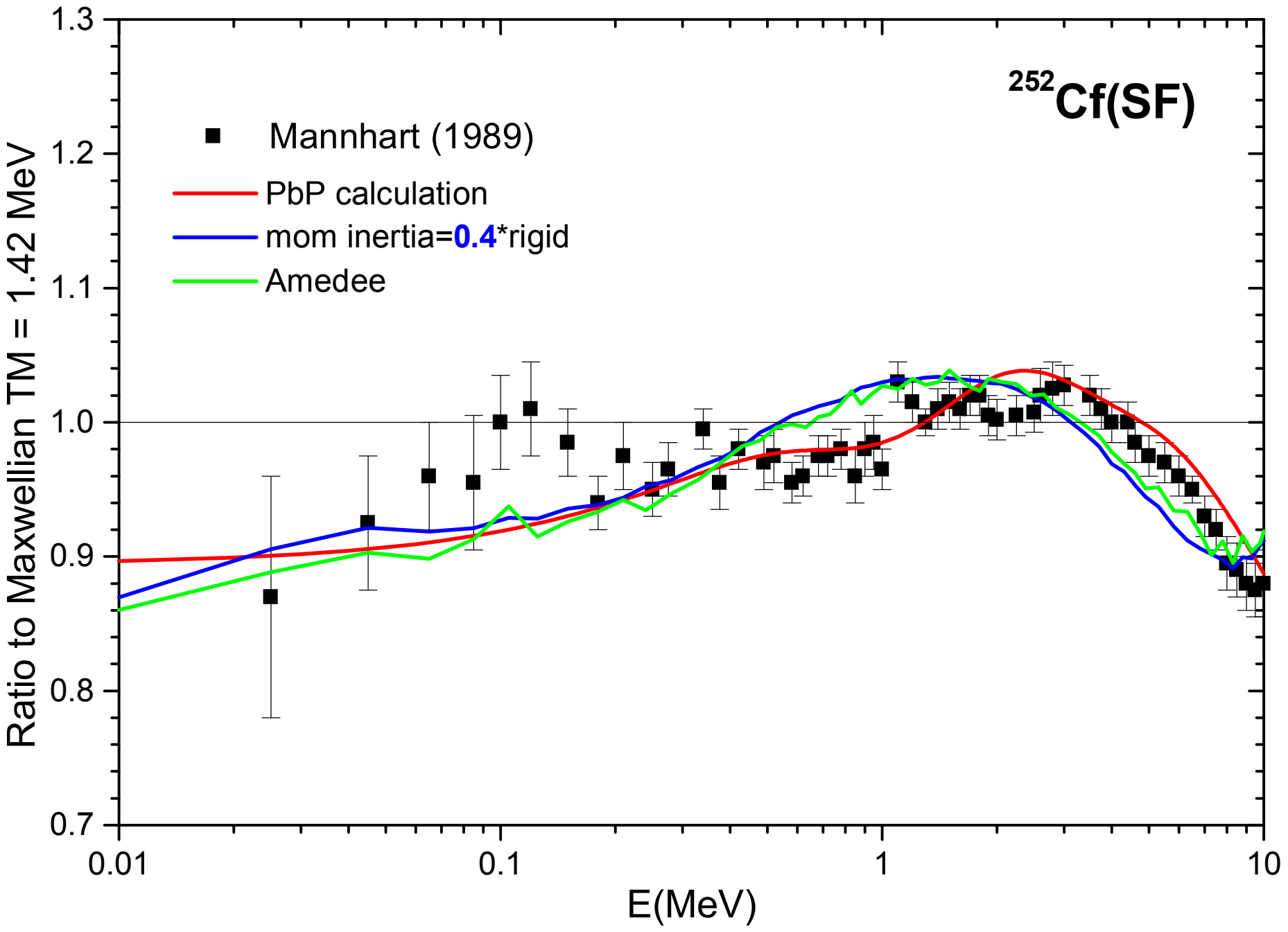

**Fig. 3.12** $^{252}$Cf(SF) Total prompt fission neutron spectrum as a ratio to a Maxwellian spectrum with $T_M$=1.42 MeV in comparison with the experimental data of Manhart (open squares) and the PbP results (red line)

# Chapter IV

## Comparative study between the PbP model and the Monte–Carlo treatment

### IV.1 Prompt neutron emission calculation of $^{236, 238, 240, 242, 244}$Pu(SF) in the frame of the PbP model and Monte Carlo treatment (FIFRELIN code)

As it was mentioned in the previous chapters both the determinist PbP model and the probabilistic Monte-Carlo treatment included in the code FIFRELIN are able to provide all quantities characterizing the prompt emission by averaging the multi-parametric matrices of different quantities as a function of fragment (Z, A) and of TKE (for instance ν(Z,A,TKE), ε(Z,A,TKE) and so on) calculated with PbP or FIFRELIN over different FF distributions, giving:

- average quantities as a function of fragment mass number (such as ν(A), <ε>(A), Eγ(A) and so on) obtained by averaging the corresponding multi-parametric matrices over the charge and TKE distribution
- average quantities as a function of TKE (such as <ν>(TKE), <ε>(TKE) and so on) by averaging the corresponding multi-parametric matrices over the charge and mass distributions
- total average quantities (such as total average prompt neutron multiplicity <νp>, spectra N(E), prompt gamma-ray energy <Eγ>, prompt neutron multiplicity distribution P(ν) and so on) by averaging over charge, mass and TKE distributions.

#### IV.1.1 Fission fragment distributions used in calculations

In both PbP and FIFRELIN calculations the same Gaussian charge distribution (given by eq.(2.25)) was used. Also in both treatments the experimental fragment distributions Y(A), TKE(A) and $\sigma_{TKE}$(A) of Demattè et al, (taken from Demattè, 1996 and from EXFOR, 2012a) were used.

For all Pu(SF) the mass distributions Y(A) measured by Demattè et al. were taken from EXFOR, 2012a, these distributions covering almost all mass fragment range (from near symmetric fission up to far asymmetric fragmentations with $A_H$ of about 155-159).

In the case of experimental distributions TKE(A) and especially $\sigma_{TKE}(A)$ taken from (Dematte, 1996) data near symmetric fission are missing for all Pu(SF) (meaning data for fragment pairs with $A_H$ from symmetric up to 124).

The re-construction of the double distribution Y(A,TKE) from the experimental single ones according to:

$$Y(A,TKE) = Y(A)\frac{1}{\sqrt{2\pi}\,\sigma_{TKE}(A)}\exp\left(-\frac{(TKE - TKE(A))^2}{2(\sigma_{TKE}(A))^2}\right) \tag{4.1}$$

was verified by comparing the TKE distribution obtained from the reconstructed Y(A,TKE) as:

$$Y(TKE) = \sum_A Y(A,TKE) \Big/ \sum_A Y(A) \tag{4.2}$$

with the available experimental Y(TKE) data of Dematte taken from (EXFOR, 2012a), see **Figs.4.1a-e**.

In the case of $^{236,238}$Pu(SF) the use of Y(A), TKE(A) and $\sigma_{TKE}(A)$ from (Dematte, 1996) in order to obtain Y(A,TKE) leads to Y(TKE) in good agreement with experimental data as it can be seen in **Figs.4.1a,b**.

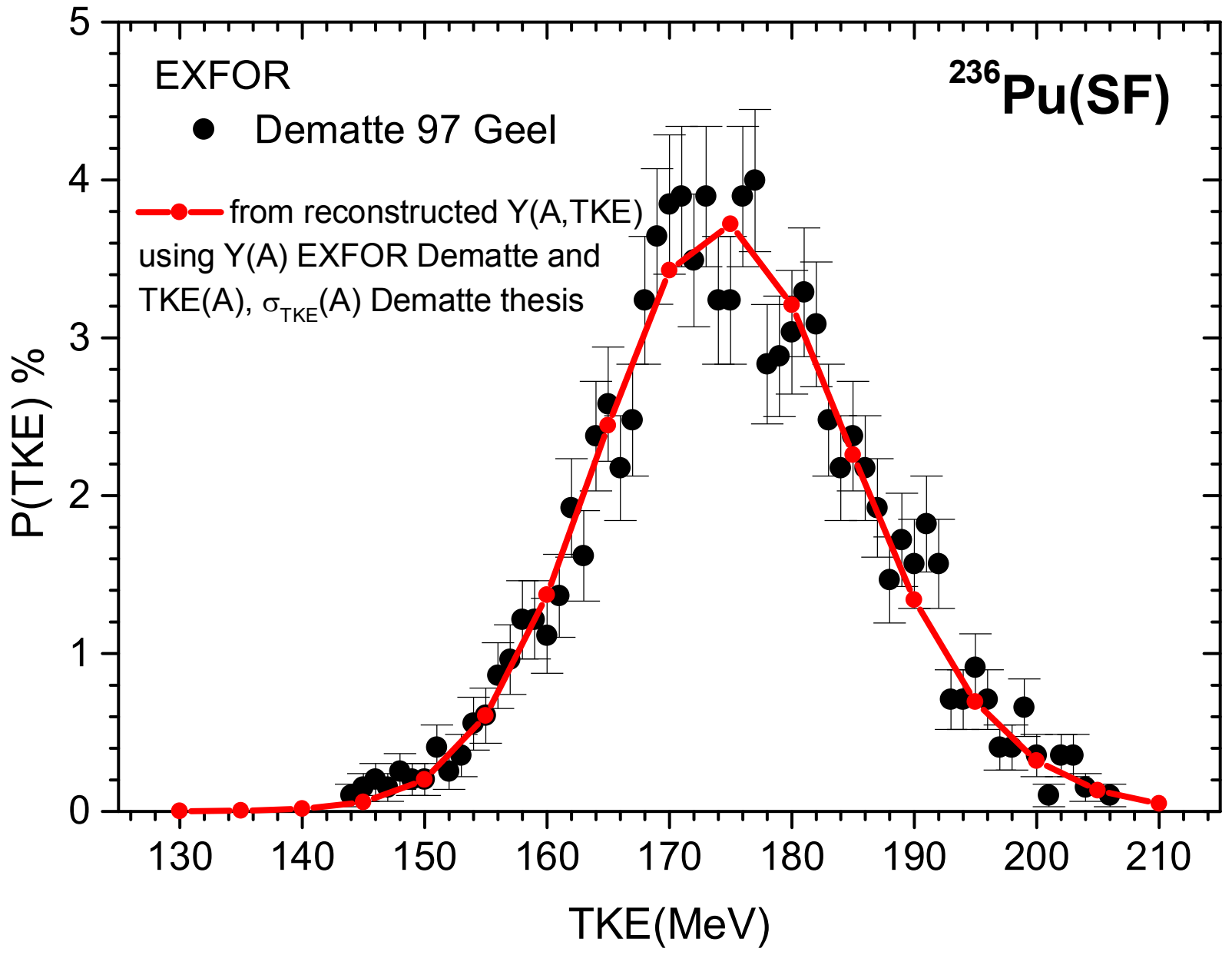

**Fig. 4.1a**: $^{236}$Pu(SF) Experimental Y(TKE) taken from EXFOR and calculated Y(TKE) (red line) by using Y(A,TKE) obtained from experimental Y(A), TKE(A) and $\sigma_{TKE}(A)$.

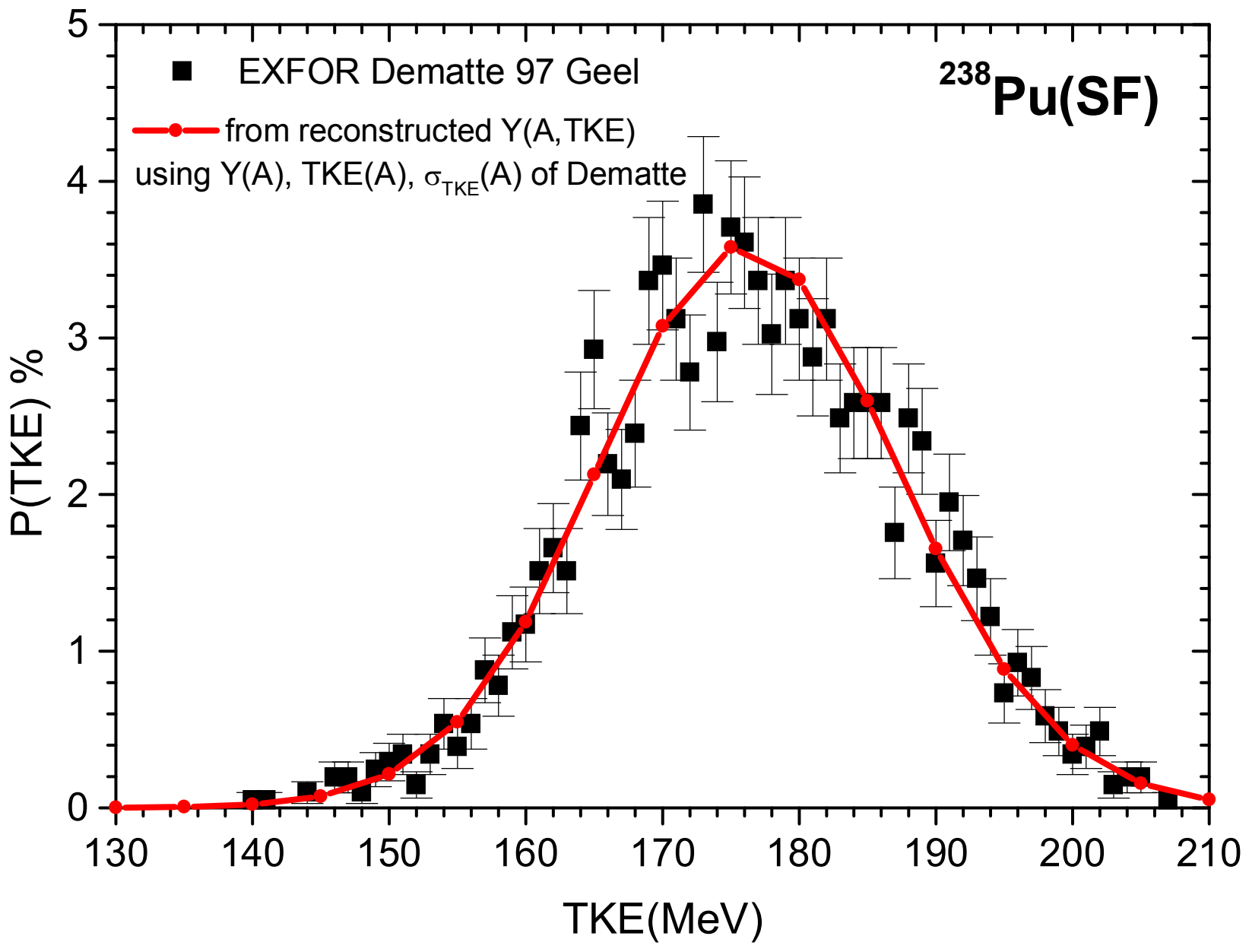

**Fig.4.1b**: $^{238}$Pu(SF) Experimental Y(TKE) taken from EXFOR and calculated Y(TKE) (red line) by using Y(A,TKE) obtained from experimental Y(A), TKE(A) and $\sigma_{TKE}$(A).

For the fissioning nucleus $^{240}$Pu experimental $\sigma_{TKE}$(A) data covering the entire fragment mass range exist (including data for near symmetric fragmentations). These data were measured by Asghar et al. (EXFOR, 2012b) for the reaction $^{239}$Pu($n_{th}$,f). The best agreement of Y(TKE) (obtained according to eqs.(4.1, 4.2)) with the experimental data was obtained by using for $\sigma_{TKE}$(A) the data of Demattè for $A_H$>125 and the data of Asghar for $A_H$=120-125 (see the red line in **Fig.4.1c**). These adjusted $\sigma_{TKE}$(A) data were used in the PbP calculation while in the case of FIFRELIN the data of Demattè were used. In **Fig.4.1c** Y(TKE) calculated with $\sigma_{TKE}$(A) of Demattè (without data for near symmetric fragmentations) is plotted with cyan line. Both Y(TKE) (red and cyan lines) are close each other and describe well the experimental Y(TKE) data. In the same figure the Gaussian fit of (Tudora, Hambsch, 2010) is also plotted with dashed blue line.

For $^{242}$Pu(SF) also other experimental Y(A) and TKE(A) data exist in EXFOR, 2012c (measured by Vorobyeva et al). In the calculation of Y(A,TKE) several combinations of Y(A), TKE(A) (from Demattè and Vorobyeva) were tested.

The best agreement of calculated Y(TKE) with experimental data was obtained in two cases: using Y(A), TKE(A) and $\sigma_{TKE}$(A) of Demattè (the red line in **Fig.4.1d**) and using Y(A) and TKE(A) of Vorobyeva and $\sigma_{TKE}$(A) of Demattè (the green line in **Fig.4.1d**). In both PbP and FIFRELIN calculations for $^{242}$Pu(SF) the experimental Y(A), TKE(A) and $\sigma_{TKE}$(A) of Demattè were used.

In the case of $^{244}$Pu(SF) the calculated Y(A,TKE) using the experimental Y(A), TKE(A) and $\sigma_{TKE}$(A) data of Demattè is in reasonable agreement with the experimental data as it can be seen in **Fig.4.1e**.

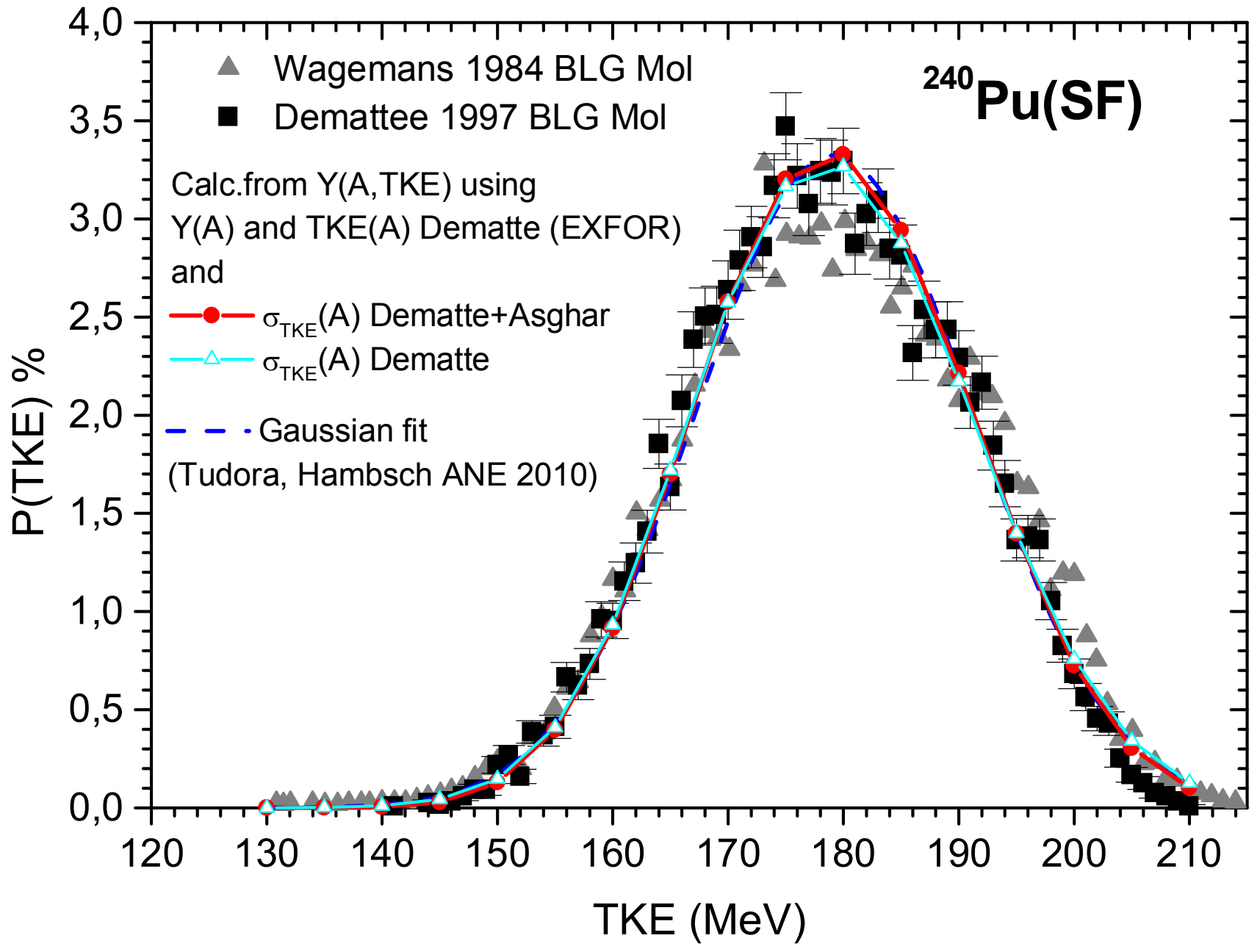

**Fig.4.1c**: $^{240}$Pu(SF) Experimental Y(TKE) taken from EXFOR and calculated Y(TKE) (solid lines) by using Y(A,TKE) obtained from experimental Y(A), TKE(A) and $\sigma_{TKE}$(A), a Gaussian fit with dashed line.

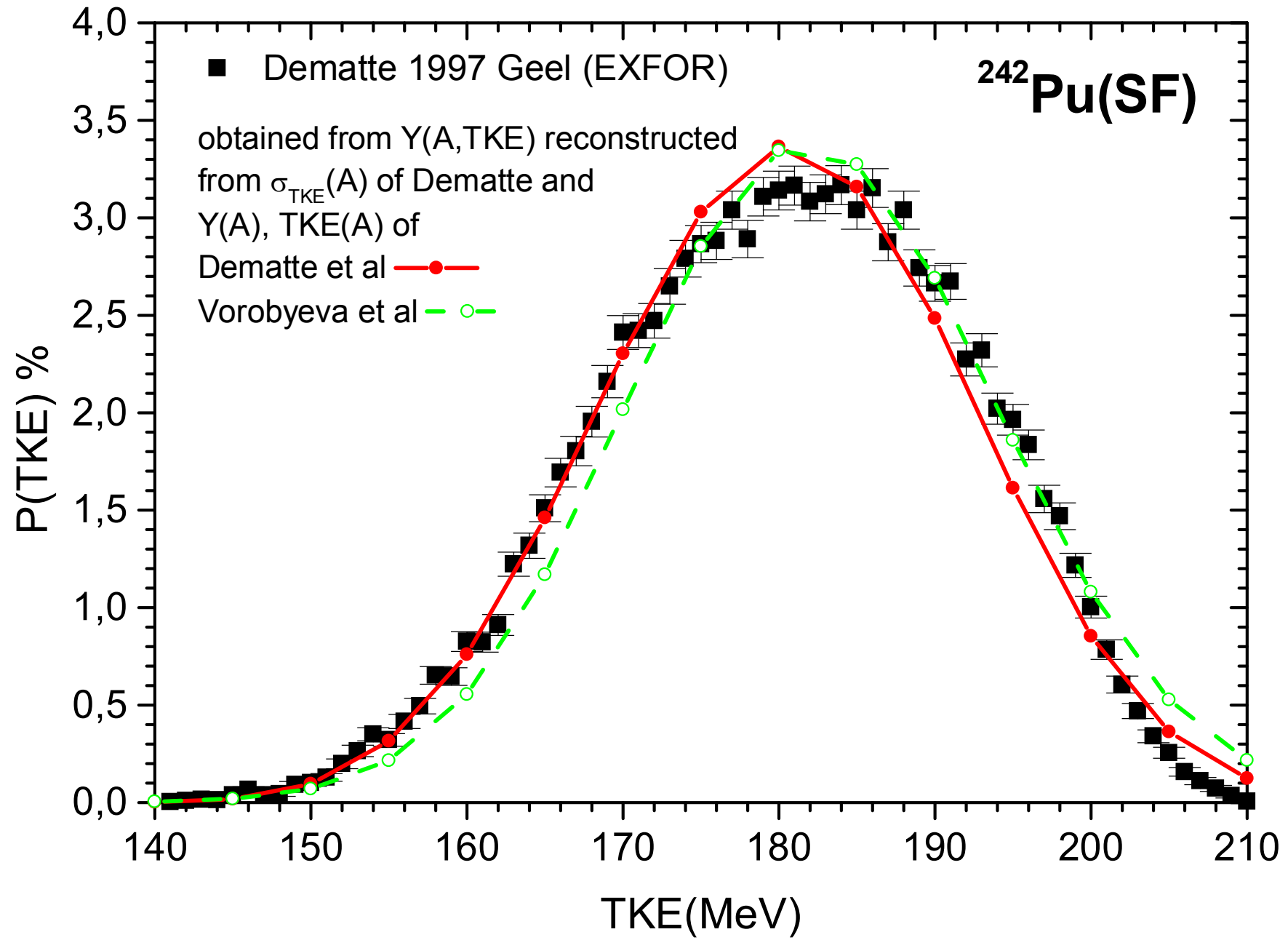

**Fig.4.1d**: $^{242}$Pu(SF) Experimental Y(TKE) taken from EXFOR and calculated Y(TKE) (solid and dashed lines) by using Y(A,TKE) obtained from different experimental Y(A), TKE(A) and $\sigma_{TKE}$(A).

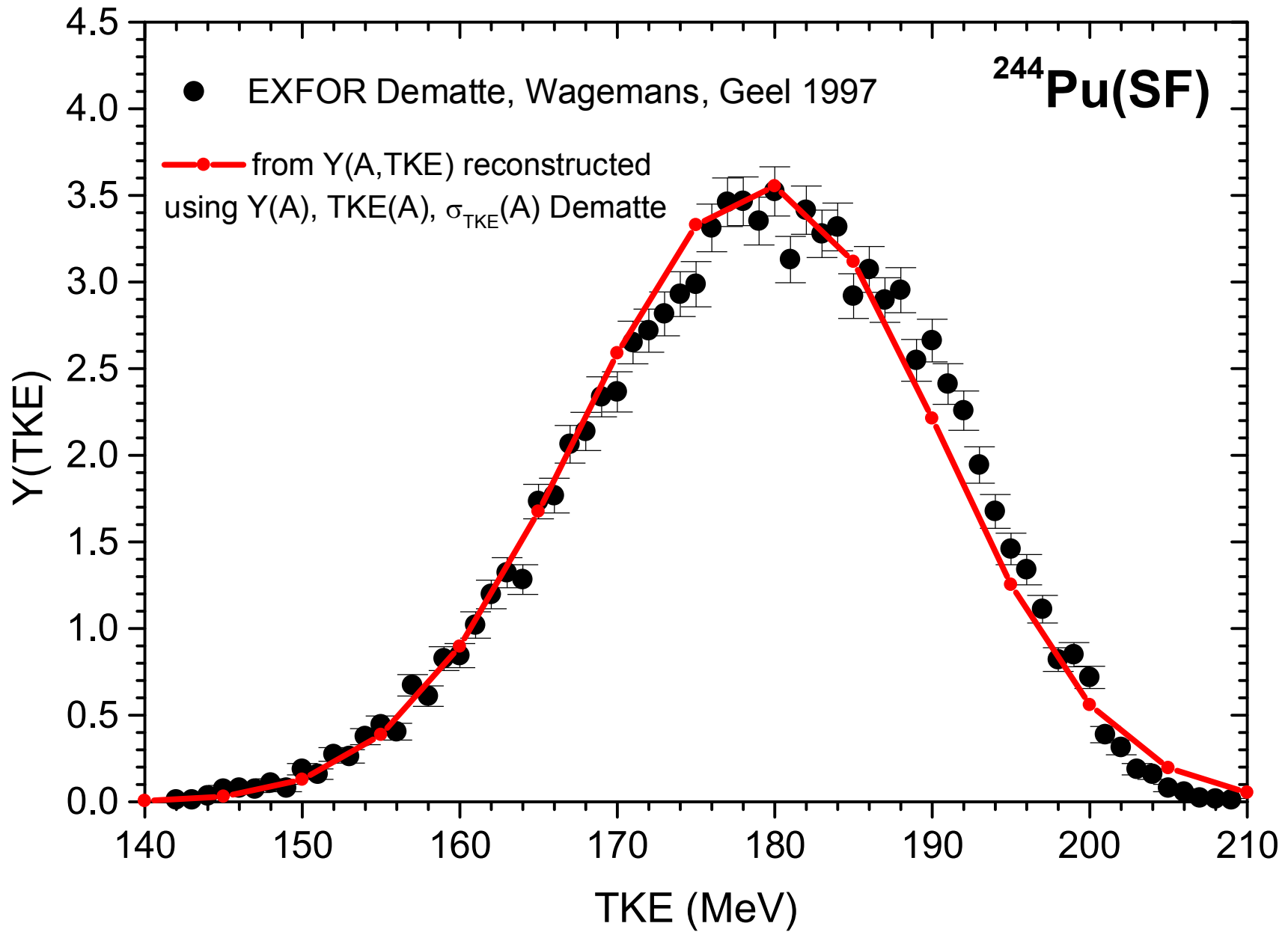

**Fig.4.1e**: $^{244}$Pu(SF) Experimental Y(TKE) taken from EXFOR and calculated Y(TKE) (solid line) by using Y(A,TKE) obtained from experimental Y(A), TKE(A) and $\sigma_{TKE}$(A) of Dematte

### IV.1.2. Comparison between FIFRELIN and PbP results

FIFRELIN calculations were performed in five cases as following:

1) considering a constant compound nucleus cross-section of the inverse process ($\sigma_c(\varepsilon)$) of neutron evaporation from fragments and a fraction of the moment of inertia equal to 0.4 of the rigid body momentum (according to eq. (3.14))

2) taking $\sigma_c(\varepsilon)$ variable with the energy, provided by optical model calculation (performed with the SCAT2 code of Bersillon, 1991 with subsequent developments (Manea, 2010)) using the phenomenological potential of Becchetti-Greenless taken from (RIPL–3, 2012a) and a fraction of the moment of inertia equal to 0.4 from the rigid body momentum.

3) considering $\sigma_c(\varepsilon)$ constant and using the moment of inertia results of theoretical HFB calculations performed at CEA-Bruyères-le-Châtel included in the AMEDEE database (AMEDEE, 2011)

4) using the moment of inertia values from the AMEDEE database and $\sigma_c(\varepsilon)$ provided by SCAT2 optical model calculations with the Becchetti-Greenless potential

5) using the moment of inertia values from the AMEDEE database and $\sigma_c(\varepsilon)$ provided by SCAT2 optical model calculations with the potential of Koning-Delaroche taken from (RIPL–3, 2012a)

For all the FIFRELIN calculations the unchanged charge distribution (UCD) was considered and the temperature ratio $RT=T_L/T_H$ as a function of $A_L$ described in Chapter III for $^{252}$Cf(SF) case was adapted according to the fragment mass range of each Pu(SF). The experimental FF distributions of Demattè, 1996 were used in all cases. The calculations were performed by sampling $10^7$ events for the calculation cases with constant $\sigma_c$ and $10^6$ events for the cases with $\sigma_c(\varepsilon)$ variable with energy.

In the case of the PbP calculations for all Pu(SF) the fragmentation range was build by taking the entire fragment mass range covered by the experimental distributions, for each mass number A two charge numbers were taken as the nearest integer values above and below the most probable charge that was taken as UCD for $^{236,238}$Pu(SF) cases and considering a charge polarization $\Delta Z$ of 0.5 (with + for light fragments and – for heavy ones) in the cases of $^{240,242,244}$Pu(SF) (according to Tudora and Hambsch, 2010 where PbP calculations for $^{240,242}$Pu(SF) were done by considering a charge polarization).

As it was mentioned in chapters II and III a major difference between the PbP and FIFRELIN consists in the treatment of the sequential emission. This is considered globally by the residual temperature distribution P(T) in the case of PbP and is taken sequentially up to the lower limit of fragment excitation energy taken as *Sn+Erot* in the Monte-Carlo treatment. Other difference maybe less significant than the sequential emission consists in different methods of TXE partition used in the PbP and FIFRELIN treatments (see chapters II and III).

The manner to calculate the energy release and the neutron separation energy from fragments is the same, by using mass excesses from Audi and Wapstra database (RIPL–3, 2012b).

Both treatments PbP and Monte-Carlo (FIFRELIN) use the generalized super-fluid model of Ignatiuk, 1998 to calculate the level density parameter of fragments. In the PbP treatment super-fluid model calculations are done two times: at the scission moment in an iterative procedure under the condition of thermodynamic equilibrium at scission and at full acceleration at the fragment excitation energies resulted from the TXE partition.

In the code FIFRELIN super-fluid model calculations are done in an iterative procedure, too, this time under the condition of fragment temperatures satisfying the RT function.

As it was mentioned in chapter III some little differences between the super-fluid model versions applied in PbP and FIFRELIN appear in the case of the dumping factor $\gamma$ and the parameterization of the asymptotic level density parameter.

Consequently the comparison between the PbP and FIFRELIN results regarding different prompt emission quantities as well as their comparison with experimental data when they exist can lead to interesting conclusions regarding the physical assumptions, the models and the procedures used in each case.

### IV.1.2.1 FIFRELIN and PbP results of P(ν)

The existing experimental P(ν) data for $^{236,238,240,242}$Pu(SF), majority given in EXFOR, 2012d and also reported by Santi et al, 2005; Santi and Miller, 2008, allow comparisons with the FIFRELIN and PbP results, as it can be seen in **Figs.4.2a-d**.

In these figures the experimental P(ν) data are plotted with different black and wine symbols. The results of FIFRELIN calculations done in the five cases mentioned above are plotted with solid lines (black or gray for case 1, orange for case 2, green for case 3, blue for case 4 and cyan for case 5). The PbP results are plotted with red circles (connected with solid lines) for P(ν) calculations done by considering a single Y(A) distribution and with magenta stars (also connected with solid lines) when a double distribution Y(A,TKE) was used.

As it can be seen in all **Figures 4.2a-d**, in the case of FIFRELIN calculations the P(ν) results of case 1 and 2 (moments of inertia taken as a fraction of 0.4 of rigid body momentum and $\sigma_c(\varepsilon)$ constant or variable) are close to each other and differ from P(ν) results of case 3, 4, 5 ((all using moment of inertia values from the database AMEDEE and $\sigma_c(\varepsilon)$ constant or given by optical model calculations with different potentials) that are also close each other. These facts reveal that for the models used in the FIFRELIN code the moment of inertia calculation has more significant influence on P(ν) results compared to $\sigma_c(\varepsilon)$.

Concerning the PbP calculations, as it can be seen in **Figs.4.2 a,b**, the P(ν) results obtained by using Y(A) and Y(A,TKE) visibly differ to each other, a better agreement with the experimental data being obtained when Y(A) is used.

In the case of $^{236}$Pu(SF), as it can be seen in **Fig.4.2a**, the FIFRELIN results using the moments of inertia from the AMEDEE database and the PbP result using Y(A) are in a reasonable agreement with the experimental data (exhibiting large error bars). In the case of $^{238}$Pu(SF) (see **Fig.4.2b**) the P(ν) result of PbP (using Y(A)) and the FIFRELIN results (cases with moment of inertia values taken as a fraction of 0.4 from the rigid body momentum) succeeded to describe rather well the experimental data of Holden and Santi.

For $^{240,242}$Pu(SF), as it can be seen in **Figs.4.2c,d**, the PbP results (plotted with red circles) already reported in (Tudora and Hambsch, 2010), describe very well the experimental data. All FIFRELIN results (five calculation cases) are this time rather close to each other and do not succeed to describe the experimental data.

A possible explanation of this visible disagreement with the experimental data is the consideration of UCD in the case of FIFRELIN while in the case of PbP (giving P(ν) in excellent agreement with experimental data) the fragmentation range was obtained by considering also a charge polarization.

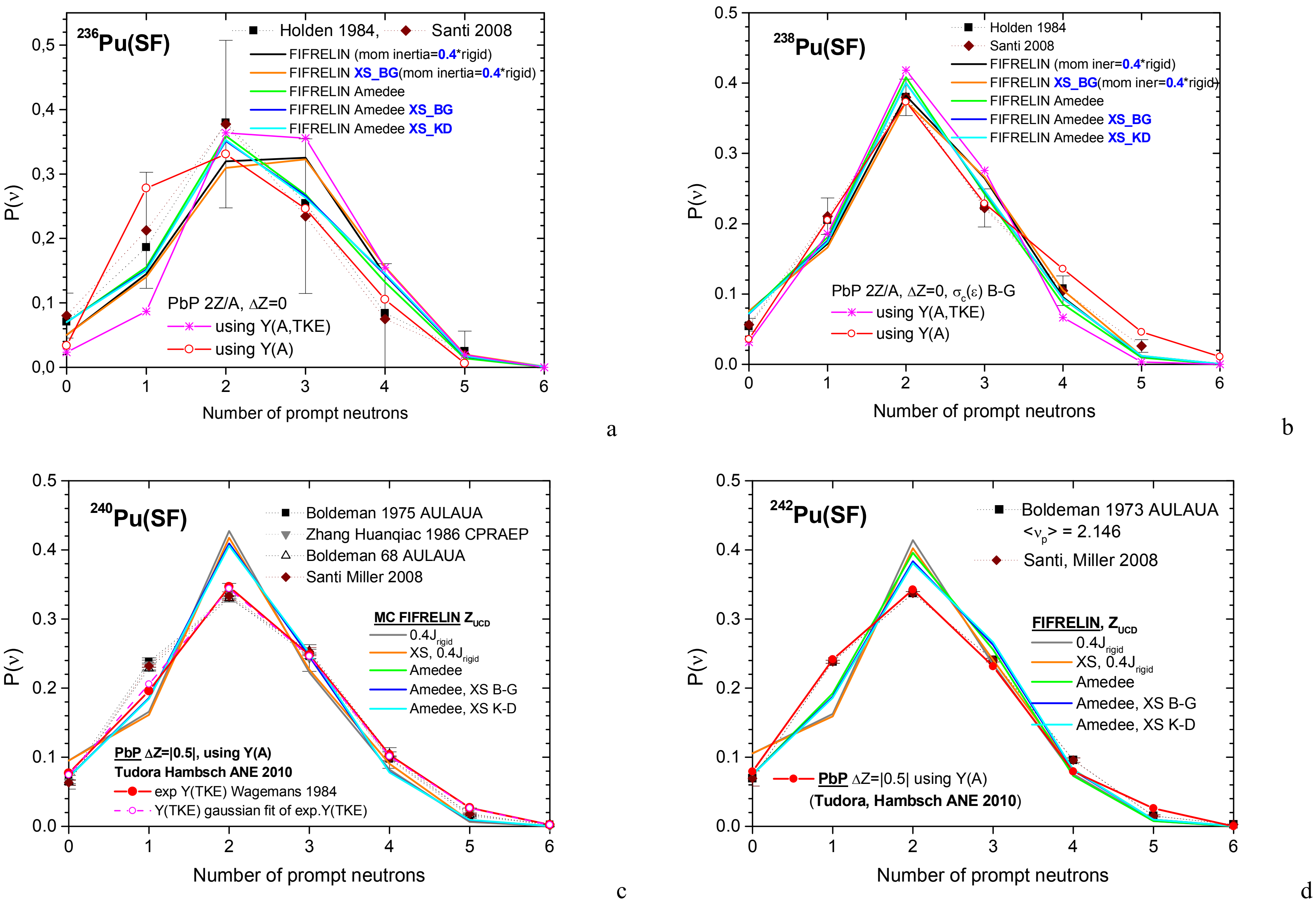

**Fig.4.2**: FIFRELIN and PbP calculations of P(ν) for $^{236,238,240,242}$Pu(SF) in comparison with experimental data

### IV.1.2.2 FIFRELIN and PbP results of <ν>(TKE)

Even if experimental <ν>(TKE) data of $^{236\text{-}244}$Pu(SF) do not exist to allow a validation of each model calculations, the comparison between the results of <ν>(TKE) provided by the two different treatments (PbP and Monte-Carlo) can reveal interesting facts, too.

<ν>(TKE) calculations with the FIFRELIN code were done for the five cases mentioned above and in all cases ν(Z,A,TKE) was averaged over the Gaussian charge distribution and over the double distributions Y(A,TKE) reconstructed from the experimental single ones (according to eq.(4.1)). It is interesting to mention that the FIFRELIN results of <ν>(TKE) of the five calculation cases are close to each other. For this reason in the figures referring to <ν>(TKE) only two relevant cases are plotted: the case considering a constant $\sigma_c(\varepsilon)$ and moments of inertia taken 0.4 from rigid body momentum and the case with $\sigma_c(\varepsilon)$ provided by optical model calculations potential and moments of inertia from the database AMEDEE). The <ν>(TKE) results of FIFRELIN for $^{236\text{ - }244}$Pu(SF) are plotted in **Figs.4.3a-e** with full blue and red open circles respectively.

In the same figures the PbP results of <ν>(TKE) are plotted with full black circles connected with thin dotted lines (to guide the eye) and the results of the most probable fragmentation approach with black solid lines. The PbP results were obtained by averaging the multi-parametric matrices ν(Z,A,TKE) over the same charge distribution as in FIFRELIN and over the double distributions Y(A,TKE) (reconstructed from the single ones as mentioned in Section IV.1.1).

An overall good agreement between the PbP and FIFRELIN results is obtained for all Pu(SF). In the case of $^{236,238}$Pu(SF), as it can be seen in **Figs.4.3a,b**, the PbP and FIFRELIN results of <ν>(TKE) are very close each other for TKE values above 160 MeV. At lower TKE values the PbP result exhibits a smooth flattening of <ν> while the FIFRELIN ones have another shape of this flat behaviour and exhibit structures at TKE values below 140 MeV. In principle similar behaviours of <ν>(TKE) can be observed for other Pu(SF), too (see **Figs.4.3c-e**) with the mention that for TKE above 160 MeV the agreement between PbP and FIFRELIN is not so good as in the case of $^{236,238}$Pu(SF). The slope dTKE/dν of PbP results is a little bit lower than the slope of FIFRELIN results. We assume that for $^{240,242,244}$Pu(SF) the less good agreement between the PbP and FIFRELIN results can be due to the consideration of a charge polarization in the case of PbP compared to FIFRELIN where only UCD is taken.

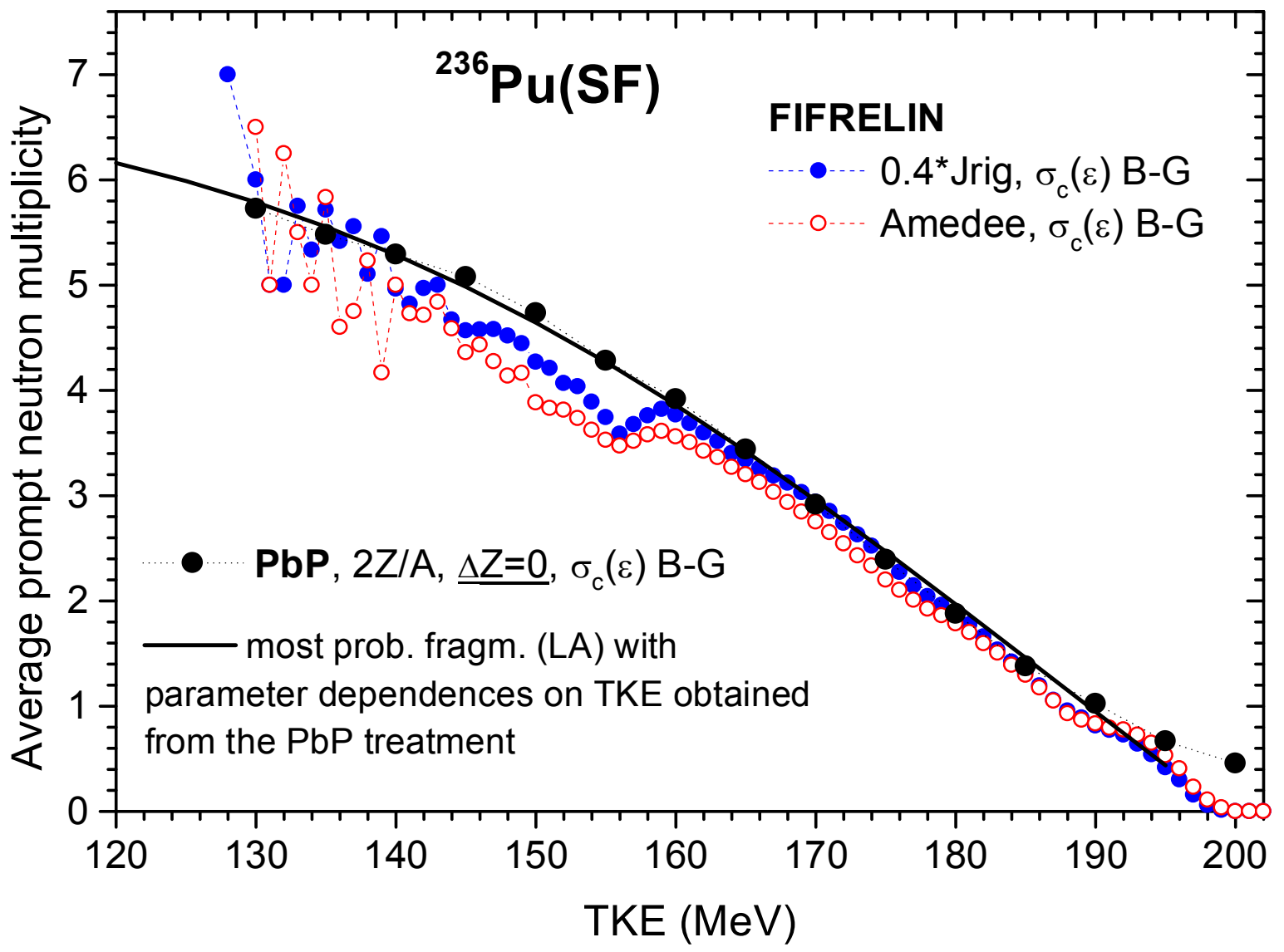

**Fig.4.3a**: $^{236}$Pu(SF): <ν>(TKE) FIFRELIN calculations using $\sigma_c(\varepsilon)$ form optical model calculation with Becchetti-Greenless potential are plotted with full blue circles ( case of moment of inertia 0.4 from rigid body momentum) and open red circles (moment of inertia from the AMEDEE database). PbP calculations with full black points, most probable fragmentation result with black solid line.

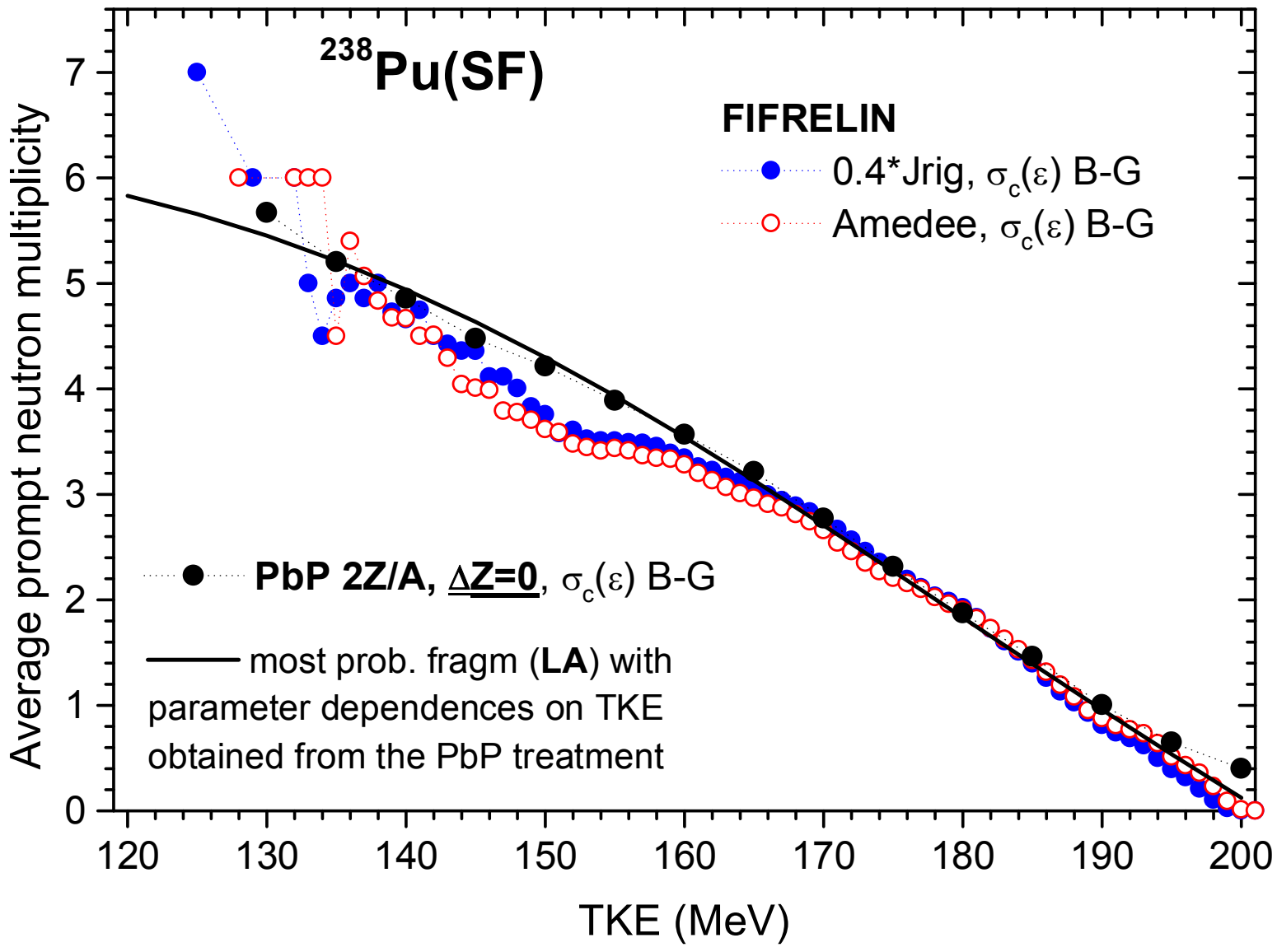

**Fig.4.3b**: $^{238}$Pu(SF): <ν>(TKE) FIFRELIN calculations using $\sigma_c(\varepsilon)$ form optical model calculation with Becchetti-Greenless potential are plotted with full blue circles ( case of moment of inertia 0.4 from rigid body momentum) and open red circles (moments of inertia from the AMEDEE database). PbP calculations with full black points, most probable fragmentation result with black solid line.

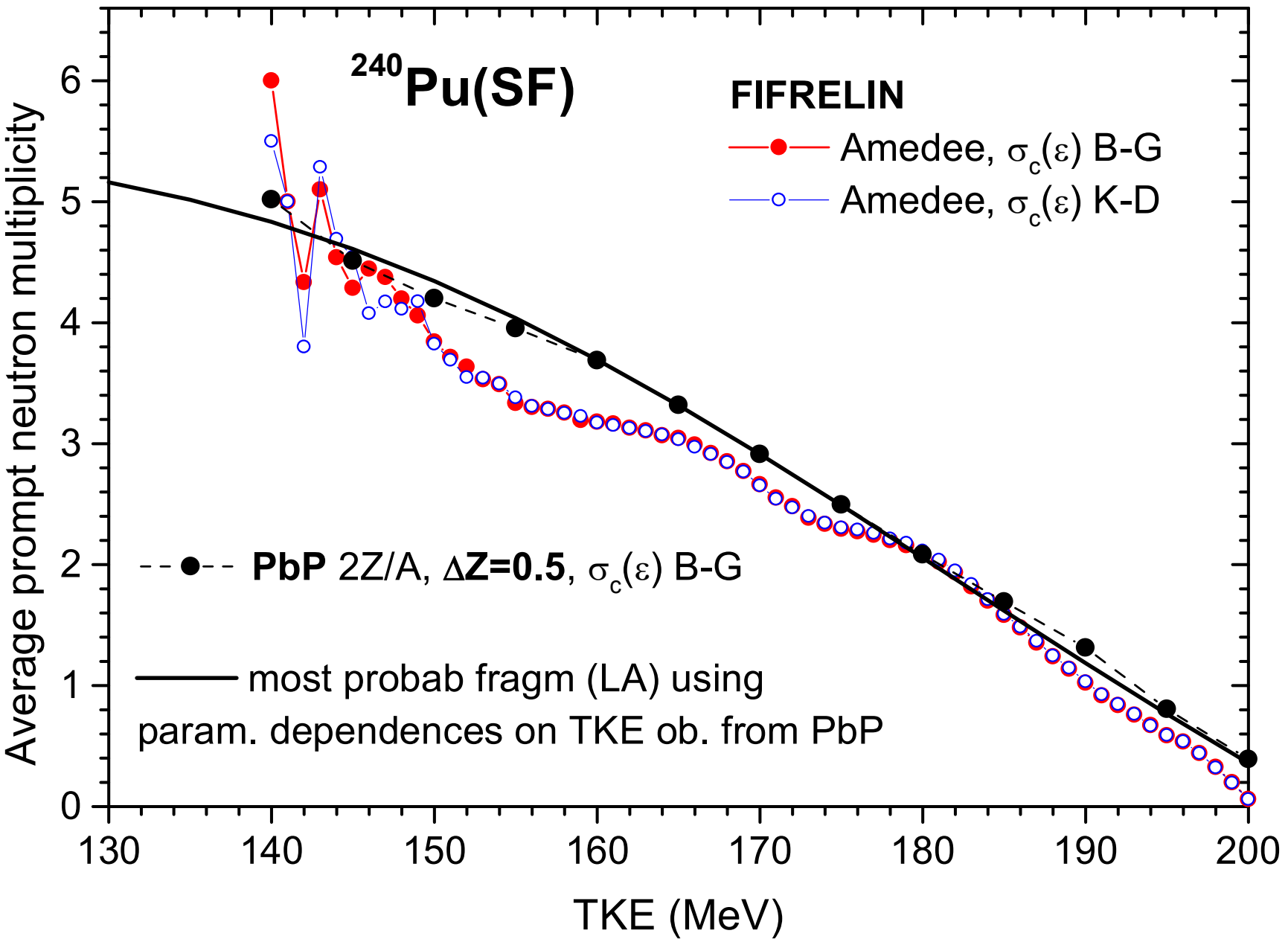

**Fig.4.3c**: $^{240}$Pu(SF): <ν>(TKE) FIFRELIN results with full blue and open red circles (cases with AMEDEE and $\sigma_c(\varepsilon)$ calculated with Becchetti-Greenless and Koning-Delaroche optical potentials). PbP calculations with full black points, most probable fragmentation result with black solid line.

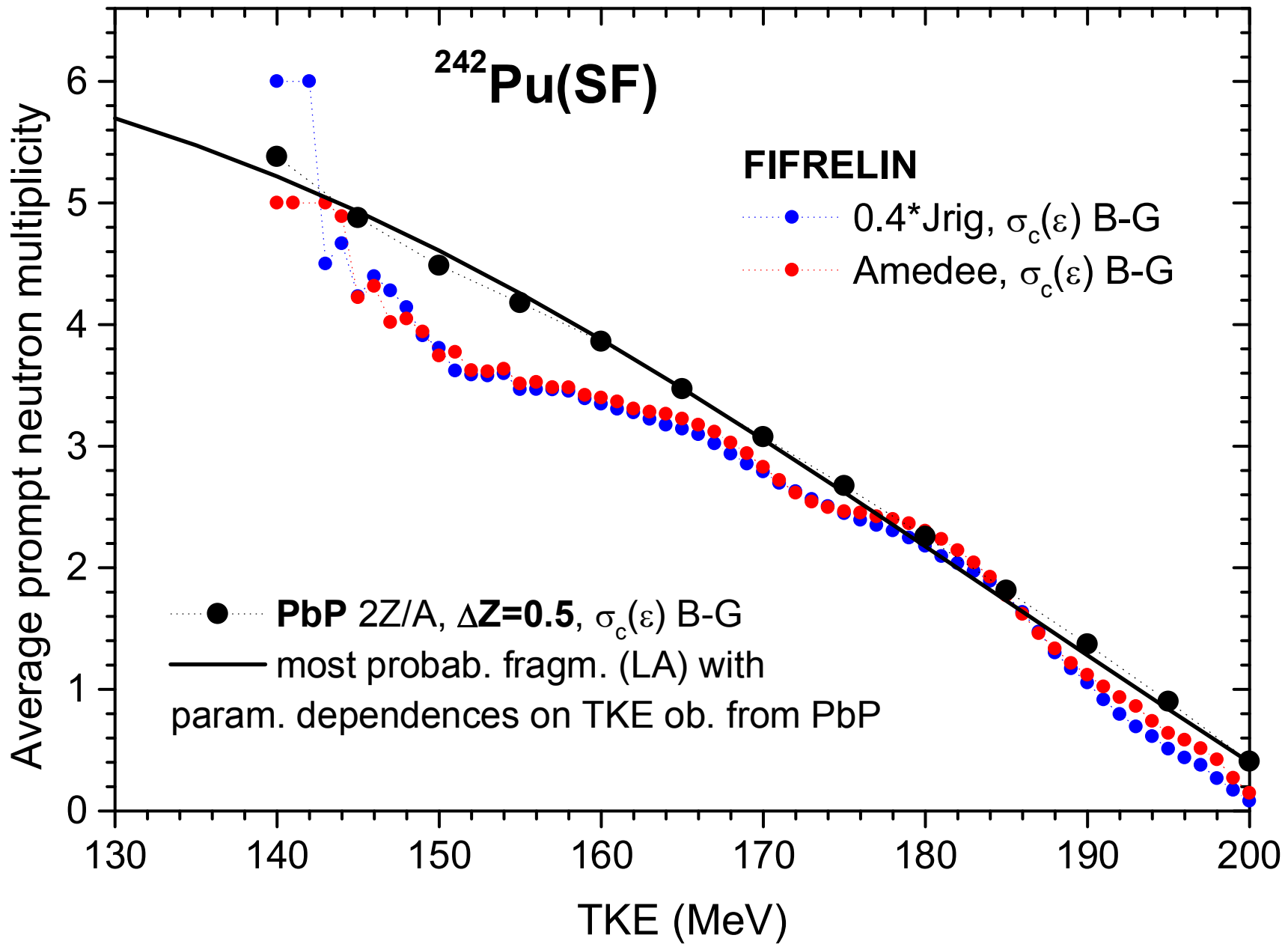

**Fig.4.3d**: $^{242}$Pu(SF): <ν>(TKE) FIFRELIN results with full blue and open red circles (same calculation cases as in Fig.4.3a,b). PbP calculations with full black points, most probable fragmentation result with black solid line.

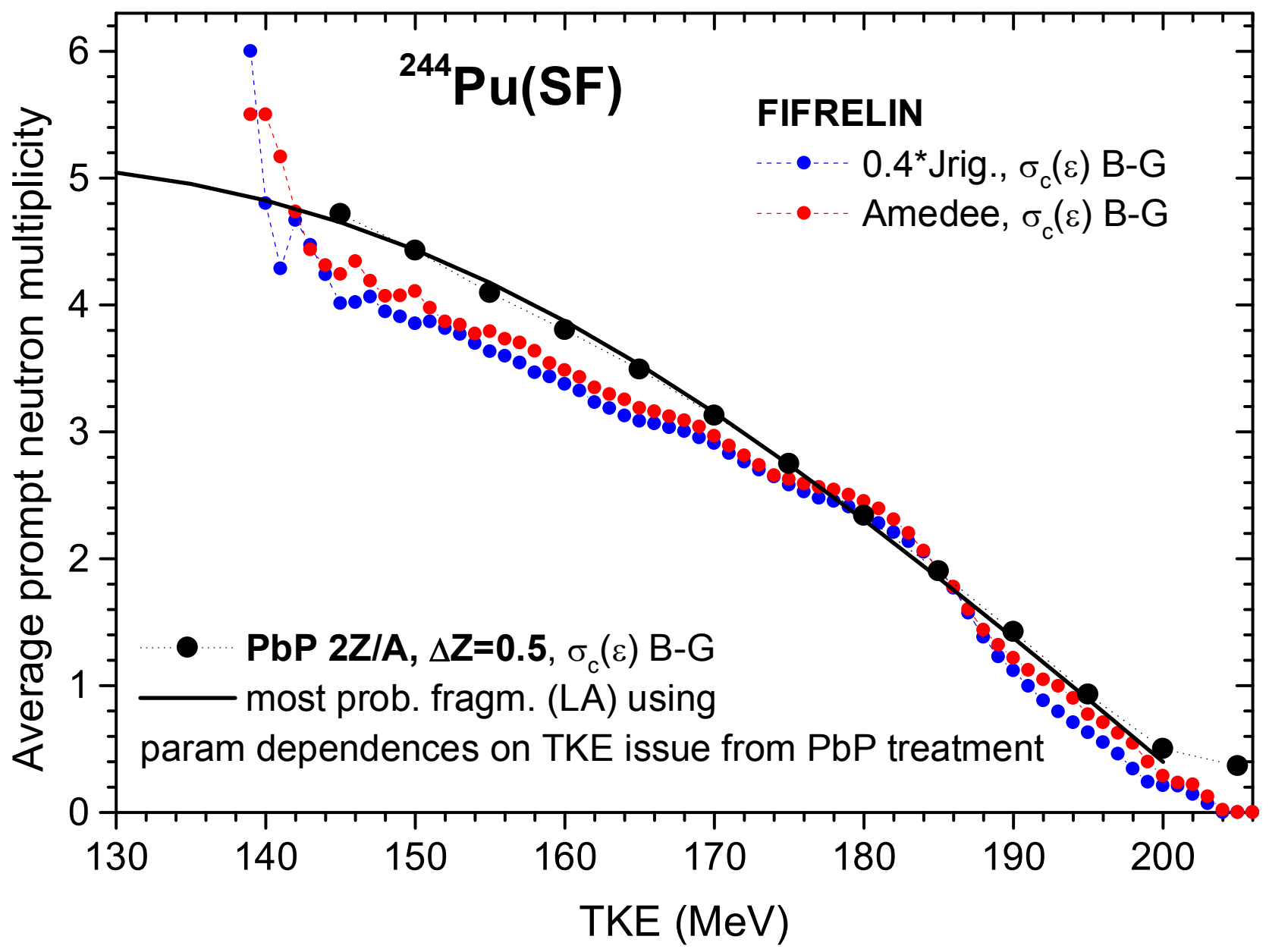

**Fig.4.3e**: $^{244}$Pu(SF): <ν>(TKE) FIFRELIN results with full blue and open red circles (same calculation cases as in Fig.4.3a,b,d). PbP calculations with full black points, most probable fragmentation result with black solid line.

PbP calculations of <ν>(TKE) by averaging ν(A,TKE) over the Y(A) distribution (independent on TKE) were done for $^{236,238}$Pu(SF). These results are plotted in **Fig.4.4a,b** (upper parts) with green stars (connected with dashed lines to guide the eye). The same behaviour as in other studied cases, such as $^{252}$Cf(SF) (discussed in detail in Regnier et al, 2012b, and Tudora, 2012a, 2012b) and $^{235}$U($n_{th}$,f) (discussed in Tudora, 2012a, 2012b) are obtained.

The effect of the approximation by averaging over a distribution Y(A) independent on TKE is an almost linear behaviour of <ν> over the entire TKE range (without a flattening of <ν> at low TKE values) and a lower slope dTKE/dν compared to the case of averaging over a TKE dependent distribution Y(A,TKE). In **Figs.4.4a,b** the values of inverse slopes dTKE/dν obtained from the linear fit of <ν>(TKE) results of FIFRELIN (both plotted calculation cases) and of PbP results by averaging over Y(A,TKE) and over Y(A) are written with the same colors as the corresponding plotted <ν>(TKE).

In **Figs.4.4c,d** the PbP calculations of <ν> as a function of TKE are plotted this time versus <TXE> using the same symbols and colors as in **Figs.4.4a,b**. A linear increase with practically the same slope is observed in both cases (by averaging over Y(A,TKE) and over Y(A)).

The values of the inverse slopes dTXE/dν obtained from the linear fit of <ν>(TXE) (full black circles and green stars) are also given with the same colors as the corresponding <ν>(TXE) plotted in **Figs.4.4c,d**. It is also interesting to observe that the values of the increasing slope dTXE/dν are very close to the absolute values of the decreasing slope dTKE/dν obtained in the approximation of averaging over the single distribution Y(A) (independent on TKE).

FIFRELIN calculations of <ν>(TKE) for the case of $^{252}$Cf(SF) recently reported in (Regnier et al, 2012b) revealed that a large number of events sampled in the Monte-Carlo treatment leads to an almost linear behaviour of <ν> over the entire TKE range without the flattening of <ν> at low TKE values.

Consequently the behaviour exhibited at low TKE values by the present FIFRELIN results of Pu(SF) (flattening of <ν> and structure below 140 MeV) can be due to the number of events sampled in the Monte-Carlo treatment.

Also it is interesting to observe that in the case of $^{252}$Cf(SF) the slope dTKE/dν recently obtained by Regnier et al, 2012b by sampling a large number of events is placed between the slopes dTKE/dν previously obtained by averaging over Y(A,TKE) (in agreement with Budtz-Jorgensen experimental data) and over Y(A) (in agreement with Zeynalov data). Therefore we can say that an almost linear behaviour of <ν> over the entire TKE range with a lower slope, obtained by sampling a very large number of events in a Monte-Carlo treatment, can explain on one side the PbP results obtained in the approximation of averaging over a distribution Y(A) independent on TKE (as reported for many fissioning systems ($^{252}$Cf(SF), $^{233,235}$U($n_{th}$,f), $^{244,248}$Cm(SF) in Tudora, 2008, 2010b, 2012a) and on the other hand the behaviour of a few experimental data sets (such as of Zeynalov, 2011 and van Aarle et al, 1994 as discussed in Tudora, 2012a).

As observation, the shape of <ν>(TKE) especially at low TKE values depends also on the function RT($A_L$) used in the code FIFRELIN, as it will be described in detail in the next part of this chapter.

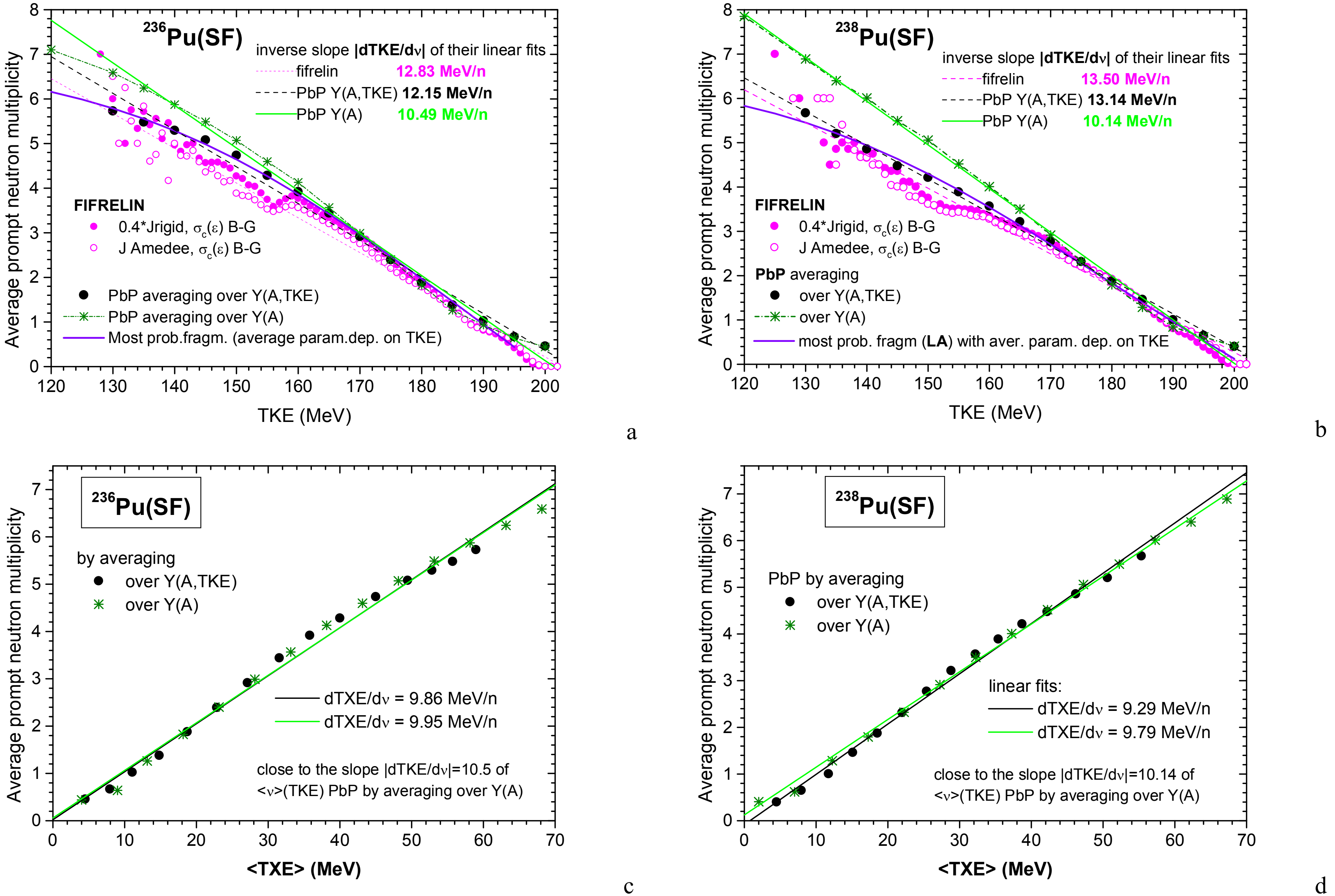

**Fig.4.4**: Inverse slopes dTKE/dν of FIFRELIN and PbP calculations of <ν>(TKE) by averaging over Y(A,TKE) and Y(A)

### IV.1.2.3 Average parameter dependences on TKE obtained from the PbP treatment used in the most probable fragmentation approach

In the frame of the PbP treatment average model parameters as a function of TKE can be easily obtained by averaging the multi-parametric matrices corresponding to the respective parameter as following (Tudora, 2012a,b, Tudora et al, 2012c):

$$< param > (TKE) = \sum_{Z,A} param(Z,A,TKE)\, p(Z,A)\, Y(A,TKE) \Big/ \sum_{Z,A} p(Z,A)\, Y(A,TKE) \qquad (4.3)$$

where "*param*" refers to one of the following quantities: energy release (*Er*), average fragment neutron separation energy (*Sn*) and fragment level density parameter (*a*).

As in previous studied cases (reported in Tudora 2012a,b) for the present studied systems $^{236\text{-}244}$Pu(SF) the obtained average parameters $<Er>$, $<Sn>$, $<a>$ (given traditionally as $<C>=A_0/<a>$ with $A_0$ the mass number of the fissioning nucleus) as a function of TKE exhibit nice and regular behaviours that can be fitted well, see the black circles in **Figs.4.5-4.9**. Their appropriate fits are also given in these figures (with solid red lines).

The obtained parameter dependences on TKE (given in the legends of **Figs.4.5-4.9**) allow the use of the "most probable fragmentation approach" (meaning the Los Alamos model with subsequent improvements), the resulted $<\nu>$(TKE) being plotted with solid black lines in **Figs.4.3a-e** and **4.4a,b**.

The use of the Los Alamos model has as major advantage the possibility to obtain $<\nu>$ at many TKE values in a very short computing time compared to the PbP and Monte Carlo (FIFRELIN) treatments.

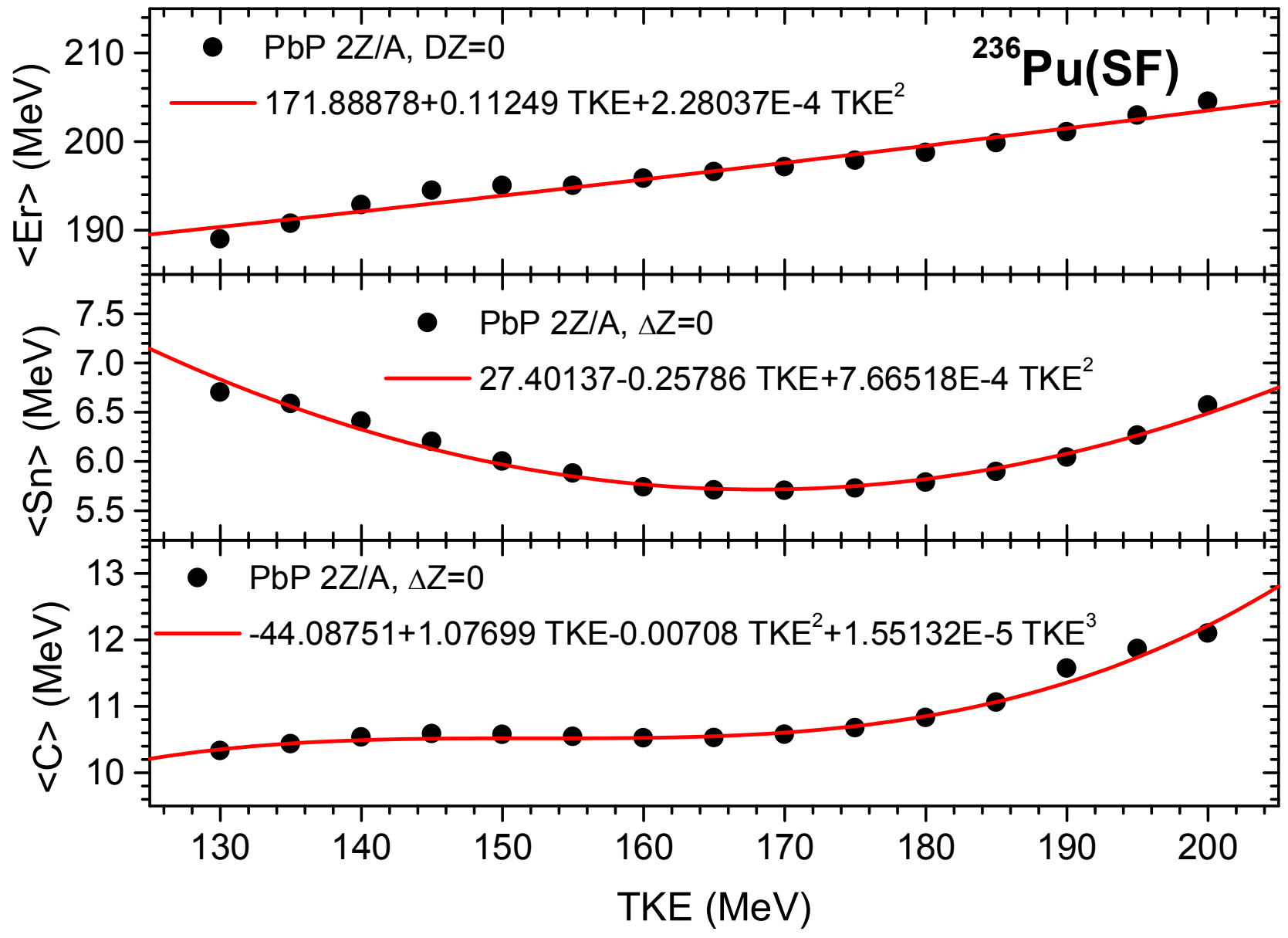

**Fig.4.5**: $^{236}$Pu(SF) Average model parameters resulted form the PbP treatment (full circles) and appropriate fits (solid lines)

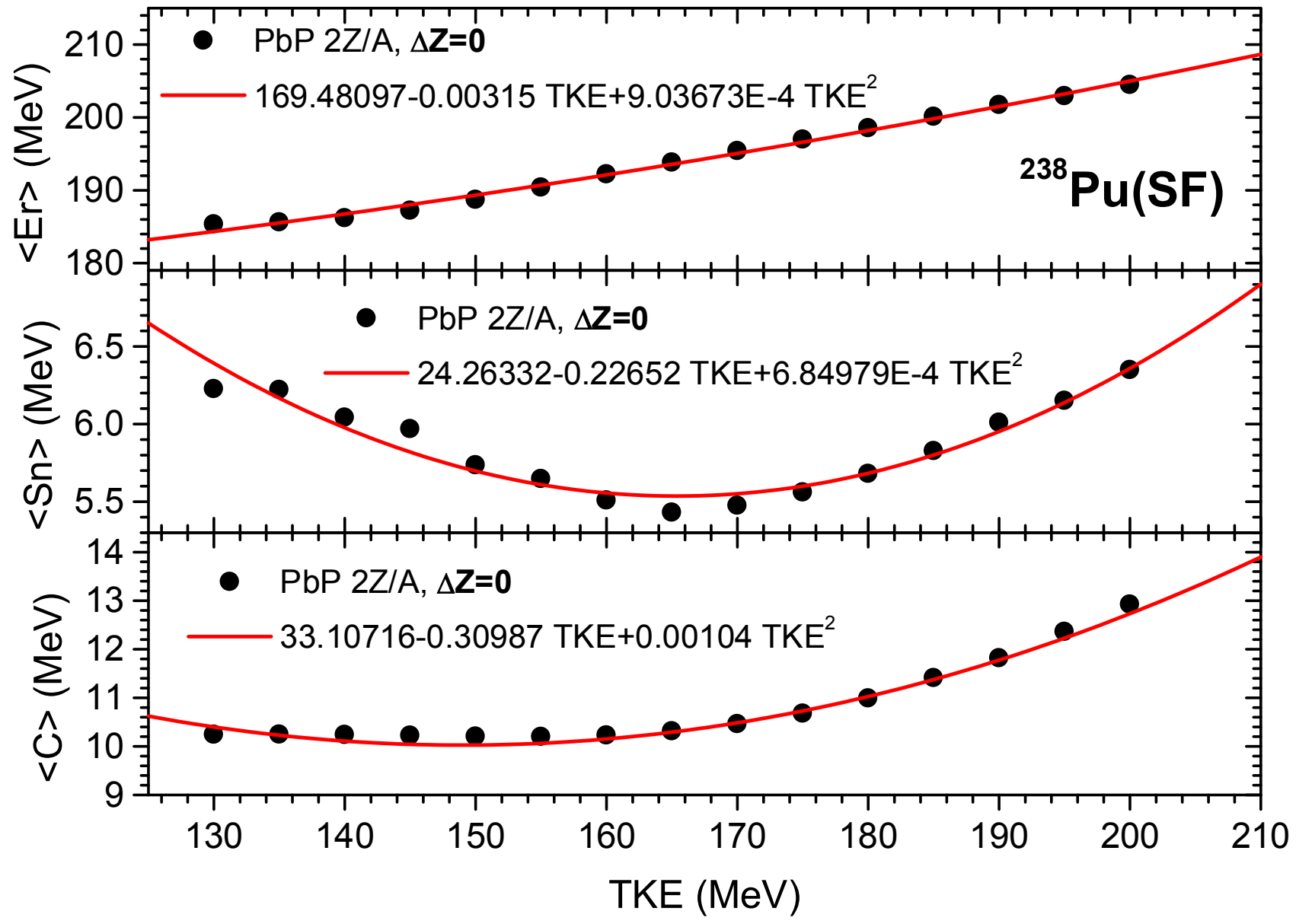

**Fig.4.6**: $^{238}$Pu(SF) Average model parameters resulted form the PbP treatment (full circles) and appropriate fits (solid lines)

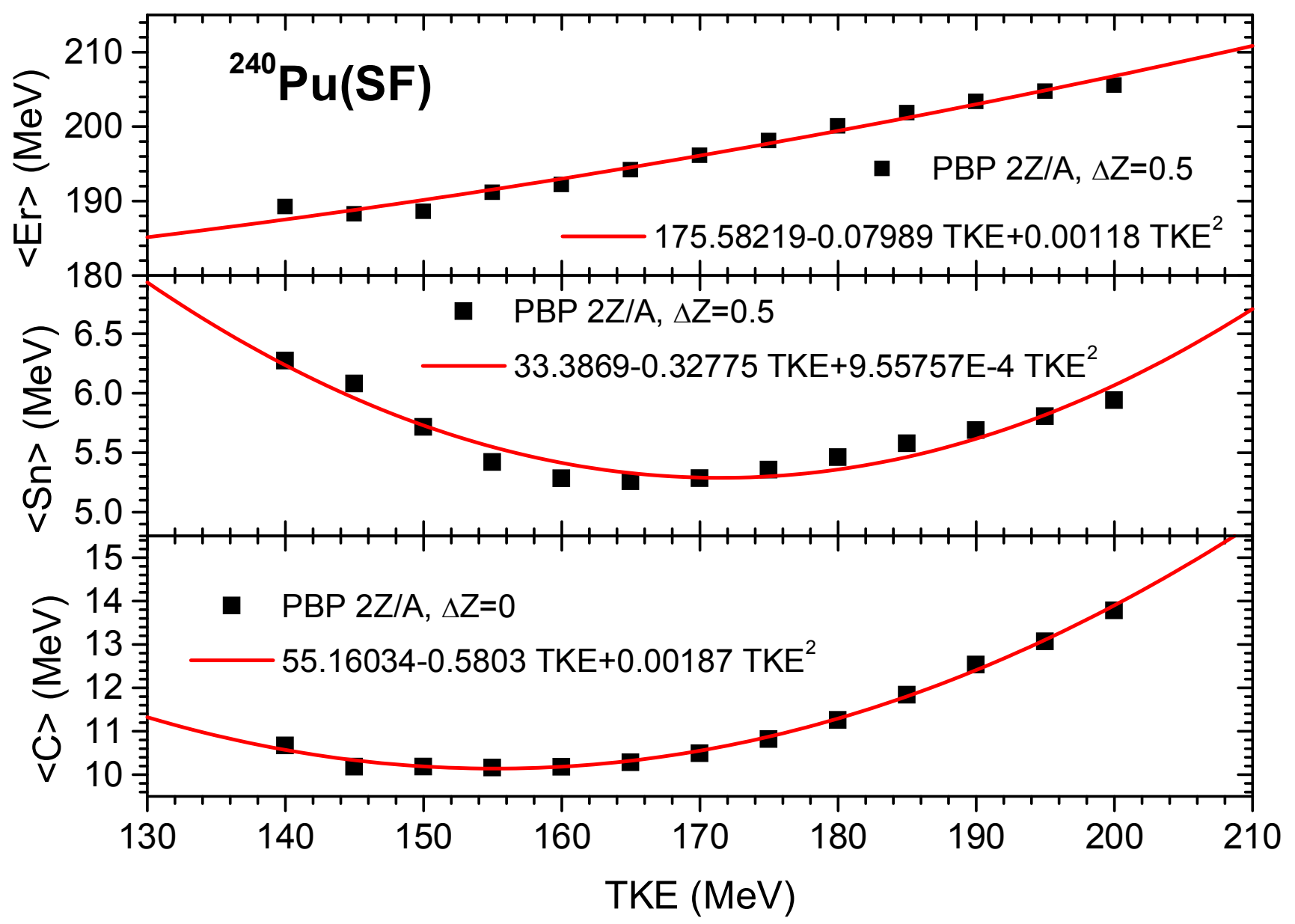

**Fig.4.7**: $^{240}$Pu(SF) Average model parameters resulted form the PbP treatment (full circles) and appropriate fits (solid lines)

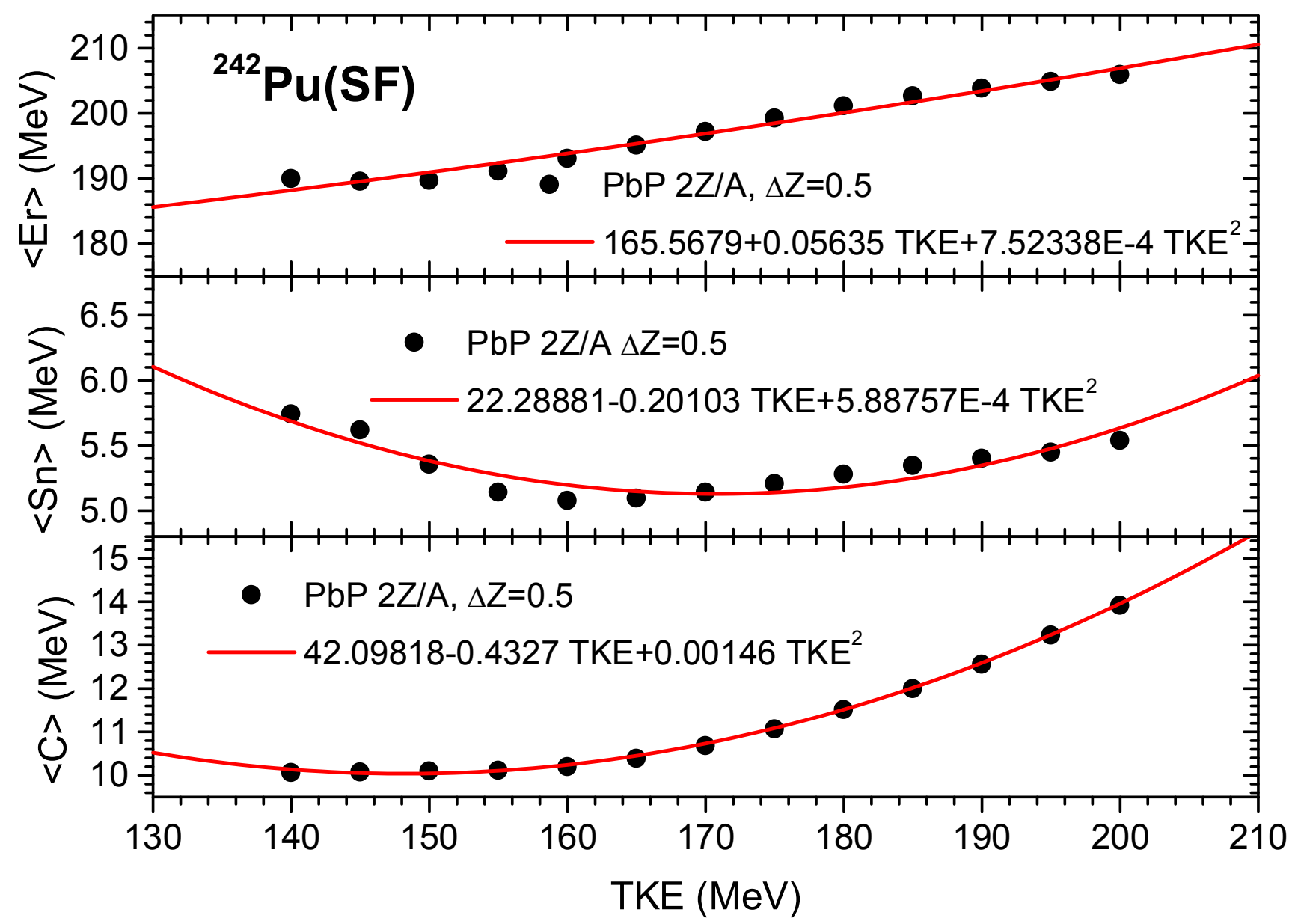

**Fig.4.8**: $^{242}$Pu(SF) Average model parameters resulted form the PbP treatment (full circles) and appropriate fits (solid lines)

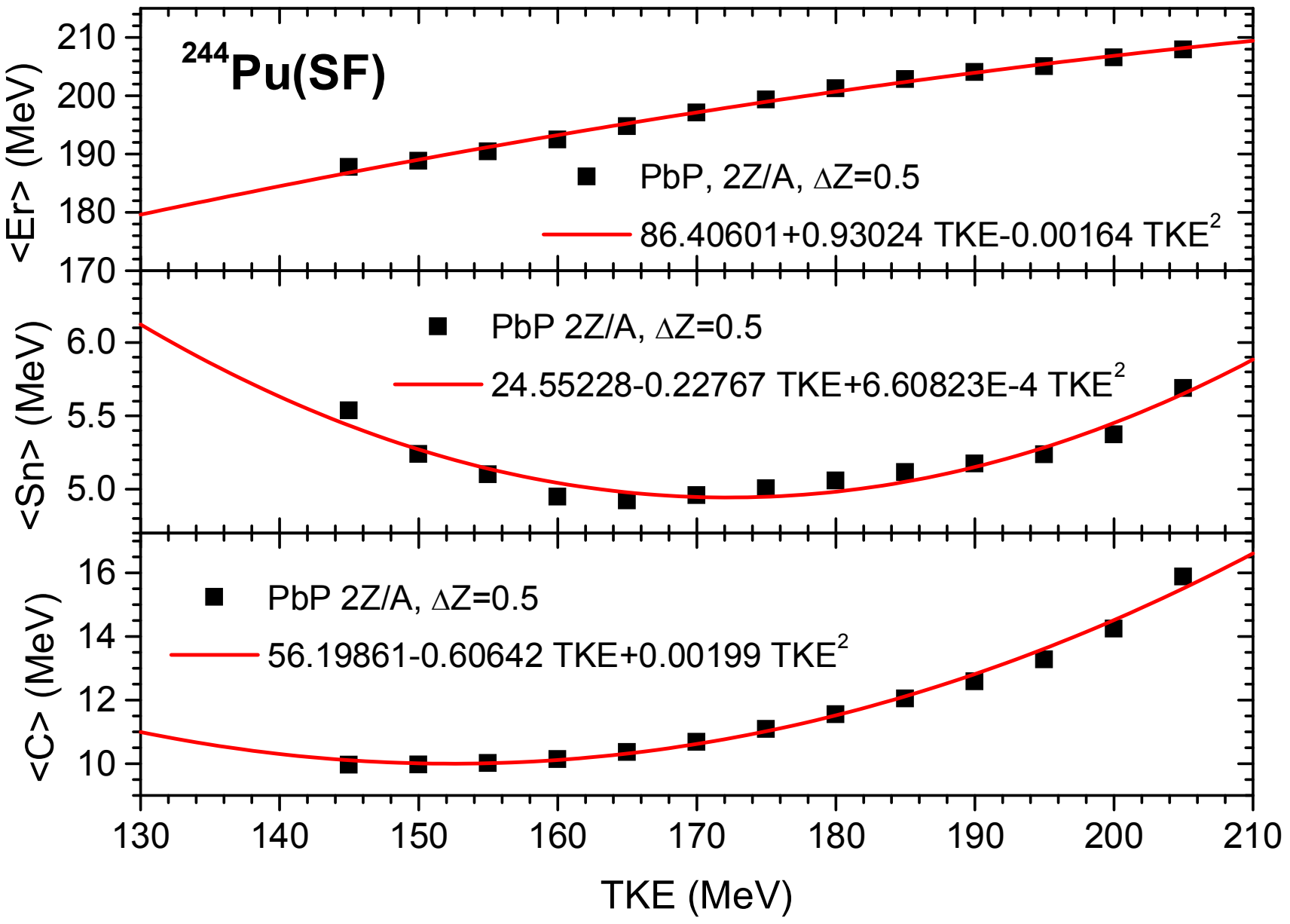

**Fig.4.9**: $^{244}$Pu(SF) Average model parameters resulted form the PbP treatment (full circles) and appropriate fits (solid lines)

### IV.1.2.4 Total average prompt emission quantities obtained by FIFRELIN and PbP calculations

For application purposes one of the most important prompt fission data is the total average prompt neutron multiplicity $<\nu_p>$.

The comparison of $<\nu_p>$ results provided by FIFRELIN and PbP for $^{236-244}$Pu(SF) with existing experimental data (EXFOR 2001, 2012e, Santi et al, 2005; Santi and Miller, 2008) is synthesized in **Fig.4.10**, where the total average multiplicity is plotted as a function of the mass number of the fissioning nucleus. The old and recent data from EXFOR are plotted with gray and black squares and the data of Santi and Miller with green triangles. In this figure only the FIFRELIN result of the calculation case 4 (moment of inertia values from the AMEDEE database and $\sigma_c(\varepsilon)$ provided by optical model calculations with the Becchetti-Greenless potential) is plotted with cyan diamonds. The PbP results are given with red circles.

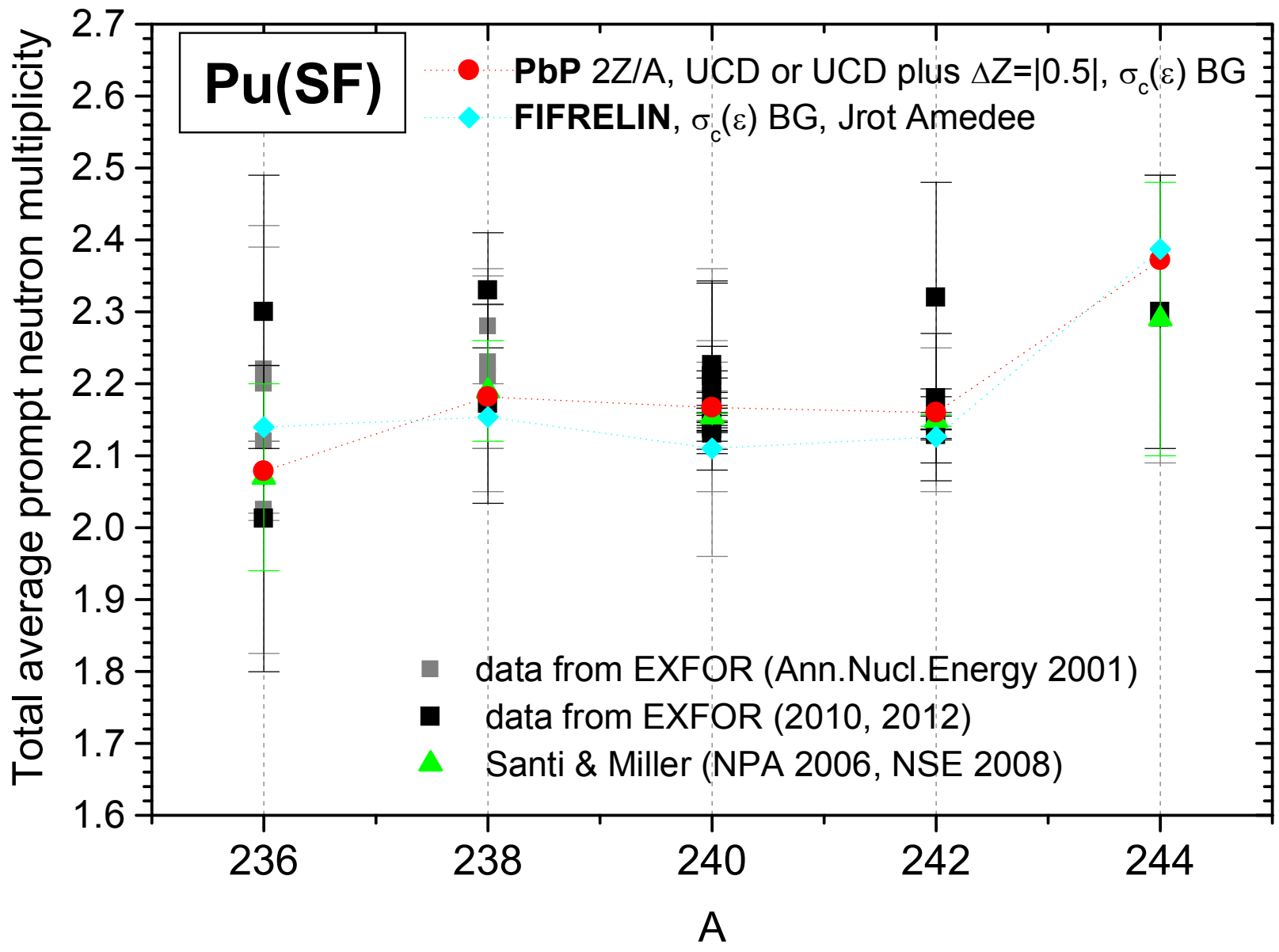

**Fig.4.10**: $^{236,238,240,242,244}$Pu(SF) total average prompt neutron multiplicity as a function of fissioning nucleus mass number: PbP results (red circles) and FIFRELIN results (cyan diamonds) in comparison with experimental data (black and gray squares and green triangles).

As it can be seen both FIFRELIN and PbP results are rather close to each other and in good agreement with the experimental data. In the case of $^{244}$Pu(SF) the PbP and FIFRELIN treatments provide practically the same result for $<\nu_p>$. The maximum difference between the PbP and FIFRELIN results is of 2.86% in the case of $^{236}$Pu(SF).

In the case of total average prompt gamma-ray energy $<E\gamma>$ significant differences between the PbP and FIFRELIN results occur, see **Fig.4.11** where $<E\gamma>$ is plotted versus the mass number of the fissioning nucleus. The PbP results are plotted with full red circles and the FIFRELIN results of the five calculation cases with different other symbols as indicated in the figure's legend. The dotted lines connecting the plotted points are only to guide the eye.

As it can be seen the PbP results of $<E\gamma>$ are visibly lower than all FIFRELIN results and exhibit a slight increase with the mass number of the fissioning nucleus while all FIFRELIN results are decreasing with the mass number, this decrease being very pronounced in the case of FIFRELIN calculations made by using the moment of inertia values from the AMEDEE database. Only for $^{244}$Pu(SF) $<E\gamma>$ results of all FIFRELIN calculations are very close each other and also close to the PbP result.

As in the case of P(ν), again the FIFRELIN results of cases 1 and 2 (using a moment of inertia of 0.4 from the rigid body momentum) are very close each other and the FIFRELIN results of cases 3, 4, 5 (using moment of inertia values from the AMEDEE database) are very close to each other, too and differ visibly from the first two cases.

These behaviours of <Eγ> prove again that in the case of the Monte Carlo treatment of prompt emission included in the FIFRELIN code the moment of inertia has a much more influence on the results of total average prompt quantities than the compound nucleus cross section of the inverse process of neutron evaporation from fragments. This fact is not surprising taking into account that the sequential emission of neutrons ends at the limit of *Sn+Erot*. Consequently the rotational energy of fragments that depends on the moment of inertia plays a very important role in the FIFRELIN calculations.

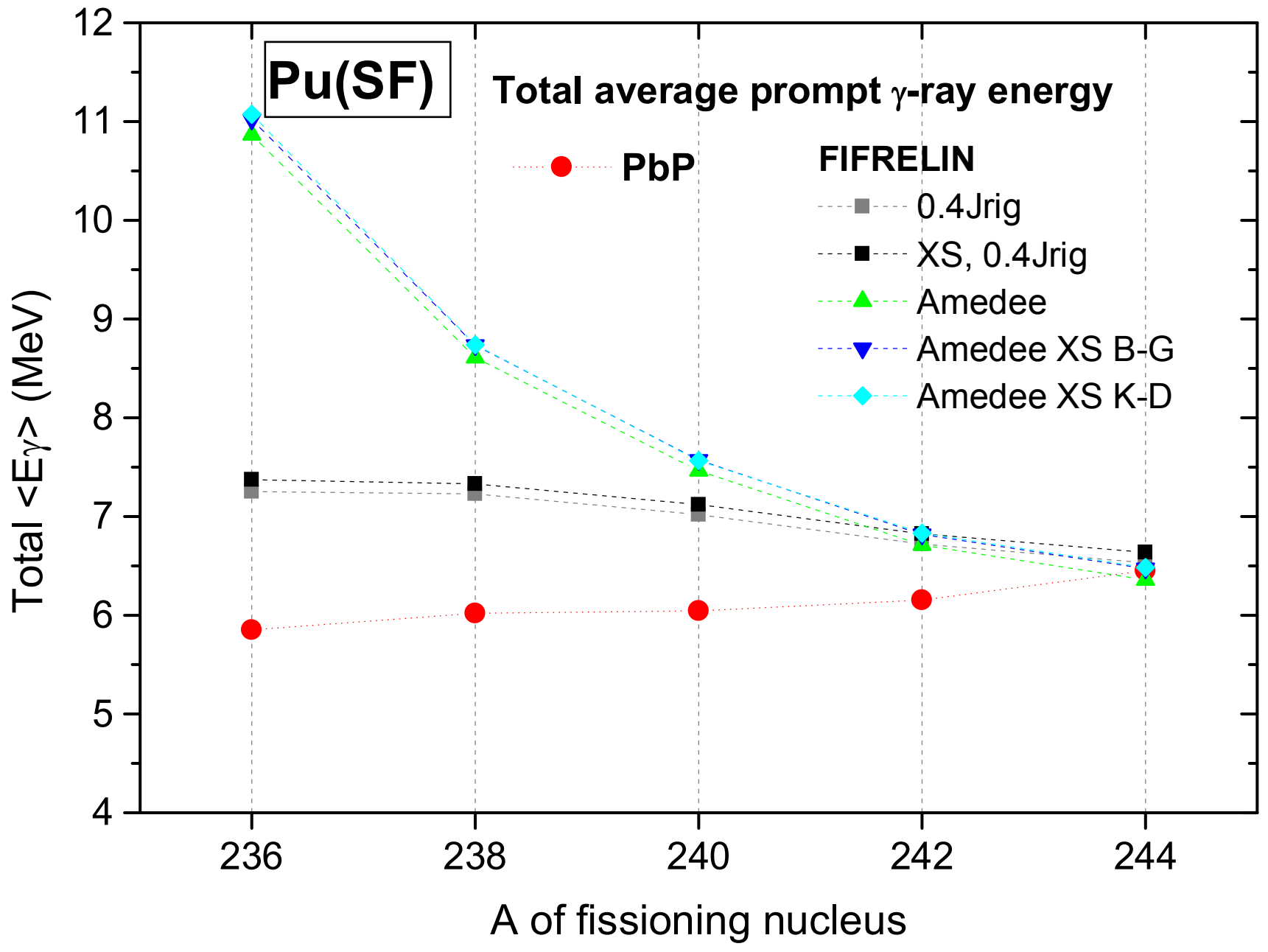

**Fig.4.11**: $^{236,238,240,242,244}$Pu(SF) total average prompt gamma-ray energy as a function of the mass number of the fissioning nucleus: PbP results with red circles, FIFRELIN results of the 5 calculation cases with different symbols as indicated in the legend.

Contrary, in the case of the PbP model, the TXE partition methods (that require the use of different assumptions and approaches, including also different calculations of β2 deformation parameters of fragments or parameterizations) have an insignificant influence on the results of total average prompt emission quantities (more details about this fact can be found in Manailescu et al, 2011 and Morariu et al, 2012).

Experimental data regarding the total average prompt gamma-ray energy were measured for a few fissioning systems. For the standard $^{252}$Cf(SF) experimental <Eγ> data are of about 7 MeV. The measurements performed by Fréhaut, 1989 for the neutron induced fission of three actinides shows values around 6 MeV at low energy neutron induced fission as following: <Eγ> = (6.72 ± 0.04) MeV for $^{235}$U(n,f) at En = 1.14 MeV, <Eγ> = (6.804 ± 0.042) MeV for $^{237}$Np(n,f) also at En = 1.14 MeV and <Eγ> = (5.834 ± 0.033) MeV for $^{232}$Th(n,f) at En = 2.35 MeV.

For this reason, in the case of Pu(SF) <Eγ> values around 6 MeV as obtained by PbP model calculations can be considered as a realistic physical result. The FIFRELIN <Eγ> results around 7 MeV (obtained in the case of a moment of inertia taken as 0.4 from rigid body momentum) seem to be realistic, too.

The high <Eγ> values provided by FIFRELIN (in the case of moment of inertia values taken form the AMEDEE database) especially for $^{236,238}$Pu(SF) (values of about 11MeV or 9 MeV) are due to the approximation contained in the FIFRELIN version used in the present calculations. This approximation consists in a prompt gamma-ray energy taken as the energy left when no further emission of neutrons is possible (in other words the energy limit of *Sn+Erot*). Important developments of the FIFRELIN code are in progress at CEA Cadarache regarding a refined statistical treatment of the prompt gamma emission from fragments that will lead certainly to significant improvements of prompt gamma-ray energy results in the near future.

Regarding the total average prompt fission neutron spectra (PFNS) a few experimental data exist only for $^{240,242}$Pu(SF). Comparisons of PbP and FIFRELIN calculated spectra with these experimental data are given in **Figs.4.12-4.15**.

As it can be seen in **Fig.4.12**, in the case of $^{240}$Pu(SF) the PbP spectrum calculations by using two TXE partition methods (described in Chapter II) are very close each other and in overall good agreement with the experimental data of Aleksandrova et al (taken from the Madland, 1998). The PFNS result obtained by using the TXE partition based on the $\nu_H/\nu_{pair}$ parameterization, plotted with red line in **Fig.4.12**, was already reported in (Tudora and Hambsch, 2010).

The other PbP spectrum result, plotted with blue line in the same figure, was obtained by using the TXE partition method described in Morariu et al, 2012. It is important to mention that this time in the calculation of the extra-deformation energy of fragments at scission, instead of β2 values taken from HFB calculations (as in Ref. Morariu et al, 2012) the following nice and simple parameterization proposed by Schmidt and Jurado, 2012 for the β2 deformation parameters of light and heavy fragments was used:

$$\beta_{LF} = 0.04(Z_{LF} - 26.6), \quad \beta_{HF} = 0.035(Z_{HF} - 48) \qquad (4.4)$$

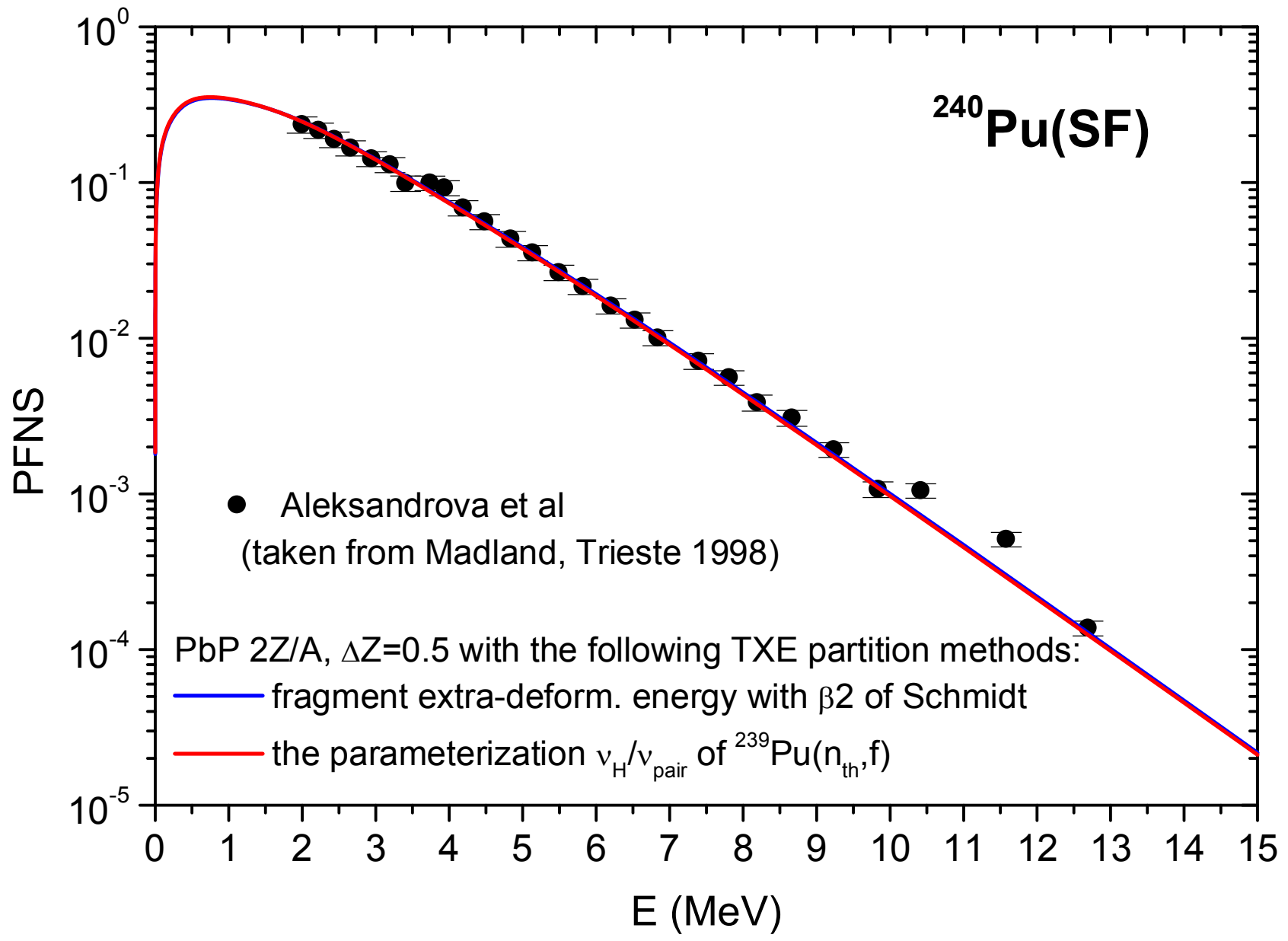

**Fig.4.12**: $^{240}$Pu(SF) PbP calculations of PFNS using the TXE partition of Manailescu et al, 2011 with red line and using the TXE partition of Morariu et al, 2012 with the β2 parameterization proposed by Schmidt and Jurado, 2012 with blue line. Experimental data (black circles) are renormalized to the spectrum calculation with the first mentioned TXE partition.

The very close PbP spectrum results obtained by using different TXE partitions (see the red and blue lines in Fig.4.12) confirm again the fact (already mentioned in Manailescu et al, 2011 and Morariu et al, 2012) that the TXE partition has not a significant influence on total average prompt emission quantities.

In order to better see the agreement of the PbP calculations with the experimental data the spectrum results are plotted as ratio to a Maxwellian spectrum (with TM = 1.35 MeV) in **Figs.4.13** as following: in red the case of TXE partition based on the $\nu_H/\nu_{pair}$ parameterization and in blue the new calculation with extra-deformation energies calculated using the β2 parameterization of eqs.(4.4). The experimental data are plotted with full and open circles in the same color as the calculated PFNS at which the data were renormalized.

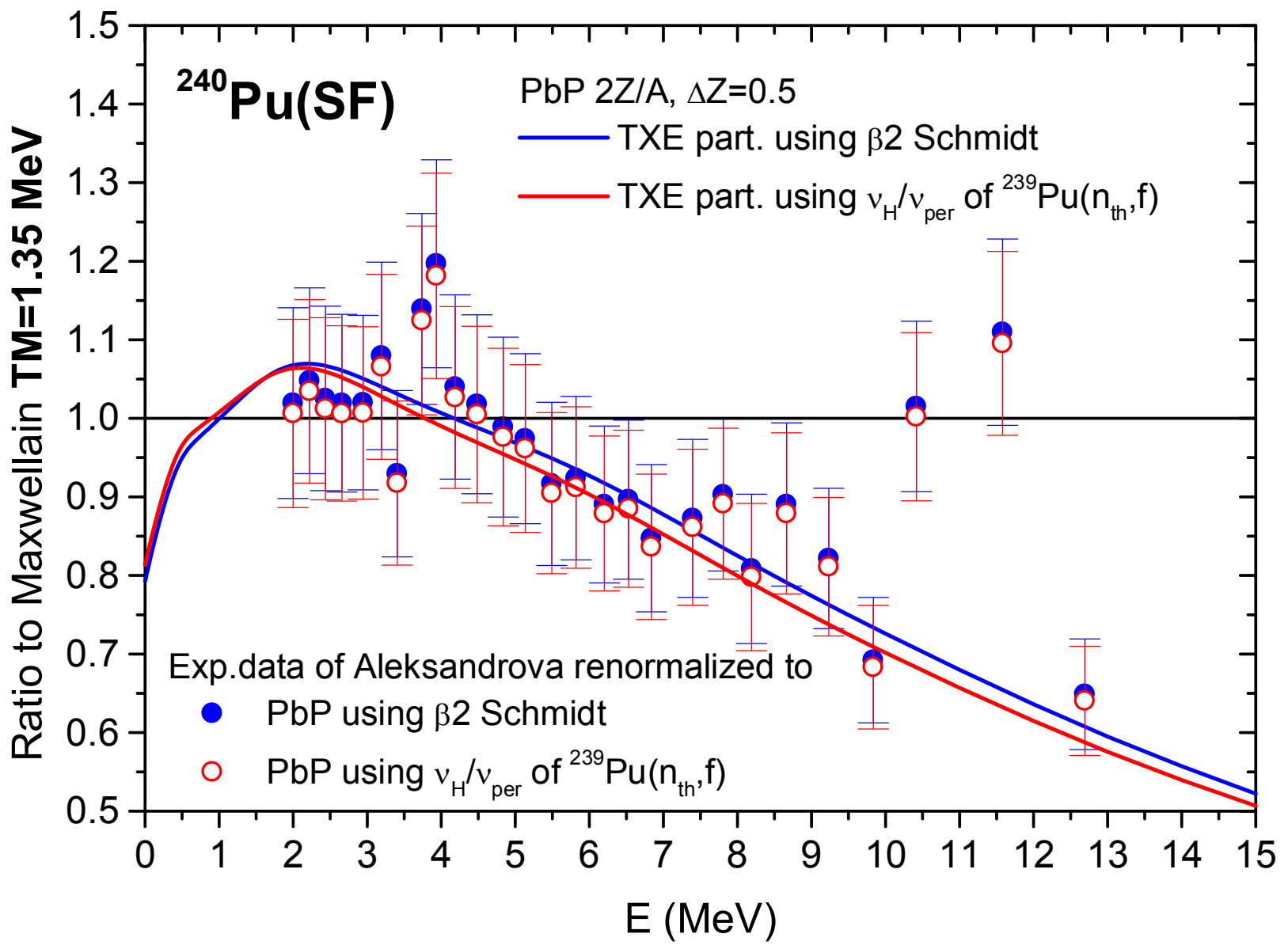

**Fig.4.13**: $^{240}$Pu(SF) PbP spectrum calculations given as ratio to Maxwellian in blue color for the case of extra-deformation energy calculated with β2 parameterization of Schmidt and Jurado and in red color for the case of TXE partition based on the $\nu_H/\nu_{pair}$ parameterization. The experimental data are plotted with circles in the same color as the calculated spectrum at which they were re-normalized.

As it is known and already reported many years ago (in the works of Madland and Nix, 1982 and Vladuca and Tudora, 2000b, 2001b) the shape of $\sigma_c(\varepsilon)$ has a great influence on the PFNS shape. As it can be seen in Fig.4.13, the use in the PbP calculation of $\sigma_c(\varepsilon)$ provided by optical model calculations with the Becchetti-Greenless potential lead to a good reproduction of the shape exhibited by experimental data.

In **Figs.4.14a-c** FIFRELIN spectrum results are plotted in comparison with the experimental data, in each figure the experimental data being re-normalized to the respective calculated spectrum. As it can be seen in Figs.4.14a,b PFNS calculated by taking $\sigma_c(\varepsilon)$constant (and using different moments of inertia) are in a reasonable agreement with the experimental data. The FIFRELIN calculation using $\sigma_c(\varepsilon)$ provided by optical model calculations with the Becchetti-Greenless potential gives a PFNS result that does not succeed to describe the experimental data (see Fig.4.14c).

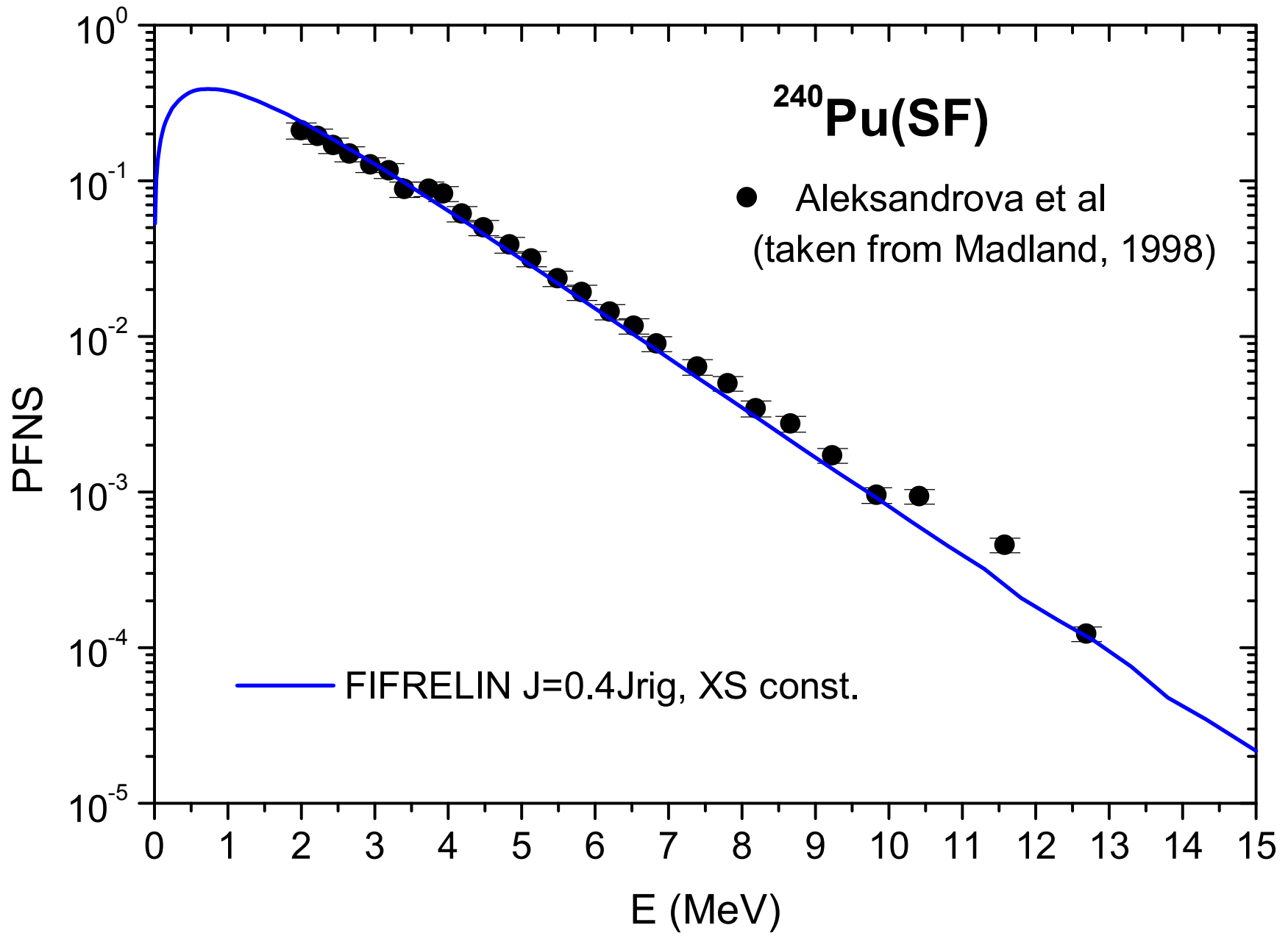

**Fig.4.14a**: $^{240}$Pu(SF): FIFRELIN spectrum calculation (case $\sigma_c(\varepsilon)$ constant and moment of inertia taken as 0.4 of rigid body momentum) in comparison with experimental data of Aleksandrova et al (full circles)

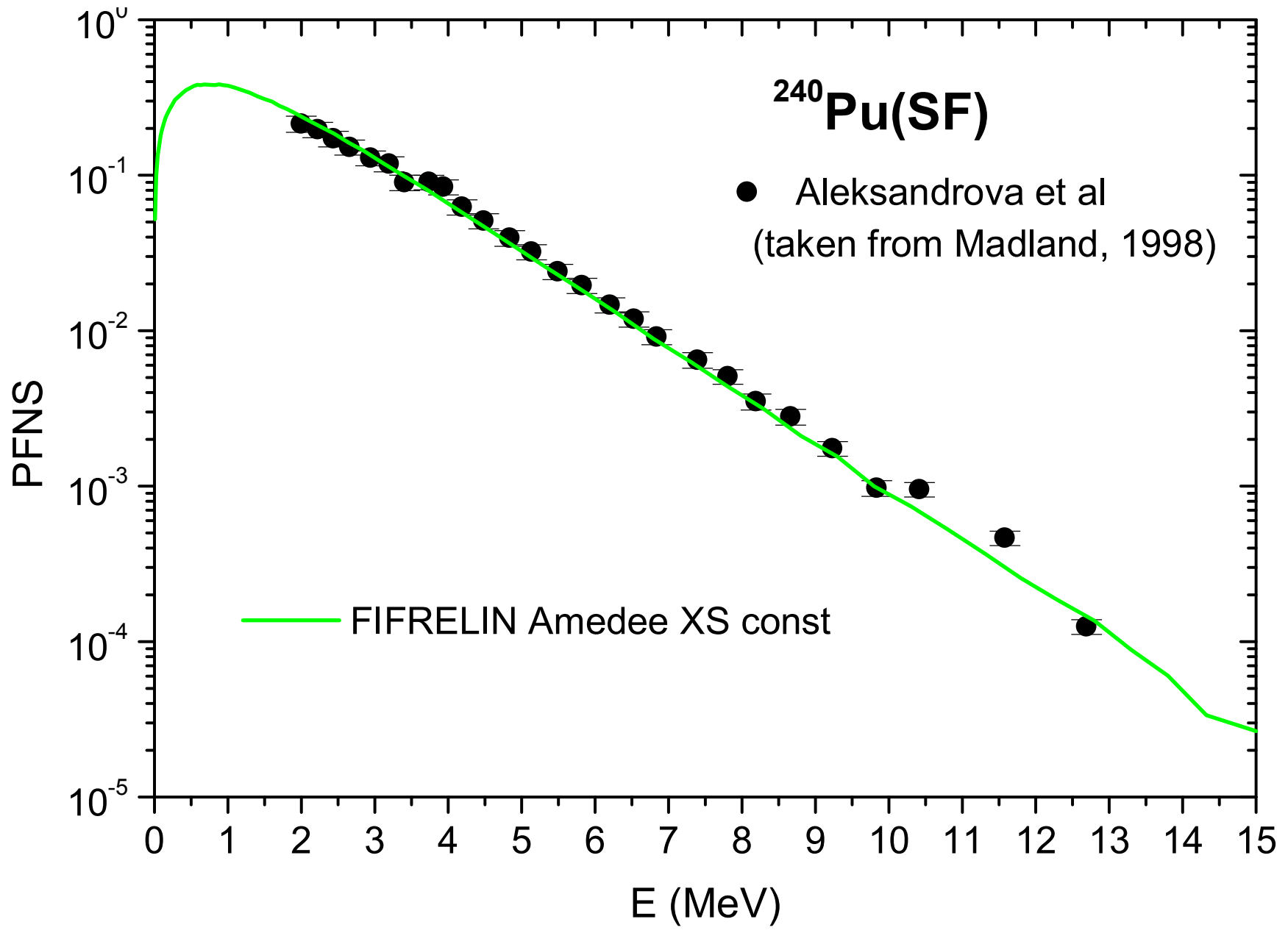

**Fig.4.14b**: $^{240}$Pu(SF): FIFRELIN spectrum calculation (case $\sigma_c(\varepsilon)$ constant and moment of inertia values from the AMEDEE database) in comparison with experimental data of Aleksandrova et al (full circles).

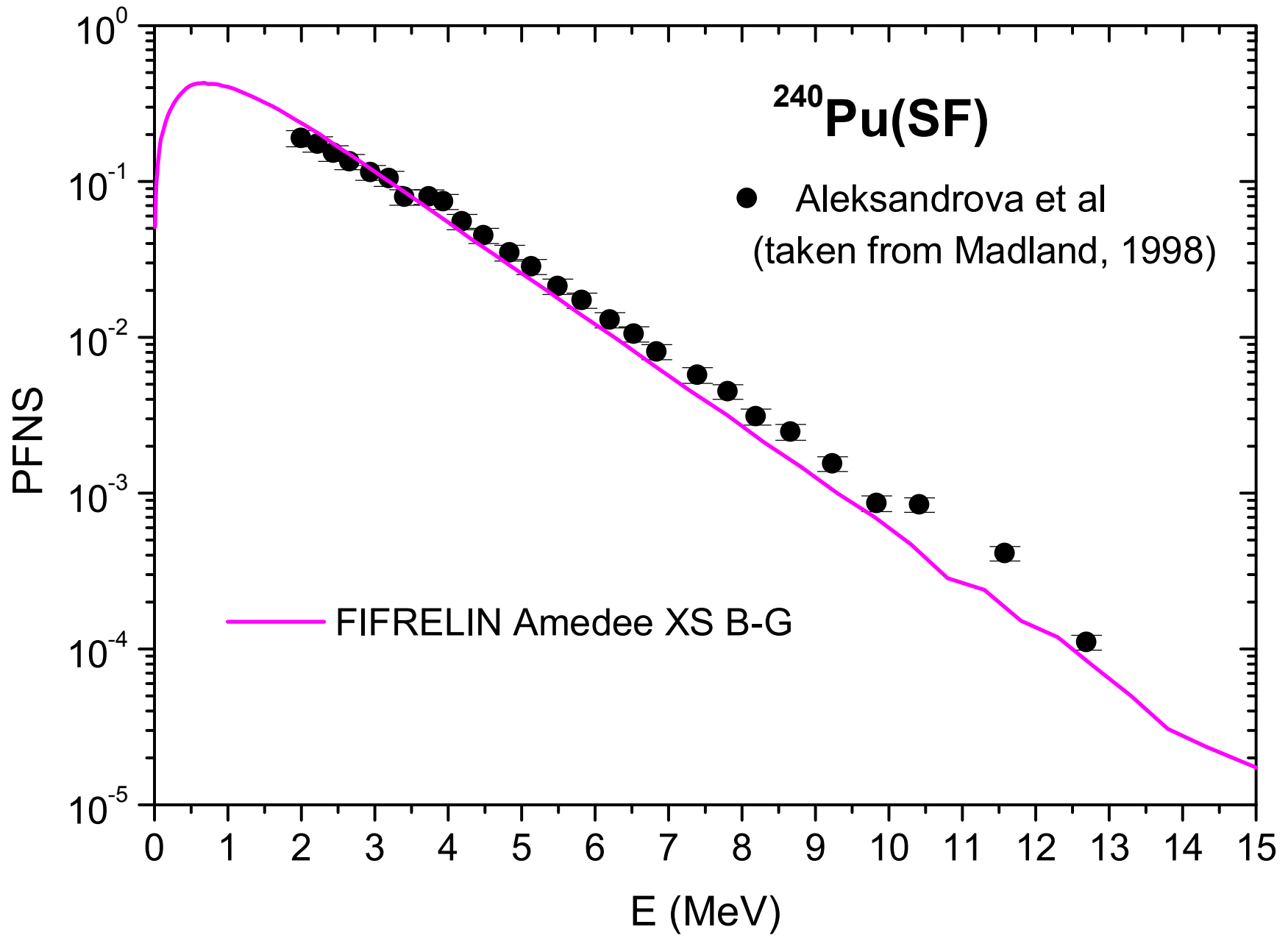

**Fig.4.14c**: $^{240}$Pu(SF): FIFRELIN spectrum calculation (case $\sigma_c(\varepsilon)$ provided by optical model calculations with Becchetti-Greenless potential and moment of inertia values from the AMEDEE database) in comparison with experimental data of Aleksandrova et al (full circles).

The FIFRELIN spectra calculated considering $\sigma_c$ constant are plotted in **Fig.4.15** as a ratio to a Maxwellian spectrum (with TM=1.32 MeV) in comparison with experimental data. The re-normalized experimental data at each calculated PFNS are plotted with the same color as the respective calculated spectrum. The fluctuations visible at the spectrum queues are due to the number of events sampled in the Monte-Carlo treatment. As it can be seen the experimental spectrum is not so well reproduced by these FIFRELIN results.

Because in the case of PFNS the comparison with experimental data requires each time a normalization of these data to the respective calculated spectrum it is difficult to compare on the same graph the PbP and the FIFRELIN spectra with the experimental data.

Anyway, looking at **Figs. 4.13** and **4.15** where PbP and FIFRELIN spectra of $^{240}$Pu(SF) are given as ratios to Maxwellian spectra, it is obvious that the spectrum shapes are different and the PbP spectrum result describe better the experimental data.

It is surprising that the PFNS result of FIFRELIN when $\sigma_c(\varepsilon)$ is taken variable with the energy (case plotted in Fig.4.14c) which is a much more physical situation than the consideration of a constant $\sigma_c$ does not succeed to reproduce the experimental data. A possible explanation of this contradictory PFNS result can be an insufficient number of events sampled in the Monte-Carlo treatment.

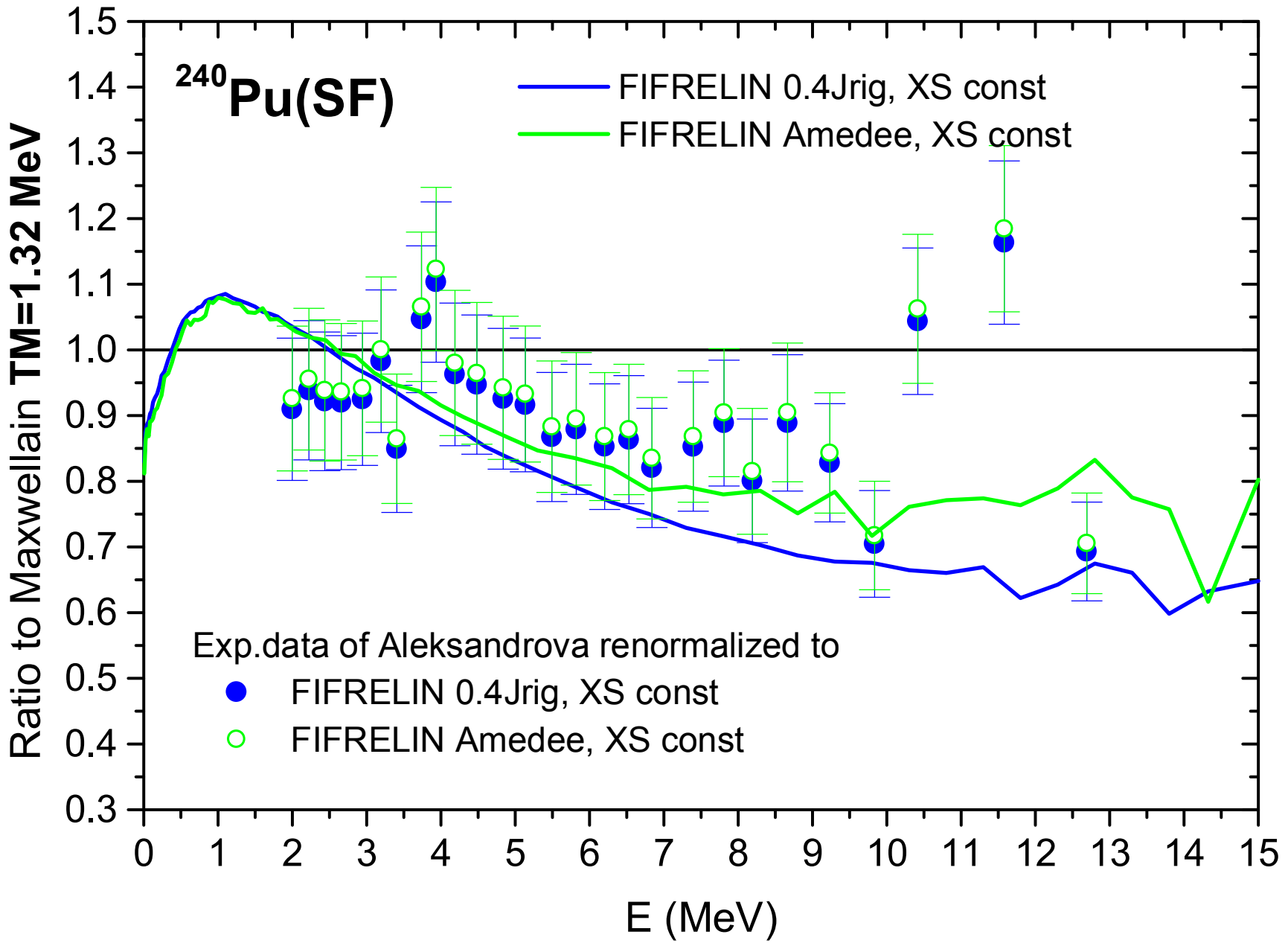

**Fig.4.15**: $^{240}$Pu(SF) FIFRELIN spectra given as ratio to a Maxwellian spectrum with TM=1,32 MeV plotted with blue line for the case $\sigma_c(\varepsilon)$ constant and moment of inertia values taken 0.4 of the rigid body momentum and with green line for the case $\sigma_c(\varepsilon)$ constant and moment of inertia values from the AMEDEE database. The experimental data are plotted with circles in the same color as the calculated spectrum at which they were re-normalized.

In the case of $^{242}$Pu(SF), as it can be seen in **Fig.4.16**, the PbP calculation (done by using $\sigma_c(\varepsilon)$ calculated with the Becchetti-Greenless potential and the TXE partition with extra-deformation energies calculated with β2 of eqs.(4.4)) are in fair agreement with the scarce experimental data taken from EXFOR. This PFNS result is rather close the previous PbP calculation reported in (Tudora and Hambsch, 2010) where the TXE partition based on the $\nu_H/\nu_{pair}$ parameterization was used.

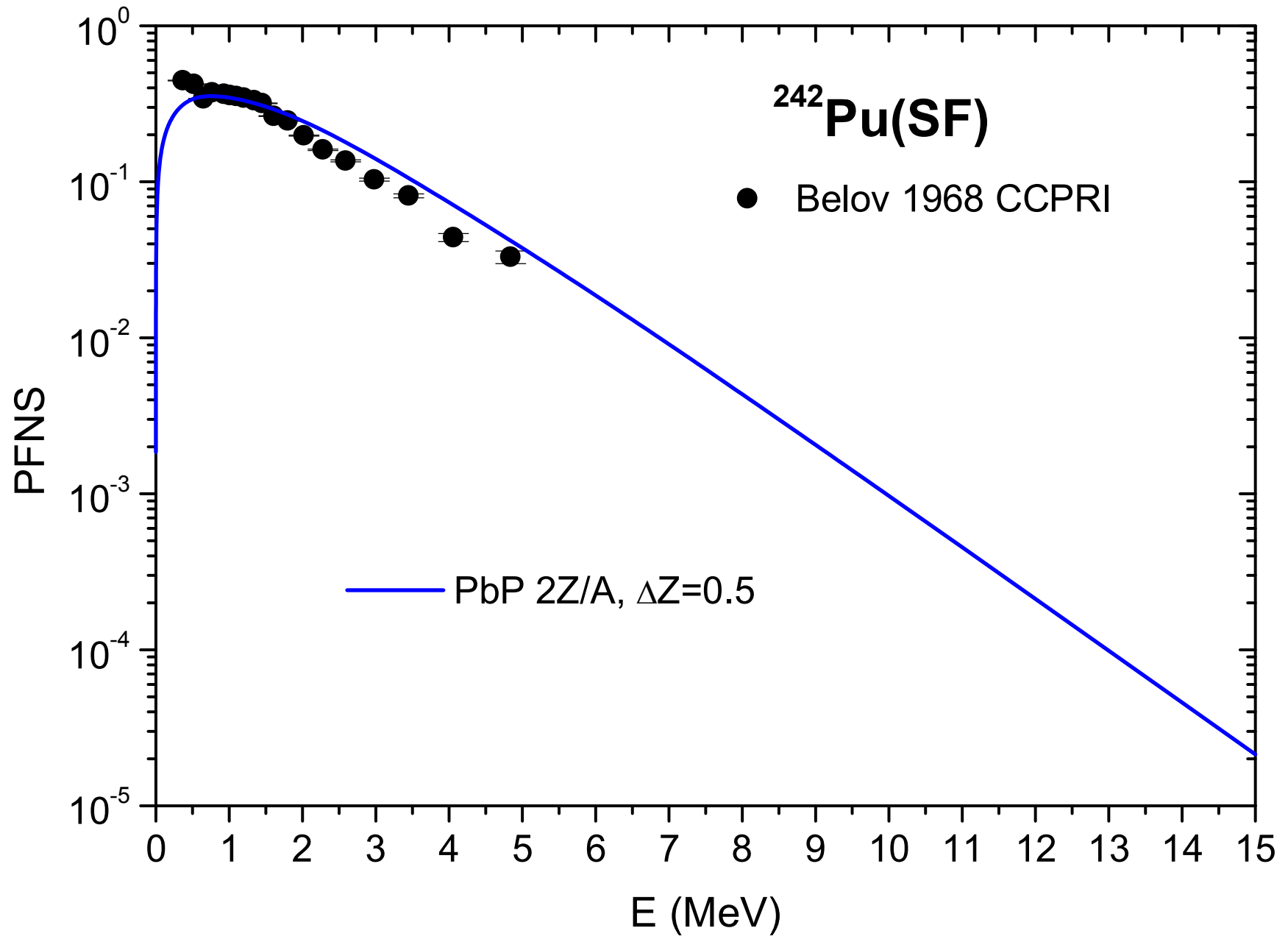

**Fig.4.16**: $^{242}$Pu(SF) PbP spectrum calculation in comparison with experimental data from EXFOR.

The $^{242}$Pu(SF) spectrum results of FIFRELIN in four calculation cases are compared with the same experimental data from EXFOR in **Figs.4.17a,b**.

In **Fig.4.17a** the PFNS calculations with moments of inertia taken 0.4 from the rigid body momentum and considering $\sigma_c(\varepsilon)$ constant (blue line) and with $\sigma_c(\varepsilon)$ provided by optical model calculations with Becchetti-Greenless potential (red line) are plotted in comparison with experimental data plotted with circles in the same color as the calculated spectrum used for re-normalization. As it can be seen in both calculation cases the FIFRELIN spectra are in good agreement with the experimental data (excepting the two experimental points below 0.5 MeV), a better description being obtained for the case of $\sigma_c(\varepsilon)$ taken variable with the energy (calculated spectrum and experimental data plotted with red color).

FIFRELIN spectrum calculations with moment of inertia values taken from the AMEDEE database and considering $\sigma_c(\varepsilon)$ constant (magenta line) and $\sigma_c(\varepsilon)$ provided by optical model calculations with Becchetti-Greenless potential (black line) are plotted in **Fig.4.17b** in comparison with experimental data renormalized to the respective calculated spectrum plotted in the same color. Both calculations describe satisfactory the experimental data (excepting the two experimental points below 0.5 MeV). Nevertheless it seems that a better agreement with experimental data is obtained in the case of $\sigma_c(\varepsilon)$ considered variable with the energy and moments of inertia taken as 0.4 from the rigid body momentum (the red line and circles in Fig.4.17a).

In the case of $^{242}$Pu(SF) the FIFRELIN spectrum results describe better the scarce experimental data of Belov et al. (taken from EXFOR, 2012f) than the PbP results.

In **Figs.4.17a,b** the influence of $\sigma_c(\varepsilon)$ on the spectrum shape is visible, spectrum calculations done by using $\sigma_c(\varepsilon)$ provided by optical model (red and black lines in Figs.4.17) lead to a better description of experimental data than spectra obtained by taking a constant $\sigma_c$.

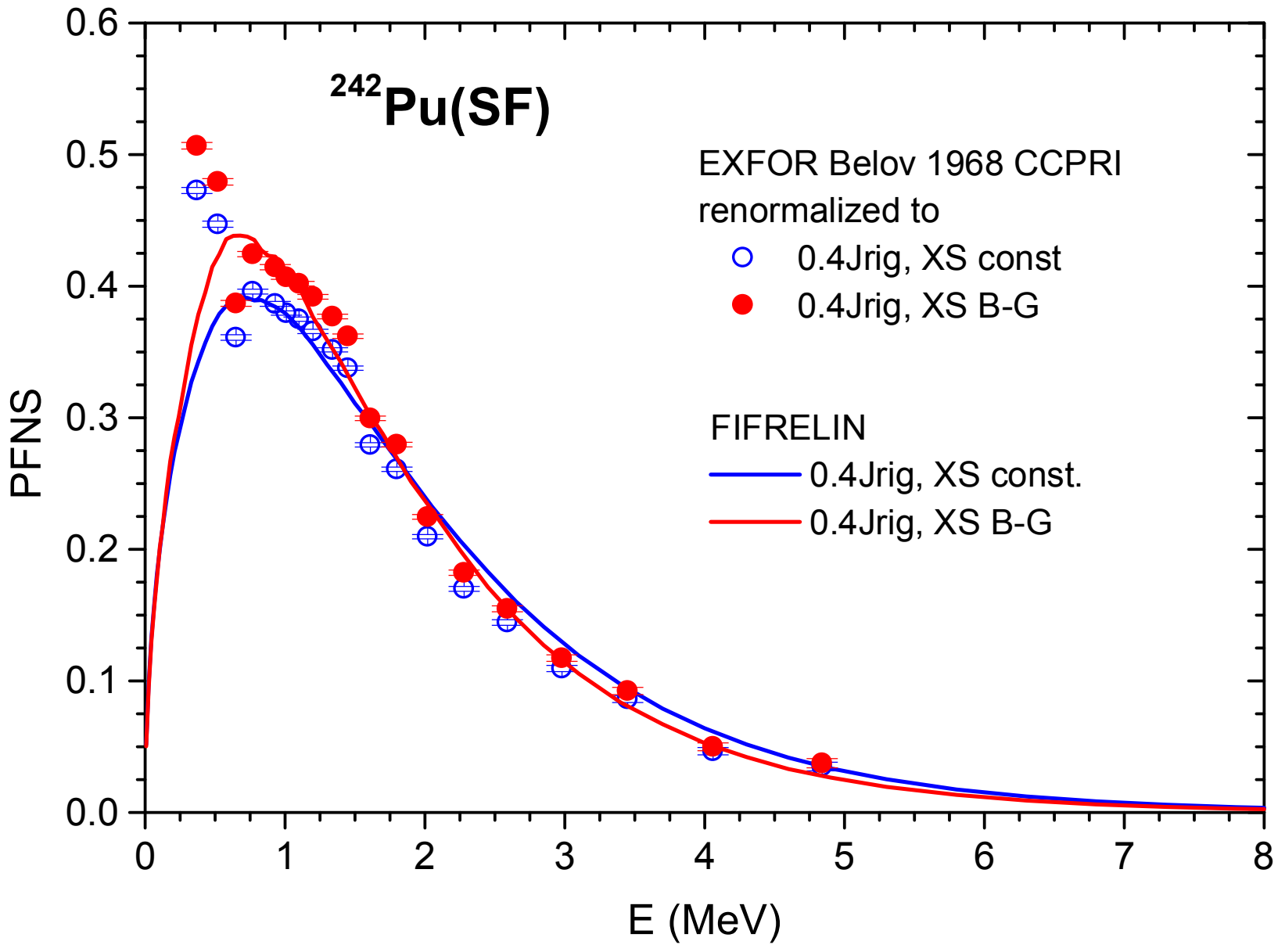

**Fig.4.17a**: $^{242}$Pu(SF) FIFRELIN spectrum calculations with moment of inertia taken as 0.4 from the rigid body momentum and with $\sigma_c(\varepsilon)$ taken constant (blue line) and provided by optical model calculations with Becchetti-Greenless potential (red line) in comparison with experimental data in the same color as the calculated spectrum used for re-normalization.

The FIFRELIN results for $^{242}$Pu(SF) confirm the general statement, available for all Los Alamos and PbP spectrum calculations, that the consideration of an energy variable compound nucleus cross-section of the inverse process is not only more physical than the consideration of a constant $\sigma_c$ but also it leads to a better description of experimental spectrum data.

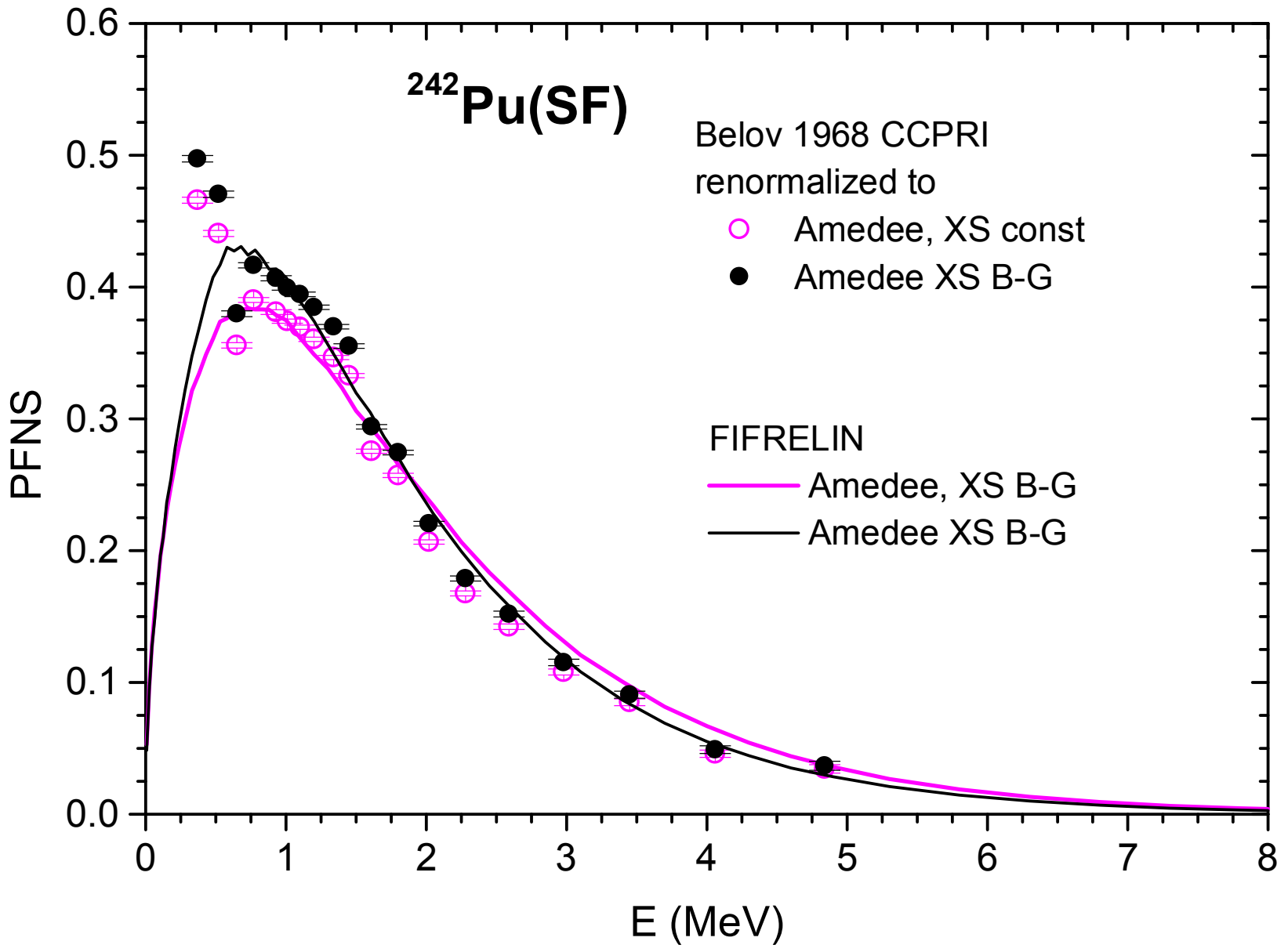

**Fig.4.17b**: $^{242}$Pu(SF) FIFRELIN spectrum calculations with moment of inertia values from the AMEDEE database $\sigma_c(\varepsilon)$ taken constant (green line) and provided by optical model calculations with Becchetti-Greenless potential (black line) in comparison with experimental data in the same color as the calculated spectrum used for re-normalization.

A test of consistency of PbP calculations consists in the comparison of total average prompt emission quantities, meaning quantities that are usually input parameters of the Los Alamos model <*Er*>, <*TKE*>, <*Sn*>, <*a*> and also prompt emission quantities such as prompt neutron multiplicity, spectrum, gamma-ray energy, obtained by averaging in two ways as following:

a) a parameter (labeled "*param*") or a prompt emission quantity (labeled "*quantity*") calculated as a function of fragment mass number A is averaged over the fragment mass distribution Y(A):

$$
\begin{aligned}
&< param >= \sum_A < param > (A)\, Y(A) \Big/ \sum_A Y(A) \\
&< quantity >= \sum_A < quantity > (A)\, Y(A) \Big/ \sum_A Y(A)
\end{aligned}
\qquad (4.5.1)
$$

b) a parameter or a prompt emission quantity calculated as a function of TKE is averaged over the TKE distribution Y(TKE):

$$
\begin{aligned}
&< param >= \sum_{TKE} < param > (TKE)\, Y(TKE) \Big/ \sum_{TKE} Y(TKE) \\
&< quantity >= \sum_{TKE} < quantity > (TKE)\, Y(TKE) \Big/ \sum_{TKE} Y(TKE)
\end{aligned}
\qquad (4.5.2)
$$

as it was described in detail in (Tudora 2012a, 2012b, Tudora et al,, 2012c).

We give as example of such test of consistency the comparison between the values of total average parameters <*Er*>, <*Sn*> and <*a*> (given as $<C>=A_0/<a>$) as well as between the total average multiplicity obtained by averaging according to eqs.(4.5.1-4.5.2) for $^{240}$Pu(SF):

**Table 4.1** $^{240}$Pu(SF) Total average quantities obtained by averaging the corresponding quantity as a function of A over Y(A) and as a function of TKE over Y(TKE)

| Total average parameter | param(A) averaged over Y(A) | param(TKE) averaged over Y(TKE) | Deviation (%) |
|---|---|---|---|
| <*Er*> (MeV) | 199.42 | 199.22 199.29 | 0.100 0.065 |
| <*Sn*> (MeV) | 5.4949 | 5.4804 5.4723 | 0.264 0.411 |
| $<C>=A_0/<a>$ (MeV) | 11.2978 | 11.3681 11.4631 | 0.622 1.463 |
| Total average prompt neutron multiplicity | ν(A) averaged over Y(A) | <ν>(TKE) averaged over Y(TKE) | |
| $<\nu_p>$ | 2.1670 | 2.1987 2.1662 | 1.463 0.037 |

Because in both PbP and FIFRELIN calculations experimental TKE(A) and KE(A) data are used, in Table 4.1 the total average <TKE> was not included being irrelevant for the test of consistency.

In the averaging over the TKE distribution the Y(TKE) resulted from eq.(4.2) and plotted with red line in Fig.4.1c was used. In the column containing total quantities obtained by averaging over Y(TKE) the first data in each cell refers to the parameters resulted from the PbP calculation (plotted with full circles in Fig.4.7) and the second data to the parameter dependences on TKE (plotted with red lines and given in the legend of Fig.4.7). Similarly in the case of <νp>, the first data is obtained by averaging the PbP result of <ν>(TKE) (plotted with black circles in Fig.4.3c) over Y(TKE) and the second data by averaging the most probable fragmentation result of <ν>(TKE) (plotted with solid black line in Fig.4.3c) over Y(TKE).

As it can be seen in the last column of Table 4.1, the deviation between the total average parameters obtained by averaging over Y(A) and Y(TKE) is very low, practically insignificant, in great part of cases less than 1%. In the case of $<\nu_p>$ the deviations are also very low.

## IV.2. Prompt neutron emission calculations for $^{239}$Pu($n_{th}$,f) in the frame of the PbP and Monte-Carlo (FIFRELIN) treatments

Prompt neutron emission data of $^{239}$Pu($n_{th}$,f) obtained by using the PbP model and the Monte-Carlo treatment (code FIFRELIN) are inter-compared and compared with existing experimental data, too.

FIFRELIN calculations were done in the same five cases as for Pu(SF) and the experimental fragments distributions of Demattè, 1996 were used. Fragment mass numbers A covering the range from 75 to 165 and the unchanged charge distribution (UCD) were used in all FIFRELIN calculations.

In the case of PbP calculations, the fragmentation range was build by taking the entire A-mass range covered by experimental fragment distributions taken from EXFOR with a step of one mass unit. For each pair of fragment masses two charge numbers were taken as the nearest integer values above and below the most probable charge considered as UCD and corrected with a charge polarization. Part of PbP results regarding $^{239}$Pu($n_{th}$,f), already reported (see for instance (Tudora, 2010b, 2010c, Tudora and Hambsch, 2010, Tudora, 2012a) were obtained by using different experimental fragment distributions (of Wagemans, Demattè, Surin, Tsuchiya and so on) taken from (EXFOR, 2012g).

In the PbP calculations given in this chapter in comparison with the FIFRELIN results, the experimental distributions Y(A) and TKE(A) of Wagemans et al (EXFOR, 2012g) and $\sigma_{TKE}$(A) of Asghar et al. (EXFOR, 2012b) were used. They are plotted with red circles in **Fig.4.18** in comparison with the experimental distributions of Demattè, 1996 (plotted with blue squares) used in the FIFRELIN calculations. As it can be seen the two sets of fragment distributions do not differ significantly, consequently they do not affect the comparison between FIFRELIN and PbP results of different average quantities characterizing the prompt neutron emission.

In both PbP and FIFRELIN calculations the same narrow Gaussian charge distribution given by eq.2.25 was used.

The FIFRELIN and PbP results of P(ν) are compared with the experimental data of Holden et al (taken from EXFOR, 2012h and plotted with full black squares) in **Fig.4.19**.

The PbP result (Tudora and Hambsch, 2010) plotted with full red circles connected with solid line is in excellent agreement with the experimental data.

The FIFRELIN results of the two calculation cases taking the moment of inertia as a fraction of 0.4 of the rigid body momentum and $\sigma_c(\varepsilon)$ constant (gray line) or provided by optical model calculation with the Becchetti-Greenless potential (black line) are close to each other and describe rather well the experimental data. P(ν) results of other three calculation cases (using momentum of inertia from the database AMEDEE and $\sigma_c(\varepsilon)$ constant or given by optical model calculations with Becchetti-Greenless and Koning-Delaroche potentials) visibly overestimate the experimental emission probability of 3 neutrons.

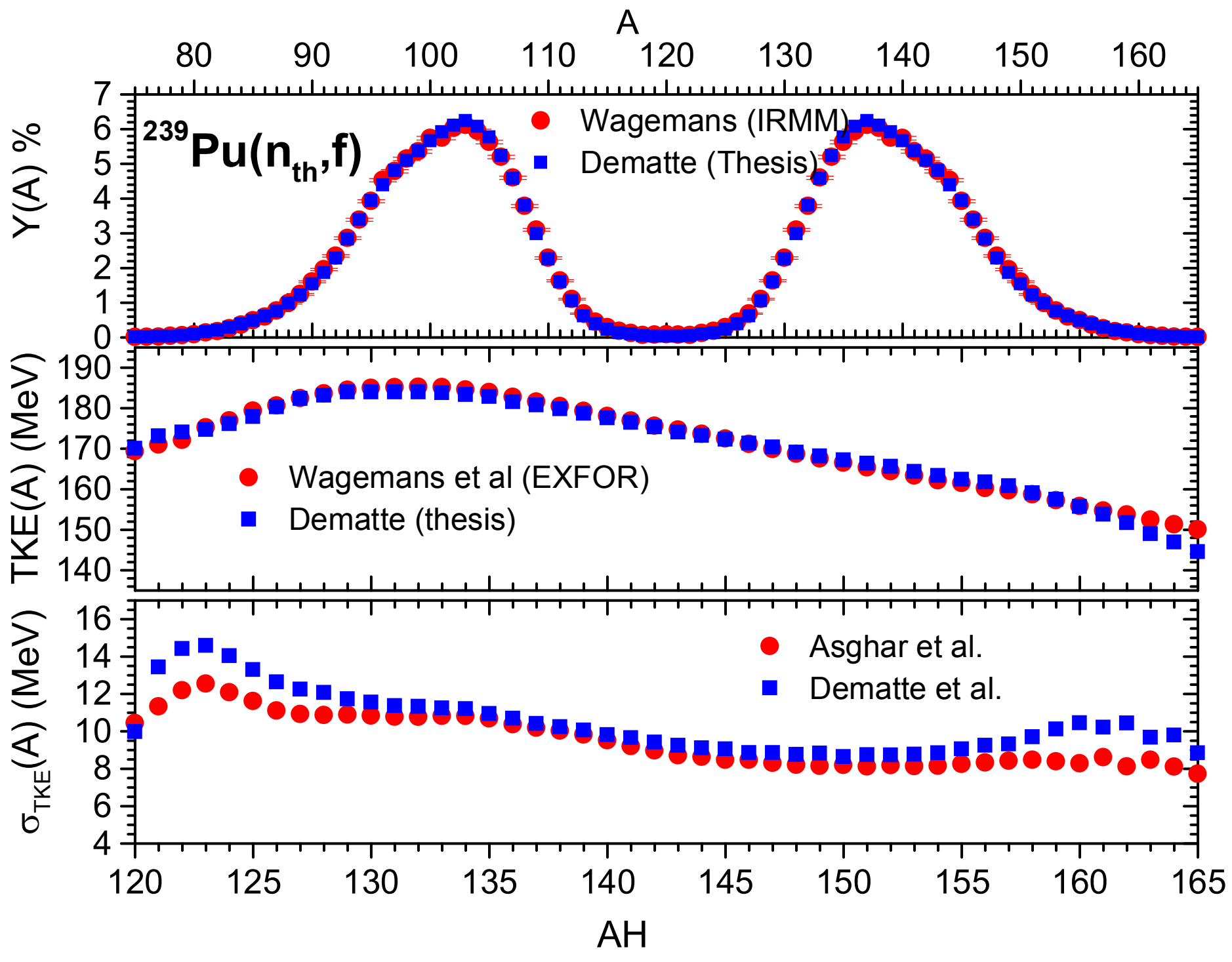

**Fig.4.18**: $^{239}$Pu($n_{th}$,f) Fission fragment distributions taken from EXFOR (red circles) used in the PbP calculations and measured by Demattè et al (blue squares) used in the FIFRELIN calculations

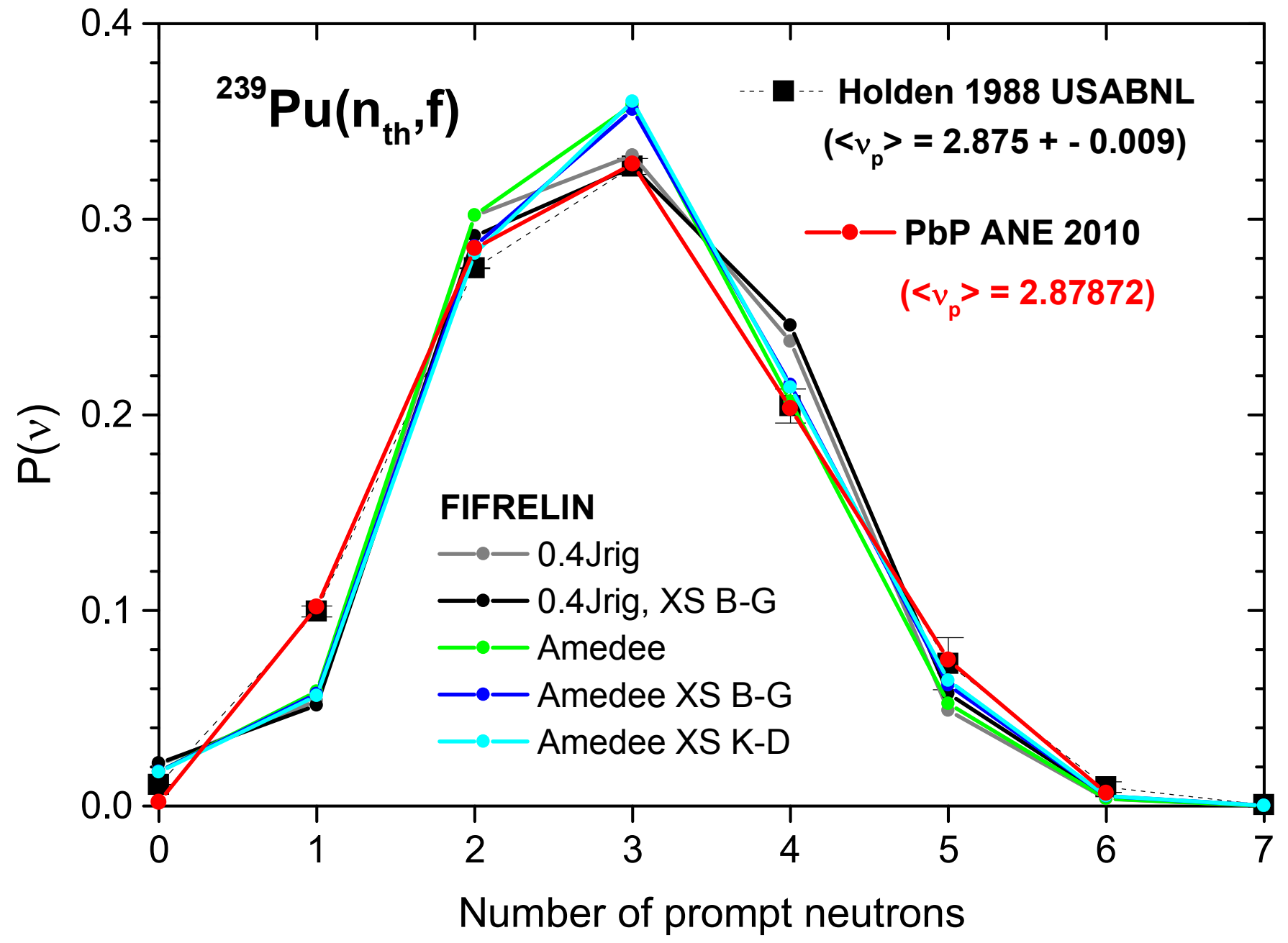

**Fig.4.19**: $^{239}$Pu($n_{th}$,f) P(ν) results of PbP calculation (red circles connected with lines) and of FIFRELIN calculations with solid lines in gray and black for cases with moments of inertia taken as 0.4 of the rigid body momentum and $\sigma_c(\varepsilon)$ constant and variable (B-G) and in green, blue and cyan for cases with moments if inertia form the database AMEDEE and $\sigma_c(\varepsilon)$ constant and variable (B-G and K-D) in comparison with the experimental data of Holden et al. (full black squares)

The PbP and FIFRELIN results of <ν>(TKE) are compared with the experimental data of Tsuchiya (taken from EXFOR, 2012i) in **Fig.4.20**.

The PbP result plotted with full red diamonds (connected with dotted lines to guide the eye) already reported in (Tudora, 2012a) was obtained by averaging the multi-parametric matrix ν(A,TKE) over the Y(A,TKE) re-constructed from the the experimental Y(A) and TKE(A) distributions of Tsuchiya and $\sigma_{TKE}$(A) of Asghar. The other PbP result plotted with open magenta diamonds, also reported in (Tudora, 2012a) was obtained by using Y(A,TKE) reconstructed from experimental Y(A) and TKE(A) distributions of Wagemans et al and $\sigma_{TKE}$(A) of Asghar. The influence on <ν> of the two reconstructed Y(A,TKE) is visible only at TKE values lower than 155 MeV. The most probable fragmentation result using the average parameter dependences on TKE resulted from the PbP treatment (see Tudora, 2012a) is plotted with wine solid line.

Because the FIFRELIN results of the five calculation cases are close each other in Fig.4.20 only the results obtained by using $\sigma_c(\varepsilon)$ from optical model calculations with the Becchetti-Greenless potential are plotted with full cyan circles (case of moments of inertia taken as 0.4 of the rigid body momentum) and open blue circles (case of moments of inertia from the database AMEDEE).

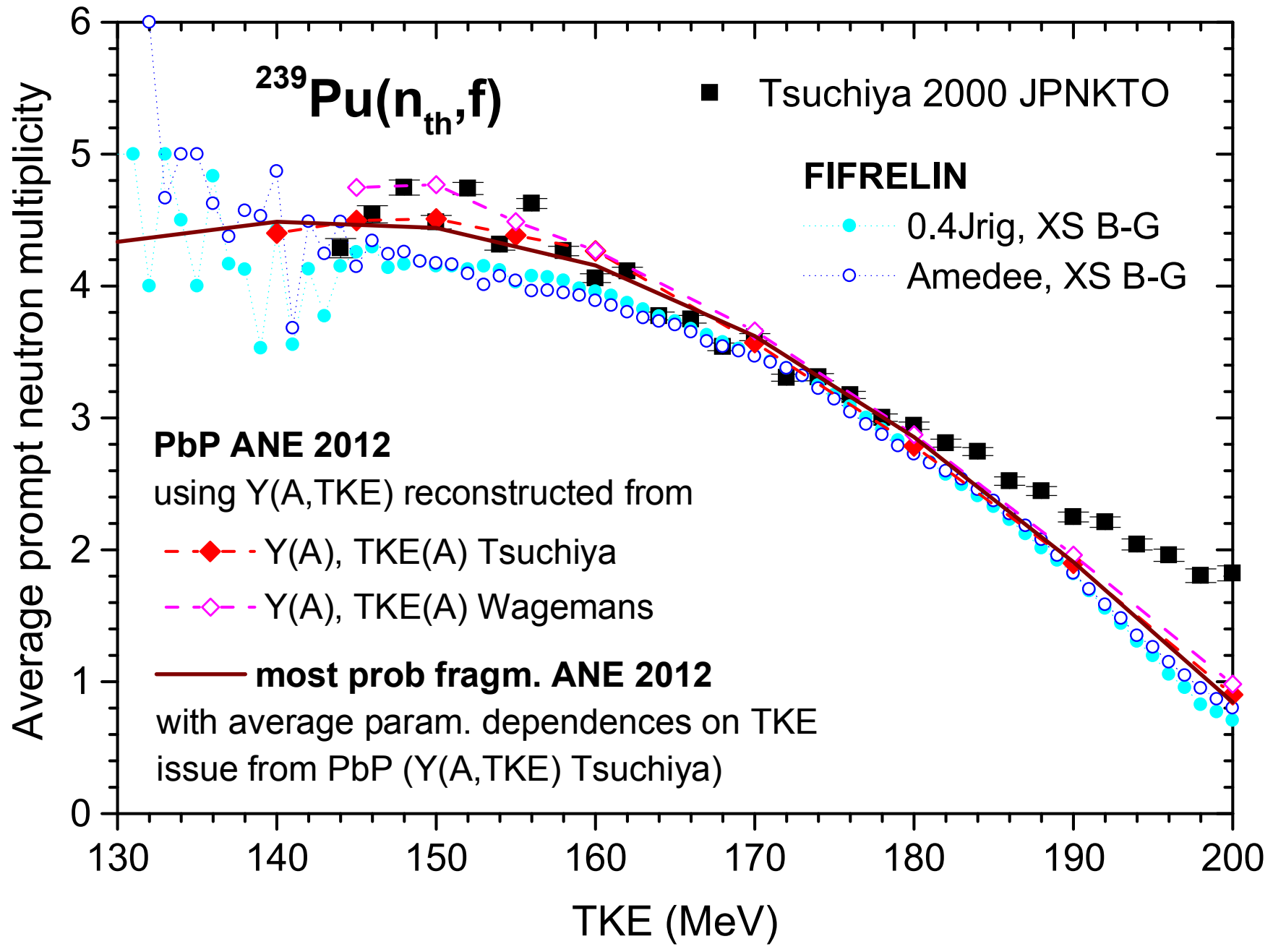

**Fig.4.20**: $^{239}$Pu(n$_{th}$,f) <ν>(TKE) results of PbP (full red and open magenta diamonds connected with thin dashed lines) and FIFRELIN results for cases with $\sigma_c(\varepsilon)$ variable (B-G) and moments of inertia taken as 0.4 from rigid body momentum (cyan full circles) and from the database AMEDEE (open blue circles) in comparison with the experimental data of Tsuchiya et al. (full black squares)

As it can be seen, for TKE values above 170 MeV the FIFRELIN results are very close to the results provided by PbP and most probable fragmentation calculations. The flattening of <ν> at lower TKE values is more pronounced in the case of FIFRELIN and a structure appears at TKE values below 145 MeV (probably due to the insufficient number of events sampled in the Monte-Carlo treatment).

It is interesting to mention that the present FIFRELIN results exhibit similar behaviour (consisting in a similar shape and a structure at TKE values below 140 MeV) as another Monte-Carlo result reported by Talou et al, 2011.

Both PbP and FIFRELIN results of <ν>(TKE) reproduce the general trend of experimental data of Tsuchiya but they do not succeed to describe the high experimental value of about 2 neutrons at TKE values of about 190-200 MeV. More comments about this high experimental value in connection with available TXE at these high TKE values can be found in Tudora, 2012a and Talou et al, 2011. At TKE values below 160 MeV the PbP results describe better the experimental data than the FIFRELIN results.

Prompt emission quantities as a function of fragment mass (such as $\nu(A)$, $\varepsilon(A)$) are very sensitive to the excitation energy partition between complementary fully-accelerated fragments. For this reason the comparison of such quantities provided by PbP and FIFRELIN with available experimental data as well as their inter-comparison can lead to interesting conclusions.

Different PbP and FIFRELIN results of $\nu(A)$ in comparison with experimental data taken from EXFOR, 2012j (different open symbols) are plotted in **Fig.4.21** as following:

a) PbP results obtained by using $\sigma_c(\varepsilon)$ provided by optical model calculations with the Becchetti-Greenless potential and the fragmentation range built by taking one or two charge numbers for each A and using different methods of TXE partition as follows:

- the TXE partition based on the $\nu_H/\nu_{pair}$ parameterization (described in Chapter II and in Manailescu et al, 2011), plotted with full magenta circles

- the TXE partition method based on the calculation of the extra-deformation energy of fragments at scission (described in Chapter II and in Morariu et al, 2012) by using β2 deformation parameter values from HFB calculations (taken from RIPL–3, 2012c), plotted with open red circles

- the same TXE partition bases on the extra-deformation energy but this time using the β2 parameterization proposed by Schmidt and Jurado, 2012 (eqs.4.4 of Section IV.1), plotted with wine stars.

b) FIFRELIN results of the five calculation cases mentioned above given with solid lines in different colors specified in the legend of Fig.4.21.

As it can be seen the three PbP results of $\nu(A)$ are rather close to each other and describe well all experimental data sets excepting only the far asymmetric regions (where data of Nishio and Apalin are underestimated in the far asymmetric heavy fragment part and the data of Tsuchiya are underestimated in the far asymmetric light fragment part).

The $\nu(A)$ results of FIFRELIN obtained by taking a fraction equal of 0.4 of the rigid body momentum and $\sigma_c(\varepsilon)$ constant or variable, plotted with violet and blue line, respectively are close each other and differ from the three FIFRELIN results (also close each other) obtained by taking moments of inertia from the database AMEDEE and different $\sigma_c(\varepsilon)$ (constant or optical model calculations with Becchetti-Greenless and Koning-Delaroche potentials) plotted with green, olive and cyan lines, respectively.

As in the previous studied cases of Pu(SF) again in the FIFRELIN treatment the moments of inertia have a more influence (impact) on ν(A) than the consideration of $\sigma_c(\varepsilon)$ variable or constant. The explanation of this fact consists in the sequential emission of neutrons ending when the fragment excitation energy becomes equal to *Sn+Erot*, the momentum of inertia entering the expression of *Erot*.

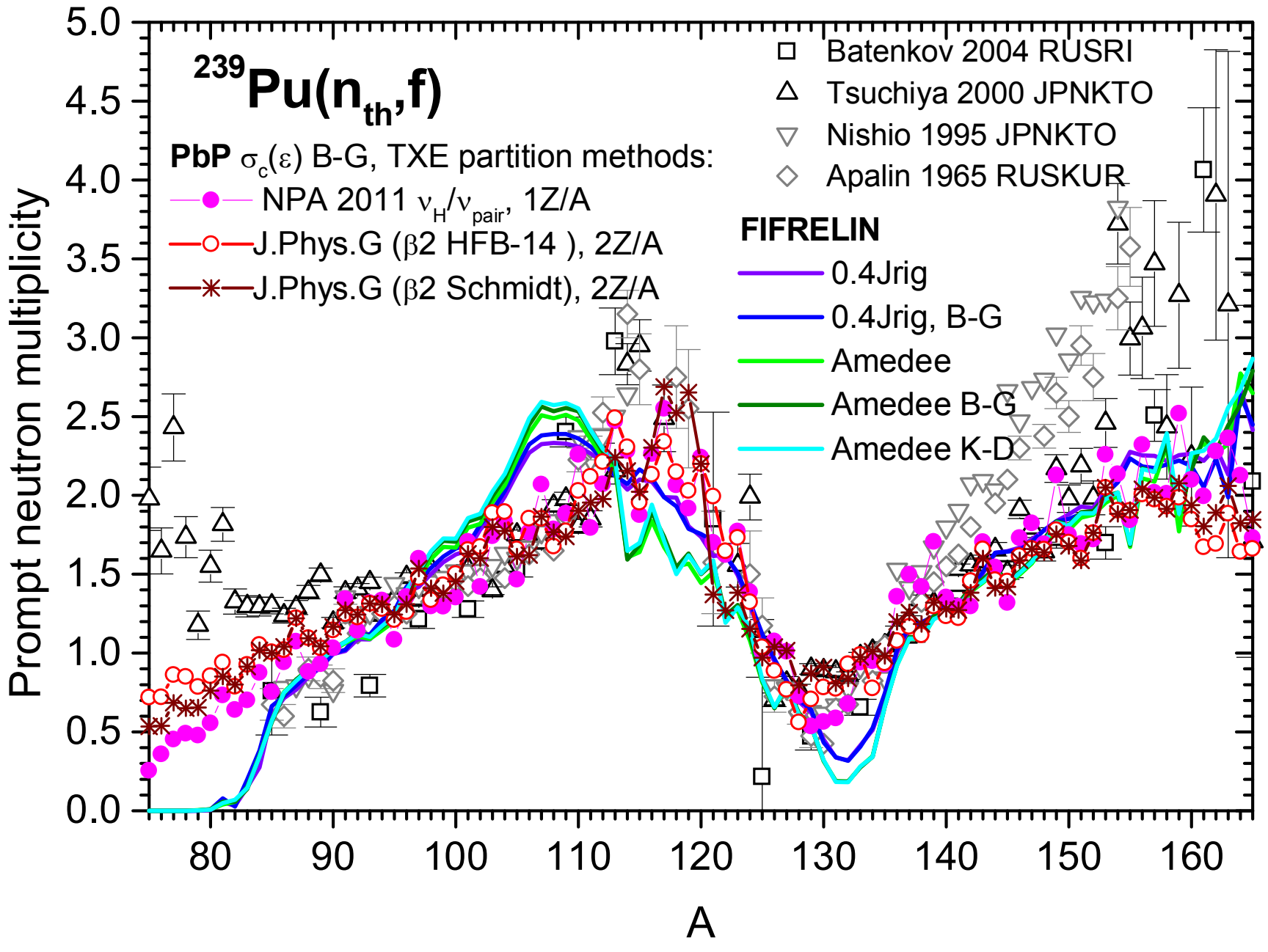

**Fig.4.21**: ν(A) of $^{239}$Pu($n_{th}$,f) : experimental data from EXFOR (different open symbols), PbP results using different TXE partition methods: based on the $\nu_H/\nu_{pair}$ parameterization with full magenta circles, based on the extra-deformation energy with β2 of the database HFB-14 of RIPL-3 with open red circles and with β2 parameterization of Schmidt and Jurado with wine stars. FIFRELIN results of 5 calculation cases with solid lines in violet and blue (moments of inertia of 0.4 from rigid body) and in green, olive and cyan (moments of inertia from the database AMEDEE).

As it can be seen in Fig.4.21 the FIFRELIN results obtained by taking 0.4 of rigid body momentum are in overall better agreement with experimental data than the results obtained by using moments of inertia from the database AMEDEE. This situation is similar with the case of another quantity <Eγ> (that is a total average quantity) of $^{236-244}$Pu(SF) when the calculations using moments of inertia from the database AMEDEE leaded to unphysical high results (also because the prompt gamma-ray energies of fragments are taken as the excitation energy limit of *Sn+Erot* when the neutron emission ends up). All FIFRELIN results of ν(A) exhibit a little bit shifted and more pronounced sawtooth shape than the experimental data.

Looking to the experimental data of average prompt neutron energy in SCM (<ε>(A)) taken from (EXFOR, 2012k) and plotted with open symbols in **Fig.4.22,** their almost symmetric behaviour suggests that maybe these data are referring to <ε> of fragment pairs.

The PbP results of <ε> corresponding to fragment pairs plotted with red circles and blue stars (obtained in the frame of the TXE partition method from Ref. Manailescu et al, 2011) succeeded to give a reasonable description of these experimental data.

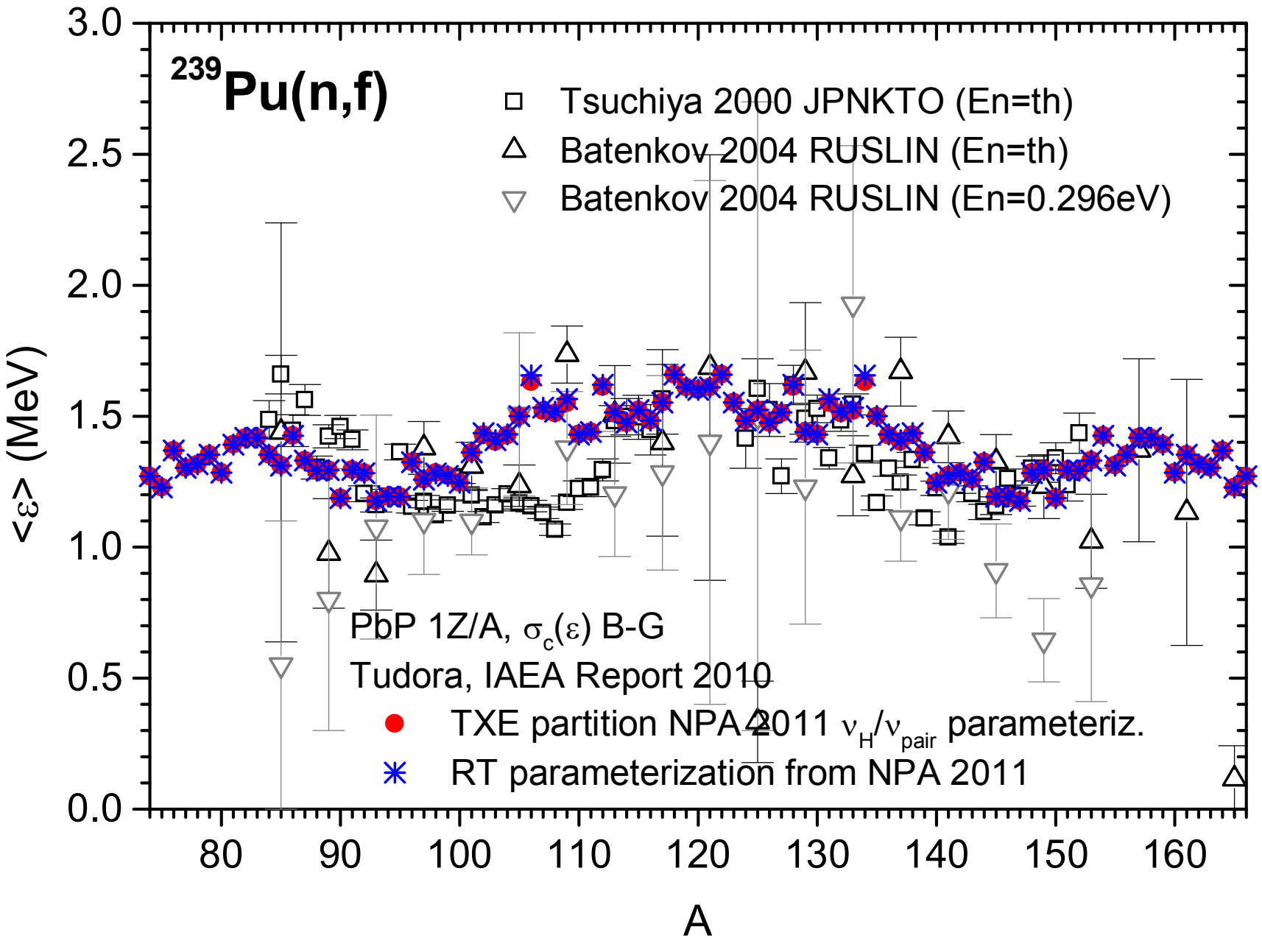

**Fig.4.22**: $^{239}$Pu($n_{th}$,f) experimental <ε>(A) data from EXFOR with different open symbols and PbP results using the TXE partition of Manailescu et al, 2011 with red circles and blue stars.

The FIFRELIN results of <ε>(A), exhibiting a sawtooth shape, are given in **Fig.4.23a**. Visible differences appears between calculation cases using $\sigma_c(\varepsilon)$ constant and variable, being known that the shape of $\sigma_c(\varepsilon)$ influences the spectrum shape and consequently the first order momentum of spectrum, too. Average <ε> of fragment pairs obtained from these results (case $\sigma_c(\varepsilon)$ constant and moments of inertia 0.4 of rigid body momentum and case $\sigma_c(\varepsilon)$ variable Becchetti-Greenless and moments of inertia from the database AMEDEE) are plotted in comparison with experimental data in **Fig.4.23b** (with the same colors as in Fig.4.6a). They visibly underestimate the experimental data.

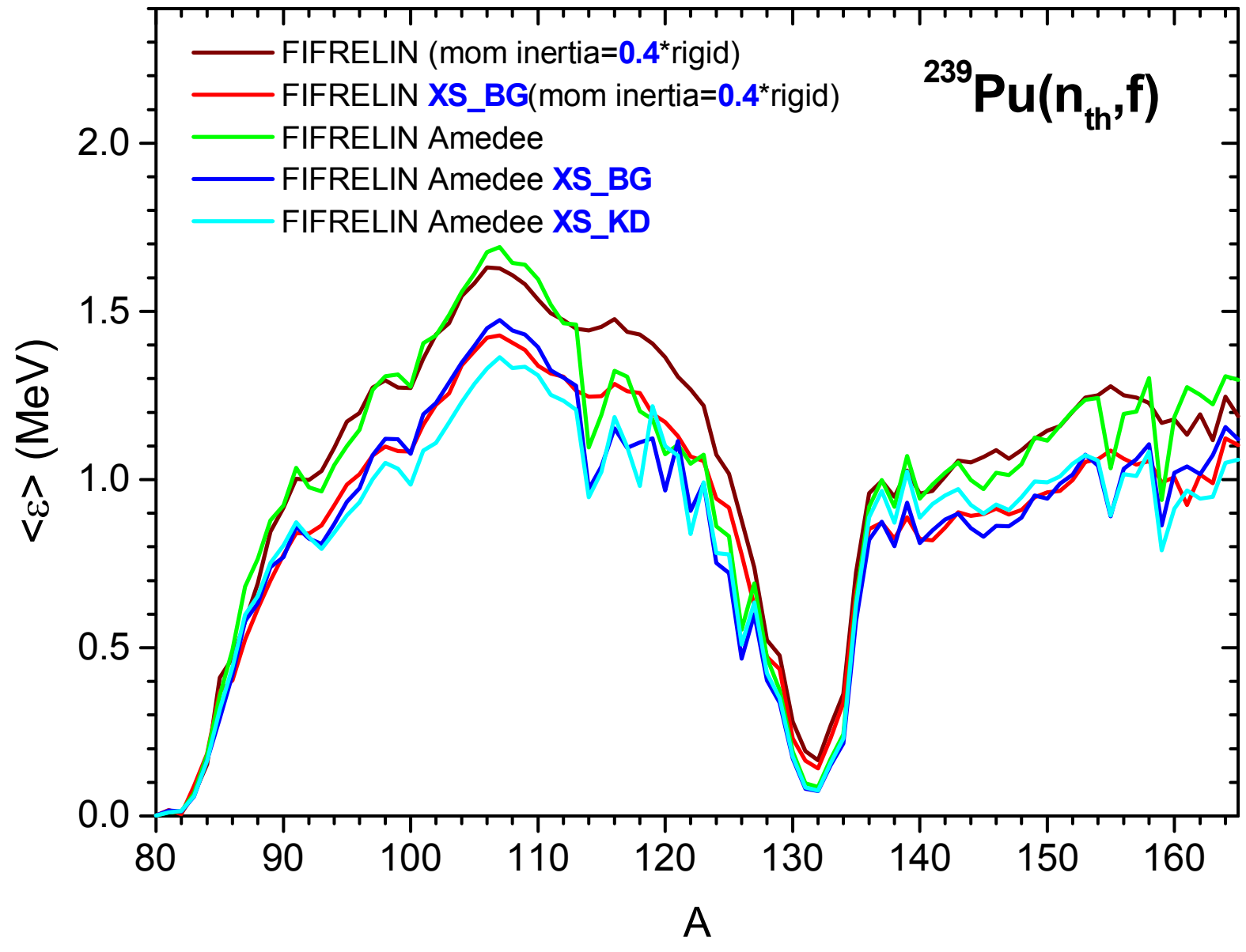

**Fig.4.23a**: FIFRELIN results of ε(A) for the 5 calculation cases

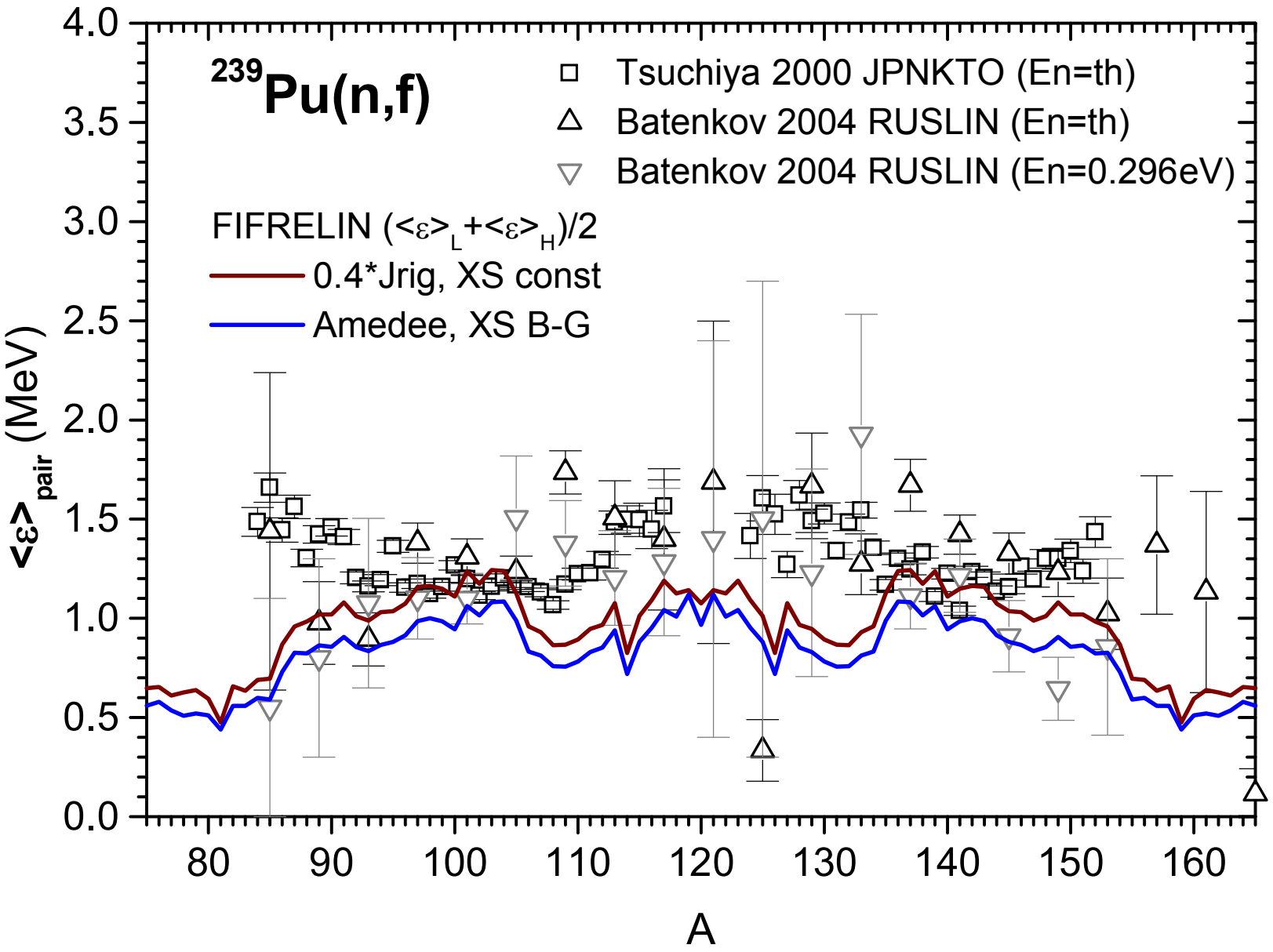

**Fig.4.23b**: FIFRELIN results of <ε> of fragment pairs in comparison with experimental data

The total average prompt neutron multiplicity results of PbP and FIFRELIN are given in **Table 4.2** together with their deviation versus the ENDF/B-VII.0 evaluation.

**Table 4.2** PbP and FIFRELIN results of $<\nu_p>$ and deviations versus ENDF/B-VII.0

| Model used and different calculation cases | Total average multiplicity result | Deviation from ENDF/B-VII.0 $<\nu_p>$=2.87245 |
|---|---|---|
| PbP 1Z/A $\sigma_c(\varepsilon)$ B-G (Tudora, 2010a) | 2.87872 | 0.22 % |
| PbP 2Z/A $\sigma_c(\varepsilon)$ B-G (Tudora, 2010c) | 2.8678 | 0.16 % |
| FIFRELIN 0.4Jrig, $\sigma_c(\varepsilon)$ constant | 2.8732 | 0.026 % |
| FIFRELIN 0.4Jrig, $\sigma_c(\varepsilon)$ B-G | 2.9155 | 1.5 % |
| FIFRELIN AMEDEE, $\sigma_c(\varepsilon)$ constant | 2.8514 | 0.73 % |
| FIFRELIN, AMEDEE, $\sigma_c(\varepsilon)$ B-G | 2.9013 | 1% |
| FIFRELIN, AMEDEE, $\sigma_c(\varepsilon)$ K-D | 2.9110 | 1.34 % |

As it can be seen in Table 4.2, both PbP results of $<\nu_p>$ as well as the FIFRELIN results for calculation cases with constant $\sigma_c$ are very close to ENDF/B-VII (with deviations less than 1%).

Total average prompt fission neutron spectrum calculated with FIFRELIN in three cases by taking $\sigma_c(\varepsilon)$ constant and moments of inertia of 0.4 from rigid body momentum and from the database AMEDEE and using $\sigma_c(\varepsilon)$ from optical model calculations with Becchetti-Greenless potential and moments of inertia from AMEDEE are plotted in comparison with the experimental data sets of Bojkov, Lajtay and Starostov (provided in January 2011 by IAEA-NDS in the frame of a coordinated research project on PFNS, CRP, 2011) in **Figs.4.24 a,b**. In the lowest parts of these figures the PbP spectrum (taken from Tudora, 2010c) is given too, in comparison with the same experimental data sets. In Figs 4.24a,b the PFNS results are given separately because the comparison with experimental data requires each time a re-normalization of experimental data to the respective calculated spectrum.

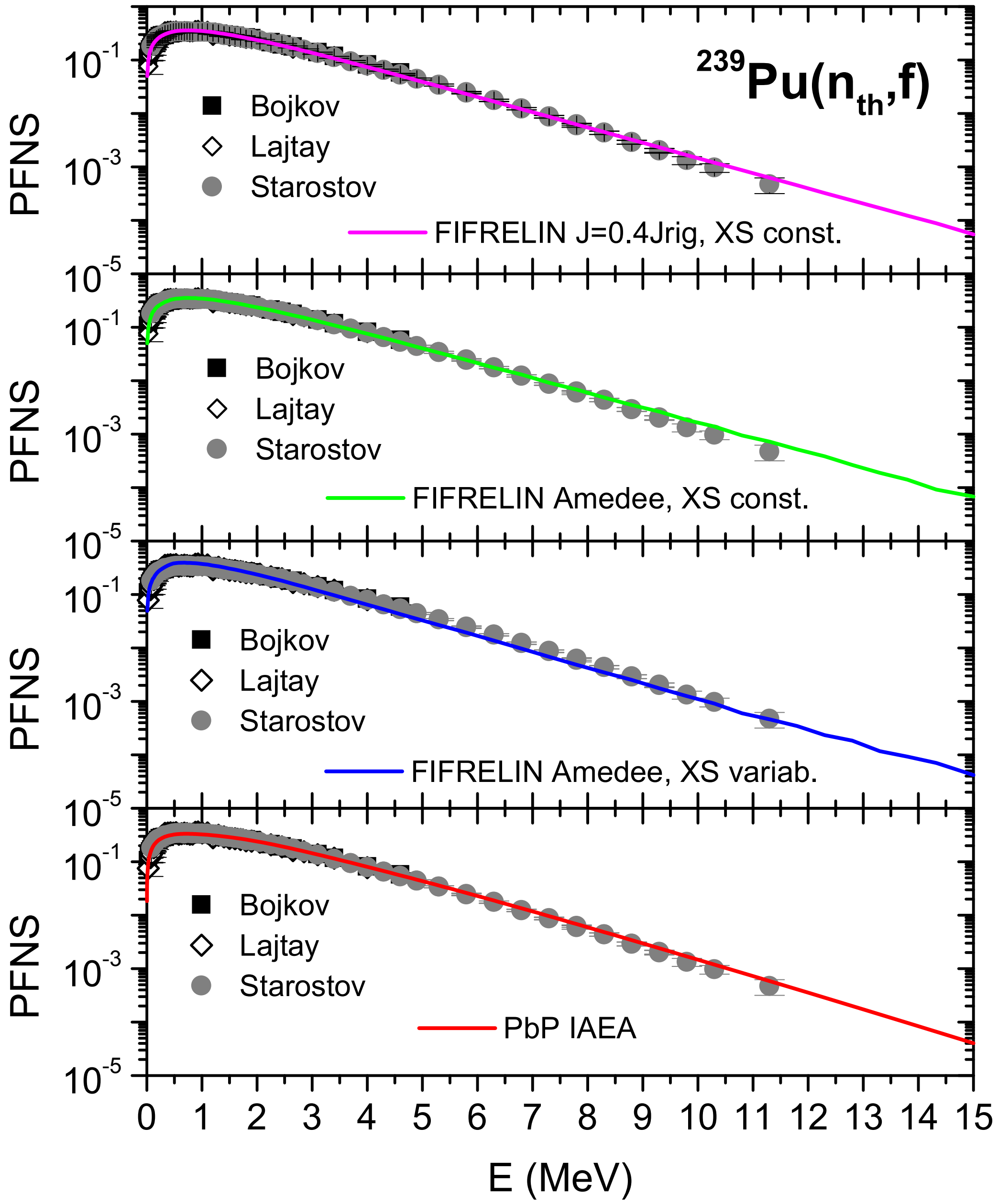

**Fig.4.24a**: $^{239}$Pu($n_{th}$,f) spectrum results of FIFRELIN (top and middle parts) and PBP bottom part) in comparison with re-processed experimental data (provided by IAEA) re-normalized to the respective calculated PFNS

Looking at **Fig.4.24a** where the representation in linear-logarithmic scales focus the high energy part of spectrum, it seems that a good agreement with experimental data in this energy range is obtained by the FIFRELIN calculations with $\sigma_c(\varepsilon)$ constant and by the PbP result. The FIFRELIN calculation with $\sigma_c(\varepsilon)$ variable underestimate the data at energies above 5 MeV.

Looking at **Fig.4.24b** where the representation in linear-linear scales focus the hard part of the spectrum, it is visible that a good description of experimental data is obtained in the same cases: the PbP spectrum and the FIFRELIN calculations with $\sigma_c(\varepsilon)$ constant.

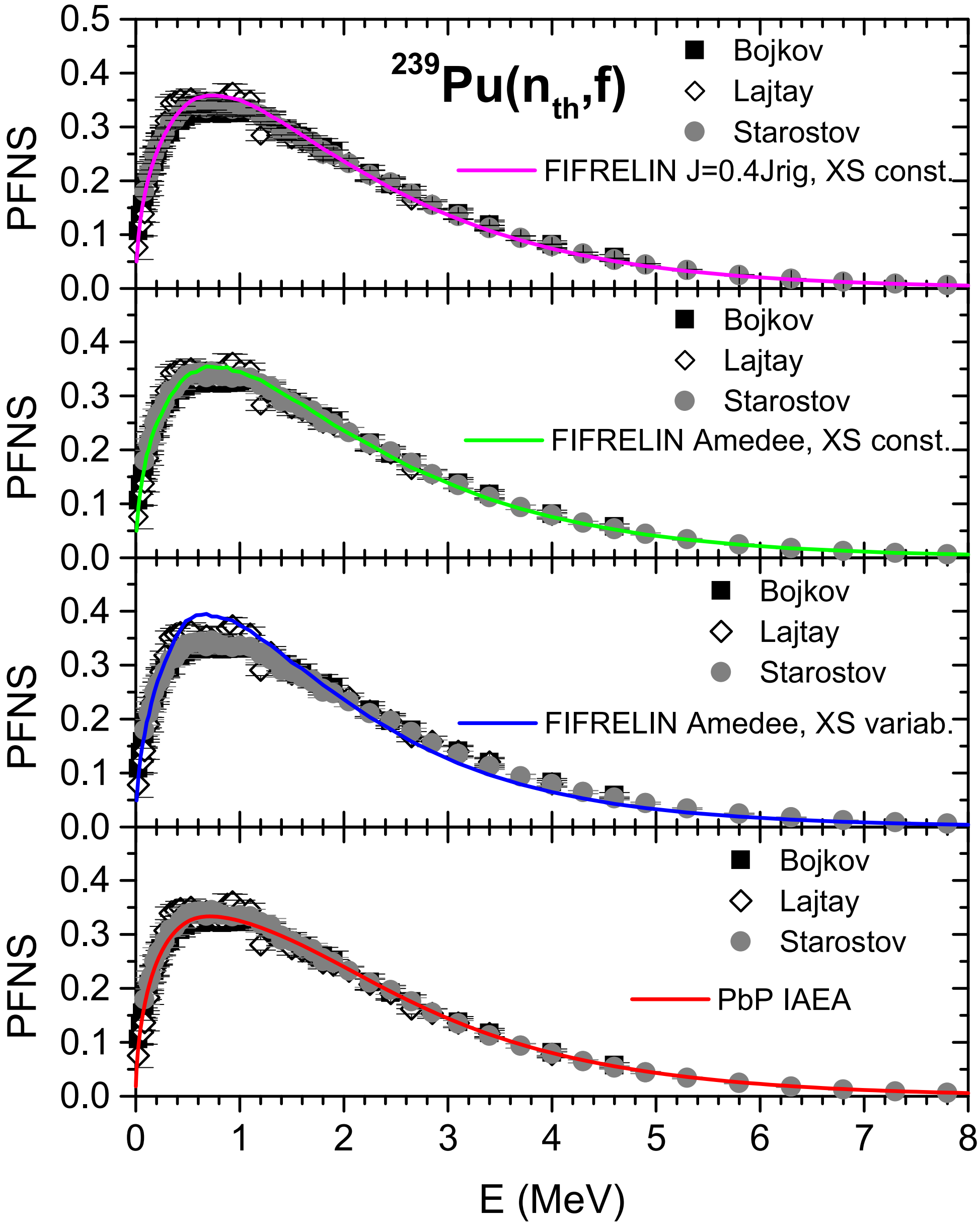

**Fig.4.24b**: $^{239}$Pu($n_{th}$,f) spectrum results of FIFRELIN (top and middle parts) and PBP bottom part) in comparison with re-processed experimental data (provided by IAEA) re-normalized to the respective calculated PFNS

The comparison with the experimental data is better visible if spectra are given as ratios to a Maxwellian spectrum. A representation as ratio to a Maxwellian spectrum with TM=1.415 MeV is plotted in **Fig.4.25**: upper part the FIFRELIN calculation with constant $\sigma_c(\varepsilon)$ and moments of inertia form the data base AMEDEE and lower part the PbP spectrum. As it can be seen both PbP and FIFRELIN spectra describe well the experimental data up to about 5-6 MeV. At high energies the PbP spectrum slightly overestimates the data of Starostov while the FIFRELIN result overestimates considerably the Starostov data at energies above 5 MeV.

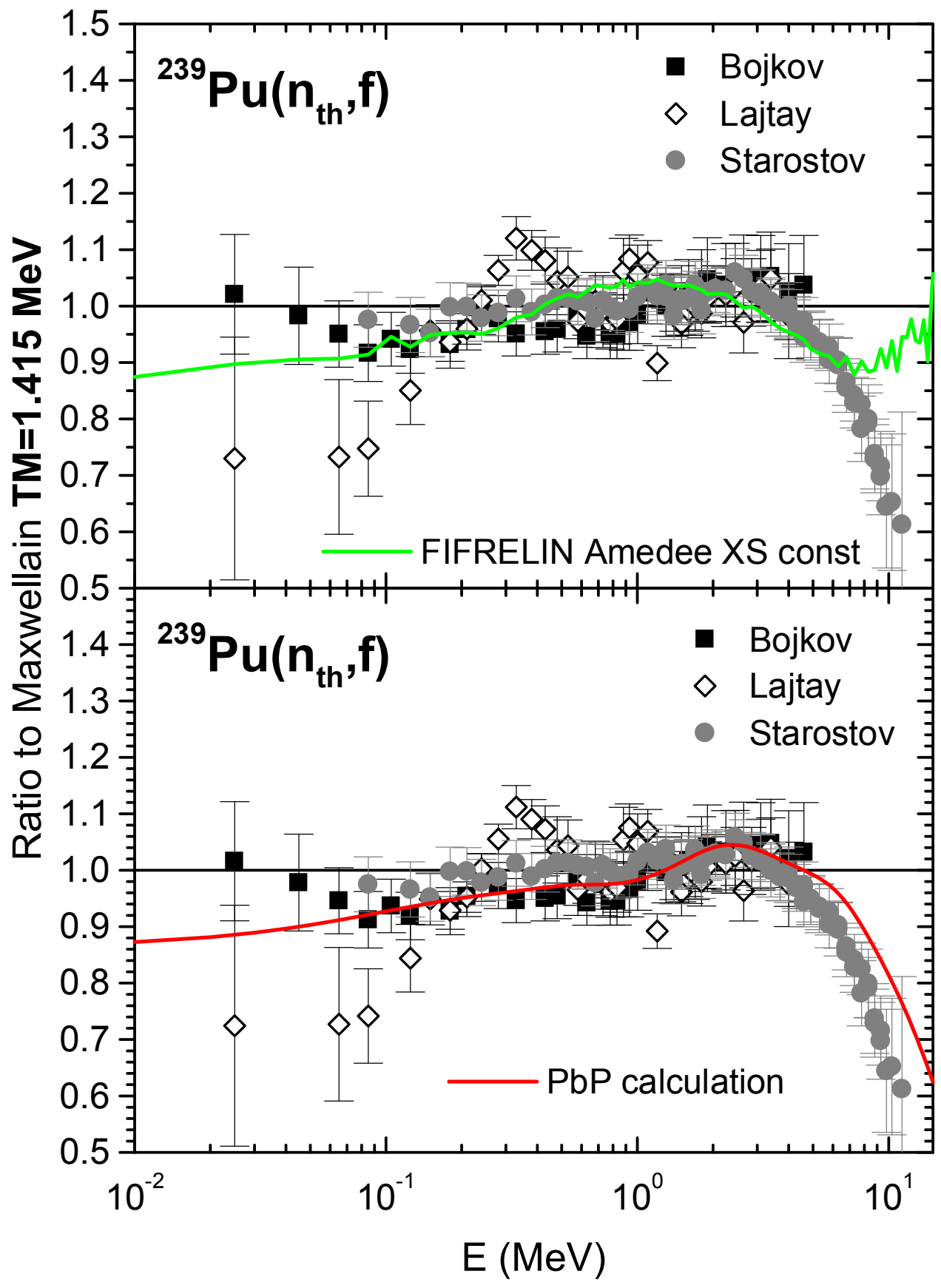

**Fig.4.25**: PFNS given as ratio to Maxwellian: FIFRELIN result in the upper part and PbP result in the lower part in comparison with the re-processed experimental data (IAEA).

It is interesting to mention that the PbP spectrum (taken from Tudora, 2010c) that is plotted in the lower parts of Fig.4.7 in comparison with the re-processed experimental data sets provided by IAEA CRP, 2011, is in good agreement at high energies with the data sets retrieved from EXFOR in 2010 (EXFOR, 2010) as it can be seen in **Fig.4.26**.

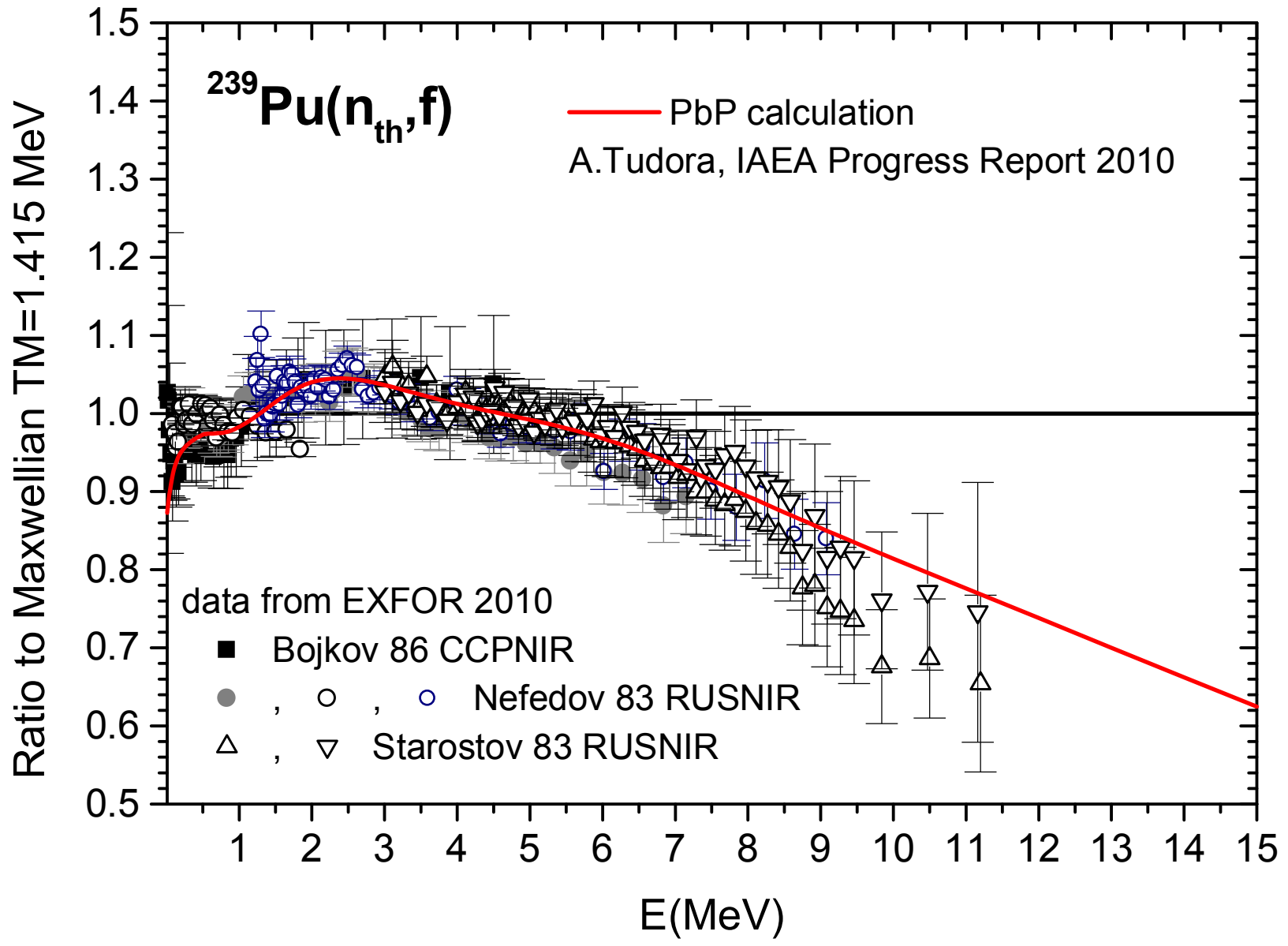

**Fig.4.26**: PbP spectrum given as ratio to Maxwellian (TM=1.415 MeV) in comparison with experimental data retrieved from EXFOR in 2010 (different sumbols).

### IV.2.1 Parameterizations of the fragment temperature ratio

Another interesting comparison between PbP and FIFRELIN treatments concerns the temperature ratios $RT=T_L/T_H$ as a function of light or heavy fragment mass number as appearing in the FIFRELIN and PbP treatments, respectively.

In the case of PbP treatment the temperature ratio RT refers the to maximum values of residual temperature distributions P(T). This RT ratio is obtained in the frame of the PbP calculation as resulting form the TXE partition and is given as a function of $A_H$. For instance examples and discussions about $RT(A_H)$ resulted from different methods of TXE partition are given in Manailescu et al, 2011, Morariu et al, 2012 and in Chapter II).

In the case of FIFRELIN, as it was described by Litaize and Serot, 2010 and it was mentioned in Chapter III, the RT ratio refers to the residual temperature entering the evaporation spectrum of fragments during the sequential emission of neutrons. This RT given as a function of $A_L$ is an input parameterization in the FIFRELIN code. RT entering FIFRELIN is defined as described in chapter III: RT=1 for the symmetric fragmentation, the maximum RT value of 1.6 is taken from fragment pairs with $A_H$=132 (driven by the double magic numbers Z=50, N=82) and a minimum RT value of 0.4 is taken for the last far asymmetric fragmentation.

Different RT functions for the $^{239}$Pu($n_{th}$,f) case are plotted in **Fig.4.27** as a function of $A_H$ as following. RT($A_H$) plotted with red points is resulted from the PbP calculation with the TXE partition method based on the parameterization $\nu_H/\nu_{pair}$ (based on the systematic behaviour of experimental ν(A) data). The RT($A_H$) parameterization proposed in (Manailescu et al, 2011) is plotted with blue line. The RT function entering the FIFRELIN code is plotted with green line. As it can be seen the linear RT parameterization resulted form the PbP treatment (blue line) and the linear parameterization used in FIFRELIN (green line) exhibit behaviours that are similar for the physical point of view: for fragment pairs from symmetric fission up to $A_H$=130 they are identical. Both RT are decreasing for pairs with $A_H$ above 132. At far asymmetric fragmentations the RT function of FIFRELIN is lower than the RT parameterization resulted from the PbP treatment.

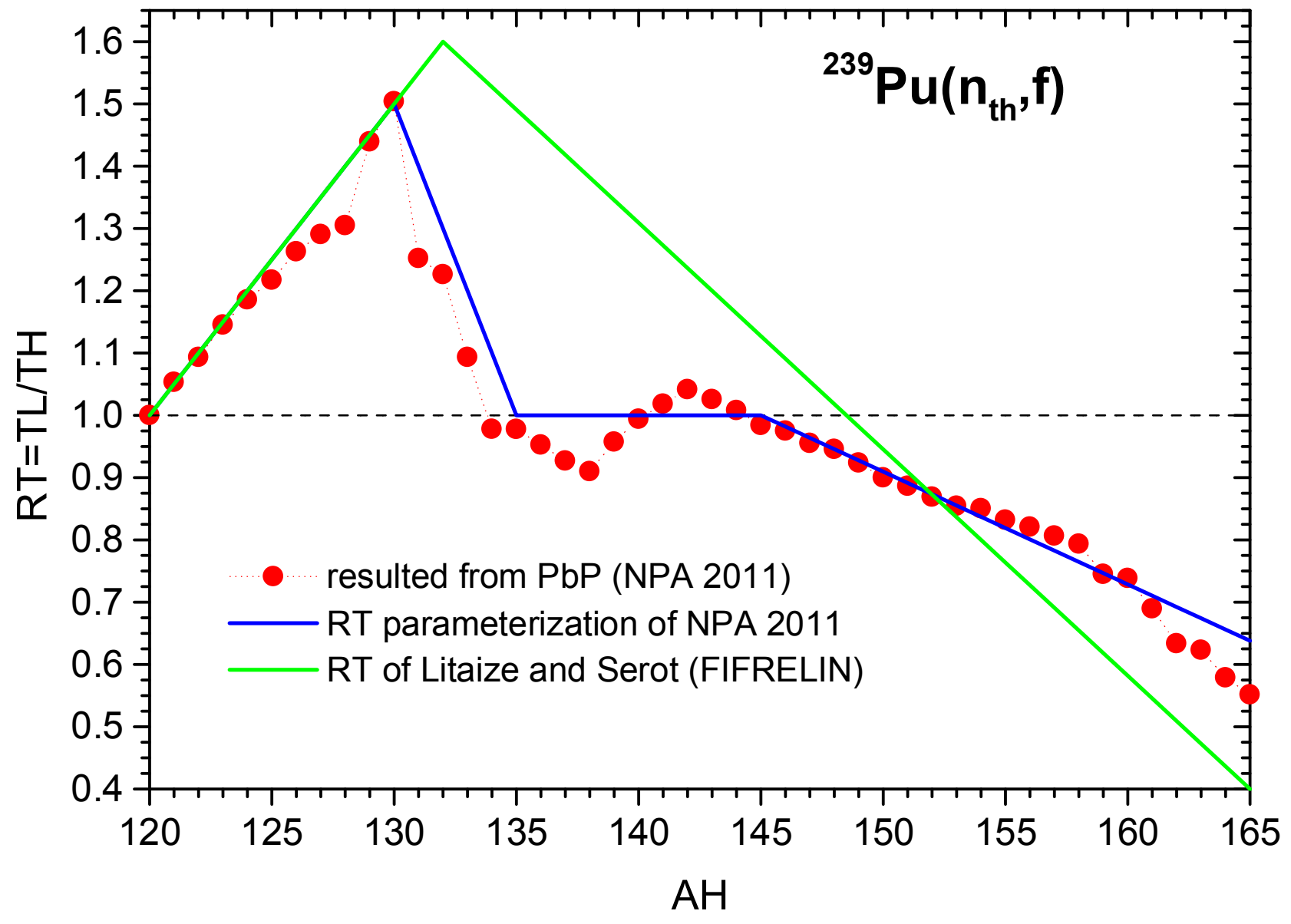

**Fig.4.27**: $^{239}$Pu($n_{th}$,f) RT ratios plotted as a function of AH : RT resulted from PbP calculation with red circles, the RT parameterization of Manailescu et al, 2011 with blue line and the RT function of Litaize and Serot, 2010 used in present FIFRELIN calculations with green line.

We suppose that the FIFRELIN results of ν(A) exhibiting in the case of $^{239}$Pu(SF) a more pronounced sawtooth shape a little bit shifted in comparison with the shape of experimental data and PbP results (see Fig.4.21) can be due to the RT function used in FIFRELIN that was initially made for the case of $^{252}$Cf(SF).

In Ref.(Manailescu et al, 2011) it is mentioned that in the case of $^{252}$Cf(SF) the use of the RT function of Litaize and Serot, 2010 in the frame of the PbP treatment leads to fragment excitation energy results E*(A) that are in very good agreement with E*(A) obtained from the TXE partition method based on the $\nu_H/\nu_{pair}$ parameterization as it can be seen in **Fig.4.28** (which is a reproduction of Fig.2e from Manailescu et al, 2011). This fact proves that in the case of $^{252}$Cf(SF) the RT function of Litaize and Serot is working very well in both treatments PbP and Monte-Carlo (FIFRELIN).

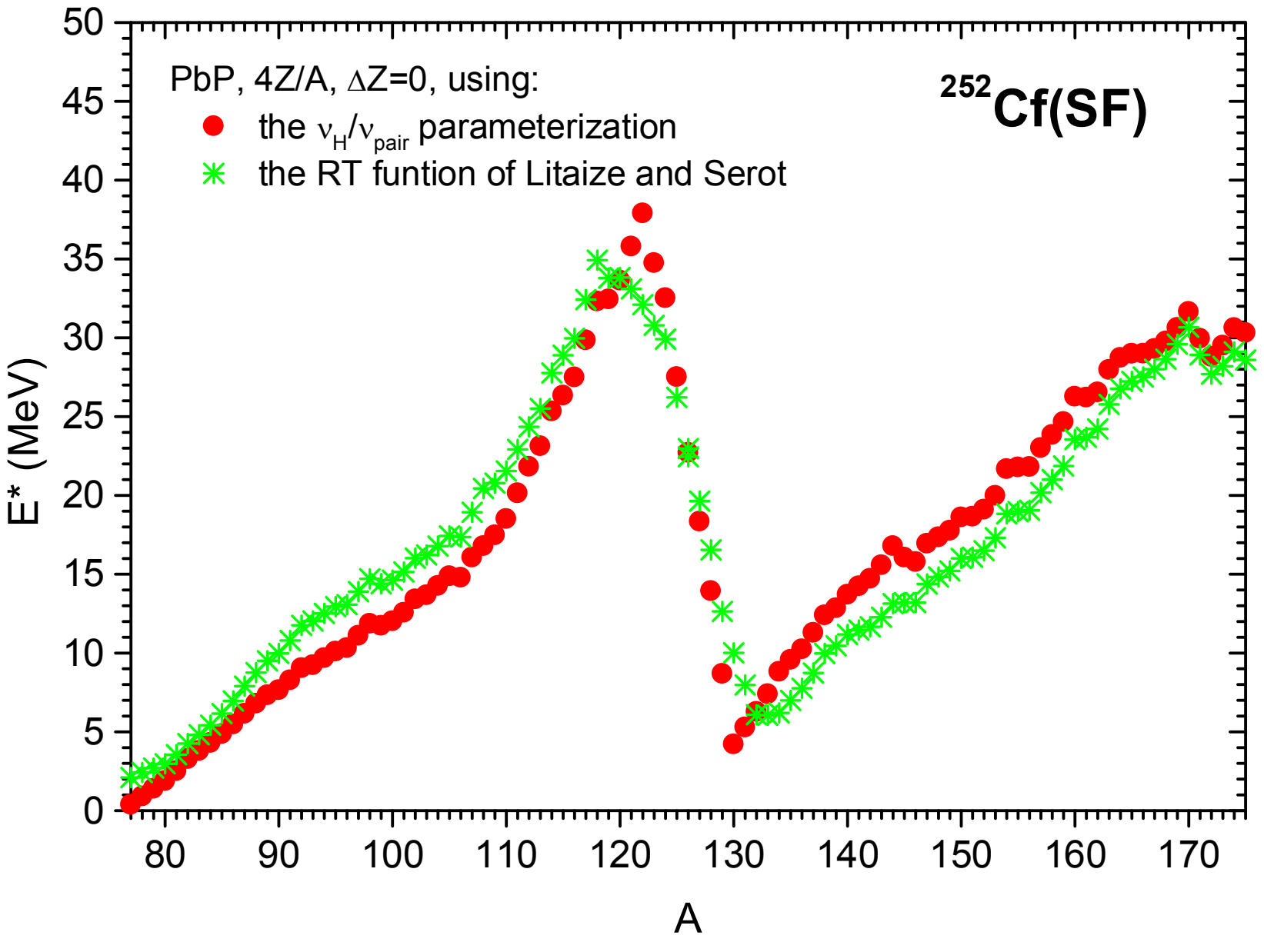

**Fig.4.28**: $^{252}$Cf(SF) Excitation energy of fully-accelerated fragments forming the fragmentation range of the PbP treatment build by taking 4 Z for each mass number A. The plotted E*(A) were averaged over the gaussian charge distribution. E*(A) resulted from the TXE partition based on the $\nu_H/\nu_{pair}$ parameterization with red circles and E*(A) obtained by using the RT function of Litaize and Serot (in an iterative procedure in the frame of the super-fluid model) with green stars. Reproduction of Fig.2e from (Manailescu et al, 2011).

In order to see the influence of different RT functions on the fragment excitation energies at full acceleration E*(A) and implicitly on the sawtooth shape of ν(A) the following simple exercise was made for the case of $^{239}$Pu($n_{th}$,f).

The fragmentation range was taken with A from 75 to 165 and for each A two charge numbers Z were taken as the nearest integers above and below the most probable charge considered one time as UCD (as in the FIFRELIN calculations) and another time as UCD corrected with a charge polarization of ±0.5 (as in the PbP calculation). TXE of fragment pairs calculated by using the experimental TKE(A) data of Wagemans in the two cases ΔZ=0 (black squares) and ΔZ=|0.5| (red circles) are plotted in **Fig.4.29**. The TXE values plotted as a function of $A_H$ were already averaged over the charge distribution

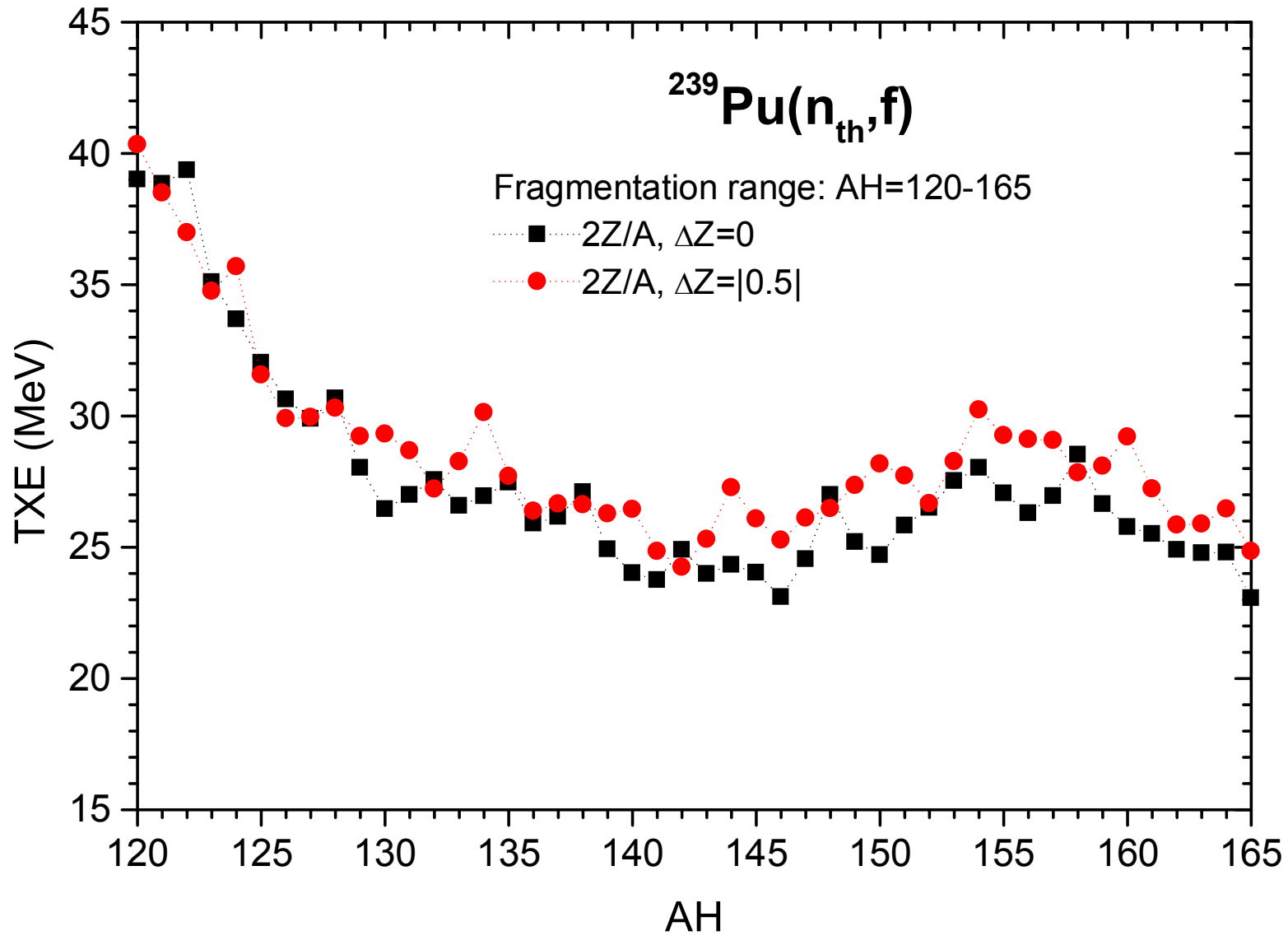

**Fig.4.29**: $^{239}$Pu($n_{th}$,f) TXE calculations considering UCD (black squares) and a charge polarization (red circles). In both cases 2Z/A were taken and the plotted TXE were averaged over the charge distribution.

As it can be seen in Fig.4.29 the differences between TXE calculated by taking ΔZ=0 (UCD) and ΔZ=|0.5| are not large.

E*(A) was calculated by an iterative procedure in the frame of the generalized super-fluid model (as described in Chapter 2 and Appendix 2) under the condition that the light and heavy fragment temperatures resulted from $T_{L.H} = \sqrt{E^*_{L,H} / a_{L,H}}$ are in the ratio given by the RT functions of Litaize and Serot (plotted with green line in Fig.4.27) or of (Manailescu et al, 2011) plotted with blue line in Fig.4.27. The obtained E*(A) (already averaged over the Gaussian charge distribution) for the cases ΔZ=0 (UCD) and ΔZ=|0.5| are plotted in **Fig.4.30** as following:

1) $\Delta Z=0$ and RT of Litaize and Serot, 2010 with full black circles (this being a simulation of the calculation conditions of FIFRELIN)

2) $\Delta Z=0.5$ and RT of Litaize and Serot with open black circles

3) $\Delta Z=0$ and RT from Ref. Manailescu et al, 2011 with magenta stars

4) $\Delta Z=0.5$ and RT from Manailescu et al. with red squares (this being the situation of the PbP calculation)

As it was expected (taking into account the TXE results plotted in Fig.4.29) the fragmentation range build by considering or not the charge polarization does not lead to significant differences in E*(A) results (see the data plotted with full and open black circles that are very close each other and the data plotted with magenta stars and red squares that are very close each other, too).

Looking at the behaviour of E*(A) plotted in **Fig.4.30** with full black circles (FIFRELIN calculation case) and red squares (PbP calculation case) it is easy to observe that different RT functions lead to visible different E*(A) results especially for A between 95-112 and 130-148.

In other words in the case of $^{239}Pu(n_{th},f)$ the two RT functions, of Litaize and Serot and of PbP, do not lead to close E*(A) as in the case of $^{252}Cf(SF)$, see comparatively Fig.4.28 and Fig.4.30.

Taking into account the great influence of the RT function on the shape of E*(A), from this exercise we can suppose that the shape of ν(A) results of FIFRELIN plotted in Fig.4.21 (exhibiting visible differences compared to PbP results for A between 98–112 and between 130–135) are mainly due to the RT function used in FIFRELIN. This statement is supported by the E*(A) result plotted in Fig.4.30 with full black circles that exhibits visibly other shape compared to the E*(A) plotted with red squares for fragmentations with $A_H$ between 130–145 (and complementary $A_L$=110–95). This behaviour of E*(A) (plotted with full black circles) being reflected in the shape of ν(A) results of FIFRELIN given in Fig.4.21.

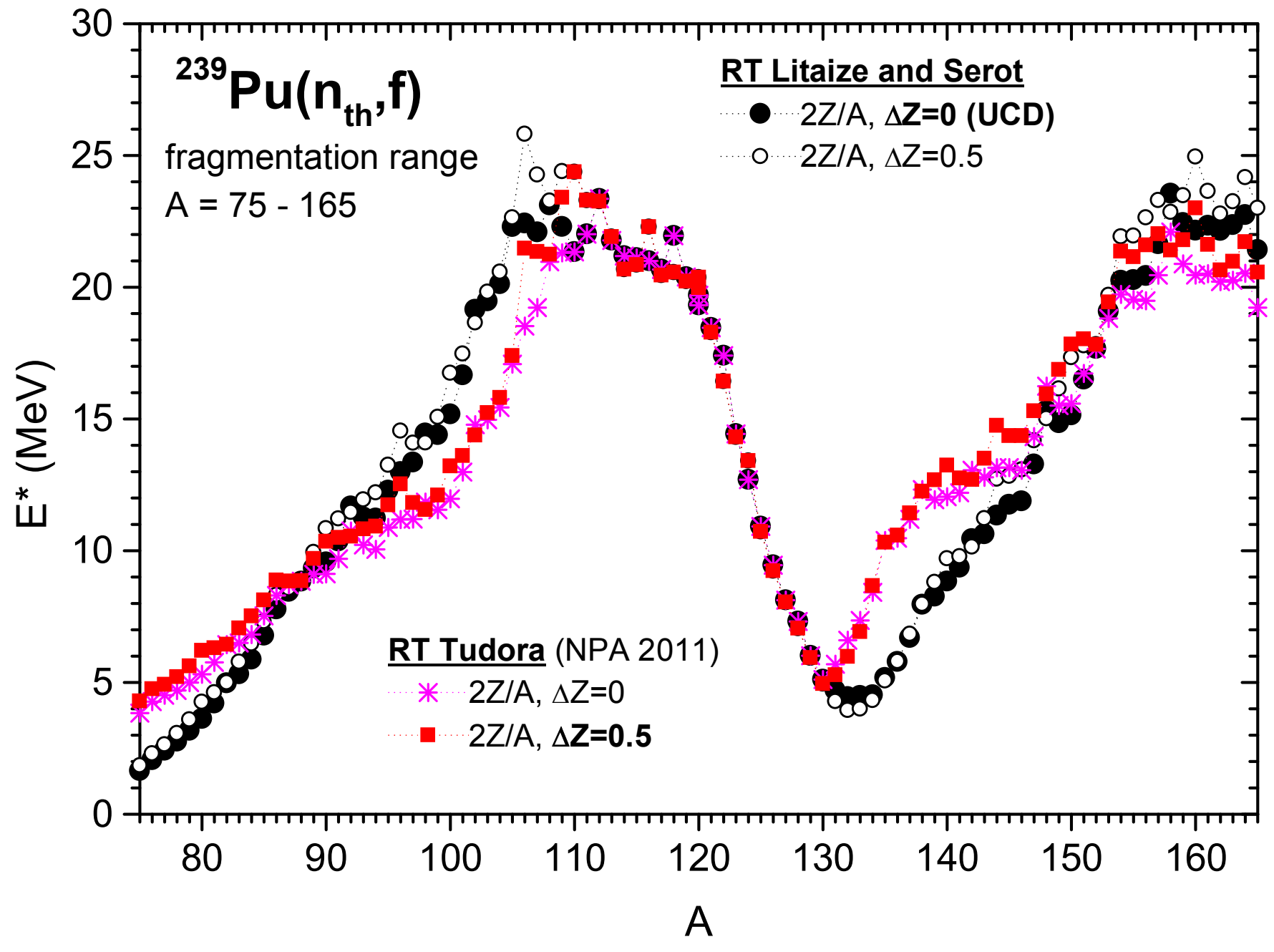

**Fig.30**: $^{239}$Pu(n$_{th}$,f) E*(A) calculations using the RT function of Litaize and Serot with full black circles for the case ΔZ=0 and with open black circles for ΔZ=0.5 and using the RT function of Manailescu et al, with magenta stars for the case ΔZ=0 and with red squares for ΔZ=0.5

Taking into account that in the frame of the super fluid model included in the codes FIFRELIN and PbP the damping parameter γ and the asymptotic level density parameterizations are a little bit different, we have tested the influence of these different parameterizations on the E*(A) results, too. For instance in **Fig.4.31** E*(A) for the case ΔZ=0 and RT of Litaize and Serot were calculated by considering the γ and $\tilde{a}$ parameterizations used in PbP calculation (black circles) and in FIFRELIN (blue open circles), in both calculations the shell corrections of Moller and Nix (RIPL–3, 2012d) being used. As it can be seen the differences due to different γ and $\tilde{a}$ parameterizations are insignificant and practically do not affect the results of this exercise.

The influence of shell-corrections taken from different databases is also insignificant, see the E*(A) result plotted with green stars in Fig.4.31 obtained by using the shell-corrections of Myer and Swiatecki (RIPL–3, 2012d).

From this exercise we conclude that in the case of $^{239}$Pu(n$_{th}$,f) the consideration of another RT function in FIFRELIN (eventually based on the systematic behaviour of experimental ν(A) data) might improve the results of prompt neutron emission quantities as a function of A (such as ν(A), ε(A)).

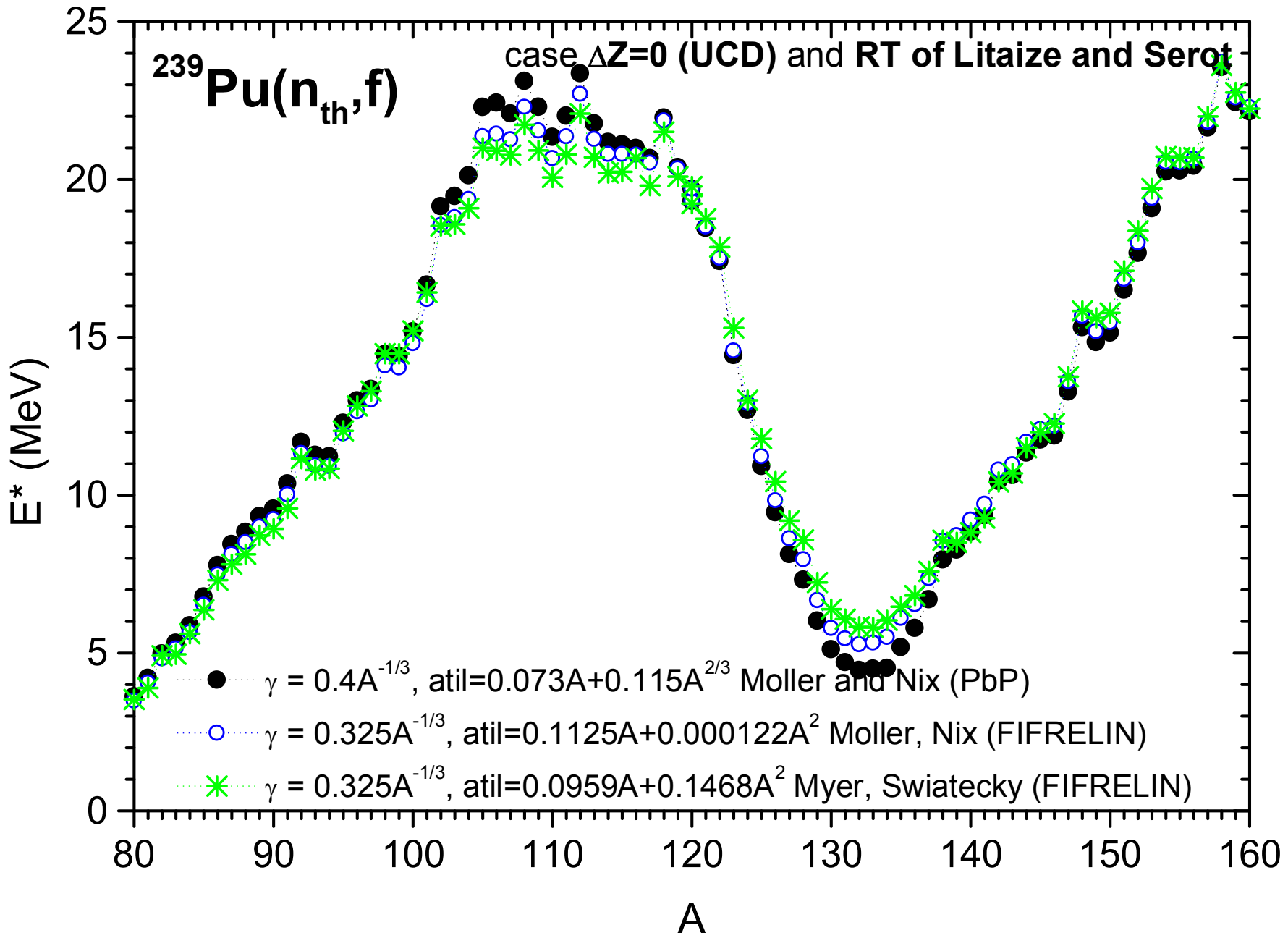

**Fig.4.31**: $^{239}$Pu(n$_{th}$,f) E*(A) for the case ΔZ=0 and RT of Litaize and Serot calculated by using the γ and $\tilde{a}$ parameterizations used in PbP calculations (full black circles) and FIFRELIN calculations (open blue circles for shell corrections of Moller and Nix and green stars for shell corrections of Myer and Swiatecki).

### IV.2.2 FIFRELIN calculations using the RT function provided by the PbP treatment

The RT function obtained from the PbP treatment (as it was described in Manailescu et al, 2011) plotted with blue line in Fig.4.27 was introduced in the code FIFRELIN. The case with moments of inertia taken 0.4 of rigid body momentum and constant $\sigma_c$ is running two times by sampling $10^6$ and $10^7$ events.

The new results of ν(A) are plotted in **Figs.4.32 a,b** with magenta stars ($10^6$ events) and red diamonds ($10^7$ events) connected with solid lines.

In Fig.4.32a the new FIFRELIN results of ν(A) are given in comparison with previous results (for the 5 studied cases). The differences between the previous and present ν(A) results are significant proving the very important role played by the RT function in the frame of FIFRELIN calculations.

As it was expected (see Fig.4.32b) the $\nu(A)$ results of FIFRELIN with the RT function obtained from the PbP treatment (red diamonds and magenta stars) are closer to PbP results (blue and green circles) than the previous FIFRELIN results, exhibiting a similar behaviour but with a little bit more pronounced sawtooth shape.

The FIFRELIN result obtained by sampling $10^6$ events describes well the experimental data and is close to the PbP results for fragments with $A_H>135$ while $\nu(A)$ obtained by sampling $10^7$ events remains in agreement with experimental data and overestimates the PbP results for $A_H>140$. For $A_L<105$ both FIFRELIN results underestimate the experimental data and the PbP results.

The total average multiplicity values provided by FIFRELIN using the RT function from PbP are $<\nu_p>=2.88052$ (case $10^7$ events) and $<\nu_p>=2.73688$ (case $10^6$ events), in good agreement with experimental data, deviating in the case of $10^7$ events only with 0.28% from the ENDF/B-VII.0 evaluation.

Multiplicities of fragment pairs as a function of $A_H$ ($\nu_{pair}(A_H)$) are plotted in **Fig.4.33** as following: experimental data with open symbols, PbP result with green circles, previous FIFRELIN results with blue triangles (for the case of moments of inertia taken 0.4 Jrig and $10^7$ events), present FIFRELIN results (using the RT function resulted from the PbP treatment) with red diamonds (case $10^7$ events) and magenta stars (case $10^6$ events).

It is very interesting to observe the very close results of $\nu_{pair}(A_H)$ provided by FIFRELIN in the same calculation conditions (regarding the moments of inertia and sampling $10^7$events) but using different RT functions: see the blue triangles and red diamonds connected with solid lines. This fact proves again that in the case of FIFRELIN the sawtooth shape of $\nu(A)$ depends practically only on the RT function used.

As it can be seen in Fig.4.33 both previous and present FIFRELIN results of $\nu_{pair}(A_H)$ obtained by sampling $10^7$ events are in agreement with the PbP result, too. As it was expected (looking ar the lower part of Fig.4.32) $\nu_{pair}(A_H)$ for the calculation by sampling $10^6$ events is a little bit lower for pairs with $A_H$ above 145 (because of the visible underestimation of complementary light fragment multiplicities for $A_L< 95$).

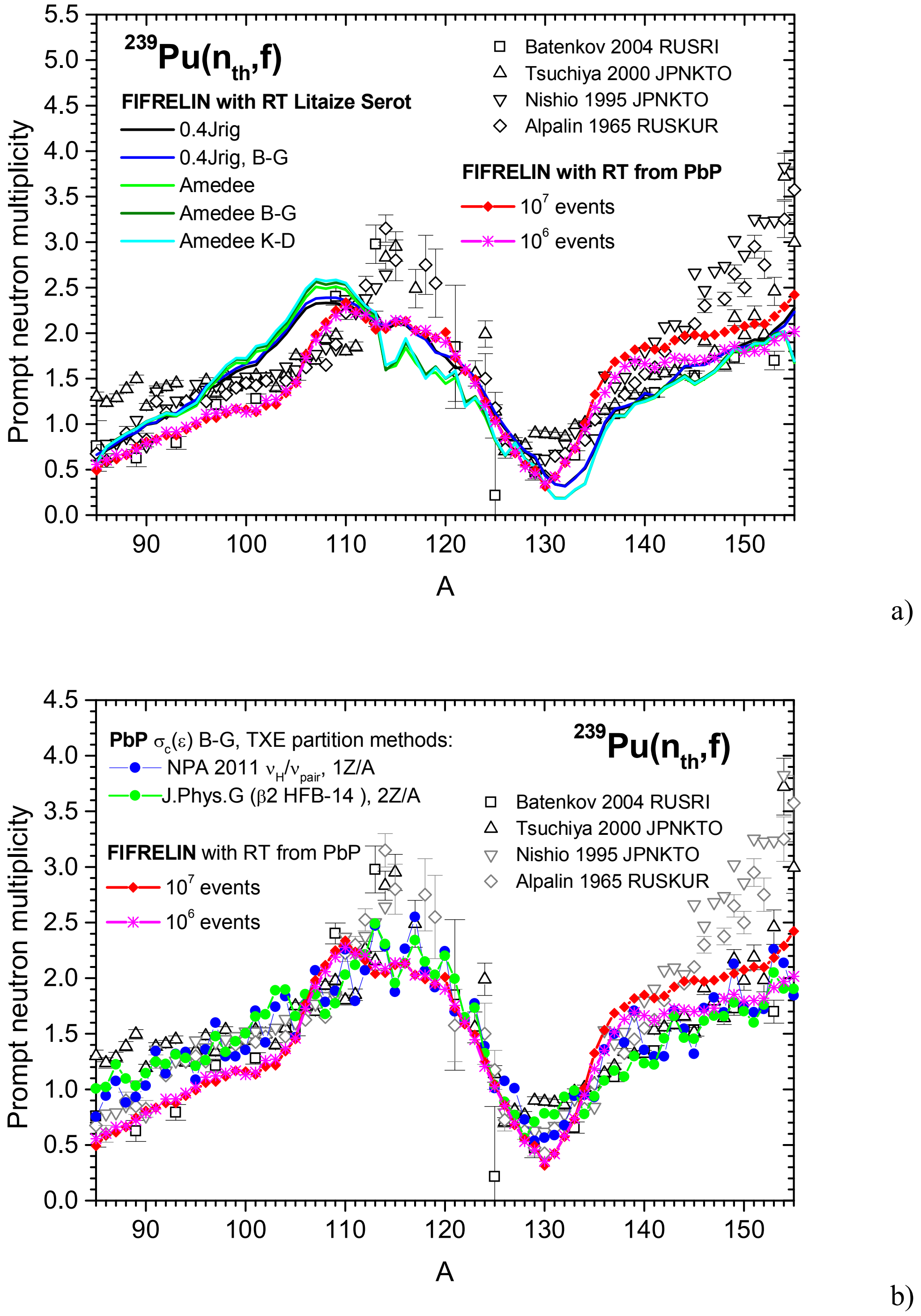

**Fig.4.32**: Different ν(A) results in comparison with experimental data (open symbols): **a)** previous FIFRELIN results (solid lines different coloured) and present results (red diamonds for the case of $10^7$events and magenta stars for the case of $10^6$ events) **b)** present results of FIFRELIN (same symbols and colours as in the upper part) and PbP results (blue and green circles).

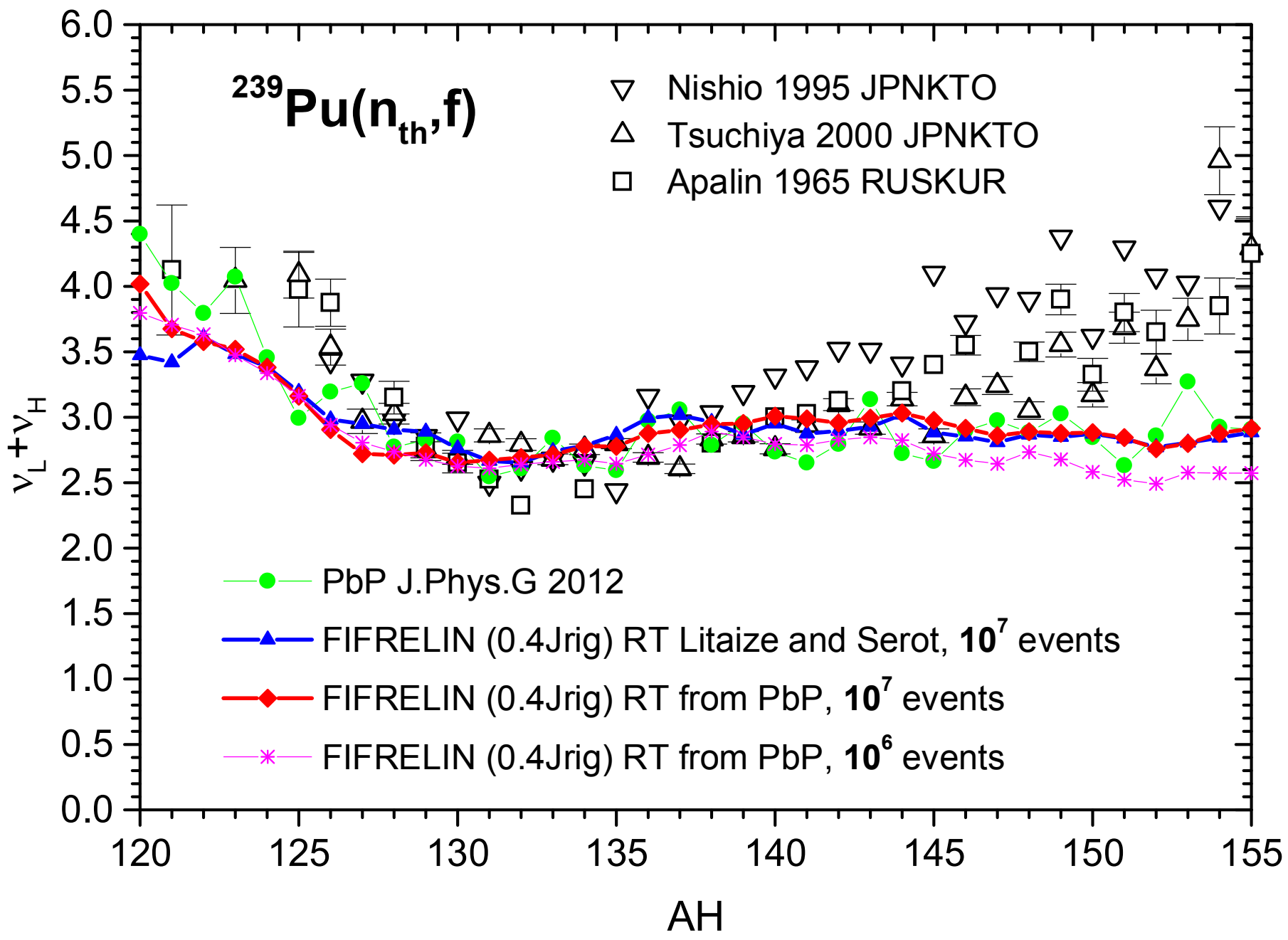

**Fig.4.33**: $\nu_{pair}(A_H)$ obtained from FIFRELIN calculations (case 0.4Jrig) using the RT function of Litaize and Serot (blue triangles) and RT provided by PbP (red diamonds by sampling $10^7$ events and magenta stars $10^6$ events) in comparison with $\nu_{pair}(A_H)$ of PbP using the TXE partition from (Morariu et al., 2012) (green circles) and with experimental data from EXFOR (different open black symbols)

The influence of the RT function on the sawtooth shape is visible in the case of ε(A), too, as it can be seen in **Fig.4.34** where the previous and present FIFRELIN results are plotted for the case of moments of inertia taken 0.4 form rigid body momentum as following: with blue triangles (using the RT function of Litaize and Serot) and with red circles and magenta stars when the RT function from PbP is used in calculations considering $10^7$ and $10^6$ events, respectively. As it can be seen the influence of the number of events sampled in the Monte Carlo treatment is visible only for fragments with $A_H$ above 140.

The present results of average <ε> of fragment pairs (plotted with the same symbols and colours as in Fig.4.34) differ from the previous ones and tend to approach a little bit the experimental data as in can be seen in **Fig.4.35**.

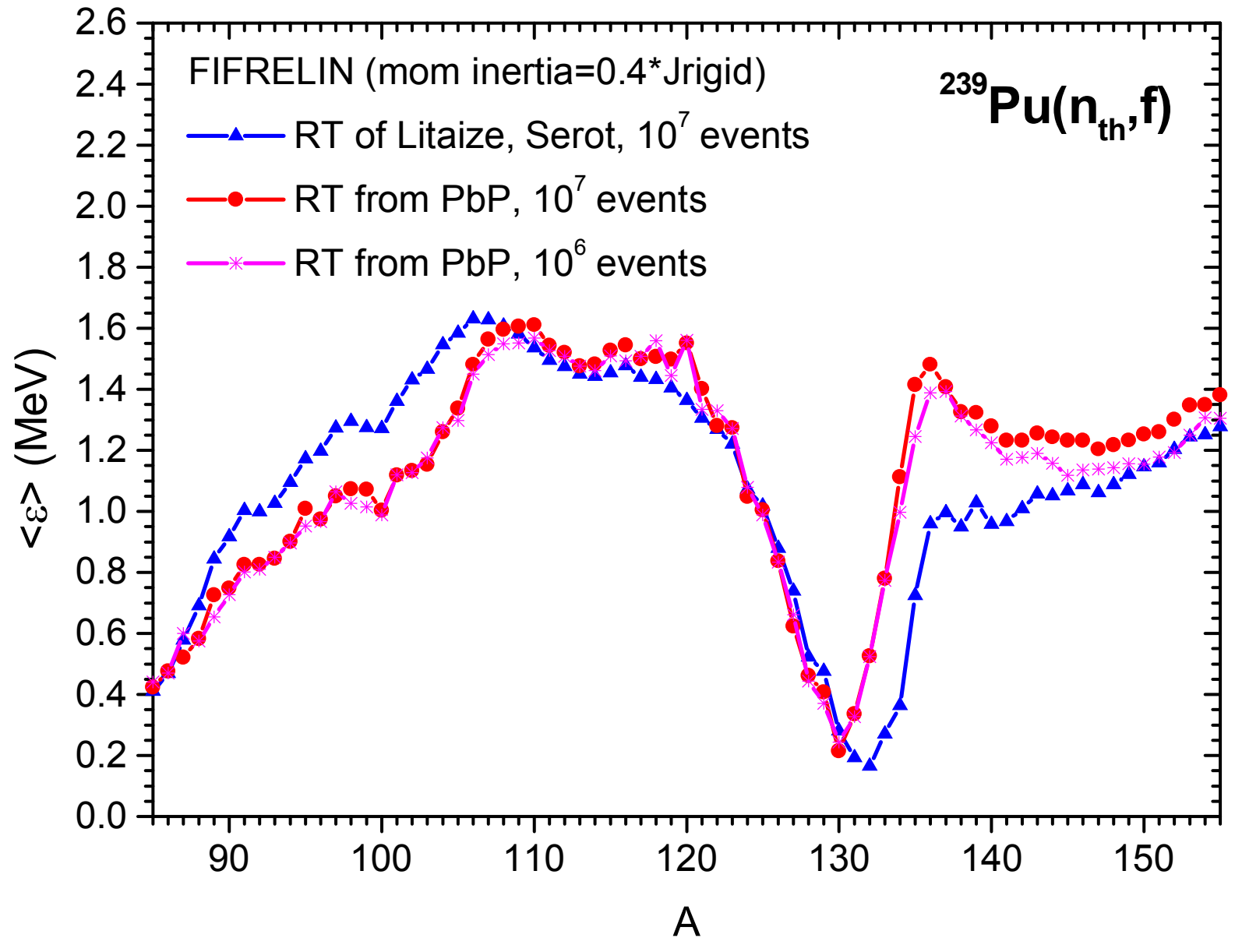

**Fig.4.34**: FIFRELIN result of ε(A) in the case on moments of inertia 0.4 of rigid body momentum: previously obtained by using the RT function of Litaize and Serot and $10^7$ events with blue triangles and the present ones using the RT function from PbP with red circles (sampling $10^7$ events) and magenta stars (sampling $10^6$ events)

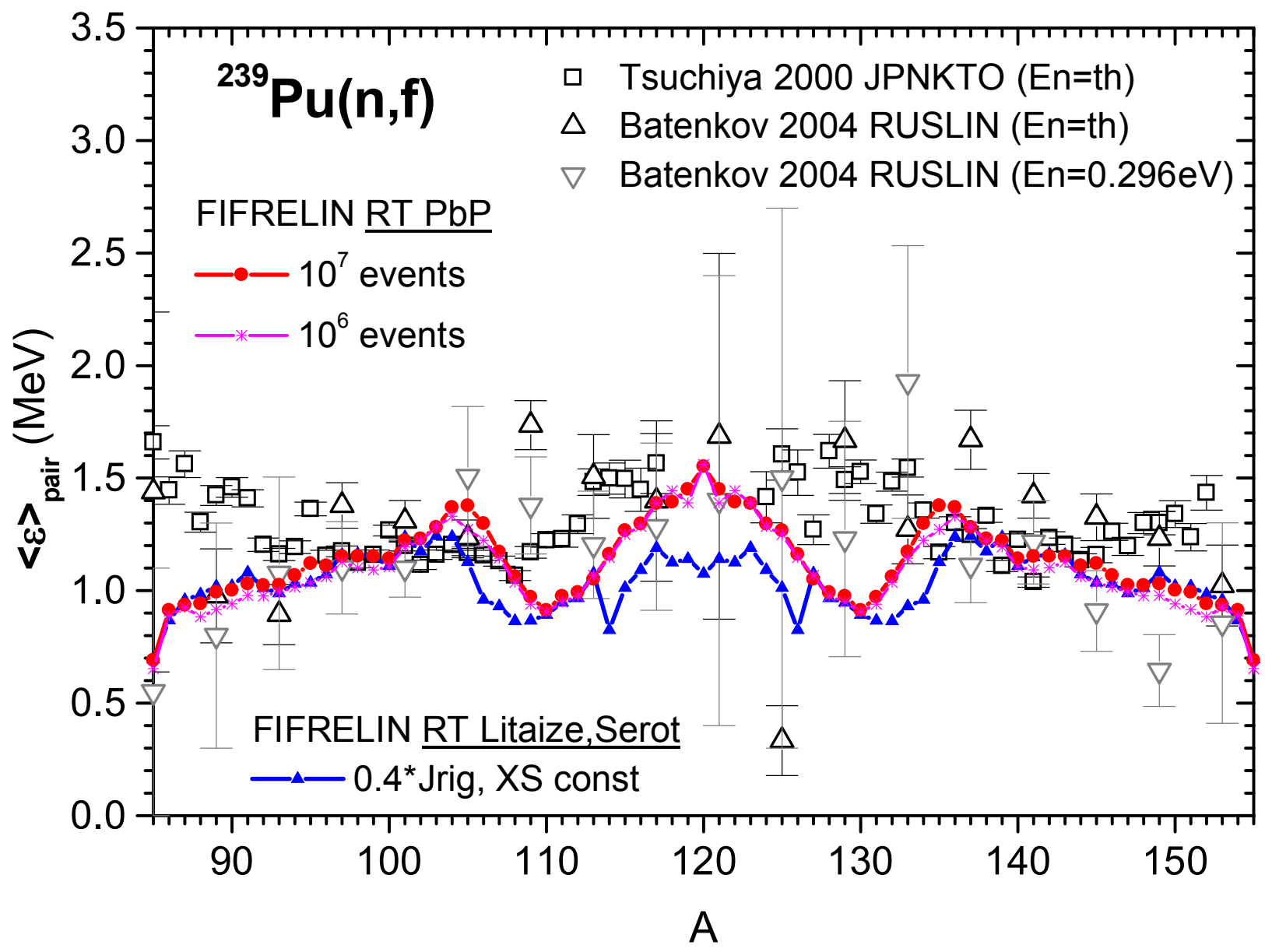

**Fig.4.35**: Previous and present FIFRELIN results of average <ε> of fragment pairs (plotted with the same symbols and colours as in Fig.4.34 in comparison with existing experimental data (open symbols)

In the case of P(ν) a much better agreement with experimental data is obtained by using the RT function from PbP as it can be seen in **Fig.4.36**. A very good description of experimental P(ν) data is obtained by sampling $10^6$ events (blue stars), this result being also very close to the PbP result (plotted with red circles).

As it can be seen in **Figs.4.37,** the inclusion of another RT function in the code FIFRELIN leads to <ν>(TKE) with a different behaviour compared to previous results. The pronounced flattening of <ν> at low TKE values exhibited by the previous FIFRELIN result (plotted in the upper part of Figs.4.37 with open blue triangles) is almost vanishing when the RT function of PbP is used, see the <ν>(TKE) results plotted in both parts of Fig.4.37 with full red circles (case of $10^7$ events) and open magenta circles (case of $10^6$ events). As it was already mentioned the structure at TKE values below 140 MeV becomes less pronounced when the number of events is increased (from $10^6$ to $10^7$).

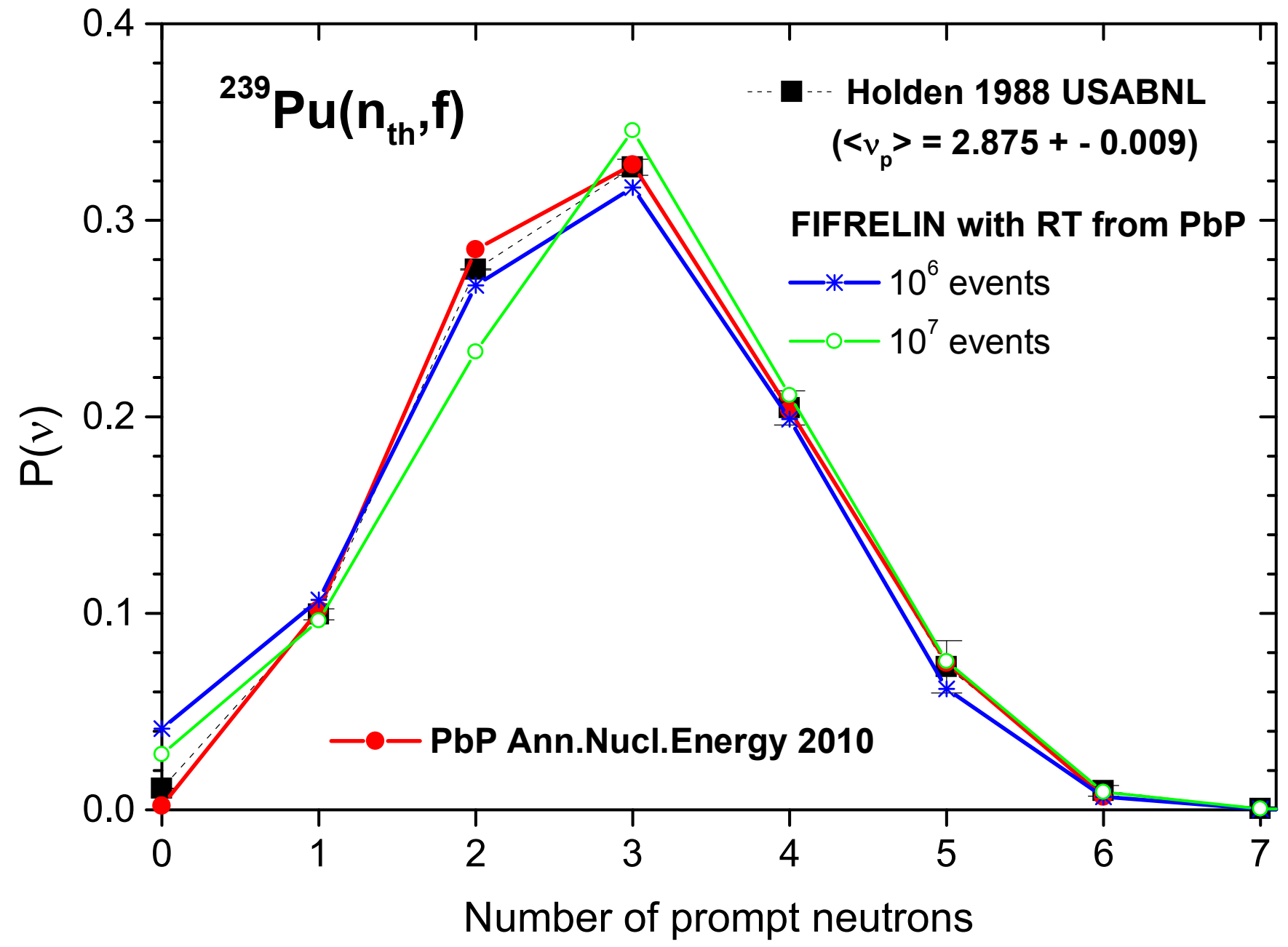

**Fig.4.36**: P(ν) results of FIFRELIN using the RT function from PbP with blue stars ($10^6$ events) and open green circles ($10^7$ events) in comparison with the PbP model result reported in (Tudora and Hambsch, 2010) (full red circles) and the experimental data (full black squares).

Also it is interesting to observe the almost linear behaviour of new <ν>(TKE) results over the entire TKE range, with an inverse slope dTKE/dν that is a little bit lower than the slope of previous <ν>(TKE) results. The present FIFRELIN results succeed to describe well the experimental data of Tsuchiya in the middle part of TKE range (between 155 – 175 MeV)

A PbP calculation of <ν>(TKE) was done by considering for each mass number A two charge numbers Z as the nearest integers below and above the most probable charge taken as UCD (as in the FIFRELIN calculations). This time the PbP multi-parametric matrix ν(A,TKE) was averaged over the single experimental distribution Y(A) of Demattè, 1996, leading to the <ν>(TKE) result plotted in the lower part of Fig.4.37 with full black diamonds (connected with solid line to guide the eye).

As it can be seen the behaviour of new FIFRELIN results tend to approach this PbP result of <ν>(TKE) obtained under the approximation of a fragment distribution independent on TKE, this fact being in agreement with the observations mentioned in (Regnier et al, 2012b).

The FIFRELIN calculations using the RT function of PbP give prompt neutron spectrum results in better agreement with experimental data as the previous ones. The PFNS obtained by sampling $10^7$ and $10^6$ events are plotted in comparison with experimental data renormalized to the respective calculated spectrum in the upper and lower parts of **Figs.4.38**, respectively**.**

Looking at Figs.**4.38a,b** in comparison with Figs.4.24a,b (both figures giving spectra in appropriate scales to focus the hard and soft spectrum parts) the improved agreement with experimental data is obvious.

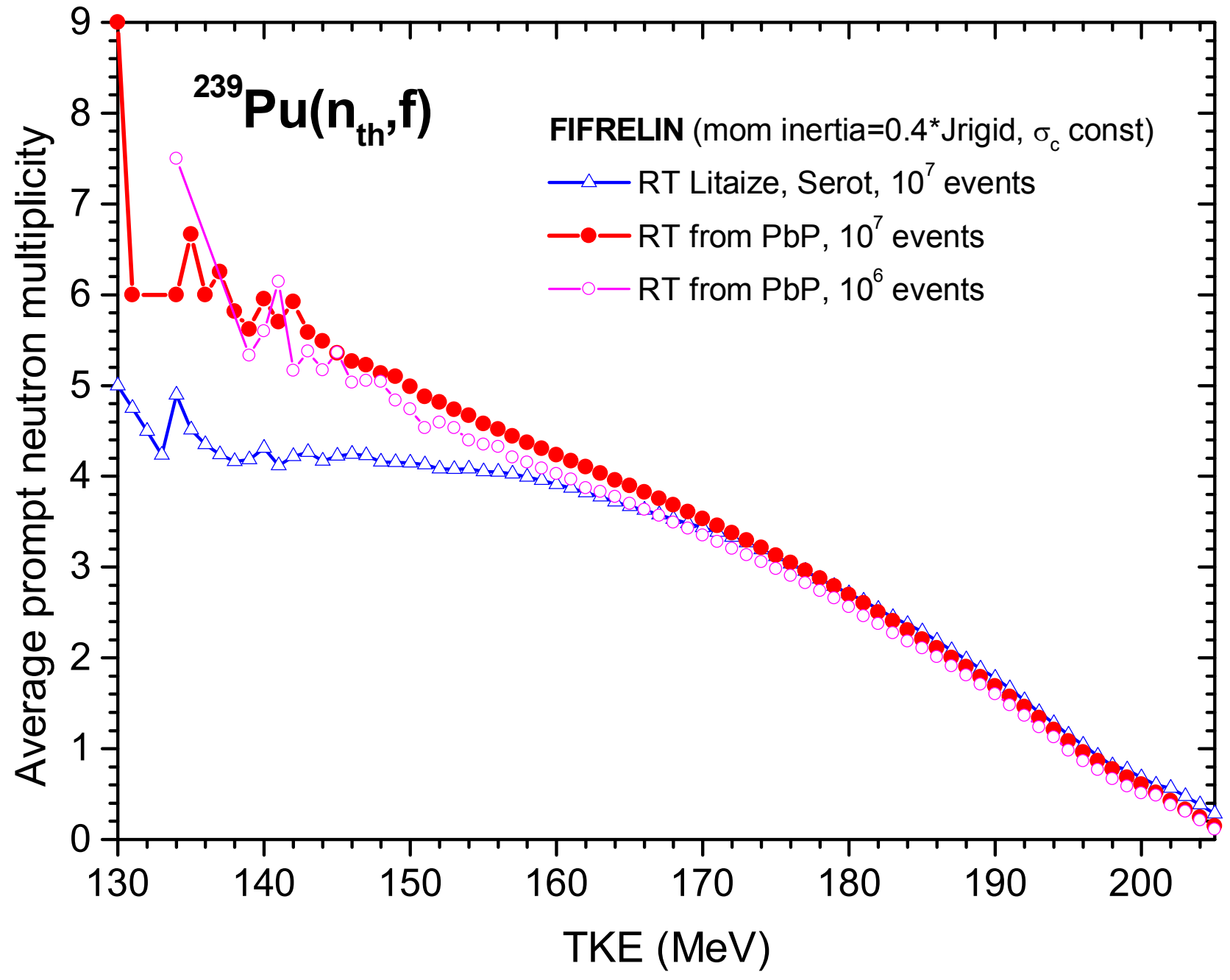

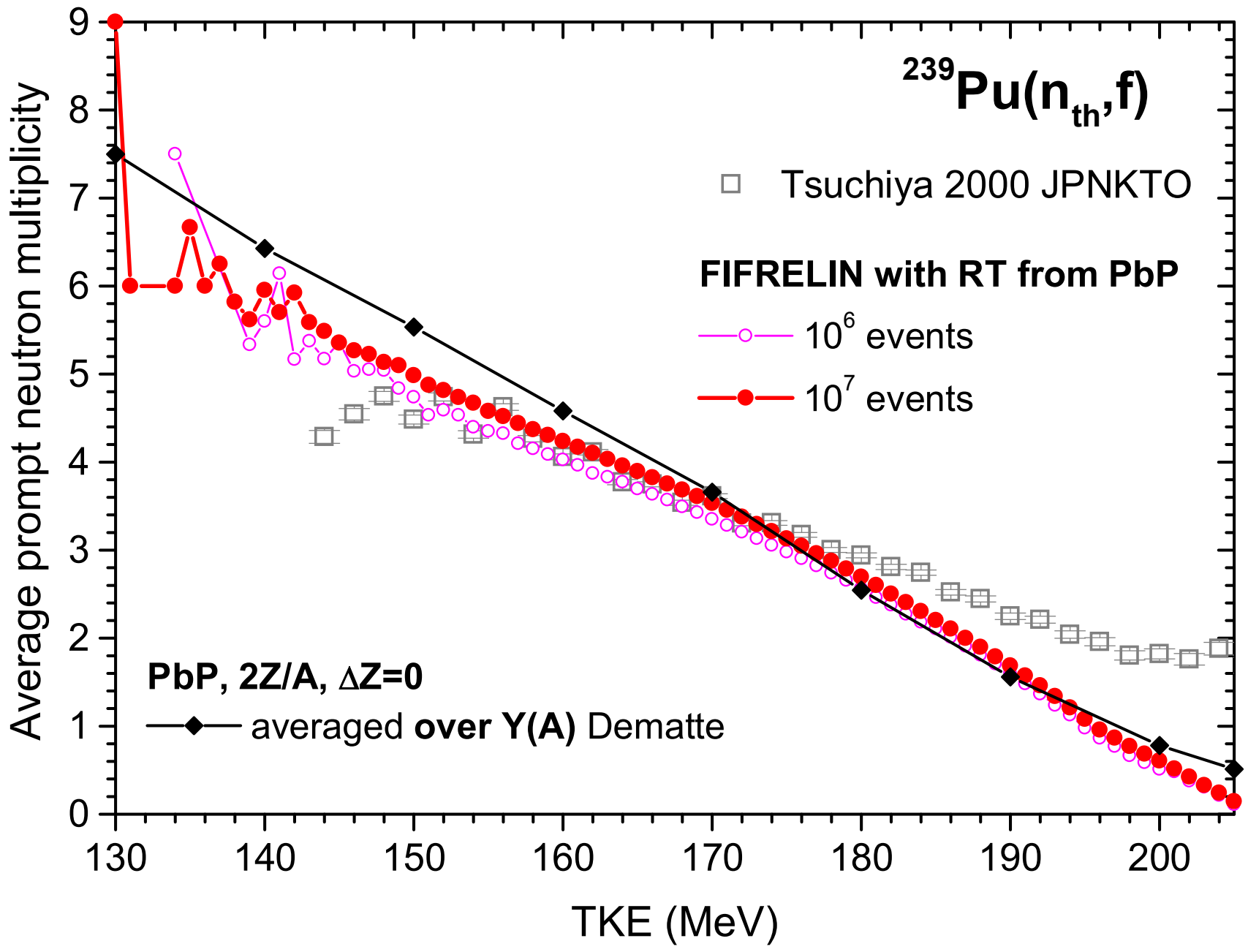

**Fig.4.37**: Upper part : present <ν>(TKE) results using the RT function from PbP plotted with full red circles and open magenta circles ($10^6$ events) in comparison with the previous result using the RT function of Litaize and Serot (open black triangles). Lower part: new FIFRELIN results (same symbols and colours as in the upper part) in comparison with the PbP result of <ν>(TKE) (plotted with full black diamonds) obtained by averaging over the single distribution Y(A) of Dematté, 1996. Experimental data of Tsuchiya taken from EXFOR are plotted with open gray squares.

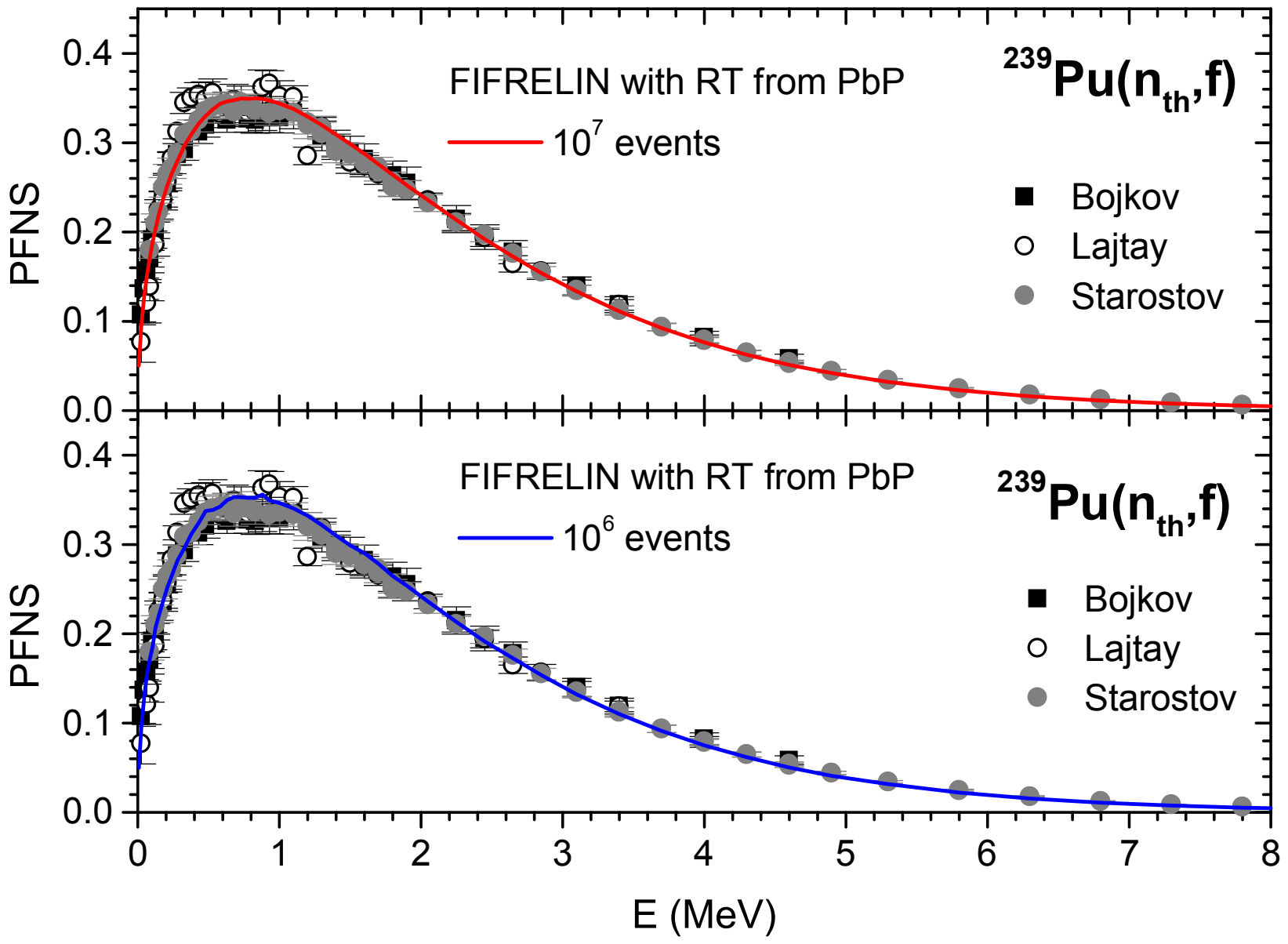

**Fig.4.38a** : PFNS provided by FIFRELIN with the RT fucntion of PbP by sampling $10^7$ events (upper part) and $10^6$ events (lower part) in comparison with experimental data. The hard spectrum part is focused.

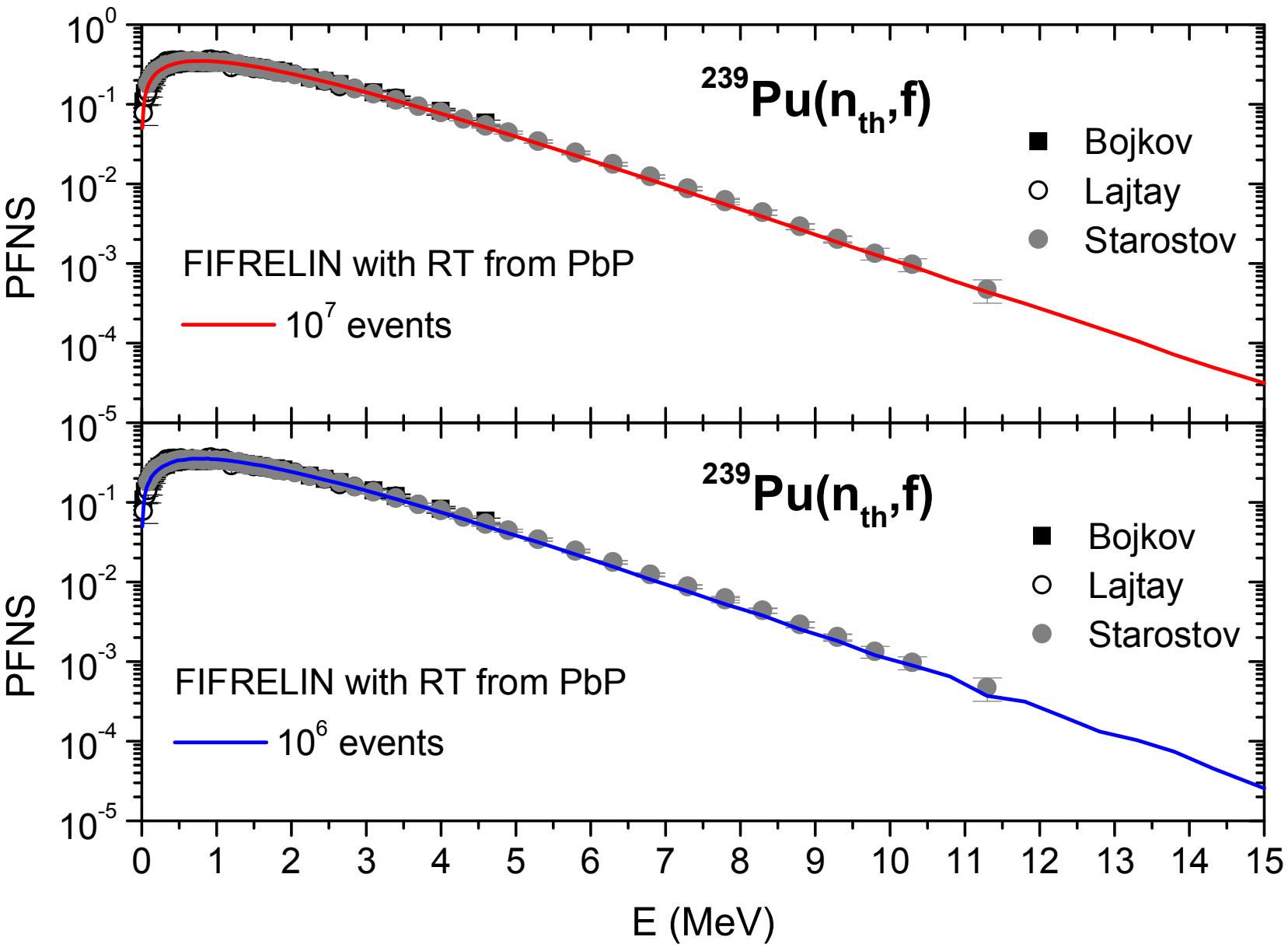

**Fig.4.38b** : PFNS provided by FIFRELIN with the RT fucntion of PbP by sampling $10^7$ events (upper part) and $10^6$ events (lower part) in comparison with experimental data. The soft spectrum part is focused

The much better description of experimental data by the present PFNS results of FIFRELIN is more visible in the representation as ratio to a Maxwellian spectrum given in **Fig.4.39** (calculations by sampling $10^7$ events in the upper part and $10^6$ events in the lower part).

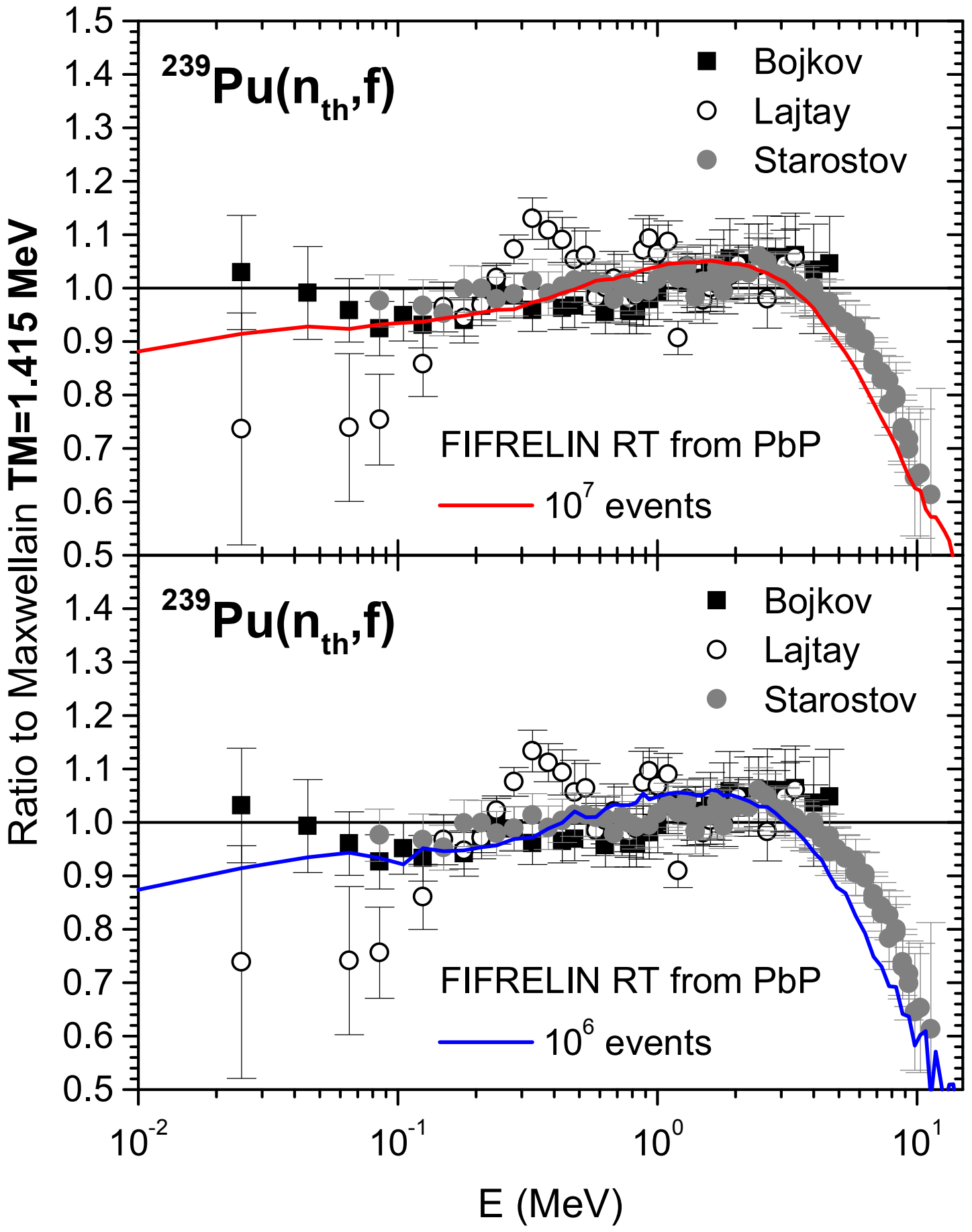

**Fig.4.39**: FIFRELIN spectrum results using the RT function from PbP in comparison with experimental data renormalized to the respective calculated spectrum plotted as ratio to Maxwellian with TM=1.415 MeV: calculation by sampling $10^7$ events in the upper part and $10^6$ events in the lower part.

Looking at both parts of Fig.4.39 in comparison with previous FIFRELIN spectrum given as ratio to a Maxwellian spectrum with the same TM=1.415 MeV in the upper part of Fig.4.25, it is easy to see the overall better agreement with experimental data obtained in the present case. The previous FIFRELIN spectrum overestimates considerably the experimental data of Starostov at prompt neutron energies above 6 MeV while the present results (especially for calculations taking $10^7$ events) are only a little bit lower than the data of Starostov.

The structure observed in the spectrum shape due –as it was already mentioned– to the insufficient number of events sampled in calculations is entirely confirmed by the present FIFRELIN results. The spectrum results plotted in the upper and lower parts of Figs.4.38b and 4.39 differ only by the number of events sampled in the Monte Carlo calculation and structures are visible only in the spectrum plotted in the lower parts (that is calculated by sampling a lower number of events).

### IV.3 Comments about the PbP and FIFRELIN results of $^{236\text{-}244}Pu(SF)$ and $^{239}Pu(n_{th},f)$

The inter-comparison of PbP and FIFRELIN results for the fissioning systems $^{236\text{-}244}Pu(SF)$ and $^{239}Pu(n_{th},f)$, as well as their comparison with existing experimental data, revealed some interesting observations that can be synthesized as following.

The overall agreement between PbP and FIFRELIN results regarding different prompt neutron emission quantities (such as average quantities a function of A, average quantities as a function of TKE and total average ones) proves that the global treatment of sequential emission in the PbP model by using the fragment residual temperature distribution P(T) can be considered as a satisfactory approximation, being indirectly validated by the sequential emission treatment in the frame of the Monte-Carlo code FIFRELIN.

Differences between the PbP and FIFRELIN results are due to the TXE partition: even if in both models the physical assumptions are the same (meaning the consideration of statistical equilibrium at scission and the Fermi-Gas description of fragment level densities with level density parameters obtained in the frame of the generalized super-fluid model) the manner to calculate the excitation energies of complementary fragments at full acceleration is different.

The ratios $RT=T_L/T_H$ (were $T_L$ and $T_H$ are referring to maximum values of residual temperature distributions) are an result of the PbP treatment (meaning of the TXE partition used in PbP) and they differ from the RT function that was used as input in the code FIFRELIN in the case of $^{239}Pu(n_{th},f)$ and $^{236\text{-}244}Pu(SF)$. The maximum and minimum values of the RT function used in FIFRELIN (placed at AH=132 and at the last far asymmetric fragmentation) were initially deduced for the case of $^{252}Cf(SF)$ and they seem to be not so adequate for other fissioning systems.

In the PbP model the shape of the compound nucleus cross-section of the inverse process of neutron evaporation from fragments has a major influence on the prompt neutron spectrum shape. The compound nucleus cross-sections $\sigma_c(\varepsilon)$ obtained by optical model calculations used in PbP with different parameterizations (giving visible different shapes of $\sigma_c(\varepsilon)$ as it was mentioned in Chapter III) lead to PFNS results with different shapes. Good descriptions of experimental spectrum shapes were obtained in the majority of cases by using the optical model parameterization of Becchetti-Greenless.

In the case of FIFRELIN, the five calculation cases showed that the manner to calculate the moments of inertia plays in many cases a much more important role than the consideration of a constant or variable $\sigma_c(\varepsilon)$. For almost all prompt emission quantities it is observed that the use of the same moments of inertia with $\sigma_c(\varepsilon)$ taken constant or variable leads to close FIFRELIN results. For instance this is the case of $P(\nu)$, $\langle\nu\rangle$(TKE), $\nu(A)$. This fact is explicable because the sequential emission of neutrons ends up at the fragment energy limit taken as *Sn+Erot,* the moments of inertia entering the rotation energy expression.

The consideration of $\sigma_c(\varepsilon)$ variable with energy is obviously a more physical assumption than to take $\sigma_c$ constant. The PbP model results with variable $\sigma_c(\varepsilon)$, describing better the experimental data in all cases, proved this fact.

FIFRELIN calculations for Pu(SF) and $^{239}$Pu($n_{th}$,f) give in some cases better agreement with experimental data when $\sigma_c$ is taken constant. For instance this is the case of the prompt neutron spectra of $^{240}$Pu(SF) and of $^{239}$Pu($n_{th}$,f) when the best agreement with experimental spectrum data is obtained with constant $\sigma_c$ and moments of inertia from the database AMEDEE. In another case, of $^{242}$Pu(SF), the scarce experimental spectrum data are very well described (better than PbP) by the FIFRELIN results considering variable $\sigma_c(\varepsilon)$ (using the Becchetti-Greenless optical potential) in both cases of moments of inertia (a fraction of 0.4 of the rigid body momentum and HFB calculations from the data base AMEDEE).

Not so important differences between FIFRELIN and PbP results can be due –in the case of $^{240,242,244}$Pu(SF) and $^{239}$Pu($n_{th}$,f) – to the consideration of a charge polarization in the PbP treatment while in FIFRELIN only UCD is taken in all calculations. But, as it was proved by the simulation exercise of Section IV.2.1, the E*(A) results are much more affected by the RT function than by the fragmentation range build by considering UCD or UCD with charge polarization.

In the case of the PbP model, that is used always with variable $\sigma_c(\varepsilon)$, total average quantities (such as prompt neutron multiplicity $\langle\nu p\rangle$ and spectra) as well as average quantities as a function of TKE (such as $\langle\nu\rangle$(TKE), $\langle\varepsilon\rangle$(TKE)) have a low sensitivity to the TXE partition. The PbP spectrum shape is very sensitive to the optical model parameterization.

In the PbP model prompt emission quantities as a function of fragment, such as $\nu(A)$, $\varepsilon(A)$ are very sensitive to the TXE partition.

In the case of the Monte-Carlo treatment of FIFRELIN, total average quantities and average quantities as a function of TKE are sensitive to different moments of inertia (affecting mainly the sequential emission by the energy limit of Sn+Erot) and also to $\sigma_c(\varepsilon)$ but the sensitivity to $\sigma_c(\varepsilon)$ is less than to moments of inertia. Quantities as a function of fragment, such as $\nu(A)$, $\varepsilon(A)$, are very sensitive to the temperature function RT that gives practically the TXE partition. In other words the sawtooth shape of $\nu(A)$, $\varepsilon(A)$ is mainly given by the RT function used as input in the code FIFRELIN.

We can say that the RT function used as input in the FIFRELIN code affects practically all prompt emission quantities: the sawtooth shapes of $\nu(A)$ and $\varepsilon(A)$, the prompt neutron spectrum shape as well as the behaviour of $<\nu>$(TKE). The results provided by the FIFRELIN code using another RT function in the case of $^{239}$Pu($n_{th}$,f) support entirely this statement.

The PbP and FIFRELIN results of different prompt emission quantities of $^{236\text{-}244}$Pu(SF) and $^{239}$Pu($n_{th}$,f) showed the following:

Total average prompt neutron multiplicities provided by FIFRELIN and PbP are close each other and in good agreement with the experimental data for all studied fissioning systems ($^{236\text{ - }244}$Pu(SF) and $^{239}$Pu($n_{th}$,f)).

Total average prompt neutron spectra provided by PbP and FIFRELIN succeeded to give a reasonable description of experimental data, the PbP results being in better agreement with the experimental data in the case of $^{239}$Pu($n_{th}$,f) and $^{240}$Pu(SF). In the case of $^{242}$Pu(SF) the FIFRELIN spectrum describes much better the experimental data.

Visible discrepancies appear between the $<E\gamma>$ results of PbP and FIFRELIN in the case of $^{236,238,240,242}$Pu(SF), especially for FIFRELIN calculations using moments of inertia taken from the database AMEDEE. Unfortunately, experimental $<E\gamma>$ of Pu(SF) are missing. Taking into account that experimental $<E\gamma>$ of other fissioning systems (spontaneous or neutron induced at low energy) are of about 6-7 MeV we can consider that $<E\gamma>$ values of Pu(SF) of about 6 MeV (as in the case of PbP results) or 7 MeV (case of FIFRELIN results with moments of inertia taken 0.4 of rigid body momentum) are physically reasonable, while high values of about 9-11 MeV (as in the case of FIFRELIN results with moments of inertia from AMEDEE) seem to be not reasonable. The explanation consists in the approximation contained in the FIFRELIN version used in the present calculations. The inclusion of a refined model for prompt gamma-ray emission is in progress at CEA-Cadarache and certainly will lead to an important improvement of <Eγ> results.

P(ν) results of FIFRELIN and PbP are in overall good agreement with existing experimental data. In the cases of $^{240,242}Pu$(SF) and $^{239}Pu(n_{th},f)$ the PbP results of P(ν) describe much better the experimental data than the FIFRELIN results.

<ν>(TKE) results of PbP and FIFRELIN exhibit practically the same decreasing slope dTKE/dν and are very close to each other for TKE values above 160 MeV (especially in the cases of $^{236,238}Pu$(SF) and $^{239}Pu(n_{th},f)$). The flattening behaviour of <ν> at low TKE values is different.

The FIFRELIN and PbP results of ν(A) for $^{239}Pu(n_{th},f)$ are different, the PbP result being in a better agreement with the experimental data. Large differences appear between the ε(A) results of PbP and FIFRELIN. The simulation exercise regarding the excitation energies of fully-accelerated fragments E*(A) obtained by using different RT functions leaded to the supposition that large differences appearing between the FIFRELIN and PbP results of ν(A) of $^{239}Pu(n_{th},f)$ are due mainly to the RT function used in the FIFRELIN calculations.

In order to verify the great impact of the RT function on the FIFRELIN results, another RT function (resulted from the PbP treatment) was included in the code FIFRELIN. The results obtained by running FIFRELIN with this RT function for the case $^{239}Pu(n_{th},f)$ differ considerably from the previous ones.

The new FIFRELIN result of ν(A) exhibits a sawtooth shape different from the previous one, being closer to the PbP result. This fact proves again the great influence of the RT function on the sawtooth shape. The new result of ε(A) also differs from the previous ones. And the agreement of <ε> of fragment pairs with experimental data is a little bit improved, too.

An interesting fact is the very close results of prompt neutron multiplicities of fragment pairs obtained by FIFRELIN calculations using different RT functions (RT of Litaize and Serot and RT provided by PbP) and sampling $10^7$events. These results are also close to $\nu_{pair}(A)$ provided by the PbP calculation.

Different RT functions used in FIFRELIN do not affect significantly the result of total average prompt neutron multiplicity. In the case of $^{239}Pu(n_{th},f)$ the differences between $\langle\nu_p\rangle$ results obtained with the RT of Litaize and Serot and of PbP (for the same cases on moments of inertia and $\sigma_c$) are only of 0.25%.

The inclusion in FIFRELIN of the RT function provided by the PbP treatment leads to $<\nu>$(TKE) results with another behaviour compared to the previous ones: the flattening of $<\nu>$ at low TKE values is practically vanished, $<\nu>$(TKE) exhibiting an almost linear behaviour over the entire TKE range with a little bit lower slope dTKE/d$\nu$. The new FIFRELIN result obtained by sampling a larger number of events (of $10^7$) tends to approach the PbP result obtained under the approximation of averaging over a single Y(A) distribution independent on TKE.

The prompt neutron spectrum provided by FIFRELIN with the RT function from PbP is in much better agreement with experimental data than the previous one.

The FIFRELIN calculations with the RT function from PbP done by sampling $10^6$ and $10^7$ events prove again that the fluctuations (structure) appearing in the spectrum and in $<\nu>$(TKE) at low TKE values (below 140 MeV) are due only to the insufficient number of events sampled in the Monte Carlo treatment.

The exercise done by changing the RT function used as input in FIFRELIN showed the major impact of this function (giving the TXE partition) on almost all prompt neutron quantities. The sawtooth shape of quantities as a function of fragment (such as $\nu$(A), $\varepsilon$(A)) is practically given by the RT function. The prompt neutron spectrum shape is influenced not only by the compound nucleus cross-section of the inverse process but also by the RT function. The behaviour of $<\nu>$(TKE) also depends on the RT function used.

The fact that the FIFRELIN results of fragment pair multiplicity as well as of total average multiplicity practically do not depend on the RT function used and are in agreement with the PbP results shows that two different treatments, one determinist (PbP) another probabilistic (Monte Carlo FIFRELIN), using in many cases the same physical assumptions and models can give results in overall good agreement for quantities as a function of fragment pair or averaged ones.

# Chapter V

# Conclusions

In the present work were presented and discussed the results of a series of calculations for several spontaneous ($^{236,238,240,242,244}Pu$(SF), $^{252}Cf$(SF)) and neutron induced fissioning systems ($^{239}Pu$(nth,f)).

The prompt emission calculations were performed considering two different treatments: the deterministic Point by Point model and the probabilistic Monte Carlo treatment included in the FIFRELIN code.

In the case of the PbP model, two methods of TXE partition were used, one based on the $\nu_H/\nu_{pair}$ parameterization, the other one based on the calculation of the extra–deformation energy of fragments at scission.

The improvements brought to the FIFRELIN code refer to the implementation of the compound nucleus cross–section of the inverse process of neutron evaporation from fragments following the procedure from the PbP model. Also it was developed a procedure to use the moment of inertia calculated with a Hartre–Fock–Bogoliubov formalism contained in the AMEDEE database.

The inter–comparison of the results obtained with both treatments for all the fissioning systems studied in this work as well as their comparison with the existing experimental data revealed some interesting aspects which can be synthesized as following.

As already mentioned, the major difference between the PbP and FIFRELIN consists in the treatment of the sequential emission, the overall agreement between the results from both models regarding different prompt neutron emission quantities proving that the global treatment of the sequential emission in the PbP model by using the fragment residual temperature distribution P(T) can be considered as a satisfactory approximation.

Even if in both models the same physical assumption are made (statistical equilibrium at scission and Fermi–Gas description of the fragment level densities with level density parameters obtained in the frame of the generalized super–fluid model), the different manner of sharing TXE between fully accelerated fission fragments lead to significant differences between the PbP and the FIFRELIN results.

The TXE partition method based on the mass dependent temperature ratio law implemented in the FIFRELIN code which was initially parameterized for the case of $^{252}$Cf(SF) give good results for this fissioning system an rather good for other systems.

It was pointed out that the RT function obtained from the PbP treatment, which was implemented in the FIFRELIN code lead to a visible improvement of the prompt emission calculations.

The insufficient number of events sampled in the FIFRELIN calculations lead to significant fluctuations in the spectrum and in the $\langle\nu\rangle$(TKE) at low TKE values. The FIFRELIN calculation performed by sampling $10^6$ and $10^7$ events and considering the RT function from PbP revealed that the fluctuation (structures) in the spectrum and in the $\langle\nu\rangle$(TKE) at low TKE values are due only to the insufficient number of events sampled. By increasing the events number and by considering the RT function from PbP, a better agreement of the FIFRELIN calculations regarding $\langle\nu\rangle$(TKE) with the PbP results (obtained by averaging over the simple distribution Y(A)) is obtained.

The very close PbP results obtained by using different TXE partition methods confirm that the TXE partition has not a significant influence on the total average prompt emission quantities.

Contrary, in the case of the FIFRELIN treatment, the TXE partition has an important influence on almost all the prompt emission quantities, excepting for the average prompt neutron multiplicity.

The strong influence of the TXE partition on the $\nu$(A) results seen for both codes, PbP and FIFRELIN is no longer visible on the $\nu_{pair}$(A), this fact being proved by the close results obtained with the FIFRELIN code considering two different RT functions.

For all the studied fissioning systems, the PbP was used only by considering an energy dependent compound nucleus cross–section of the inverse process provided by optical model calculations with the Becchetti–Greenless potential in most of the cases, the obtained results giving a better description of the experimental data.

It is surprising that in the case of the FIFRELIN code, some of the results obtained by considering a constant compound nucleus cross–section of the inverse process are in a better agreement with the experimental data than the results obtained by considering an energy dependent one.

Both models fail to describe well the scarce experimental data regarding the average prompt neutron energy in the center of mass in the case of $^{239}Pu(n_{th},f)$

FIFRELIN and PbP results showed that two different treatments can give results in good agreement for quantities as a function of the fragment pair and averaged ones.

# Appendix 1

### Simplified $\sigma_c(\varepsilon)$ formula proposed by Iwamoto

A very simple formula for the calculation of the inverse cross–sections was proposed in (Iwamoto, 2008)

$$\sigma(\varepsilon) = \sigma_0 + \sigma_s(\varepsilon) \quad \text{(iw_1)}$$

where

$$\sigma_0 = \pi R^2 \text{ with } R = r_0 A^{1/3} \quad \text{(iw_2)}$$

is a constant term, and

$$\sigma_s(\varepsilon) = \frac{\pi}{k^2} T_0 = \frac{\pi\hbar^2 S_0}{m\sqrt{\varepsilon}} \quad \text{(iw_3)}$$

is a term that depends on $1/\sqrt{\varepsilon}$, with $S_0$ the s- wave strength fucntion

Combining equation (iw_2) and (iw_3) the Iwamoto formula is obtained as:

$$\sigma(\varepsilon) = \sigma_0 \left(1 + \frac{\alpha}{\sqrt{\varepsilon}}\right) \quad \text{(iw_4)}$$

where

$$\alpha = \frac{\pi\hbar^2 S_0}{m r_0^2 A^{2/3}} \quad \text{(iw_5)}$$

indicates the magnitude of the s–wave term relative to the constant one.

# Appendix 2

## Level density parameter

### Gilbert–Cameron composite formula

The Gilbert and Cameron systematic of level density parameter $a$ is :

$$a/A = \begin{cases} 0.00917 \cdot S + 0.142 & (spherical\, nuclei) \\ 0.00917 \cdot S + 0.120 & (deformed\, nuclei) \end{cases} \qquad \text{(Gilb_1)}$$

where $S = S(Z) + S(N)$ is the shell correction.

### Generalized super–fluid model of Ignatiuk

In the frame of the generalized super–fluid model of Ignatiuk, the level density parameter $a$ of each fission–fragment with the excitation energy $E^*$ is calculated according to:

$$a(Z, A, E^*) = \begin{cases} \tilde{a}(A)\left(1 + \dfrac{\delta W(Z, A)}{U^*}\left(1 - \exp(-\gamma U^*)\right)\right) & for\ U^* \geq U_{cr} \\ a_{cr} \quad for\ U^* < U_{cr} \end{cases} \qquad \text{(Ig_1)}$$

with

$U^* = E^* - E_{cond}$, and the shell corrections $\delta W$ usually taken from the RIPL-3 database.

The $\tilde{a}$ parameter is the asymptotic level density parameter for which different parameterizations can be used as following:

- in PbP: $\tilde{a}(A) = 0.073A + 0.115A^{2/3}$
- in FIFRELIN: $\tilde{a}(A) = 0.1125A + 0.000122A^{2/3}$ for $\delta W$ of Moller and Nix

$\tilde{a}(A) = 0.0959A + 0{,}1468A^{2/3}$ for $\delta W$ of Myers and Swiatecki

The damping parameter is:

- in PbP: $\gamma = \frac{0.4}{A_f^{1/3}}$

- in FIFRELIN: $\gamma = \frac{0.325}{A_f^{1/3}}$

The condensation energy given by:

$$E_{cond} = 3a_c \frac{\Delta_0^2}{2\pi^2} - n\Delta_0 \qquad \text{(Ig_2)}$$

where $n=0,1$ and $2$ for the even–even, odd and odd–odd nuclei, respectively, and $a_c$ is the critical value of the level density parameter obtained from:

$$a_c\left(E_{cr}^*, Z, A\right) = \tilde{a}(A)\left(1 + \frac{\delta W(Z,A)}{E_{cr}^*}\left(1 - \exp\left(-\gamma E_{cr}^*\right)\right)\right) \qquad \text{(Ig_3)}$$

with

$$E_{cr}^* = aT_c^2$$

where the critical temperature is:

$$T_c = 0.567\Delta_0 \qquad \text{(Ig_4)}$$

The correlation function is given by:

$$\Delta_0 = \frac{12}{\sqrt{A_f}} \qquad \text{(Ig_5)}$$